\documentclass[a4paper,oneside,12pt]{book}
\usepackage{fancyhdr}
\usepackage{tikz}
\usepackage{sectsty}
\usepackage{lipsum}
\usepackage{geometry}                		                  	    		
\usepackage{graphicx}													
\usepackage{amsmath,amssymb,array}
\usepackage{amscd}
\usepackage{mathrsfs}
\usepackage{dsfont}
\usepackage{amsthm}
\usepackage{mathrsfs}
\usepackage[english]{babel}
\usepackage{enumitem} 
\usepackage{setspace}
\usepackage[]{latexsym}
\usepackage{subfigure}
\usepackage{a4}
\usepackage{yfonts}
\usepackage[T1]{fontenc}
\usepackage{braket}
\usepackage{caption}
\usepackage{float}
\usepackage{floatflt} 
\usepackage{wrapfig}
\usepackage[Lenny]{fncychap}
\usepackage{hyperref}
\hypersetup{
pdftitle={},%
pdfauthor={},%
pdfsubject={},%
pdfkeywords={},%
colorlinks=true,%
linkcolor=blue,%
citecolor=red,%
linktocpage=true,%
hyperfootnotes=true,%
pageanchor=true
}
\newenvironment{varpmatrix}[1][\small] 
  {\left(\mbox\bgroup#1$\begin{matrix}} 
  {\end{matrix}$\egroup\right)}

\theoremstyle{plain} 
\newtheorem{thm}{Theorem}[chapter]

\newtheorem{prop}[thm]{Proposition} 

\theoremstyle{definition} 
\newtheorem{defn}{Definition}[chapter] 

\theoremstyle{remark}

\input xy
\xyoption{all}

\begin{document}

{\thispagestyle{empty}
\vskip 0.8cm {\Large\centerline {\bf Universit\`a degli studi di Napoli ``Federico II''}}
\vskip 0.8cm {\Large\centerline {Scuola Politecnica e delle Scienze di Base}}
\vskip 0.2cm {\Large\centerline {Area Didattica di Scienze Matematiche Fisiche e Naturali}}  
\vskip 0.8cm {\Large\centerline {\bf Dipartimento di Fisica ``Ettore Pancini''}}
\vskip 1cm
\begin{center}
\includegraphics[height=5cm]{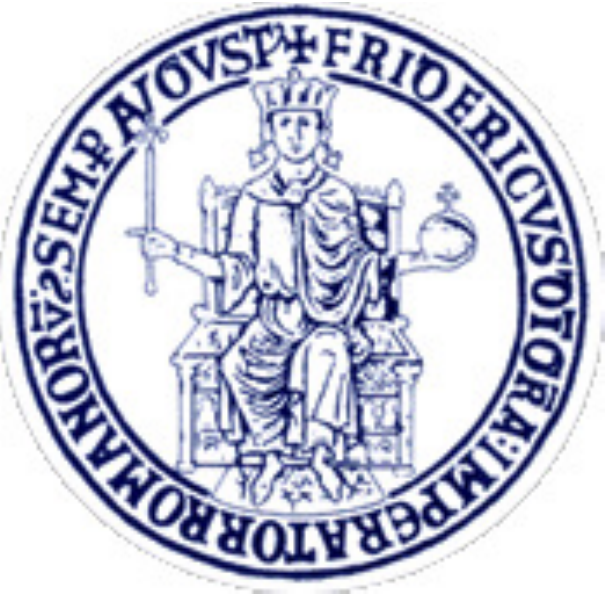}
\end{center}
\vskip 0.3cm {\Large\centerline { Laurea Magistrale in Fisica}} 
\vskip 1.2cm \center{\bf{\LARGE  Quantum Metric and Entanglement\\ on Spin Networks}} 
\vskip 1.5cm

\large
\begin{minipage}[t]{8cm}
\textbf{Relatori:}

Dr. Goffredo Chirco (AEI, Potsdam)\\Dr. Daniele Oriti (AEI, Potsdam)\\Prof. Patrizia Vitale

\vspace{0.2cm}

%
\end{minipage}
\hfill
\begin{minipage}[t]{5cm}
\hfill \textbf{Candidato:}

\hfill Fabio Maria Mele

\hfill N94/244
\end{minipage}

\vskip 1.3cm \large\centerline {A.A. $2015/2016$} \vfill
\eject}

\clearpage

%
\newpage
\thispagestyle{plain}

\begin{flushright}
\null\vspace{\stretch{1}}
\textit{To my parents Pasquale and Rosaria,\\
and to my uncle Gaspare}
\vspace{\stretch{2}}\null
\end{flushright}

\newpage

\tableofcontents
\newpage
\thispagestyle{plain}


\pagestyle{fancy}
\setlength{\headheight}{15pt}
\fancyhead[L]{\nouppercase{\textsl{\bf\leftmark}}}
\fancyhead[R]{}
\cfoot{\bfseries \thepage}%

\chapter*{\textbf{Introduction}}
\phantomsection
\addcontentsline{toc}{chapter}{Introduction}
\markboth{Introduction}{}

Quantum Mechanics (QM) and General Relativity (GR) are the two main revolutions in the Physics of the last century and provide the pillars of our current knowledge of Nature. If on the one hand Einstein's theory rules the large scales phenomena of the astrophysical and cosmological regimes and has shed a completely new light on the nature of gravity, space and time, on the other hand QM rules the small scales of the atomic and subatomic world. However, despite of their empirical success, these theories are conceptually incompatible. Indeed, QM is defined on a fixed background spacetime and its interpretation as well as its mathematical coherence deeply rely on an external time variable. As such it is incompatible with GR, according to which space and time must be treated on the same level and spacetime itself is a dynamical entity.\\ Such an incompatibility makes the attempt to unify QM and GR into a consistent theory of Quantum Gravity (QG) one of the main challenges at the foundations of modern theoretical physics.\\ \\Why then QG in first place. There are many arguments in favour of a positive answer coming from different sides \cite{intr0}:

\begin{itemize}
\item At a conceptual level, the main lesson of GR is that (the geometry of) spacetime is a dynamical object identified with the gravitational field, while QM tells us that dynamical fields are quantum objects, i.e., they have a discrete and probabilistic nature. Therefore, we conclude that spacetime itself must have a fundamental discrete quantum structure as can be synthesized in the following sillogism: ``\textit{spacetime is a dynamical field, any dynamical field is quantum, then also spacetime is quantum}'';

\item In the case of gravity, ordinary QFT perturbation techniques lead to a non renormalizable theory;

\item The fact that three of the fundamental interactions are successfully described within a quantum framework may be regarded as an hint that also the gravitational interaction should be quantized. Moreover, the success of unification in the history of science may also suggest that gravity cannot be quantized alone but we need to include also the other interactions (this is for istance the more ambitious ideology behind string theory, but we will not deal with it in this dissertation); 

\item The presence of singularities in GR and of UV divergences in QFT tells us that the both theories break down at the very small scales;

\item Horizon entropy (of black holes or more generally of any causal horizon) \cite{intr1,intr2} should be explained in terms of some fundamental degrees of freedom characterizing the microscopic structure of the spacetime region hidden behind the horizon surface;

\item Classical gravity seems to have a thermodynamic nature. Indeed, Einstein's equations can be derived by using only thermodynamic concepts \cite{intr3}. But the laws of thermodynamics are statistical, i.e., they are a macroscopic average manifestation of the microscopic behaviour, and so the statistical nature of gravity suggests that spacetime may be a macroscopic approximation of  more fundamental costituents.

\end{itemize}

\noindent
Being a still unsolved issue, QG is not just a single theory waiting for experimental tests but rather a plenary of approaches based on different motivations and techniques characterized by their own successes and internal difficulties (see \cite{LQG0} for a report). The proliferation of various paths is also due to the lack of an empirical guidance as a consequence of the fact that the physical regimes where quantum gravitational effects should become relevant are outside the present observational reach.\\
In this dissertation, we take the point of view according to which, in order to accomodate the main features of GR and QM, a theory of quantum gravity should be a background independent (i.e., a diffeomorphism invariant) quantum field theory describing in a fully relational way the fundamental costituents of space and time, in terms of which a quantized (discrete) spacetime geometry is made of. 
\\Most of the main background-independent candidates to a full theory of QG, such as Loop Quantum Gravity, Spin Foam Models and Group Field Theories, propose a radical picture of the Planck scale structure of spacetime. According to them, at the very small scales space and time dissolve into non-geometric, combinatorial and algebraic objects. Such a microscopic description is given in terms of spin networks, roughly graphs coloured by irreducible representations of the local gauge group (usually $SU(2)$). Quantum spin network states represent elementary excitations of spacetime. The properties of these quanta of the gravitational field are determined by the spectral properties of the operators representing the quantities involved in our interaction with the system.\\Specifically, geometric quantities such as area and volume correspond to quantum operators and spin networks are their eigenstates. The heuristic picture of quantum space provided by spin networks is therefore that of ``grains of space'' with discrete volumes and areas which describe the building blocks of a quantized three-dimensional geometry.\\However, it is not yet clear how ordinary spacetime would emerge from such a discrete description. The issues come from the diffeomorphism invariance of the theory and the consequent difficulty to define a notion of locality in space. There are hints that entanglement and tools from information theory should play a crucial role both in the characterization of the intrinsic properties of the quantum texture of spacetime and in the reconstruction of its geometry. For istance, even if based on completely different grounds both at the conceptual and technical level, recent developments in AdS/CFT have shown that the entanglement of particles on the boundary is directly related to the connectivity of the bulk regions thus suggesting that our three-dimensional space is held together by quantum entanglement \cite{F4}. Later works based on the so-called \textit{Ryu-Takayanagi formula}, which relates the entanglement entropy in a conformal field theory to the area of a minimal surface in its holographic dual \cite{intr4}, has also shown that the stress-energy tensor near the boundary of a bulk spacetime region can be reconstructed from the entanglement on the boundary \cite{intr5, intr6}.\\Also on the spin network side, there are proposals to use quantum information to reconstruct geometrical notions such as distance in terms of the entanglement on spin network states \cite{intr7,F3,F17}. The idea is that in a purely relational background independent context only correlations have a physical meaning and it seems reasonable to regard spin networks themselves as networks of quantum correlations between regions of space and then try to derive geometrical properties from the intrinsic informational content of the theory.\\ \\With this motivations in mind, in this thesis we make a preliminary attempt to grasp some insights on this transition ``\textit{from  pregeometry to geometry}'' within the framework of a geometric formulation of quantum mechanics (GQM). Indeed, the usual Hilbert space of a quantum mechanical system can be equipped with a K\"ahler manifold structure inheriting both a Riemannian metric tensor and a symplectic structure from the undelying complex projective space of rays. This enables us to import the powerful machinery of differential geometry also in QM and, in particular, to characterize the entanglement properties of a composite system in a purely tensorial fashion \cite{GQM3,GQM7}.\\The idea is therefore to use these quantum tensors to characterize the entanglement on spin network states. The advantages of this formalism are both computational and conceptual. Indeed, unlike the calculations involving entanglement entropy, it does not require the explicit knowledge of the Schmidt coefficients. Moreover, the key structures of the formalism are built purely from the space of states without introducing additional external structures.\\ \\The dissertation is organized as follows. In the first chapter we give a review of the role of spin networks in QG and of their interpretation as quanta of space which, because of diffeomorphism invariance, offer a realization of the pregeometric scenario in terms of building blocks of combinatorial and algebraic nature. Then we discuss how the possibility to have open spin network states allows to relate the entanglement on these states with the gluing of their links (or surfaces in the dual picture). Therefore in chapter 2, after recalling some basic notion of quantum information theory and entanglement, we perform some explicit calculations of  the entanglement entropy for  simple examples of spin network graphs and show how the gauge-invariance requirement at the gluing nodes implies a locally maximally entangled state.\\Chapter 3 instead introduces the basics of Geometric Quantum Mechanics. The space of rays is recognized to be the proper setting for the description of quantum systems and its main geometric structures are pointed out. In particular, we focus on the structures on the Hilbert space which can be regarded as the pull-back of the tensors defined on the ray space. This is the so-called Fubini-Study Hermitian tensor whose real and imaginary parts provide us with a metric and a symplectic structure, respectively. Such a pull-back procedure turns out to be very useful in the analysis of bipartite composite systems. Indeed, the pulled-back Hermitian tensor on orbit submanifolds of quantum states related by unitary transformations decomposes in block matrices which encode all the information about the separability or entangled nature of the fiducial state of the given orbit. Of particular interest are the off-diagonal blocks of the metric part which encode the information on quantum correlations between the subsystems and define an entanglement measure interpreted as a distance with respect to the separable case.\\Finally, in chapter 4 we set up a dictionary correspondence between the GQM formalism and spin networks by focusing on the most simple case of a single Wilson line. The analysis of the tensorial structures now available on the space of Wilson line states leads us to rethink of the link itself as resulting from the entanglement of its endpoint states. The entanglement measure involving the off-diagonal blocks of the Fubini-Study metric tensor can be therefore interpreted as a measure of the existence of the link and ultimately as a measure of graph connectivity. Even more interesting, in the maximally entangled case (corresponding to the gauge-invariant Wilson loop state), we find that such entanglement measure depends only on the spin $j$ labelling the $SU(2)$-representation associated to the link through a power of the area eigenvalue for the corresponding dual surface. This seems to suggest a further connection between entanglement and geometry. 

\chapter{\textbf{Quantum States of Geometry: Spin Networks}}

In this chapter we introduce the main features (both formal and conceptual) of one of the main contenders for a quantum theory of gravity, that is Loop Quantum Gravity (LQG). Our aim is to show the microscopic quantum structure of spacetime shared by some of the main background-independent approaches to Quantum Gravity \cite{LQG0} (e.g., LQG, Spin Foam Models and Group Field Theories (GFT)) which realize the pregeometric scenario described before in terms of non-spatiotemporal fundamental building blocks called spin networks. There are several ways to introduce spin networks \cite{LQG1,LQG2,LQG3}, but we choose to follow the path of the canonical quantization program for General Relativity which seems to us the most pedagogical one. Hence, we start by recalling some elements of General Relativity, in particular the ADM formalism, in order to introduce the Ashtekar variables. Then we describe the quantization process and the construction of the kinematical Hilbert space of the theory. A basis for such space is provided by spin network states which, roughly speaking, correspond to a superposition of graphs labelled by group or Lie algebra elements. We will also focus on the interpretation of these states as quanta of geometry. To enforce their geometric characterization we also study classical and quantum aspects of the tetrahedron, its duality relation with a 4-valent spin network vertex and the construction of the Hilbert space by quantizing the classical phase space obtained after the implementation of geometric constraints via a symplectic reduction procedure \cite{LQG4,LQG5}. The chapter closes with a summary on the pregeometric microscopic quantum structure of spacetime provided by the theory.\\However, it should be kept in mind that there is no room here for an exhaustive treatment of the subject and so this chapter must be intended just as an introduction of the key ideas and tools which will be used in the rest of the work. For a more detailed discussion of the foundations of LQG we refer to the following references by which the present chapter has been inspired: \cite{LQG6,LQG7,LQG8,LQG9,LQG10} for review articles, \cite{LQG11,LQG12} for introductory books, \cite{LQG13} for a detailed book and \cite{LQG14} for a more rigorous presentation of the mathematical formalism on which the theory is based.

\section{Canonical Formulation of General Relativity}

In order to understand how the quantization of General Relativity (GR) can take place in LQG, we need to recall some notions about the classical theory. In particular, we start from the Hamiltonian formulation of GR which is the starting point of the canonical formalism of the theory \cite{LQG15,LQG16} and then we show how to introduce the connection formalism. This essentially prepares the stage to perform the canonical quantization approach on which LQG is based.

\subsection{Hamiltonian Formalism}

Einstein's theory of General Relativity can be derived from the so-called Einstein-Hilbert action
\begin{equation}\label{qg1}
S_{EH}=\frac{1}{16\pi G}\int d^4x\sqrt{-g}R\;,
\end{equation}
where $g$ is the determinant of the metric and $R=g^{\mu\nu}R_{\mu\nu}$ is the scalar curvature. Varying this action with respect to the metric it is actually possible to derive the (vacuum) Einstein's equations \cite{LQG17}.\\The Einstein-Hilbert action (\ref{qg1}) is the starting point of the Hamiltonian formalism but in order to put it into a canonical form we need to identify a set of canonically conjugated variables for the theory, and then perform a Legendre transformation. Indeed, given a classical system described by a set of generalized configuaration variables $q^i\in Q$ and their time derivatives $\dot q^i$, the Lagrangian $\mathcal L(q^i,\dot q^i)\in\mathcal F(TQ)$ is a function on the tangent bundle $TQ$ to the configuration manifold $Q$ and the associated action is given by \cite{LQG18}:
\begin{equation}\label{qg2}
S=\int dt\mathcal L(q^i,\dot q^i)\;.
\end{equation}
By defining the canonical momentum $p_i=\frac{\partial\mathcal L}{\partial\dot q^i}=\frac{\delta S}{\delta\dot q^i}$, we can perform a Legendre transformation which maps $TQ$ into $T^*Q$ and rewrite the action as
\begin{equation}\label{qg3}
S=\int dt\bigl(p_i\dot q^i-H(q,p)\bigr)\;,
\end{equation}
where $H(q,p)\in\mathcal F(T^*Q)$ is the Hamiltonian function of the system defined by $H(q,p)\equiv p_i\dot q^i-\mathcal L$. On the phase space $T^*Q$ the equations of motion are given by:
\begin{equation}\label{qg4}
\begin{cases}
\dot{\vec q}=\{\vec q, H\}\\
\dot{\vec p}=\{\vec p, H\}
\end{cases}
\end{equation}
where $\{\cdot,\cdot\}$ is the Poisson bracket defined by $\{f_1,f_2\}=\frac{\partial f_1}{\partial q^i}\frac{\partial f_2}{\partial p_i}-\frac{\partial f_1}{\partial p_i}\frac{\partial f_2}{\partial q^i}\;,\;\forall f_1,f_2\in\mathcal{F}(T^*Q)$. Therefore, in the Hamiltonian formalism the system is described by the generalized configurations $q$ and their conjugate momenta $p$ at a given instant of time $t$ thus providing a natural $3+1$ splitting of space and time. Even if such a separation between space and time is not really appropriate from the GR point of view according to which we should treat space and time in the same way, it is required by Quantum Mechanics where a notion of time is needed in order to compute expectation values for istance.\\To take this into account we assume that the spacetime manifold $\mathbb M$ has the topology $\mathbb M\simeq\mathbb R\times\Sigma$ where $\Sigma$ is a fixed three-dimensional spatial manifold of arbitrary topology. This means that $\mathbb M$ foliates into a one-parameter family of (spatial) hypersurfaces $\Sigma_t=X_t(\Sigma)$ embeddings of $\Sigma$ in $\mathbb M$ and this allows to identify $t\in\mathbb R$ as a time parameter.\\ \\\textbf{Remark:} It should be stressed however that this time has no absolute meaning but it is only a parameter. This is due to the diffeomorphism invariance of the action (\ref{qg1}) which essentially implies that there is no preferred foliation (i.e., no preferred time) in a diffeomorphism invariant theory like GR. Indeed, a transformation $\phi\in Diff(\mathbb M)$ maps a given foliation $X$ into a new one $X'=X\circ\phi$ with a new time parameter $t'$. Conversely, we can write $\phi$ as the composition $\phi=X'\circ X^{-1}$ of different foliations and thus we can work with a chosen foliation but the physical quantities will not depend on this choice \cite{LQG13,LQG7}.\\ \\The meaning of this foliation can be understood by introducing the so-called ADM variables proposed by Arnowitt, Deser and Misner in 1960 \cite{LQG19}. First of all, given a foliation $X_t$, we define the time flow vector

\begin{equation}\label{qg5}
\tau^\mu(x)\equiv\frac{\partial X^\mu(t)}{\partial t}=(1,0,0,0)\qquad s.t. \qquad g_{\mu\nu}\tau^\mu\tau^\nu=g_{00}\;,
\end{equation}

\noindent
and the unit normal vector to $\Sigma$

\begin{equation}\label{qg6}
n^\mu\qquad s.t. \qquad g_{\mu\nu}n^\mu n^\nu=-1\;.
\end{equation}

\noindent
Such vectors in general are not parallel and $\tau^\mu$ can be then decomposed into a normal and a tangential part as follows:

\begin{equation}\label{qg7}
\tau^\mu(x)=N(x)n^\mu(x)+N^\mu(x)\;.
\end{equation}

\begin{figure}[h!]
\centering
\includegraphics[scale=0.40]{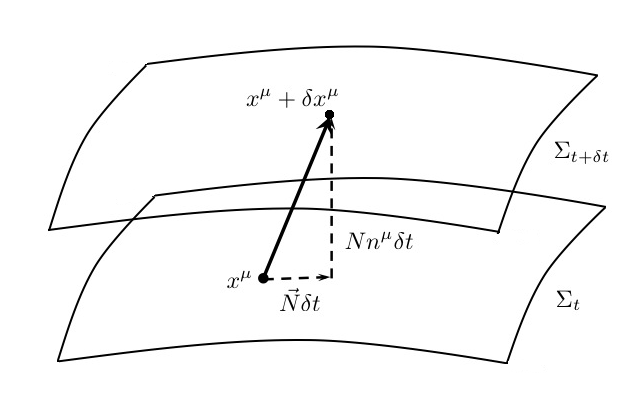}
\caption{\textit{Geometrical interpretation of the shift vector $\vec N$ and of the lapse function $N$ in the foliation induced by the $3+1$ decomposition.}}
\label{adm}
\end{figure}

\noindent
By parametrizing $n^\mu=(\frac{1}{N}, -\frac{N^a}{N})$, so that $N^\mu=(0,N^a)$, we identify the so-called Shift vector $N^a$ and Lapse function $N$ where we are using the notation according to which Greek indices denote space-time tensorial indices and Latin ones denote space indices. Thus, if $x^\mu$ is a point on $\Sigma_t$, then $x'^\mu=x^\mu+N n^\mu\delta t$ is a point on $\Sigma_{t+\delta t}$ and we can understand the Lapse and the Shift from a geometrical point of view as describing respectively the normal and tangential evolution with respect to the time parameter $t$ (see Fig. \ref{adm}).

\noindent
The metric tensor written in terms of lapse and shift is given by
\begin{equation}\label{qg8}
ds^2=g_{\mu\nu}dx^\mu dx^\nu=-(N^2-N_aN^a)dt^2+2N_adtdx^a+g_{ab}dx^adx^b\;,
\end{equation}
as can be easily derived by noticing that

\begin{equation}\label{qg9}
\tau_\mu\tau^\mu=g_{\mu\nu}\tau^\mu\tau^\nu=g_{00}=-N^2+g_{ab}N^aN^b\;,
\end{equation}

\noindent
and

\begin{equation}\label{qg10}
\tau_\mu N^\mu=g_{0b}N^b=g_{\mu\nu}(Nn^\mu+N^\mu)N^\nu=g_{ab}N^aN^b\quad\Rightarrow\quad g_{0b}=g_{ab}N^a=N_b\;.
\end{equation}

\noindent
The three-dimensional metric $g_{ab}$ given by the spatial part of $g_{\mu\nu}$ is not in general the intrinsic induced metric $q_{ab}$ on the hypersurface $\Sigma_t$ but is related to it by means of the relation $q_{\mu\nu}=g_{\mu\nu}-n_\mu n_\nu$. However, since for every tensor on $\Sigma_t$ the scalar product with the unit normal vector $n^\mu$ vanishes, such tensors can be equivalently contracted with $g$ or $q$. In other words, we have a projector on the spatial slice $\Sigma_t$ provided by $q^\mu_\nu=g^{\mu\sigma}q_{\sigma\nu}$ and we can inherit the tensorial calculus on $\Sigma_t$ from that on $\mathbb M$.\\The action (\ref{qg1}) can be therefore written in the form

\begin{equation}\label{qg11}
S_{EH}=\frac{1}{16\pi G}\int dt\int_{\Sigma}d^3x\,\sqrt{q}\,N\,[\mathcal R-K^2+\text{Tr}(KK)]\;,
\end{equation}

\noindent
where

\begin{equation}\label{qg12}
K_{\mu\nu}=q^\rho_\mu q^\sigma_\nu\nabla_\rho n_\sigma
\end{equation}

\noindent
is the extrinsic curvature of the hypersurface $\Sigma_t$, and

\begin{equation}\label{qg13}
\mathcal R^\mu_{\nu\rho\sigma}=q^{\mu'}_\mu q^{\nu'}_\nu q^{\rho'}_\rho q^{\sigma'}_\sigma R^{\mu'}_{\nu'\rho'\sigma'}-K_{\nu\sigma}K^\mu_\rho-K_{\nu\rho}K^\mu_\sigma
\end{equation}

\noindent
is the Riemann tensor of $\Sigma_t$ whose relation with the Riemann tensor $R$ of $\mathbb M$ is exactly given by Eq. (\ref{qg13}). The important feature of the action (\ref{qg11}) is that it does not involve the time derivatives of $N$ and $N^a$. This means that $N$ and $N^a$ play the role of Langrange multipliers with null conjugate momenta (i.e., $\delta\mathcal L/\delta\dot N=0$ and $\delta\mathcal L/\delta\dot N^a=0$) and ultimately that the true canonically conjugate variables of the theory are given by $q_{ab}$ and momenta

\begin{equation}\label{qg14}
\pi^{ab}\equiv\frac{\delta\mathcal L}{\delta\dot q_{ab}}=\sqrt{q}\,(K^{ab}-Kq^{ab})\;.
\end{equation}

\noindent
Then, by computing the Legendre transform, we get:

\begin{equation}\label{qg15}
S_{EH}(q_{ab},\pi^{ab},N,N^a)=\frac{1}{16\pi G}\int\,dt\int\,d^3x\,\bigl[\pi^{ad}\dot q_{ab}-N^aH_a-NH\bigr]\;,
\end{equation}

\noindent
where

\begin{equation}\label{qg16}
H_a=-2\sqrt{q}\,\nabla_b\biggl(\frac{\pi^b_a}{\sqrt{q}}\biggr)\;,
\end{equation}

\begin{equation}\label{qg17}
H=\frac{1}{\sqrt{q}}G_{abcd}\pi^{ab}\pi^{cd}-\sqrt{q}\,\mathcal R\qquad(G_{abcd}=q_{ac}q_{bd}+q_{ad}q_{bc}-q_{ab}q_{cd})\;.
\end{equation}

\noindent
The phase space of GR is thus parametrized by the variables $(q_{ab},\pi^{ab})$ with canonical Poisson brackets

\begin{equation}\label{qg18}
\{\pi^{ab}(t,x),q_{cd}(t,x')\}=\delta^a_{(c}\delta^b_{d)}\delta(x-x')\;,
\end{equation}

\noindent
and, as we can see from Eq. (\ref{qg15}), the Hamiltonian is given by:

\begin{equation}\label{qg19}
\mathbb H=\frac{1}{16\pi G}\int\,d^3x\,(N^aH_a+NH)\;.
\end{equation}

\noindent
Moreover, the variation of the action (\ref{qg15}) with respect to the Lagrange multipliers $N^a$ and $N$ gives the following equations
\begin{equation}\label{qg20}
H_a(q,\pi)=0\qquad,\qquad H(q,\pi)=0
\end{equation}
respectively known as the space-diffeomorphism and the Hamiltonian constraints. Such constraints are of the first class type as can be seen by computing their Poisson brackets which vanish on the constraint hypersurface identified by the above equations\footnote{An explicit computation of the constraint algebra can be found for istance in Sec.1.3 of \cite{LQG7}.} (i.e., on-shell). First class constraints generate gauge transformations on the constraint surface and in fact it is possible to show that the constraints (\ref{qg20}) are respectively the generators of space-diffeomorphisms on $\Sigma$ and time-diffeomorphisms which have to be satisfied by physical configurations.\\One last crucial observation is that the Hamiltonian (\ref{qg19}) is proportional to the Lagrange multipliers and thus it vanishes on the constraint surface. This is coherent with the already stressed diffeomorphism-invariance which implies that $t$ is just a parameter with no physical meaning and therefore there is no physical evolution in time as a direct consequence of a vanishing Hamiltonian function. In other words, the whole dynamical content of GR is fully encoded in the four constraints $H^\mu=(H,H^a)$. 

\subsection{Tetrad Formalism and Ashtekar Variables}

As pointed out in \cite{LQG7}, the application of the quantization procedure \'a la Dirac to the constrained theory provided by the ADM formulation of GR encounters a number of difficulties which are essentially related to the lack of a Lebesque measure on the space of metrics (modulo diffeomorphisms) that can be used to define a scalar product thus resulting into a ill-defined Hilbert space even before the implementation of constraints. However, the key step towards LQG is to reformulate GR in a way that allows to perform the Dirac's quantization procedure. In some sense, hence, the starting point of LQG is surprisingly simple consisting not of changing the gravitational theory or the quantization paradigm, but instead of choosing appropriate fundamental variables to describe gravity. In order to define such variables, let us first introduce the tetrad formalism.\\A tetrad is a quadruple of 1-forms $e_\mu^I(x)$, $I=0,1,2,3$, implicitly defined by \cite{LQG20,LQG21}
\begin{equation}\label{qg21}
g_{\mu\nu}(x)=e_\mu^I(x)e_\nu^J(x)\eta_{IJ}\;,
\end{equation}
where $\eta_{IJ}$ is the flat Minkowski metric. By definition, the tetrads provide an isomorphism between a general reference frame and an inertial one\footnote{Being a map from the tangent space $T_x\mathbb M$ at $x$ to Minkowski space, tetrads capture the intuition that spacetime locally appears like Minkowski flat space.}. The definition is invariant under Lorentz transformations and the capital Latin indices thus carry representations of the Lorentz group $SO(3,1)$. Contracting vectors and tensors with tetrads, we also obtain objects that transform under the Lorentz group. In other words, tetrads realize a mapping from the tangent bundle $T\mathbb M$ of the spacetime $\mathbb M$ to a Lorentz principal bundle $\mathbb F(\mathbb M, SO(3,1))$. The $\mathfrak{so}(3,1)$-valued connection 1-form $\omega_\mu^{IJ}$ on this bundle defines a covariant derivative of fibers
\begin{equation}\label{qg22}
D_\mu v^I(x)=\partial_\mu v^I(x)+\omega_{\mu J}^I(x)e_\nu^J(x)-\Gamma_{\mu\nu}^\rho(x)e^I_\rho(x)\;,
\end{equation}
and a derivative for objects with both kind of indices

\begin{equation}\label{qg23}
\mathcal D_\mu e^I_\nu(x)=\partial_\mu e^I_\nu(x)+\omega_{\mu J}^I(x)e_\nu^J(x)-\Gamma_{\mu\nu}^\rho(x)e_\rho^I(x)\;,
\end{equation}

\noindent
where $\Gamma_{\mu\nu}^\rho(x)$ is the Levi-Civita connection. Being the latter metric-compatible (i.e., $\nabla_\mu g_{\nu\rho}=0$), we need to require that $\omega_\mu$ is tetrad-compatible, i.e., $\mathcal D_\mu e^I_\nu=0$. In this case, $\omega$ becomes a 1-form with values in the Lie algebra of the Spin group and we will call it a spin connection. The curvature associated to this connection is \cite{LQG22}
\begin{equation}\label{qg24}
F^{IJ}=d_\omega\omega^{IJ}=d\omega^{IJ}+\omega^I_K\wedge\omega^{KJ}\;,
\end{equation}
with components
\begin{equation}\label{qg25}
F^{IJ}_{\mu\nu}=\partial_\mu\omega^{IJ}_\nu-\partial_\nu\omega^{IJ}_\mu+\omega^I_{\mu K}\omega^{KJ}_\nu-\omega_{\mu K}^J\omega_\nu^{KI}\;.
\end{equation}
The Einstein-Hilbert action can be thus written in its tetrad formulation as the so-called Palatini action \cite{LQG7}
\begin{equation}\label{qg26}
S(e)=\frac{1}{2}\,\varepsilon_{IJKL}\int_\mathbb M e^I\wedge e^J\wedge F^{KL}(\omega(e))\;,
\end{equation}
where we set $16\pi G=1$. We can also promote the connection to be an independent variable since, if the tetrad is non-degenerate, the field equation obtained by varying the action w.r.t. $\omega$ just gives the structure of the spin connection and does not add any new element to the physics of spacetime. The action (\ref{qg26}) thus becomes:
\begin{equation}\label{qg27}
S(e,\omega)=\frac{1}{2}\,\varepsilon_{IJKL}\int_\mathbb M e^I\wedge e^J\wedge F^{KL}(\omega)\;.
\end{equation}

\noindent
On the other hand, when varied with respect to $e$, this action gives the equation of motion of GR\footnote{Since only first derivatives appear in the action (\ref{qg27}), it provides a first order formulation of GR.} and, apart from the usual diffeomorphism-invariance, it presents also a gauge symmetry under local Lorentz transformations thus showing that GR is a gauge theory with local gauge group given by the Lorentz group.\\Moreover, by considering the connection as an independent variable, an additional term, which does not affect the equations of motion (in absence of fermions) and is compatible with all symmetries, can be added to Eq. (\ref{qg27}) thus leading to the so-called Holst action \cite{LQG23}
\begin{equation}\label{qg28}
S(e,\omega)=\biggl(\frac{1}{2}\,\varepsilon_{IJKL}+\frac{1}{\gamma}\,\delta_{IJKL}\biggr)\int_\mathbb M e^I\wedge e^J\wedge F^{KL}(\omega)\;,
\end{equation}
where $\delta_{IJKL}\equiv\delta_{I[K}\delta_{L]J}$ and $\gamma$ is the so-called Immirzi parameter which will play a key role in the quantum theory\footnote{Indeed, whether this parameter is chosen to be real or complex has huge consequences in the construction of the quantum constraints \cite{LQG24}. Here we will consider only the real case.}.\\Assuming now a $3+1$ splitting $\mathbb M\cong\mathbb R\times\Sigma$, from the expression (\ref{qg8}) of the metric tensor in ADM variables we can derive the expression of tetrads in terms of shift and lapse
\begin{equation}\label{qg29}
e_0^I=e_\mu^I\tau^\mu=N n^I+N^ae_a^I\qquad,\qquad g_{ab}=e_a^i e_b^j\delta_{ij}
\end{equation}
where $i,j=1,2,3$ are flat Minkowski spatial indices and the spatial part of the tetrad $e^i_a$ is called triad. As previously done for the ADM case, we should now identify canonically conjugated variables and perform a Legendre transform to explicit the Hamiltonian. However, the dependence of these variables on $e$, $\omega$ and their time derivatives together with the presence of an additional local gauge symmetry under the Lorentz group will lead to a more complicated structure of the constraint algebra which will be now of the second class. Such difficulties may be circumvented by introducing the famous Ashtekar variables \cite{LQG25}:
\begin{equation}\label{qg30}
E_i^a=\frac{1}{2}\varepsilon^{abc}\varepsilon_{ijk}e^j_be^k_c\qquad(\textbf{densitized triad}\;or\;\textbf{electric field})\;,
\end{equation}
\begin{equation}\label{qg31}
A^i_a=\gamma\omega_a^{oi}+\frac{1}{2}\varepsilon_{jk}^i\omega_a^{jk}\qquad(\textbf{Ashtekar-Barbero connection})\,.
\end{equation}

\noindent
In terms of these new variables the action (\ref{qg28}) can be rewritten as \cite{LQG14}
\begin{equation}\label{qg32}
S(A,E,N,N^a)=\frac{1}{\gamma}\int\,dt\int_\Sigma\,d^3x[\dot{A}^i_aE^a_i-A_0^iD_aE^a_i-NH-N^aH_a]\;,
\end{equation}

\noindent
where

\begin{equation}\label{qg33}
G_i\equiv D_aE^a_i=\partial_aE^a_i+\varepsilon_{ijk}A^j_aE^{ak}\;,
\end{equation}
\begin{equation}\label{qg34}
H=\bigl[F^i_{ab}-(\gamma^2+1)\varepsilon_{ijk}K_a^jK_b^k\bigr]\varepsilon_{i\ell m}\frac{E^a_\ell E^b_m}{\det{E}}+\frac{1+\gamma^2}{\gamma}\,G^i\partial_a\frac{E_i^a}{\det{E}}\;,
\end{equation}
\begin{equation}\label{qg35}
H_a=\frac{1}{\gamma}F^i_{ab}E_i^b-\frac{1+\gamma^2}{\gamma}K_a^iG_i\;,
\end{equation}
with $F_{ab}^i=\partial_aA_b^i-\partial_bA^i_a+\varepsilon_{jk}^iA_a^jA_b^k$ the curvature associated to the connection (\ref{qg31}). The action (\ref{qg32}) shows that $(A,E)$ are canonically conjugated variables while Lapse and Shift are still Lagrange multipliers, and so $H$ and $H_a$ again give rise to the Hamiltonian and space-diffeomorphism (first class) constraints:
\begin{equation}\label{qg36}
H(A,E)=0\quad,\qquad H_a(A,E)=0\;.
\end{equation}
Moreover, there is a new additional constraint
\begin{equation}\label{qg37}
G_i(A,E)=D_aE^a_i=0\;,
\end{equation}

\noindent
which resembles the familiar Gauss constraint of gauge theories. The presence of this extra constraint is actually related to the new symmetry introduced with the tetrad formalism. The conjugate pair $(A,E)$ indeed transforms respectively as $SU(2)$-vector and a $\mathfrak{su}(2)$-valued connection 1-form, and the Gauss constraint generates the local $SU(2)$ gauge transformations \cite{LQG7} which essentially codify the rotational symmetries of the local reference frame associated with the triad on fixed-time slices. This is confirmed by the Poisson brackets for these new variables
\begin{equation}\label{qg38}
\begin{split}
&\{A_a^i(x),A_b^j(x')\}=0=\{E_i^a(x),E_j^b(x')\}\\
&\{A_a^i(x),E_j^b(x')\}=\gamma\,\delta_a^b\,\delta^i_j\,\delta^{(3)}(x,x')
\end{split}
\end{equation}
where the internal index $i$ corresponds to the adjoint representation of $SU(2)$. Therefore, the key feature of this formalism is that with the new pair of canonical variables the phase space of GR acquires the same structure of a $SU(2)$ Yang-Mills gauge theory and as such it is more suitable for quantization. However, unlike gauge theories in which after imposing the Gauss law there is a physical Hamiltonian, here we have a fully constrained system and, as required for a diffeomorphism-invariant geometric theory of spacetime, the physical dynamical evolution is not in terms of any distinguished time variable.

\section{The Canonical Quantization Program: LQG}

Even if we have reformulated GR in a way as close as possible to a gauge theory, there are still some troubles that need to be solved to realize the full quantum implementation. In order to understand both conceptual and technical issues that one has to face in quantizing the classical theory, let us briefly recall the main steps in the quantization of usual gauge theories and point out the key differences with respect to the case of gravity.\\ \\The usual (canonical) quantization procedure of a gauge theory goes through the following steps:

\begin{itemize}

\item[\bf{1)}] Promote the canonical variables to quantum operators and the Poisson brackets to commutators;

\item[\bf{2)}] Use the Minkowski background metric to define a Gaussian measure $\delta A$ on the space of connections modulo gauge transformations and thus define a scalar product on wave functionals of the connection variables $\psi[A]$ which realizes the Hilbert space as $L^2(A, \delta A)$;

\item[\bf{3)}] Select gauge-invariant states by imposing the quantum operator version of the Gauss constraint, i.e., $\Hat G_i\psi[A]=0$;

\item[\bf{4)}] Once the physical states have been selected, study their dynamics generated by the Hamiltonian operator.

\end{itemize}

\noindent
We have already argued that in a diffeomorphism-invariant theory like GR, there is no physical Hamiltonian and the dynamics is fully enbodied in the constraints. However, troubles begin even before arriving at point 4). Indeed, one has first to address the following points:

\begin{itemize}
\item One of the main difficulties concerning the promotion of Poisson brackets to commutators is that the brackets of the canonical variables show singularities, i.e., they are distributional Poisson brackets. Therefore, as in Yang-Mills theory, in order to get rid off the delta functions we need to smear the fields $A_a^i, E^a_i$ by integrating them against properly chosen test functions;
\item In the case of GR we do not have any background metric that can be used to define the integration measure, since the metric itself is a fully dynamical quantity of the theory. Hence, we need to define a measure on the space of connections without relying on any fixed background metric.
\end{itemize}

\noindent
In the next sections we will see how these issues lead us to introduce the so-called holonomy-flux algebra and the notion of cylindrical functions. This essentially allows to construct the kinematical unconstrained Hilbert space $\mathcal H_{kin}$ of the theory. The strategy then will be to impose the constraints one by one thus constructing the physical Hilbert space $\mathcal H_{phys}$ by means of the following reduction process:

\begin{equation}\label{qg39}
\mathcal H_{kin}\xrightarrow{\hat G_i=0}\mathcal H_{kin}^0\xrightarrow{\hat H^a=0}\mathcal H_{Diff}\xrightarrow{\hat H=0}\mathcal H_{phys}\;.
\end{equation}

\noindent
We will focus on the first two steps thus limiting us to the kinematical structure of the LQG theory. The reason is that on the one hand the basis of the resulting Hilbert space will be the object of interest in the rest of the work, and on the other hand (despite the considerable effort) nowadays it is not yet clear how to construct a well-defined quantum operator implementing the Hamiltonian constraint and how to extract its solutions.

\subsection{Holonomy-Flux Algebra}

In the smearing of the algebra (\ref{qg38}) a key role is played by the different tensorial nature of $A_a$ and $E^a$. Indeed, following what is usually done for gauge theories on flat spacetime, we may be tempted to regularize the connection and the electric field by integrating them over the whole space with the same type of test functions. Nevertheless, the connection is a 1-form and so it is natural to smear it along a one-dimensional path. Hence, recalling that a connection defines a notion of parallel transport of the fiber over the base manifold \cite{LQG22,LQG26}, we can associate to the given Lie algebra-valued connection 1-form $A=A_a^i\tau_i dx^a$ \footnote{Here $\tau$ denotes the generators of $SU(2)$.} an element of the group $h_\gamma(A)$ called holonomy and defined by

\begin{equation}\label{qg40}
h_\gamma(A)\equiv\mathcal P\exp{\biggl(\int_\gamma A\biggr)}\;,
\end{equation}

\noindent
where $\gamma : [0,1]\longrightarrow\Sigma$ is a path in the spatial hypersurface $\Sigma$ parametrized by some coordinate functions $x^a(s)$ with $s\in[0,1]$ an affine parameter on the three-dimensional manifold such that

\begin{equation}\label{qg42}
\int_\gamma A=\int_0^1\,ds\,A_a^i(x(s))\frac{dx^a(s)}{ds}\,\tau_i\;,
\end{equation}

\noindent
and $\mathcal P$ stands for the path-ordered product

\begin{equation}\label{qg43}
\mathcal P\exp{\biggl(\int_\gamma A\biggr)}=\mathds1_{SU(2)}+\sum_{n=1}^{\infty}\int_0^1ds_1\int_{s_1}^1ds_2\dots\int_{s_{n-1}}^1ds_n\,A(\gamma(s_1))\dots A(\gamma(s_n))\;.
\end{equation}

\noindent
The holonomy of the connection $A$ along the path $\gamma$ is the unique solution of the equation:
\begin{equation}\label{qg44}
\begin{cases}
\frac{d}{ds}h_{\gamma}(A(\gamma(s)))=h_{\gamma}(A(\gamma(s)))\,A(\gamma(s))\\
h_\gamma(A(\gamma(0)))=\mathds1_{SU(2)}
\end{cases}
\end{equation}
Physically, a holonomy gives a measure of how data fail to be preserved when parallel transported along a closed curve \cite{LQG22}. Moreover, let us notice that performing the change of variables $A(x)\mapsto h_\gamma(A)$ there is no loss of information. Indeed, considering all possible paths $\gamma$ in $\Sigma$ and regarding holonomies as functionals of the path, they contain exactly the same information as specifying the connection at each point \cite{LQG27}.\\ \\Let us list some important properties of the holonomy \cite{LQG7,LQG8}:

\begin{itemize}
\item The definition (\ref{qg40}) of $h_\gamma(A)$ does not depend on the parametrization of the path $\gamma$;
\item The holonomy is a representation of the groupoid of oriented paths. Indeed:
\begin{itemize}
\item[a)] the holonomy of a degenerate path (i.e., a single point) is the identity;
\item[b)] given two oriented paths $\gamma_1$ and $\gamma_2$ such that $\gamma_1(1)\equiv\gamma_2(1)$, we have:
\begin{equation}\label{qg45}
h_{\gamma_1\circ\gamma_2}(A)=h_{\gamma_1}(A)\cdot h_{\gamma_2}(A)\;;
\end{equation}
\item[c)] combining a) and b) we also have that
\begin{equation}\label{qg46}
h_{\gamma^{-1}}(A)=h_\gamma^{-1}(A)\;.
\end{equation}
\end{itemize}
\item By using the fact that the connection transforms under a gauge transformation $g\in SU(2)$ as
\begin{equation}\label{qg47}
A\longmapsto A_g=g^{-1}dg+g^{-1}Ag\;,
\end{equation}
we find that the holonomy locally transforms as
\begin{equation}\label{qg48}
h_\gamma(A)\longmapsto h_{\gamma}(A_g)=g(\gamma(0))h_\gamma(A)g^{-1}(\gamma(1))\;,
\end{equation}
i.e., a gauge transformation localized in the bulk of the path $\gamma$ won't change the holonomy but the holonomy will change homogeneously under a gauge transformation on the endpoints of $\gamma$;
\item Under the action of spatial diffeomorphisms $\phi\in Diff(\Sigma)$, the holonomy transforms as
\begin{equation}\label{qg49}
h_\gamma(\phi^*A)=h_{\phi\circ\gamma}(A)\;,
\end{equation}
i.e., acting on the connection with a diffeomorphism is equivalent to move the path $\gamma$.
\end{itemize}

\noindent
Coming back now to the smearing, the connection variables are only a half of the phase space and we need to smear also the conjugate variables. By looking at the definition (\ref{qg30}) of the densitised triad, we see that it is a 2-form\footnote{Indeed $E_i^a$ has a natural dual pseudo-($D-1$)-form (pseudo-2-form for $D=3$)
$$
(*E)^i_{bc}:=\varepsilon_{abc}E_i^a=sgs(\det{e})\varepsilon_{ijk}e^j_be^k_c
$$
that can be integrated along submanifolds of codimension one, namely analytic 2-surfaces.}. Hence, it is natural to smear it on a two-dimensional surface $S\subset\Sigma$, i.e.

\begin{equation}\label{qg50}
E_i(S)\equiv\int_Sd^2\sigma\,n_aE^a_i=\int_S(*E_a)n^a\;\in\;\mathfrak{su}(2)\;,
\end{equation}

\noindent
where $n_a=\varepsilon_{abc}\frac{\partial x^b}{\partial\sigma_1}\frac{\partial x^c}{\partial\sigma_2}$ denotes the normal to the surface. The quantity (\ref{qg50}) is called the flux of the electric field $E$ through the surface $S$.\\The above regularization (\ref{qg40},\ref{qg50}) by means of paths and surfaces instead of the all of space as in traditional smearings leads us to a smeared version of the Poisson algebra (\ref{qg38}) in terms of $h_\gamma(A)$ and $E_i(S)$ which is called \textbf{holonomy-flux algebra}. Such an algebra is in general very complicated but, in the simple case in which there is only one intersection between $\gamma$ and $S$, the Poisson brackets for these variables are given by:
\begin{equation}\label{qg51}
\begin{split}
&\{h_\gamma(A),h_{\gamma'}(A)\}=0\\
&\{E_i(S),E_j(S)\}=-\varepsilon_{ijk}E_k(S)\\
&\{E_i(S),h_\gamma(A)\}=\tau_ih_\gamma(A)
\end{split}
\end{equation}
and we see that no delta functions appear now. Furthermore, we have that the $E_i(S)$ play the role of conjugate momenta but the above Poisson brackets tell us that they do not commute. This is the reason why in the traditional representation of LQG holonomies are chosen as configuration variables thus avoiding the difficulties related to a non-commutative flux algebra that come out from the quantum perspective\footnote{However, in more recent works such a non-commutative momentum space has been defined and this has opened the exploration of new connections between background-independent approaches to quantum gravity and non-commutative geometry \cite{LQG28,LQG29,LQG30}.}.

\subsection{Cylindrical Functions and Structure of the Kinematical Hilbert Space}

We are now ready to make the first step of the Dirac quantization procedure outlined before, that is to construct the unconstrained kinematical Hilbert space of the theory $\mathcal H_{kin}$. Hence, let us introcuce the notion of cylindrical function which will play a crucial role in the definition of the configuration space of the theory.\\Generically, a cylindrical function is a functional of a field that depends only on some subset of components of the field itself. In our case, the field is the connection, and the cylindrical functions are functionals that depend on the connection only through the holonomies along some finite set of paths. More precisely, let $\Gamma\subset\Sigma$ be a graph, i.e., a finite and ordered collection of smooth oriented paths $\gamma_\ell\in\Sigma$ with $\ell=1,\dots,L$ meeting at most at their endpoints (such paths will be called the \textit{links} or the \textit{edges} of the graph, while the intersection points will be called \textit{nodes} or \textit{vertices}), and let $f: SU(2)^L\rightarrow\mathbb C$ be a smooth function $f=f(h_1(A),\dots,h_L(A))$ of the $L$ group elements given by the holonomies $h_\ell(A)\equiv h_{\gamma_\ell}(A)$ of the connection A along the links $\gamma_\ell$ of the graph $\Gamma$. Then, we have the following:

\begin{defn}\textbf{(Cylindrical Function)}
A cylindrical function is defined as the functional of the connection indentified by the couple $(\Gamma, f)$:
\begin{equation}\label{qg52}
\psi_{(\Gamma,f)}[A]=f(h_1(A),\dots,h_L(A))\;.
\end{equation}
\end{defn}

\noindent
The linear space of cylindrical functionals w.r.t. a given graph $\Gamma\subset\Sigma$

\begin{equation}\label{qg53}
Cyl_\Gamma=\bigl\{\mathcal C_\Gamma: A\longmapsto\mathcal C_\Gamma(A)\in\mathbb C\;|\;\mathcal C_\Gamma(A):=\psi_{(\Gamma,f)}[A]\bigr\}
\end{equation}

\noindent
can be turned into a Hilbert space by equipping it with the following scalar product between two functionals constructed on the same graph

\begin{equation}\label{qg54}
\braket{\psi_{(\Gamma,f)}|\psi_{(\Gamma,f')}}\equiv\int\prod_{\ell=1}^Ldh_\ell\,\overline{f(h_1(A),\dots,h_L(A))}\,f'(h_1(A),\dots,h_L(A))\;,
\end{equation}

\noindent
where $dh_\ell$ are $L$ copies of the Haar measure of $SU(2)$. The space (\ref{qg53}) equipped with the scalar product (\ref{qg54}) defines a Hilbert space $\mathcal H_\Gamma\cong L^2(SU(2)^L)$ associated to a given graph $\Gamma$. Notice the crucial role played in this definition by the previous switch from connections to holonomies. Indeed, the holonomy is an element of the group $SU(2)$ and the integration over $SU(2)$ is well-defined. In particular, there is a unique gauge-invariant normalized measure $dh$ on $SU(2)$ called the Haar measure \cite{LQG31}, i.e., such that:

\begin{equation}\label{qg55}
dh=d(gh)=d(hg)=dh^{-1}\qquad\forall g\in SU(2)\;,
\end{equation}

\noindent
and

\begin{equation}\label{qg56}
\int_{SU(2)}dh=\mathds1\;.
\end{equation}

\noindent
\textbf{Remarks:}
\begin{itemize}

\item[i)] The inner product (\ref{qg54}) is invariant under $SU(2)$ gauge transformations. This is a direct consequence of the invariance of the Haar measure (\ref{qg55}) and of the transformation law for a cylindrical functional

\begin{equation}\label{qg57}
\begin{split}
\psi_{(\Gamma,f)}[A]\mapsto&f(g(\gamma_1(0))h_1g^{-1}(\gamma_1(1)),\dots,g(\gamma_L(0))h_Lg^{-1}(\gamma_L(1)))\\
&=\psi_{(\Gamma,f)}[A_g]\\
&=\psi_{(\Gamma,f_g)}[A]
\end{split}
\end{equation}

\item[ii)] The inner product (\ref{qg54}) in also invariant under (spatial) diffeomorphisms $\phi\in Diff(\Sigma)$ as can be easily deduced by using the transformation law (\ref{qg49}) and noticing that the integrand in (\ref{qg54}) does not depend explicitly on the graph.

\end{itemize}

\noindent
At this point, next step is to define the unconstrained Hilbert space $\mathcal H_{kin}$ as the space of all cylindrical functions for all graphs $\Gamma\subset\Sigma$. Indeed, coming from the canonical quantisation of GR in the continuum, this involves the holonomies associated to all paths embedded in the canonical hypersurface and the fluxes across all surfaces similarly embedded. This essentially means to consider all graphs $\Gamma$ embedded in the spatial manifold $\Sigma$, the holonomies along the edges of these graphs, and the set of surfaces dual to them (i.e., such that each surfaces is pierced by one and only one edge of the graph), that is

\begin{equation}\label{union}
\bigcup_{\Gamma\subset\Sigma}\mathcal H_\Gamma\;.
\end{equation}

\noindent
To turn it into a Hilbert space, we need to define a scalar product for cylindrical functions based on different graphs. Such a scalar product can be deduced from that on $\mathcal H_\Gamma$ as follows. The construction is based on the introduction of the so-called \textit{cylindrical equivalence relations} which reflects properties of the underlying continuum connection field. Essentially, we define equivalence classes of graphs that can be regarded as subgraphs of a bigger one, i.e.:

\begin{equation}\label{qg59}
[\Gamma]=\bigl\{\Gamma_1\sim\Gamma_2\quad\text{iff}\quad\exists\,\Gamma\subset\Sigma\;:\;\Gamma\supset\Gamma_1,\Gamma_2\bigr\}\;.
\end{equation}

\noindent
In this way in fact the inner product can be reconduced to that in (\ref{qg54}) for such bigger graph by defining

\begin{equation}\label{qg60}
\braket{\psi_{(\Gamma_1,f_1)}|\psi_{(\Gamma_2,f_2)}}\equiv\braket{\psi_{(\Gamma,f_1)}|\psi_{(\Gamma,f_2)}}\;,
\end{equation}

\noindent
where $\Gamma=\Gamma_1\cup\Gamma_2$ and $f_1,f_2$ are trivially extended on $\Gamma$ by setting them constant over the links which do not belong to $\Gamma_1,\Gamma_2$, respectively\footnote{Obviously, this new product reduces to the previous one if $\Gamma_1$ and $\Gamma_2$ coincide.}. The (unconstrained) kinematical Hilbert space wil be therefore given by:

\begin{equation}\label{qg58}
\mathcal H_{kin}=\frac{\bigcup_{\Gamma\subset\Sigma}\mathcal H_\Gamma}{\sim}\;.
\end{equation}

\noindent
It is possible to prove that the space (\ref{qg58}) equipped with the inner product (\ref{qg60}) can be realized as a Hilbert space

\begin{equation}\label{qg61}
\mathcal{H}_{kin}\cong L^2(A,d\mu_{AL})\;,
\end{equation}

\noindent
over ``generalized'' connections on $\Sigma$ with the so-called Ashtekar-Lewandowski measure $d\mu_{AL}$ \cite{LQG32,LQG33,LQG34}.\\Finally, let us introduce a basis for the kinematical Hilbert space. According to the Peter-Weyl theorem \cite{LQG31}, a cylindrical function $\psi_{(\Gamma, f)}[A]\in\mathcal H_\Gamma\cong L^2(SU(2)^L, d\mu_{Haar})$ can be decomposed as

\begin{equation}\label{qg62}
\psi_{(\Gamma,f)}[A]=\sum_{j_\ell,m_\ell,n_\ell}f_{m_1,\dots,m_L,n_1,\dots,n_L}D^{(j_1)}_{m_1n_1}(h_1(A))\dots D^{(j_L)}_{m_Ln_L}(h_L(A))\;,
\end{equation}

\noindent
where $D^{(j_\ell)}_{m_\ell n_\ell}(h_\ell(A))$ are the Wigner matrices which give the spin-$j$ irreducible representations of the group elements $h_\ell(A)\in SU(2)$. An orthonormal basis for the Hilbert space $\mathcal H_\Gamma$ is thus provided by 

\begin{equation}\label{qg63}
\braket{A|\Gamma; \vec j, \vec m, \vec n}\equiv D^{(j_1)}_{m_1n_1}(h_1(A))\dots D^{(j_L)}_{m_Ln_L}(h_L(A))\;,
\end{equation}

\noindent
where we have used a compact vectorial notation $\vec j, \vec m, \vec n$ to denote the spin labels of the unitary irreducible representations of $SU(2)$ associated with each link of the graph. Remarkably, as proven in \cite{LQG35, LQG36}, on this basis there is a unique representation of the holonomy-flux algebra respectively acting by multiplication (the holonomy) and through the derivative $-i\hbar\frac{\delta}{\delta A}$ (the flux).\\ \\\textbf{Remark:} In the spin representation, the equivalence condition $\sim$ which allows us to define the scalar product (\ref{qg60}) amounts to take all spins zero on the ``virtual'' links of the extended graph which do not belong to the starting one \cite{LQG14}. This implies that the Hilbert space (\ref{qg58}) can be recasted as a direct sum of single graph-based Hilbert spaces

\begin{equation}\label{lqgkin}
\mathcal H_{kin}=\frac{\bigcup_{\Gamma\subset\Sigma}\mathcal H_\Gamma}{\sim}=\bigoplus_{\Gamma\subset\Sigma}\tilde{\mathcal H}_\Gamma\;,
\end{equation}

\noindent
but the individual graph-based Hilbert spaces $\tilde{\mathcal H}_\Gamma$ correspond to $\mathcal H_\Gamma$ without zero modes, i.e., where the spins $j_\ell$ never take the value zero.

\subsection{Gauge-invariant States and Spin Network Basis}

Now that a well-behaved unconstrained kinematical Hilbert space has been constructed, the reduction procedure (\ref{qg39}) can take place, at least in principle. First of all we have to determine the solutions of the quantum Gauss constraint. These solutions will form the Hilbert space $\mathcal H_{kin}^0$ of $SU(2)$-gauge invariant states, i.e.:

\begin{equation}\label{qg64}
\mathcal H_{kin}^0\equiv \text{Inv}_{SU(2)}\bigl[\mathcal H_{kin}\bigr]\;.
\end{equation}

\noindent
\\Recalling the transformation law (\ref{qg48}) of the holonomy under a gauge transformation, we have the following transformation for a generic irreducible spin-$j$ representation

\begin{equation}\label{qg65}
\begin{split}
D^{(j_\ell)}_{m_\ell n_\ell}(h_\ell)\mapsto D^{(j_\ell)}_{m_\ell n_\ell}(h'_\ell)&=D^{(j_\ell)}_{m_\ell n_\ell}\bigl(g(\gamma_\ell(0))h_\ell g^{-1}(\gamma_\ell(1))\bigr)\\
&=\sum_{\alpha_\ell,\beta_\ell=-j_\ell}^{j_\ell}D^{(j_\ell)}_{m_\ell \alpha_\ell}\bigl(g(\gamma_\ell(0))\bigr)D^{(j_\ell)}_{\alpha_\ell \beta_\ell}(h_\ell)D^{(j_\ell)}_{\beta_\ell n_\ell}\bigl(g^{-1}(\gamma_\ell(1))\bigr)
\end{split}
\end{equation}

\noindent
from which follows that a gauge transformation acts only on the nodes of the graph. Therefore, the gauge-invariance requirement for cylindrical functions translates into the requirement of invariance under the action of the group at the nodes, i.e.:

\begin{equation}\label{qg66}
f_0(h_1,\dots,h_L)=f_0\bigl(g(\gamma_1(0))h_1 g^{-1}(\gamma_1(1)),\dots,g(\gamma_L(0))h_L g^{-1}(\gamma_L(1))\bigr)\;.
\end{equation}

\noindent
For a generic cylindrical function $f\in Cyl_\Gamma$, the above invariance can be implemented by means of the group averaging

\begin{equation}\label{qg67}
f_0(h_1,\dots,h_L)=\int\prod_{v=1}^Vdg_v\,f\bigl(g(\gamma_1(0))h_1 g^{-1}(\gamma_1(1)),\dots,g(\gamma_L(0))h_L g^{-1}(\gamma_L(1))\bigr)\;,
\end{equation}

\noindent
where $V$ is the number of nodes (vertices) of the graph $\Gamma$. This corresponds to inserting on each node $v$ of the graph the following projector\footnote{For the proof that (\ref{qg68}) is a projector we refer to \cite{LQG7}, (p. 47).}$^,$\footnote{In order to visualize this statement, we report in Appendix A the explicit computation of the gauge invariant state for the simple example of the theta graph.}:

\begin{equation}\label{qg68}
\mathcal I_v=\int dg\,\prod_{\ell\in v}D^{(j_\ell)}(g)\;.
\end{equation}

\noindent
But

\begin{equation}\label{qg69}
\prod_{\ell\in v}D^{(j_\ell)}_{m_\ell n_\ell}(g)\,\in\,\bigotimes_{\ell\in v}\mathcal V^{(j_\ell)}
\end{equation}

\noindent
where $\mathcal V^{(j_\ell)}$ denote the $SU(2)$ irreducible spin-$j_\ell$ representation spaces. Therefore, by using the decomposition of the tensor product $\bigotimes_{\ell\in v}\mathcal V^{(j_\ell)}$ into irreducible representations

\begin{equation}\label{qg70}
\bigotimes_{\ell\in v}\mathcal V^{(j_\ell)}=\bigoplus_i\mathcal V^{(j_i)}\,,
\end{equation}

\noindent
we find that $\mathcal I_v$ projects onto the gauge invariant part of $\bigotimes_{\ell\in v}\mathcal V^{(j_\ell)}$, namely the singlet space $\mathcal V^{(0)}$, i.e.:

\begin{equation}\label{qg71}
\mathcal I_v\;:\;\bigotimes_{\ell\in v}\mathcal V^{(j_\ell)}\longrightarrow\mathcal V^{(0)}\,.
\end{equation}

\noindent
Moreover, being $\mathcal I_v$ a projector, it can be decomposed in terms of a basis $\{i_\alpha\}$ of $\mathcal V^{(0)}$ and its dual basis $\{i_\alpha^*\}$ as

\begin{equation}\label{qg72}
\mathcal I_v=\sum_{\alpha=1}^{\text{dim}\,\mathcal V^{(0)}}i_\alpha i_\alpha^*\;\in\;\mathcal V^{(0)}\otimes\mathcal V^{(0)*}\;,
\end{equation}

\noindent
from which, together with the decomposition of $\bigotimes_{\ell\in v}\mathcal V^{(j_\ell)}=(\bigotimes_{\ell\,in}\mathcal V^{(j_\ell)*})\otimes(\bigotimes_{\ell\,out}\mathcal V^{(j_\ell)})$ between ingoing and outgoing links of the vertex $v$, it follows that $\mathcal I_v$ is the invariant map between the representation spaces associated with the edges joined at the node $v$, i.e.:

\begin{equation}\label{qg73}
\mathcal I_v\;:\;\bigotimes_{\ell\,in}\mathcal V^{(j_\ell)}\longrightarrow\bigotimes_{\ell\,out}\mathcal V^{(j_\ell)}\;.
\end{equation}

\begin{figure}[h!]
\centering
\includegraphics[scale=0.40]{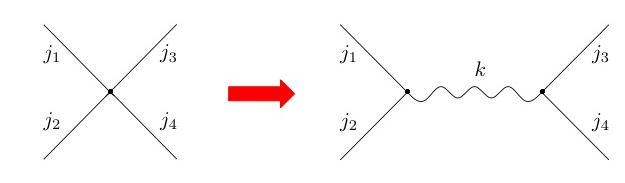}
\caption{\textit{Construction of an intertwiner for a 4-valent vertex.}}
\label{inter}
\end{figure}

\noindent
Such invariants are called \textit{intertwiners}. Hence, if we have an $p$-valent node, the intertwiner is an element of the invariant subspace $\text{Inv}_{SU(2)}\bigl[\mathcal V^{(j_1)}\otimes\dots\otimes\mathcal V^{(j_p)}\bigr]$ of the tensor product space between the $p$ irreducible representations associated to the links joining that node. However, such a procedure is possible only if some conditions necessary to have an invariant subspace are satisfied. For instance, in the case of a 3-valent node, there exists an intertwiner space only if the spin numbers $j_1,j_2,j_3$ labelling the representations associated to the three links satisfy the Clebsch-Gordan condition:
\begin{equation}\label{qg74}
|j_1-j_2|\leq j_3 \leq j_1+j_2\;.
\end{equation}
As explicitly computed in Appendix A, in this case dim$\mathcal V^{(0)}=1$ and the unique intertwiner is given by the Wigner's $3j$ symbols (see Eq. (\ref{A7})). For a $p$-valent node (with $p>3$) the space $\mathcal V^{(0)}$ can have a larger dimension and the construction consists of adding first two irreducible representations, then the third, and so on, thus giving rise to a decomposition in virtual 3-valent nodes in which virtual links are labelled by spins $k$ satisfying the condition (\ref{qg74}) and represent the intertwiners as shoved in Fig. \ref{inter} for a 4-valent node.

\noindent
By the way, since the projector (\ref{qg72}) acts only on the nodes of the graph that label the basis of $\mathcal H_{kin}$, we can write the result of the action of $\mathcal I_v$ on elements of $\mathcal H_{kin}$ as a linear combination of products of representation matrices $D^{(j_\ell)}_{m_\ell n_\ell}(h_\ell(A))$ contracted with intertwiners. This leads us to give the following

\begin{defn}\textbf{(Spin Networks):} A triplet $(\Gamma, \vec j, \vec i)$ representing a graph $\Gamma$ embedded in $\Sigma$ whose $L$ links are colored by the spins $\vec j=(j_1, \dots, j_L)$ and whose $V$ nodes are labelled by intertwiners $\vec i=(i_1,\dots,i_V)$ is called a \textbf{spin network} $S$ embedded in $\Sigma$ associated with the graph $\Gamma$. A \textbf{spin network state} $\ket{S}\equiv\ket{\Gamma;\vec j,\vec i}$ is defined as the cylindrical function over the spin network $S$ associated with the graph $\Gamma$ which can be written as
\begin{equation}\label{qg75}
\braket{A|\Gamma;\vec j,\vec i}=\psi_{\Gamma,\vec j,\vec i}[A]=\bigotimes_\ell D^{(j_\ell)}(h_\ell(A))\cdot\bigotimes_vi_v\;,
\end{equation}
where $D^{(j_\ell)}(h_\ell(A))$ are the spin irreducible representations of the holonomy along each link and $\cdot$ denotes the contraction with the intertwiners whose indices (hidden for simplicity) can be reconstructed from the connectivity of the graph.
\end{defn}

\noindent
One of the fundamental results in LQG is that spin network states form a complete orthonormal basis for the kinematical Hilbert space of solutions of the quantum Gauss constraint associated with a given graph \cite{LQG37}. Indeed, using the Peter-Weyl theorem according to which the Wigner matrices form an orthonormal basis of $L^2(SU(2))$, and the definition of the scalar product (\ref{qg60}), we have:
\begin{equation}\label{qg76}
\braket{\Gamma';\vec j\,',\vec i\,'|\Gamma;\vec j,\vec i}\equiv\braket{\psi_{\Gamma';\vec j\,',\vec i\,'}|\psi_{\Gamma;\vec j,\vec i}}=\delta_{\Gamma',\Gamma}\,\delta_{\vec j\,',\vec j}\,\delta_{\vec i\,',\vec i}\;.
\end{equation}
The $SU(2)$ constraint is implemented by choosing an intertwiner at each node as discussed before. Furthermore, the gauge-invariant Hilbert space on a fixed graph $\Gamma$ with $L$ links and $V$ nodes is given by
\begin{equation}\label{qg77}
\mathcal H^0_\Gamma=\bigotimes_{j_\ell}\bigl(\bigotimes_v Inv_{SU(2)}\bigl[\bigotimes_{\ell\in v}\mathcal V^{(j_\ell)}\bigr]\bigr)\cong L^2(SU(2)^L/SU(2)^V)\;,
\end{equation}
and we construct the kinematical Hilbert space of the theory $\mathcal H_{kin}^0$ as a direct sum of spaces over all possible graphs
\begin{equation}\label{qg78}
\mathcal H_{kin}^0=\bigoplus_{\Gamma\subset\Sigma}\mathcal H_{\Gamma}^0\;.
\end{equation}

\section{Geometric Operators on Spin Networks}

In the previous section we have constructed an orthonormal basis for the kinematical Hilbert space and we have shown how the gauge invariance is implemented. Such a basis is provided by spin network states which essentially correspond to ``colored'' graphs with links labelled by the spin numbers of $SU(2)$ irreducible representations and nodes labelled by invariant intertwiners. Now we are interested in constructing (geometric) operators acting on such states and require them to be gauge invariant. The operators which satisfy the gauge invariance condition, i.e., characterized by vanishing Poisson brackets with the constraints, are called Dirac's observables.\\As we have seen in the previous sections, the starting point of the theory was the choice of the pair of canonical variables given by the connection $A^i_a$ and its conjugate momentum $E^a_i$. We can then define two field operators associated with each one of them:

\begin{equation}\label{qg79}
\hat A_a^i\psi_{\Gamma,\vec j, \vec i}[A]=A^i_a\psi_{\Gamma,\vec j, \vec i}[A]\;,
\end{equation}

\begin{equation}\label{qg80}
\hat E_i^a\psi_{\Gamma,\vec j, \vec i}[A]=-i\hbar 8\pi G\frac{\delta}{\delta A^a_i}\psi_{\Gamma,\vec j, \vec i}[A]\;.
\end{equation}

\noindent
However, we need to find a suitable smearing of the variables which satisfies the gauge invariance requirement. We have already seen that in order to smear the Poisson algebra associated to the canonically conjugate variables $(A_a^i,E^a_i)$, we are led to introduce a new pair of canonical variables given by the holonomies of the connection and fluxes. The holonomy itself defines an operator which acts as a multiplicative operator but, in order to implement gauge invariance, we need to take the trace of the holonomy along a closed curve $\gamma$ thus defining the following Dirac observable called Wilson loop

\begin{equation}\label{qg81}
\hat W[\gamma]=-\text{Tr}\bigl(\hat h_\gamma(A)\bigr)\qquad s.t. \qquad \hat W[\gamma]\psi[A]=-\text{Tr}\bigl(h_\gamma(A)\bigr)\psi[A]\;.
\end{equation}

\noindent
The operator which acts as a derivation operator is more difficult to construct. We have seen that in order to regularize the distributional character of the densitized triad it is necessary to smear it by integrating against a two-dimensional surface $S$ thus defining the so-called electric fluxes as:

\begin{equation}\label{qg82}
E_i(S)=\int_S\ast E_i\;.
\end{equation}

\noindent
At the quantum level, the flux operator acts through the functional derivative\cite{LQG7}
\begin{equation}\label{qg83}
\hat E_i(S)h_\ell(A)=-i\hbar\gamma\int_S d^2\sigma\,n_a\frac{\delta h_\ell(A)}{\delta A^i_a(x(\sigma))}=\pm i\hbar\gamma h_{\ell_1}(A)\tau_ih_{\ell_2}(A)\;,
\end{equation}
where $\ell_1,\ell_2$ are the two new edges defined by the point at which the triad acts and the sign depends on the relative orientation of $\ell$ and $S$. In particular, the action vanishes when $\ell$ is tangential to the surface $S$ or $\ell\cap S=0$. However, such a quantity does not transform nicely under gauge transformations, but it can be used to construct more complicated gauge-invariant operators.\\The simplest operator that can be constructed in LQG is given by the following limit \cite{LQG7,LQG13}
\begin{equation}\label{qg84}
\hat{\mathcal A}(S)=\lim_{N\rightarrow\infty}\hat{\mathcal A}_N(S)\equiv\lim_{N\rightarrow\infty}\sum_{I=1}^N\sqrt{\hat E_i(S_I)\hat E^i(S_I)}
\end{equation}
where we have decomposed the surface $S$ in $N$ 2-dimensional cells $S_I$, $I=1,\dots, N$, such that $\lim_{N\rightarrow\infty}S_I=0$ and $\bigcup_IS_I=S$. Such an operator is gauge invariant and self-adjoint \cite{LQG13}. We can now evaluate the action of the operator (\ref{qg84}) on a generic spin network state $\psi_{\Gamma}$ as follows. Using the result (\ref{qg83}), we can evaluate the action of the scalar product $\hat E_i(S)\hat E^i(S)$ of two flows acting inside a link:
\begin{equation}\label{qg85}
\hat E_i(S)\hat E^i(S)h_\ell(A)=-\hbar^2\gamma^2h_{\ell_1}(A)\tau_i\tau^ih_{\ell_2}(A)=-\hbar^2\gamma^2 C^2h_\ell(A)\;,
\end{equation}
where we have used the fact that $\tau_i\tau^i=C^2$ is the Casimir operator of the algebra that commutes with all group elements. The action (\ref{qg85}) can be extended to a generic basis element $D^{(j_\ell)}_{m_\ell n_\ell}(h_\ell(A))$ by simply replacing $\tau_i$ with the generator $J_i$ in the arbitrary irreducible spin-$j_\ell$ representation. In this case the Casimir operator is $C^2=-j_\ell(j_\ell+1)\mathds1_{2j_\ell+1}$ and we get:

\begin{equation}\label{qg86}
\hat E_i(S)\hat E^i(S)D^{(j_\ell)}_{m_\ell n_\ell}(h_\ell(A))=\hbar^2\gamma^2j_\ell(j_\ell+1)D^{(j_\ell)}_{m_\ell n_\ell}(h_\ell(A))\;.
\end{equation}

\noindent
The action can be naturally extended by linearity over the whole space $\mathcal H_{kin}$. We then see that $\hat E_i(S)\hat E^i(S)$ gives zero if the surface $S$ is not intersected by any link of the graph $\Gamma$. Therefore, going back to the operator (\ref{qg84}), we have that once the decomposition of $S$ is sufficiently fine so that each surface $S_I$ is punctured once and only once, taking a further refinement has no consequences. Therefore, the limit (\ref{qg84}) simply amounts to sum the contributions of those links $\ell$ of $\Gamma$ which intersect $S$ and, using Eq. (\ref{qg86}), we finally get:

\begin{equation}\label{qg87}
\hat{\mathcal A}(S)\psi_{\Gamma,\vec j,\vec i}[A]=\sum_{\ell\in S\cap\Gamma}\hbar\sqrt{\gamma^2j_\ell(j_\ell+1)}\psi_{\Gamma,\vec j,\vec i}[A]\;.
\end{equation}

\noindent
This means that spin networks are eigenstates of this operator. As we will discuss in the next sections, this result has a very important physical meaning and enters in a crucial way the interpretation of spin networks as quanta of geometry. The key point is that the operator $\hat{\mathcal A}(S)$ can be interpreted as the area operator. Indeed, at the classical level, the expression of the area of a surface $S$ embedded in the three-dimensional space $\Sigma$ is

\begin{equation}\label{qg88}
\mathcal A(S)=\int_S d^2\sigma\sqrt{\det{\biggl(q_{ab}\frac{\partial x^a}{\partial\sigma^\alpha}\frac{\partial x^b}{\partial\sigma^\beta}\biggr)}}=\int_S d^2\sigma\sqrt{q^{(S)}}\qquad(\alpha,\beta=1,2)\;,
\end{equation}

\noindent
where $q^{(S)}$ is the determinant of the metric $q^{(S)}_{ab}$ on $S$ induced by that on $\Sigma$ ($q_{ab}$) given by
\begin{equation}\label{qg89}
q^{(S)}_{ab}=q_{ab}-\frac{1}{n^2}n_an_b
\end{equation}
with
\begin{equation}\label{qg90}
n_a=\varepsilon_{abc}\frac{\partial x^b}{\partial\sigma^1}\frac{\partial x^c}{\partial\sigma^2}
\end{equation}
the normal vector to $S$ and $\vec\sigma=(\sigma^1,\sigma^2)$ denoting local coordinates on $S$. But

\begin{equation}\label{qg91}
q\,\underset{q^{(S)}/q}{\underbrace{q^{ab}n_an_b}}=E_i^aE_i^bn_an_b\qquad\Longrightarrow\qquad q^{(S)}=E_i^aE_i^bn_an_b\;.
\end{equation}

\noindent
The classical area can be thus written in terms of the densitized triad as:

\begin{equation}\label{qg92}
\mathcal A(S)=\int_Sd^2\sigma\sqrt{n_aE_i^an_bE_j^b\delta^{ij}}\;.
\end{equation}

\noindent
By introducing now a decomposition of $S$ into small pieces $S_I$, the integral (\ref{qg92}) can be written as the limit of the Riemann sum

\begin{equation}\label{qg93}
\mathcal A(S)=\lim_{N\rightarrow\infty}\sum_{I=1}^N\sqrt{E_i(S_I)E^i(S_I)}
\end{equation}

\noindent
from which we see that the operator $\hat{\mathcal A}(S)$ defined in (\ref{qg84}) is the quantum area operator obtained by replacing the classical flux $E_i(S_I)$ with the corresponding operator $\hat E_i(S_I)$. This shows two key results:

\begin{itemize}
\item[\textbf{1)}] The geometric notion of area is now quantized and its spectrum is completely known (for the case in which the surface $S$ intersects a node of the graph we refer to \cite{LQG13,LQG14,LQG38});
\item[\textbf{2)}] The area can only take discrete values and there is a minimal value of area corresponding to the smallest eigenvalue which, restoring the Newton's constant, is proportional to the squared Planck length $\ell_p^2=\hbar G$. To be precise, the minimal planckian area corresponds to the minimal eigenvalue $j=1/2$ if one has imposed the (cylindrical) equivalence relations on quantum states, discussed in the previous section, and if one has used a symmetric quantization map in defining the product of flux operators. If one does not impose cylindrical equivalence, $j=0$ is an allowed eigenvalue \cite{LQG60}, and if for istance the Duflo map is used, then the spectrum becomes $j+1/2$ \cite{duflo}.
\end{itemize}

\noindent
Finally, a similar procedure can be done with the volume operator and the length operator which we do not discuss here but we refer to \cite{LQG7,LQG39,LQG40,LQG41,LQG42}. However, let us stress that the definition of these operators is more complicated and not unique. In particular, for the length operator, the physical interpretation is far less intuitive. Moreover, while the area operator acts on the links of a spin network, the two volume operators known in the literature (one proposed by Rovelli and Smolin \cite{LQG40}, and the other by Ashtekar and Lewandowski \cite{LQG41}) act on the nodes and more precisely on the intertwiners. Both of them annihilate a 3-valent node and coincide (up to a proportionality constant) in the 4-valent case \cite{LQG7}. Being every intertwiner space finite dimensional, the volume also has a discrete spectrum with a minimal value proportional to $\ell_p^3$.

\section{The Dual Picture: Tetrahedron (Classical and Quantum)}

\begin{figure}[h!]
\centering
\includegraphics[scale=0.4]{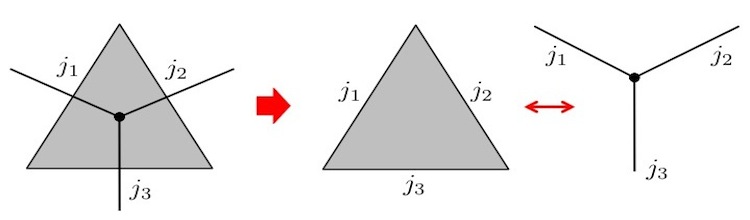}
\caption{\textit{A triangle and its dual 3-valent node.}}
\label{tria}
\end{figure}

In order to understand the geometrical interpretation of a spin network, in this section we will discuss the duality between 4-valent nodes and tetrahedra showing that the intertwiner state for a 4-valent node can be be understood as the state of a quantum tetrahedron. The role played by $SU(2)$ representations in the description of geometric objects was known from the sixties in the context of the so-called Ponzano-Regge (PR) model for 3-dimensional quantum gravity in which spins $j_\ell$ label the edges of a 3-dimensional simplicial complex which triangulates the manifold and they are interpreted as the quantized lengths of the edges of a tetrahedron, i.e., of a building block of 3-dimensional simplicial complexes \cite{LQG43}. However, it was Barbieri in his seminal work \cite{LQG4}, who realized that quantum states of LQG can be understood (and heuristically derived) by applying the quantization procedure to the usual geometric tetrahedron in 3 spatial dimensions thus suggesting that quantum gravity may be about quantizing (discrete, e.g., simplicial) geometric structures. Such a dual picture was then generalized to the case of $k$-valent nodes whose intertwiners correspond to polyhedra states and a spin network is interpreted as the dual skeleton of an abstract $k$-complex \cite{LQG44}.\\Following \cite{LQG5}, let us consider as a preliminary example the case of a triangle in 3D gravity and show how the dual 3-valent node can be obtained by applying a symplectic reduction and geometric quantization argument\footnote{For a brief review of these procedures see Appendix B.}. As in the PR model, we associate three spins $j_1,j_2,j_3$ (that is three $SO(3)$ irreducible representations) to the edges of the triangle. A quantum state of geometry of the triangle will be an element of $\bigotimes_{\ell=1}^3\mathcal V^{(j_\ell)}$ satisfying the closure condition $j_1+j_2+j_3=0$ whose quantum version translates the invariance of the state under the action of $SO(3)$. If the $j$'s satisfy the triangles inequalities, there exists a unique invariant element called vertex. As showed in Fig.\ref{tria}, this object can be interpreted as a 3-valent node of a spin network dual to the triangulation. This reflects the fact that the geometry of a Euclidean triangle is completely specified by its edge lengths.

\noindent
The uniqueness can be understood by means of the following argument \cite{LQG5}. Any irreducible representation of $SO(3)$ can be realized by the action of $SU(2)$ on the space of holomorphic sections of some line bundle over $S^2$. Therefore, elements of the space $\mathcal V^{(j_1)}\otimes\mathcal V^{(j_2)}\otimes\mathcal V^{(j_3)}$ correspond to holomorphic sections of the tensor product line bundle on $S^2\times S^2\times S^2$. A vertex corresponds to a $SU(2)$-invariant section of this tensor product line bundle. In more general terms, invariant holomorphic sections of a line bundle over a K\"ahler manifold correspond to sections of a line bundle over the symplectic reduction of this manifold by the group action \cite{LQG45}. By noticing that symplectic reduction eliminates d.o.f. in conjugate pairs, we understand that the reduction of the 6-dimensional manifold $S^2\times S^2\times S^2$ by the action of the 3-dimensional group $SU(2)$ gives a $6-(3\cdot 2)=0$ dimensional reduced space and, in fact, when the spins satisfy the tringular inequality, the reduced space is just a single point. A vertex corresponds to a section of a line bundle over this point, i.e., the spaces of vertices is 1-dimensional.

\subsection{Classical Phase Space via Symplectic Reduction}

\begin{figure}[t!]
\centering
\includegraphics[scale=0.33]{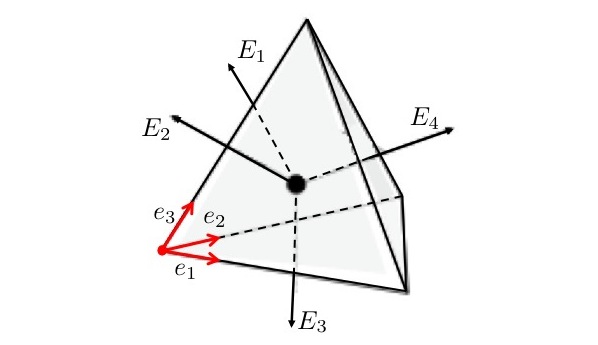}
\caption{\textit{Descriptions of the classical geometry of a tetrahedron in 3-dimensions.}}
\label{tetrahedron}
\end{figure}

A tetrahedron can be understood as the convex envelope of four points in the 3-dimensional Euclidean space. Its geometry, modulo translations, is determined by three vectors $e_1, e_2, e_3$ which are the edge vectors for the three edges pointing out from a common vertex \cite{LQG4,LQG5}. As discussed in \cite{LQG4}, the classical geometry of a tetrahedron can be equivalently described by means of the bivectors associated to the triangular faces
\begin{equation}\label{qg94}
\begin{cases}
&E_1=e_3\wedge e_2\;,\;E_2=e_1\wedge e_3\;,\;E_3=e_2\wedge e_1\\
&E_4=-E_1-E_2-E_3
\end{cases}
\end{equation}
where the last equation, which implements the \textit{closure condition}, shows that only three of these bivectors are independent. In three dimensions $\Lambda^2\mathbb R^3\cong\mathbb R^3$ and the $\wedge$ operation is the usual vector cross product. The bivectors (\ref{qg94}) can be then interpreted as the ``vectorial areas'' of the faces of the tetrahedron and, somewhat improperly, they are called normals (see Fig. \ref{tetrahedron}).

\noindent
 With this identification we have also that:
\begin{equation}\label{qg95}
-E_1\cdot(E_2\times E_3)=36V^2>0\;,
\end{equation}
where $V=\frac{1}{6}[e_1\cdot(e_2\times e_3)]$ is the volume of a non-degenerate tetrahedron. Eq.(\ref{qg95}) is called the \textit{positivity constraint}\footnote{It can be written in different equivalent ways obtained by means of a even permutation of $\{1,2,3,4\}$ which correspond to different choises of the starting vertices in which the triad $e_1,e_2,e_3$ is centered.}.\\Each bivector $E_i$ is an element of $\mathfrak{so}(3)^*$ and so, in order to obtain a Poisson structure on the space of geometries of a tetrahedron in 3 dimensions, we start by taking the product of four copies of $\mathfrak{so}(3)^*$, i.e.:

\begin{equation}\label{qg96}
\bigl(\mathfrak{so}(3)^*\bigr)^4=\mathfrak{so}(3)^*\times\mathfrak{so}(3)^*\times\mathfrak{so}(3)^*\times\mathfrak{so}(3)^*\;.
\end{equation}

\noindent
Then we can construct the classical phase space of the tetrahedron by performing a Poisson reduction w.r.t. the closure constraint as follows \cite{LQG5}:
\begin{itemize}
\item The constraint submanifold is idetified as:
\begin{equation}\label{qg97}
\mathcal C=\{E_1+E_2+E_3+E_4=0\}\subset(\mathfrak{so}(3)^*)^4\;;
\end{equation}
\item The closure constraint generates the diagonal action of $SO(3)$ on $(\mathfrak{so}(3)^*)^4$ preserving $\mathcal C$;
\item The reduced space is given by the quotient of $\mathcal C$ w.r.t. this action, i.e.:
\begin{equation}\label{qg98}
\mathcal T=\mathcal C / SO(3)\;.
\end{equation}
Such a space $\mathcal T$ is half of the phase space of a tetrahedron in three dimensions.
\end{itemize}

\noindent
To be honest one should also impose the positivity constraint on $\mathcal T$ to get the phase space of the tetrahedron. However, the resulting space is hard to quantize and so the strategy is to quantize $\mathcal T$ first and then impose the positivity constraint directly at the quantum level.\\Any $SO(3)$-invariant function on $(\mathfrak{so}(3)^*)^4$ determines a function on $\mathcal T$. In particular, from a geometrical point of view, we have the following relevant functions on $\mathcal T$:
\begin{equation}\label{qg99}
A_i=|E_i|\qquad,\qquad A_{ij}=|E_i+E_j|\qquad,\qquad U=E_1\cdot(E_2\times E_3)
\end{equation}
which are respectively interpreted as twice the area of the $i$th face of the tetrahedron ($A_i$), four times the area of the parallelogram whose vertices are given by the midpoints of the tetrahedron edges contained in either the $i$th or the $j$th face but not both ($A_{ij}$), and $U<0$ implements the positivity constraint for the tetrahedron described by the bivectors $E_i$\footnote{The case $U>0$ corresponds to the tetrahedron with opposite orientations $-E_i$.}. The map from $SO(3)$-invariant functions on $(\mathfrak{so}(3)^*)^4$ to functions on $\mathcal T$ preserves Poisson brackets. Moreover, since $|E_i|$ are constant on the symplectic leaves of $(\mathfrak{so}(3)^*)^4$, it follows that the functions $A_i$ have vanishing Poisson brackets with all functions on $\mathcal T$. The functions $A_{ij}$ instead have non-vanishing Poisson brackets and this leads to an uncertainty relation which, as pointed out in \cite{LQG4}, implies that at the quantum level the geometry of the tetrahedron exists only in the sense of ``mean geometry''.\\The symplectic leaves of $\mathcal T$ can be obtained from those of $(\mathfrak{so}(3)^*)^4$ by symplectic reduction w.r.t. the closure constraint. Indeed, considering a symplectic leaf $\Lambda=\{|E_i|=r_i\}$ in $(\mathfrak{so}(3)^*)^4$, the corresponding leaf on $\mathcal T$ is given by $\Lambda_\mathcal T=(\Lambda\cap\mathcal C)/SO(3)\cong S^2$ \cite{LQG5}.

\subsection{Quantization and 4-Valent Vertex Hilbert Space}

There are two strategies for constructing the space of states of the quantum tetrahedron in 3-dimensions:

\begin{itemize}
\item[\textbf{i)}] Geometrically quantize $(\mathfrak{so}(3)^*)^4$ and impose the closure constraint at the quantum level;

\item[\textbf{ii)}] Impose the closure constraint at the classical level and geometrically quantize the resulting reduced space $\mathcal T$.
\end{itemize}

\noindent
In general, these strategies may be not equivalent. Indeed, quantization commutes with reduction only if the action of $SO(3)$ generated by the closure constraint is free, in which case the two strategies give naturally isomorphic vector spaces \cite{LQG45}. However, even if generically $SO(3)$ acts freely on the symplectic leaves of $(\mathfrak{so}(3)^*)^4$, there are also cases in which the reduced leaf has singular points where it is not a K\"ahler manifold and the action of $SO(3)$ is not free. This complicates the geometric quantization procedure thus making the second strategy more delicate. For this reason and being the first strategy sufficient for our present purposes, we will not deal with the second strategy for which we refer to \cite{LQG5}.\\Since geometric quantization takes products to tensor products, the Hilbert space corresponding to $(\mathfrak{so}(3)^*)^4$ will be the tensor product of four copies of the Hilbert space of a quantum bivector in 3 dimensions that, according to the results of Appendix B, is given by:

\begin{equation}\label{qg100}
\mathscr H=\mathcal H^{\otimes4}\qquad,\qquad\text{with}\quad\mathcal H=\bigoplus_j\mathcal V^{(j)}\;.
\end{equation}

\noindent
The coordinate functions $E_1^i,\dots,E_4^i$ on $(\mathfrak{so}(3)^*)^4$ will then correspond to the following operators on $\mathcal H^{\otimes4}$

\begin{equation}\label{qg101}
\begin{split}
&\hat E^i_1=\hat E^i\otimes\mathds1\otimes\mathds1\otimes\mathds1\\
&\hat E^i_2=\mathds1\otimes\hat E^i\otimes\mathds1\otimes\mathds1\\
&\hat E^i_2=\mathds1\otimes\mathds1\otimes\hat E^i\otimes\mathds1\\
&\hat E^i_2=\mathds1\otimes\mathds1\otimes\mathds1\otimes\hat E^i
\end{split}
\end{equation}

\noindent
with $i=1,2,3$. By imposing the closure constraint at the quantum level, we get the subspace of $\mathcal H^{\otimes4}$ consisting of states $\ket\psi$ such that:
\begin{equation}\label{qg102}
\bigl(\Hat{\vec{E}}_1+\Hat{\vec{E}}_2+\Hat{\vec{E}}_3+\Hat{\vec{E}}_4\bigr)\ket\psi=0\;.
\end{equation}
This equation implements at the quantum level the fact that the set of the faces must form the boundary of the tetrahedron and as such it must have no boundary according to the topological principle that ``the boundary of a boundary is zero'' usually encoded in the well-known Stokes Theorem. It essentially corresponds to the requirement of invariance under rotations of the tetrahedron which as a whole must be invariant even if its faces, carrying non-zero spins, transform non trivially. In other words, the states which satisfy the condizion (\ref{qg102}) are those invariant under the action of $SU(2)$. They form the $SU(2)$-invariant subspace of $\mathcal H^{\otimes4}$ which identifies the Hilbert space of the quantum tetrahedron in 3-dimensions:
\begin{equation}\label{qg103}
\mathcal H_\mathcal T=\text{Inv}_{SU(2)}\bigl(\mathcal H^{\otimes4}\bigr)\cong\bigoplus_{j_1,\dots,j_4}\text{Inv}_{SU(2)}\biggl(\bigotimes_{\ell=1}^4\mathcal V^{(j_\ell)}\biggr)\;.
\end{equation}
Just for completeness let us mention that, since we have not imposed the positivity constraint, the phase space discussed so far includes both genuine tetrahedra and their negatives. In \cite{LQG4} it is showed that at the quantum level $U$ corresponds to an operator $\hat U$ which induces a symmetry of the Hilbert space w.r.t. the interchanges of its eigenvalues with their opposites. Such a parity symmetry translates the fact that the positivity constraint is not affected by taking a right-handed or a left-handed tetrahedron.\\Finally, as discussed in \cite{LQG4}, the condition (\ref{qg102}) corresponds in the LQG framework to the compatibility condition for the spins associated to the irreducible representations which label the links adjacent to a 4-valent vertex, the eigenvalues of $A_i$ correspond to the values of the area of the $i$th face and $\hat U$ is related to the volume operator restricted to 4-valent vertices.

\section{Interpretation of Spin Networks: Quanta of Geometry}

\begin{figure}[h!]
\centering
\includegraphics[scale=0.40]{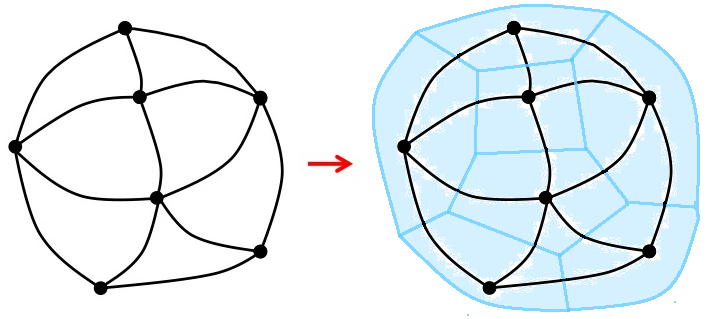}
\caption{\textit{Heuristic picture of the quantum geometry of space described by spin network states.}}
\label{CHUNK}
\end{figure}

The main point of the theory developed so far is the exhibition of a basis for the kinematical Hilbert space provided by spin network states. The key feature of these states is that they diagonalize geometric operators such as area and volume. In particular, we have mentioned that only nodes contribute to the spectrum of the volume operator implying that the volume of a given region of a spin network is actually the sum of $v$ terms each of which is associated with one of the nodes inside the region. This, together with the identification of spin networks with states of quantum polyhedra dual to the nodes of the graph, leads us to interpret spin networks as the quantum states of space geometry \cite{LQG6}. By this we mean that the algebraic data associated to a spin network define a notion of quantum geometry where each face dual to a link $\ell$ has an area proportional to the spin label $j_\ell$, and each region around a node $v$ has a volume determined by the intertwiner $i_v$ as well as the spins of the links sharing that node.\\We may translate such an interpretation into the heuristic picture of Fig. \ref{CHUNK} where space is represented by a collection of ``chunks'' (given by the polyhedra dual to the nodes) with quantized volume. Neighbouring chunks share surfaces whose area is determined by the spin carried by the dual link which intersects it.

\noindent
This essentially shows that in LQG the space geometry is discrete at the Planck scale. It is important to stress that this is not a built-in discretization as in lattice approaches to quantum gravity but a result of the quantum theory \cite{LQG7}. More precisely, during the process of quantization, we have encountered two kinds of dicreteness which are conceptually different. The first one is a discretization of space related to the use of spin networks but it is not a quantum discreteness and in some sense it can be compared to the analysis of a single mode in a plane wave expansion. This first type of discretization is actually not very different (if not at the interpretation level) from the one used in lattice approaches to QG. The second one instead concerns the discrete nature of the spectra of area and volume operators and corresponds to a discretization of geometry which occurs at the quantum scale. Moreover, if after the imposition of the diffeomorphism and Hamiltonian constraints these results still hold, then they might be considered as predictions of the theory about the microscopic structure of quantum spacetime. The fundamental discrete structure whose minimal geometric exitations are proportional to the Planck length would then imply that the theory is expected to be free from UV divergences and resolve the classical singularities of GR such as black holes and the Big Bang \cite{LQG46,LQG47,LQG48,LQG49}.\\Summarizing, spin network graphs are interpreted as quanta of geometry whose algebraic data carry spatial volume and area degrees of freedom. However, it should be stressed that the quantum geometry described by a spin network $(\Gamma, j_\ell, i_v)$ is radically different from the classical notion of geometry. The key differences are basically related to the following aspects \cite{LQG7}:

\begin{itemize}
\item the discreteness of geometric quantities encoded by the quantized nature of the spectra of geometric operators;
\item the non-commutativity of some geometric operator which is a consequence of the non-commutativity of the flux operators and leads to visualize the quantum geometry of LQG in terms of fuzzy polyhedra stuck together;
\item the distributional nature of states which capture only a finite number of components of the original fields (the values along paths and surfaces respectively for the connection and the triad).
\end{itemize}

\noindent
However, if the theory is correct, it must admit both a continuum and a semiclassical regime in which a smooth continuous geometry should emerge and whose dynamics should be ruled by GR, at least in some approximation. The \textit{problem of the continuum limit} is still an open problem and represents one of the main challenges for all current research programs in quantum gravity. It is studied in different ways depending on the specific approach to quantum gravity. For istance, from the LQG side, the notion of a continuous surface would be related to a refinement of the graph with an increasing number of the links intersecting the surface and the continuum spectra are recovered in the large spin limit $j_\ell\rightarrow\infty$. As for the other two points listed before, the proposals deal with the construction of \textit{coherent states} which are namely linear superpositions of spin network states peacked on a classical geometry which minimizes the uncertainty of flux operators \cite{LQG50,LQG51,sct,COH,LQG52, LQG53, F11}. In more recent approaches like Group Field Theories, there are hints supporting the idea that the continuum effective physics shoulde emerge from a condensate phase of the microscopic degrees of freedom of the full quantum theory, at least in cosmological regimes \cite{gftcosm,gftcosm1}.

\section{A Digression on Diffeomorphism Invariance and Relationalism}

The quantum geometry picture outlined in the previous section is heuristic and only helps to visualize in the sense that we do not actually deal with chunks of space as it could be understood in a classical way, but rather with specific modes of an interaction involving areas and volumes. In other words, these ``chunks'' must not be intended as objects in space but as encoding space itself. This is due to a key feature of both classical and quantum gravity that is diffeomorphism invariance. Indeed, in the Dirac quantization program (\ref{qg39}), we have not yet imposed the diffeomorphism constraint $\hat H^a=0$. What we would like to do now is to sketch the construction of $\mathcal H_{Diff}$ and point out the main conceptual consequences of diffeomorphism invariance.\\ \\Let us start from the classical theory. The main features of Einstein's theory of General Relativity are background independence and diffeomorphism invariance. The lesson Einstein taught us is that the gravitational field carries the information about spacetime geometry. This means that a theory of the gravitational field is a theory of spacetime itself. Therefore, its geometrical properties are fully dynamical and the construction of the theory (both at the classical and quantum level) cannot rely on any background metric structure. Background independence is deeply related to diffeomorphism invariance. It is important to distinguish passive and active diffeomorphisms \cite{LQG54,LQG6}. Passive diffeomorphisms essentially connect the same object in different coordinate systems, while active diffeomorphisms relate an object to another one in the same coordinate system. Every theory can be formulated in a passive diffeomorphism-invariant way thus implying that passive diffeomorphism invariance is not an intrinsic property of the theory but only a consequence of the formulation. Active diffeomorphism invariance instead is a property of the theory itself and requires a background-independent framework. The peculiarity of GR which makes it different from other dynamical field theories is precisely its invariance under active diffeomorphisms. As discussed in \cite{LQG54}, the main consequence is that the localization of dynamical objects (i.e., positions and motions) is a fully relational notion in GR. This means that the mathematical notion of individual points in a manifold has no intrinsic physical significance and a physical state is not located somewhere in a absolute sense but only relative localization with respect to other parts of space is relevant.\\At the quantum level, this will reflect in the fact that our theory of quantum gravity should be a background-independent quantum field theory without an \textit{a priori} space-time localization and hence its quantum states should not describe quantum excitations on space-time but rather quantum exitations of space-time. Let us then briefly discuss how this is actually realized in LQG just giving the main steps of the construction (for a more detailed discussion see for istance \cite{LQG55}). In section 1.2.2. we argued that the Hilbert space $\mathcal H^0_{kin}$ carries also a unitary representation of the diffeomorphism group $Diff(\Sigma)$ of the 3-manifold $\Sigma$ via $\psi[A]\mapsto\psi[\phi^*A],\phi\in Diff(\Sigma)$. The unitarity of the representations was actually due to the invariance of the scalar product (see remarks in Sec. 1.2.2). Therefore, the action of a diffeomorphism $\phi$ on the holonomy naturally induces an operator $\hat\phi$ on the space of cylindrical functions such that
\begin{equation}\label{qg104}
\hat\phi\;:\;Cyl_\Gamma\longrightarrow Cyl_{\phi\circ\Gamma}\qquad by\qquad\psi_\Gamma[A]\longmapsto\hat\phi\psi_\Gamma[A]=\psi_{\phi\circ\Gamma}[A]\;.
\end{equation}
The diffeomorphism-invariace requirement can be implemented by group averaging as we did for the Gauss constraint. However to do this there are two main differences that might be taken into account. First of all, we should remove the trivial diffeomorphisms $\tilde\phi\in TDiff\Gamma$ which only shift points inside the links without changing them and consider only those which change links among themeselves without changing the graph $\Gamma$. Second, being $Diff$ a non-compact group, the diffeomorphism-invariant space turns out to be not a subspace of $\mathcal H^0_{kin}$ but of its topological dual $\mathcal H^{0*}_{kin}$\footnote{We can understand this point by considering a very simple example of a function $f(x)\in L^2(\mathbb R,dx)$ and requiring invariance under the non-compact group of translations. The resulting invariant function will be constant $f_0(x)=c\notin L^2(\mathbb R,dx)$, but it defines a linear functional on $L^2(\mathbb R,dx)$ since $\int dx cf(x)=c\tilde f(0)$, with $\tilde f$ the Fourier transform of $f$.}. The solutions of the diffeomorphism constraint are thus described by linear functionals $\xi\in\mathcal H^{0*}_{kin}$ such that
\begin{equation}\label{qg105}
\xi(\hat\phi\psi)=\xi(\psi)\qquad\forall\psi\in\mathcal H^{0}_{kin}\;,
\end{equation}
and we denote the space of such functionals $\mathcal H^*_{Diff}$. The Hilbert space $\mathcal H_{Diff}$ of diffeomorphism-invariant states will be then constructed by duality that, implementing group averaging, amounts to define a projector $\mathcal P_{Diff}$ on $\mathcal H_{Diff}$ such that
\begin{equation}\label{qg106}
\braket{\psi'|\psi}\equiv\braket{\psi'|\mathcal P_{Diff}|\psi}=\sum_{\phi\in Diff/TDiff}\braket{\hat\phi\psi'|\psi}\;,
\end{equation}
where we sum over all diffeomorphisms mapping $\Gamma$ into $\Gamma'$ except for those trivial ones in $TDiff$. What we get are equivalence classes of graphs in $\Sigma$ under diffeomorphisms called \textit{knots} and the resulting basis states of $\mathcal H_{Diff}$ are usually called \textit{knotted spin networks} or simply \textit{s-knots}.\\Moreover, since the action of the diffeomorphism group can change the way a graph is embedded in $\Sigma$ (but not the presence of knots within the graph), the direct consequence of imposing diffeomorphism invariace is that what is really physically relevant is only the the combinatorial structure of the graph not the way in which it is embedded. s-knots are thus non-embedded graphs carrying the information about the quantized space geometry. This means that s-knots represent quanta of space which do not live on a manifold but define themselves the structure of space. As such, all geometrical notions such as distance should be reconstructed from spin networks themselves. However, being the fundamental structure of space a quantum superposition of abstract non-embedded entities each of which has a different connectivity (i.e., a different graph structure), what is local in one term of the superposition will in general not be local in others \cite{LQG56,LQG57}. In particular, the lack of background implies that adjacent regions of a spin network do not necessarily correspond to close space regions. Indeed, there is no metric structure that allows us to define a notion of distance and there is no absolute position at all. But a given region of a spin network can be localized with respect to other parts of the graph. Therefore, the notions of ``close'' and ``far'' should be understood in terms of relations between parts of a spin network.\\The picture of the microscopic structure of space(-time) proposed by LQG nicely realizes Penrose's original idea to find a description of quantum geometry which is at the same time discrete and relational (that is, built up from purely combinatorial and algebraic (i.e., the group data) structures without any reference to background notions of space, time or geometry) \cite{LQG58} whose roots go back to Leibniz's relational phylosophy of space and time \cite{LQG59}. Spin networks thus provide relations between different regions of space and geometry should be extracted from such an intrinsic informational content whose understanding may be useful to derive a not \textit{a priori} assumed embedding for spin networks. This is the ideological motivation which has led people to investigate for istance the possibility that a notion of distance might be derived from correlations on spin network states which translates at the quantum level in the attempt to use entanglement as a measure of distance \cite{F3}.

\section{Pregeometric Structure of Quantum Spacetime}

In the light of the previous disscusion we see that, even pursuing the most conservative road of strictly quantize canonical GR, one ends up at the Plank scale with fundamental structures of a completely different type and the geometry of space should come out of their combinatorial and algebraic nature. However, in the construction of the Hilbert space there are still some reminiscences of the continuum theory we are coming from which reflect in the imposition of cylindrical equivalence relations between quantum states associated to different graphs allowing us to give a continuum interpretation in terms of a (generalized) connection field. Following \cite{LQG60}, we close this chapter by showing a recently developed perspective where a new Hilbert space is defined organizing graph-based states in a different way which does not make any use of contiuum notions thus making perfect sense in an abstract, non-embedded context and realizing the pregeometric picture of a microscopic structure of quantum spacetime made of non-spatiotemporal building blocks of combinatorial and algebraic nature\footnote{For a general conceptual discussion on the meaning of ``pregeometry'' see for istance \cite{LQG61}.}. Even if this direction is less explored in the canonical LQG literature, there are some works on the possibility of defining spin network states in a more abstract, combinatorial way \cite{absn1,absn2}.\\The key point of the construction is to understand how LQG states can be reformulated as ``many-particles'' states. Let us consider a closed graph $\Gamma$ with $V$ $d$-valent vertices labelled by the index $i=1,\dots,V$ and denote the set of its edges by
\begin{equation}
L(\Gamma)=(\{1,\dots,V\}\times\{1,\dots,d\})^2
\end{equation}
such that
\begin{equation}
[(ia)(ia)]\notin L(\Gamma)\qquad,\qquad [(ia)(jb)]\in L(\Gamma)
\end{equation}
where the last condition specifies the connectivity of the graph telling us the existence of a directed edge connecting the $a$-th link at the $i$-th node to the $b$-th link at the $j$-th node, with source $i$ and target $j$. A generic cylindrical function based on the graph $\Gamma$ will be a function of the group elements $h_{ij}^{ab}\in\mathbb G$ ($\mathbb G\equiv SU(2)$ in LQG) assigned to each link $\ell:=[(ia)(jb)]\in L(\Gamma)$\footnote{Since in what follows it is important to distinguish different nodes and the links joining them, we sacrify a little simplicity of notation using more indices to label links ($ab$) and their source and target nodes ($ij$).}

\begin{equation}
\psi_\Gamma(h_{12}^{11},h_{13}^{21},\dots)=\psi_\Gamma(\{h_{ij}^{ab}\}\equiv\{h_\ell\})\;\in\;\mathcal H_{\Gamma}\cong L^{2}(\mathbb G^L/\mathbb G^V)\;,
\end{equation}

\noindent
with $h_{ij}=h_{ji}^{-1}$ and we impose gauge invariance at each vertex $i$ of the graph, i.e.:

\begin{equation}
\psi_{\Gamma}(\{h_{ij}\})=\psi_{\Gamma}(\{g_ih_{ij}g_j^{-1}\})\qquad\forall g_i\in\mathbb G\;.
\end{equation}

\noindent
Consider now a new Hilbert space given by

\begin{equation}
\mathcal H_{V}\cong L^2(\mathbb G^{d\times V}/\mathbb G^V)\;,
\end{equation}

\noindent
whose generic element will be a function of $d\times V$ group elements

\begin{equation}
\varphi(\{g_i^a\})=\varphi(g_1^1,\dots,g_d^1,\dots,g_1^V,\dots,g_d^V)\;\in\;\mathcal H_V
\end{equation}

\noindent
satisfying the gauge invariance at the vertices of the graph, i.e.:
\begin{equation}
\varphi(\dots,g_a^i,\dots,g_b^j,\dots)=\varphi(\dots,\alpha_ig_a^i,\dots,\alpha_jg_b^j,\dots)\quad,\quad\forall\alpha\in\mathbb G\;.
\end{equation}

\begin{figure}[h!]
\centering
\includegraphics[scale=0.55]{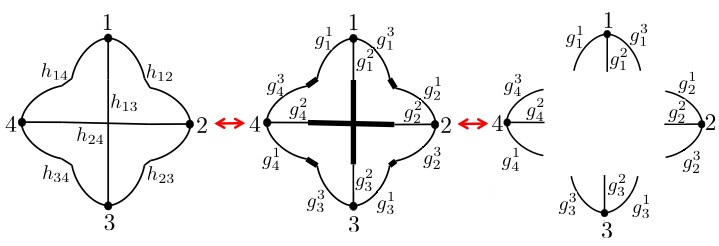}
\caption{\textit{Gluing of open spin network vertices to form a spin network closed graph.}}
\label{glu}
\end{figure}

\noindent
As in LQG, the measure of the Hilbert space is taken to be the Haar measure. The interpretation of such functions is that each $\varphi$ is associated to a $d$-valent graph formed by $V$ disconnected components, each corresponding to a single $d$-valent vertex and $d$ 1-valent vertices, which are called \textit{open spin network vertices}.\\Given a closed $d$-valent graph $\Gamma$ with $V$ vertices specified by $L(\Gamma)$, a cylindrical function $\psi_\Gamma$ can be obtained by group averaging a wave function $\varphi$

\begin{equation}\label{gluope}
\psi_\Gamma(\{h_{ij}^{ab}\})=\int_\mathbb G\prod_{[(ia)(jb)]\in L(\Gamma)}d\alpha_{ij}^{ab}\;\varphi(\{g_i^a\alpha_{ij}^{ab};g_j^b\alpha_{ij}^{ab}\})=\psi_{\Gamma}(\{g_i^a(g_j^b)^{-1}\})
\end{equation}

\noindent
in such a way that each edge is associated with two group elements $g_i^a,g_j^b\in\mathbb G$. The integrals over $\alpha$ operate a ``gluing'' of the open spin network vertices corresponding to $\varphi$, pairwise along common links, thus forming the closed spin network represented by the closed graph $\Gamma$. Such a gluing can be interpreted as a symmetry requirement. Essentially, what we are saying is that we impose the function $\varphi$ to depend on the group elements $g_i^a,g_j^b$ only through the combination $g_i^a(g_j^b)^{-1}=h_{ij}^{ab}$ which is invariant under the group action, by the same group element, at the endpoint of two open edges to which these group elements are associated as showed in Fig. \ref{glu} for the simple example of the tetrahedral graph.

\noindent
This shows that, only using functions $\varphi$, it is always possible to construct a generic function $\psi$ with all the right variables and symmetry properties, i.e., the space of functions $\psi$ is a subset of the space of functions $\varphi$.\\Moreover, using the Peter-Weyl decomposition theorem, we can give the corresponding formula in the spin representation which expresses the gluing of open spin network vertices and defines cylindrical functions for closed graphs as special cases of functions associated to a given number of them. Indeed, a cylindrical function $\psi_{\Gamma}$ can be decomposed as

\begin{equation}\label{114}
\begin{split}
\psi_{\Gamma}(\{h_{ij}^{ab}\})&=\sum\psi_{\{m_{ij}^{ab}k_{ij}^{ab}\}}^{J_{ij}^{ab}}\prod_i\overline{C^{J_{ij}^{ab}\mathcal I_i}_{m_{ij}^{ab}}}C^{J_{ij}^{ab}\mathcal I_i}_{n_{ij}^{ab}}\prod_{[(ia)(jb)]}D^{(J_{ij}^{ab})}_{s_{ij}^{ab}n_{ij}^{ab}}\bigl(\{h_{ij}^{ab}\}\bigr)\\
&=\sum_{\{J\},\mathcal I}\tilde\psi^{\{J_{ij}^{ab}\},\mathcal I_i}\prod_iC^{J_{ij}^{ab}\mathcal I_i}_{n_{ij}^{ab}}\prod_{[(ia)(jb)]}D^{(J_{ij}^{ab})}_{s_{ij}^{ab}n_{ij}^{ab}}\bigl(\{h_{ij}^{ab}\}\bigr)
\end{split}
\end{equation}
where
\begin{itemize}
\item $J_{ij}^{ab}$ label the representations of the group $\mathbb G$ and $D^{(J)}$ are the corresponding representation matrices whose indices refer to the start and end vertex of the edge $[(ia)(jb)]$ to which the group element $h_{ij}^{ab}$ is attached;
\item $C^{\{J\}, \mathcal I}$ are the normalized intertwiners for the group $\mathbb G$, attached in pairs to the vertices, resulting from the gauge-invariace requirement, a basis of which is labelled by additional quantum numbers $\mathcal I$. These intertwiners contract all indices of both nodes and of the representation functions, leaving a gauge-invariant function of spin variables only. 
\end{itemize}
By using a similar decomposition for the function $\varphi$, the group averaging expression of $\psi$ in terms of $\varphi$ can be written as
\begin{equation}\label{115}
\begin{split}
\psi_{\Gamma}(\{h_{ij}^{ab}=g_i^a(g_j^b)^{-1}\})&=\int\prod_{[(ia)(jb)]}d\alpha_{ij}^{ab}\sum_{\{\vec J_i,\vec m_i\},\mathcal I_i}\varphi_{\vec m_i}^{\vec J_i,\mathcal I_i}\prod_i\biggl(\prod_{j\neq i}D^{(J_i^a)}_{m_i^an_i^a}(g_i^a\alpha_{ij}^{ab})\biggr)C_{\vec n_j}^{\vec J_j,\mathcal I_i}\\
&=\sum_{\{J_{ij}^{ab}\},\{m_i^j\},\mathcal I_i}\varphi_{\vec m_i}^{\vec J_i,\mathcal I_i}\prod_iC_{n_{ij}^{ab}}^{J_{ij}^{ab},\mathcal I_i}\prod_{[(ia)(jb)]}\delta_{J^a_i,J_j^b}\delta_{m^a_i,m_j^b}D^{(J_{ij}^{ab})}_{s_{ij}^{ab}n_{ij}^{ab}}(g_i^a(g_j^b)^{-1})
\end{split}
\end{equation}
from which, comparing with (\ref{114}), we get the gluing formula in spin representation
\begin{equation}\label{gsr}
\psi^{\{J_{ij}\},\mathcal I_i}=\sum_{\{\vec m\}}\varphi_{\vec m_i}^{\vec J_i,\mathcal I_i}\prod_{[(ia)(jb)]}\delta_{J^a_i,J_j^b}\delta_{m^a_i,m_j^b}\;.
\end{equation}
This means that LQG states can be regarded as linear combinations of disconnected open spin network states with additional conditions enforcing the gluing and encoding the connectivity of the graph. Explicitly, Eq. (\ref{gsr}) shows that such conditions basically correspond to insert intertwiners given by the identity map at the bivalent vertices where the open links are pairwise glued.\\Now, in order to deal with graphs with an arbitrary number of vertices, we consider the Hilbert space
\begin{equation}\label{new}
\mathcal H=\bigoplus_{V=0}^\infty\mathcal H_V\;.
\end{equation}
Eq. (\ref{gluope}), or equivalently (\ref{115}), shows that there is a correspondence between LQG states and states in $\mathcal H$. This is actually more than a correspondence at the level of sets of states since it is possible to prove that the scalar product in $\mathcal H_V$ for the special class of states corresponding to closed graphs induces the standard LQG kinematical scalar product for cylindrical functions $\psi_\Gamma\in\mathcal H_\Gamma$ based on a fixed graph (see \cite{LQG60} for details). This means that, assuming that the graph $\Gamma$ has $V$ vertices, $\mathcal H_\Gamma$ can be embedded into $\mathcal H_V$ faithfully, i.e., preserving the scalar product.\\ \\It is important to stress the main differences between the new Hilbert space and the LQG one:

\begin{itemize}
\item[\textbf{1)}] The Hilbert space $\mathcal H$ in (\ref{new}) is defined by taking the direct sum over all the Hilbert spaces $\mathcal H_V\supset\mathcal H_\Gamma$ with fixed number of vertices without introducing any cylindrical equivalence class. As such, unlike the LQG case, zero modes are now included in the Hilbert space.

\item[\textbf{2)}] In the new Hilbert space, states associated to different graphs are organized in a different way w.r.t. the LQG space. Indeed, states associated to graphs with different number of vertices are orthogonal, but those associated to different graphs but with the same number of vertices are not orthogonal.
\end{itemize}

\noindent
Therefore, the new Hilbert space decreases the importance of the graph structure while maintaining the number of vertices as the key purely combinatorial feature of the quantum states\footnote{We may say that, since this new viewpoint does not deal with equivalence classes of objects reminiscent of the smearing performed at the classical level coming from GR, these states are totally abstract and their combinatorial algebraic data are not affected by any influence of the continuum theory which one would like to extract from the fundamental degrees of freedom characterizing the full quantum theory.}.\\Finally, the functions $\varphi(\vec g_1,\dots\vec g_V)$ can be understood as ``many-body'' wave functions for $V$ quanta corresponding to the $V$ open spin network vertices to which the function refers. Indeed, each state can be decomposed into products of ``single-particle''/``single-vertex'' states
\begin{equation}
\ket{\varphi}=\sum_{\{\vec\chi_i\}_{i=1,\dots,V}}\varphi^{\vec\chi_1\dots\vec\chi_V}\ket{\vec\chi_1}\otimes\dots\otimes\ket{\vec\chi_V}\;,
\end{equation}
which in the group representation reads as
\begin{equation}
\varphi(g)\equiv\braket{g|\varphi}=\sum_{\{\vec\chi_i\}}\varphi^{\vec\chi_1\dots\vec\chi_V}\braket{\vec g_1|\vec\chi_1}\dots\braket{\vec g_V|\vec\chi_V}
\end{equation}
where the complete basis of single-vertex wave functions is given by wave functions for individual spin network vertices, i.e.
\begin{equation}
\ket{\vec\chi}=\ket{\vec J,\vec m, \mathcal I}\quad:\quad\psi_{\vec\chi}(\vec g)=\braket{\vec g|\vec\chi}=\biggl(\prod_{\ell=1}^dD^{(J_\ell)}_{m_\ell n_\ell}\biggr)C^{J_1\dots J_d,\mathcal I}_{n_1\dots n_d}\;.
\end{equation}
The normalization condition for the $\varphi$ is provided by
\begin{equation}
\int\prod_{v=1}^Vd\vec g_v\bar\varphi(\vec g_1,\dots\vec g_V)\varphi(\vec g_1,\dots\vec g_V)=\sum_{\{\chi_v\}}\bar\varphi^{\{\chi_v\}}\varphi^{\{\chi_v\}}\;,
\end{equation}
where we have used the normalization condition of single-particle wave functions
\begin{equation}
\int d\vec g\,\bar\psi_{\vec\chi\,'}(\vec g)\psi_{\vec\chi}(\vec g)=\delta_{\vec\chi\,',\vec\chi}\;.
\end{equation}
The functions $\varphi$ are exactly the many-body wave functions for point particles living on the group manifold $\mathbb G^d$, whose classical phase space is $(T^*\mathbb G)^d\cong(\mathbb G\times\mathcal G^*)^d$ which is also the classical phase space of a single polyhedron dual to a $d$-valent spin network vertex. The resulting picture of the microstructure of spacetime is thus based on glued pregeometric fundamental building blocks. This is even more evident in the 2nd quantized GFT formalism whose starting point is to consider the Hilbert space $\mathcal H$ as a ``first quantized'' space and rephrase it in the Fock representation where building blocks are created and annihilated and their gluing corresponds to interactions of combinatorial nature \cite{LQG60,LQG27}. We will not treat GFT here but we have introduced this perspective since it is actually the gluing of open spin networks which will allow us to introduce a notion of entanglement on spin network states as we will discuss in the next chapter. 

\chapter{\textbf{Entanglement for Spin Network States}}

Now that the main aspects of the theory have been set in our background knowledge, next step is to introduce a notion of entanglement for spin network states. To this aim, in the first part of the chapter we recall some important notions of quantum information theory concerning composite systems, Schmidt decomposition and entanglement entropy. Then, in the second part of the chapter we will specify such notions for the case of spin network states. In particular, in order to emphasize the connection between entanglement and gauge-invariance resulting from the gluing of open spin networks, we will consider some explicit computations of entanglement entropy for simple spin network states.

\section{Quantum Information: Basics}

Quantum systems exhibit properties that are radically different from classical ones. Already at the level of a single particle system, effects like superposition of quantum states, interference, or tunneling are purely quantum-mechanical effects which are unknown for classical systems. Further differences between classical and quantum systems become manifest when one considers composite systems, i.e., systems that are formed by several subsystems (at least two). Indeed, such systems are characterized by non-classical correlations, that is, correlations which cannot be described in terms of classical probabilities. States that display such peculiar quantum-mechanical feature are referred to as \textit{entangled states}.\\Entanglement was discovered since the early days of Quantum Mechanics and, leading at apparent paradoxes suggesting, at first glance, the existence of remote actions in Quantum Mechanics, it rapidly captured a great interest \cite{QI1}. Now we know that there is no superluminal or ``at distance'' exchange of information violating causality and the paradox is solved by observing that due to the common preparation of two subsystems, the property we want to measure has a well defined meaning only when analyzed for the whole system while the same property for the parts individually remains undefined. Therefore, if similar measurements are being performed on the two entangled subsystems, there will always be a correlation between the outcomes resulting in a well defined global outcome, i.e., for both subsystems together. The investigation of entanglement is crucial for shedding light on foundational aspects of the quantum theory but many of them remain to be understood. However, within the constantly growing field of Quantum Information Theory, a lot of success has been achieved in the study of the nature of entangled states and quantitative tools have been developed to evaluate their correlations \cite{QI2}.\\In this section we will review some basic definitions and tools in quantum information\footnote{To avoid technical complications, in what follows we will always assume finite-dimesional Hilbert spaces.}. For a more detailed discussion of the subject we refer to \cite{QI2,QI3,QI4}.

\subsection{Composite Systems and Entanglement}

Let us first recall that, for a given quantum system, a quantum state $\rho$ is a non-negative Hermitian operator of trace 1 acting on the Hilbert space $\mathcal H$, i.e., such that:
\begin{equation}\label{qi1}
\rho\geq0\qquad,\qquad\text{Tr}\rho=1\qquad,\qquad\rho^\dagger=\rho\;.
\end{equation}
The components of $\rho$ in some basis are called the \textit{density matrix} even if people usually refer to $\rho$ with the same nomenclature. A quantum state is called \textit{pure} if it can be written as an outer product
\begin{equation}\label{qi2}
\rho_{pure}=\ket\psi\bra\psi\;,
\end{equation}
where $\ket\psi\in\mathcal H$ with unit norm $\braket{\psi|\psi}=1$. To avoid confusions, we will refer to $\rho$ as the quantum state of the system and to $\ket\psi$ as the state vector of the system. We will give more details about the geometrical structure of the space of pure states in Chapter 3.\\A state which is not pure is called a \textit{mixed state}. Any mixed state $\rho$ can be interpreted, not necessarily uniquely, as a classical probability distribution over a set of mutually orthogonal pure states. By this we mean that $\rho$ describes an ensemble (i.e., a mixture) of pure states and can therefore be written as
\begin{equation}\label{qi3}
\rho_{mixed}=\sum_i p_i\rho_i=\sum_i p_i \ket{\psi_i}\bra{\psi_i}\;,
\end{equation}
where $p_i$ denotes the probability weight or fractional population associated to the pure state $\rho_i$ satisfying the normalization condition $\sum_i p_i=1$. More generally, it is always possible to find a basis in which $\rho$ is diagonal, and the diagonal elements in that basis can be interpreted as the classical probabilities for the system being in the various pure states eigenvectors of the density matrix \cite{QI5}. In the case of a pure state,  there is only one non-zero diagonal element which is equal to 1 and we have ``complete'' knowledge (modulo global phases) of the state of the system. The system can be then equivalently described in terms of a density operator or of a state vector.\\A \textit{composite} or \textit{multipartite} system is a system that naturally decomposes into two or more subsystems each of which is itself a proper quantum system. Formally, the Hilbert space of a composite system is given by the tensor product of the Hilbert spaces corresponding to each of the subsystems, i.e.:
\begin{equation}\label{qi4}
\mathcal H=\mathcal H_1\otimes\dots\otimes\mathcal H_N\;.
\end{equation}
In what follows, we shall focus on (finite-dimensional) \textit{bipartite} quantum systems, i.e., systems composed of two distinct subsystems $A, B$ and hence described by the tensor product Hilbert space
\begin{equation}\label{qi5}
\mathcal H=\mathcal H_A\otimes\mathcal H_B
\end{equation}
of the two subsystems spaces $\mathcal H_A$ and $\mathcal H_B$. Nevertheless, many of the concepts and results we are going to introduce can be easily generalized to multipartite systems. The Hilbert space of a bipartite system is spanned by state vectors of the form $\ket{e_A}\otimes\ket{e_B}\equiv\ket{e_{AB}}$, where $\ket{e_A}$ and $\ket{e_B}$ denote complete bases for $\mathcal H_A$ and $\mathcal H_B$, respectively. On a bipartite Hilbert space we can define product operators
\begin{equation}\label{qi6}
\hat O_A\otimes\hat O_B\in\mathcal O(\mathcal H)
\end{equation}
which satisfy the following properties
\begin{equation}\label{qi7}
(\hat O_A\otimes\hat O_B)(\hat O_A'\otimes\hat O_B')=\hat O_A\hat O_A'\otimes\hat O_B\hat O_B'\;,
\end{equation}
\begin{equation}\label{qi8}
\text{Tr}(\hat O_A\otimes\hat O_B)=\text{Tr}(\hat O_A)\text{Tr}(\hat O_B)\;,
\end{equation}
\begin{equation}\label{qi9}
(\hat O_A\otimes\hat O_B)^\dagger=\hat O_A^\dagger\otimes\hat O_B^\dagger\;.
\end{equation}
In particular, we are often interested in measuring product operators which act nontrivially only on one subsystem, say $A$, i.e., of the form
\begin{equation}\label{qi10}
\hat O_A\otimes\mathds1_B\;,
\end{equation}
where $\hat O_A$ is the Hermitian operator acting on the subsystem $A$ and $\mathds1_B$ is the identity operator on $\mathcal H_B$ (analogous expression holds for the subsystem $B$).\\Let us now consider a bipartite system with each subsystem prepared in a pure state $\ket{\psi_i}$, $i=A,B$. The state of the composite system $\ket\psi$ is then given by the direct product
\begin{equation}\label{qi11}
\ket\psi=\ket{\psi_A}\otimes\ket{\psi_B}\;.
\end{equation}
Suppose that we can have access to only one of the subsystems at a time, i.e., we can perform only local measurements on the system. Then, after a measurement of a local observable $\hat O_A\otimes\mathds1_B$ on the first subsystem, the state of the subsystem $A$ will be projected onto an eigenstate of $\hat O_A$, but the state of the subsystem $B$ remains unchanged. If we now perform a second local measurement, this time on the second subsystem, the result will not depend on the result of the first measurement. Therefore, for a system prepared in the state $(\ref{qi11})$, the outcomes of a measurement on different subsystems are uncorrelated and depend only on the state of the respective subsystem. On the other hand, if we consider a general pure state in $\mathcal H$ given by a superposition of pure states of the form $(\ref{qi11})$, the outcome of a measurement of a local observable $\hat O_A\otimes\mathds1_B$ on the first subsystem will be
\begin{equation}\label{qi12}
\begin{split}
\braket{\hat O_A}&=\braket{\psi|\hat O_A\otimes\mathds1_B|\psi}\\
&=\text{Tr}[\rho(\hat O_A\otimes\mathds1_B)]\\
&=\text{Tr}[\ket\psi\bra\psi(\hat O_A\otimes\mathds1_B)]\\
&=\text{Tr}_A[(\text{Tr}_B\rho)\hat O_A]\\
&=\text{Tr}_A(\rho_A\hat O_A)\;,
\end{split}
\end{equation}
where $\text{Tr}_{A}\,(\text{Tr}_{B})$ denotes the partial trace over the subsystem $A\,(B)$, and $\rho_{A,B}=\text{Tr}_{B,A}\rho$ is the so-called \textit{reduced density matrix} or \textit{reduced state} of the subsystem $A,B$. Since Eq.(\ref{qi12}) holds for any local operator $\hat O_A$, we may conclude that the state of the subsystem $A$ alone is given by the reduced density matrix $\rho_A$, and similarly the state of the subsystem $B$ is described by $\rho_B$. In other words, for the purposes of measuring any observables acting nontrivially only on a subsystem, the reduced states $\rho_A$ and $\rho_B$ can be thought of as the quantum states of the subfactors $\mathcal H_A$ and $\mathcal H_B$, respectively. However, the state of the composite system is not equal to the tensor product of its reduced state, i.e.:
\begin{equation}\label{qi13}
\rho=\ket\psi\bra\psi\neq\rho_A\otimes\rho_B\;.
\end{equation}
Moreover, the local measurement on one subsystem induces a reduction of the state of the entire system, not only of the subsystem on which the measurement is performed. This means that the measurement outcomes on different subsystems will be now correlated, i.e., the probability for measuring an outcome on one subsystem will be influenced by prior measurements on the other subsystem.\\ \\Summarizing the above considerations, we give the following:

\begin{defn}
\textbf{(Separable vs Entangled States)} A state $\ket\psi\in\mathcal H$ is said to be \textit{separable} if it can be written as a product of pure states, i.e.:
\begin{equation}\label{qi14}
\textbf{separable}\quad\,\rightarrow\,\quad\ket\psi=\ket{\psi_A}\otimes\ket{\psi_B}\quad,\quad\ket{\psi_i}\in\mathcal H_i\;(i=A,B)\;.
\end{equation}
If, on the contrary, there are no local states $\ket{\psi_A}\in\mathcal H_A$ and $\ket{\psi_B}\in\mathcal H_B$ such that the state $\ket\psi$ can be written as a product thereof, then $\ket\psi$ is called an \textit{entangled state}, i.e.:
\begin{equation}\label{qi15}
\textbf{entangled}\quad\rightarrow\quad\nexists\,\ket{\psi_i}\in\mathcal H_i\,(i=A,B)\;s.t.\;\ket\psi=\ket{\psi_A}\otimes\ket{\psi_B}\;.
\end{equation}
\end{defn}

\noindent
In real experiments, however, one usually deal with mixed states rather than pure states\footnote{Indeed, in real situations, a quantum system cannot be completely isolated from the surrounding environment and so the state of the system is given by the partial trace over the environment which, in general, gives a mixed reduced state.}. In this case, the previous definitions are slightly different.\\Similarly to the case of pure states, a \textit{mixed product state}, i.e., a state $\rho$ such that

\begin{equation}\label{qi16}
\rho=\rho^{(A)}\otimes\rho^{(B)}
\end{equation}

\noindent
does not exhibit correlations. A \textit{separable mixed state}, i.e., a mixed state which can be written as a convex combination of different product states

\begin{equation}\label{qi17}
\rho=\sum_ip_i\rho^{(A)}_i\otimes\rho^{(B)}_i\qquad\text{with}\qquad p_i\geq0\;,\;\sum_ip_i=1\;,
\end{equation}

\noindent
will exhibit some correlations, that is, there are local observables $\hat O_A,\hat O_B$ such that

\begin{equation}\label{qi18}
\text{Tr}[\rho(\hat O_A\otimes\hat O_B)]\neq\text{Tr}[\rho(\hat O_A\otimes\mathds1_B)]\text{Tr}[\rho(\mathds1_A\otimes\hat O_B)]=\text{Tr}_A(\rho_A\hat O_A)\text{Tr}_B(\rho_B\hat O_B)\;.
\end{equation}

\noindent
However, such correlations can be described by means of classical probabilities $p_i$ and are hence considered as classical correlations. In contrast, if does not exist any decomposition into product states, i.e.

\begin{equation}\label{qi19}
\nexists\;\rho_i^{(A)},\rho_i^{(B)}\;\text{and}\;p_i\geq0\qquad s.t. \qquad\rho=\sum_ip_i\rho^{(A)}_i\otimes\rho^{(B)}_i\;,
\end{equation}

\noindent
then the system will exhibit correlations of measurements on different subsystems which cannot be described in terms of only classical probabilities (i.e., quantum correlations), and in this case we say that the system is in a \textit{mixed entangled state}. In this thesis work we will deal only with pure states.

\subsection{The Schmidt Decomposition}

Even if the above definitions of separable and entangled states appear very intuitive and simple on a first sight, the explicit check of separability of a given state can turn out to be much more involved than one might expect. Indeed, the definition of separability relies on the existence of a decomposition of a state into product states which in general may be difficult to be identified. We therefore need potentially simpler criteria to discriminate separable and entangled states which do not require an explicit search. These criteria descend from the so-called \textit{Schmidt decomposition} whose key observation is that it is always possible to represent a quantum state in a basis that allows to reveal its entanglement properties \cite{QI6}.\\Given two arbitrary local bases $\{\ket{e_{\,i}^{(A)}}\}$ and $\{\ket{e_{\,i}^{(B)}}\}$ in the spaces $\mathcal H_A$ and $\mathcal H_B$, any pure state $\ket\psi\in\mathcal H=\mathcal H_A\otimes\mathcal H_B$ can be decomposed into the corresponding product basis as

\begin{equation}\label{qi20}
\ket\psi=\sum_{ij}d_{ij}\,\ket{e_{\,i}^{(A)}}\otimes\ket{e_{\,j}^{(B)}}\;,
\end{equation}

\noindent
with coefficients

\begin{equation}\label{qi21}
d_{ij}=\bigl(\bra{e_{\,i}^{(A)}}\otimes\bra{e_{\,j}^{(B)}}\bigr)\ket\psi=\braket{e_{\,ij}^{(AB)}|\psi}\;.
\end{equation}

\noindent
Let us consider now a change of bases

\begin{equation}\label{qi22}
\ket{\tilde e_{\,i}^{(A)}}=U\ket{e_{\,i}^{(A)}}\qquad,\qquad\ket{\tilde e_{\,i}^{(B)}}=V\ket{e_{\,i}^{(B)}}
\end{equation}

\noindent
with $U$ and $V$ arbitrary local unitary transformations on $\mathcal H_A$ and $\mathcal H_B$, respectively. The corresponding expansion coefficients in the new basis will be then given by

\begin{equation}\label{qi23}
\begin{split}
\tilde d_{ij}&=\bigl(\bra{\tilde e_{\,i}^{(A)}}\otimes\bra{\tilde e_{\,j}^{(B)}}\bigr)\ket\psi\\
&=\bigl(\bra{e_{\,i}^{(A)}}U^\dagger\otimes\bra{e_{\,j}^{(B)}}V^\dagger\bigr)\ket\psi\\
&=\sum_{k\ell}\braket{e_{\,i}^{(A)}|U^\dagger|e_{\,k}^{(A)}}\braket{e_{\,j}^{(B)}|V^\dagger|e_{\,\ell}^{(B)}}\bigl(\bra{e_{\,k}^{(A)}}\otimes\bra{e_{\,\ell}^{(B)}}\bigr)\ket\psi\\
&=\sum_{k\ell}u_{ik}d_{k\ell}v_{\ell j}\\
&=(udv)_{ij}\quad,
\end{split}
\end{equation}

\noindent
where in the third line we inserted a resolution of the identity on each subsystem, say $\mathds1_A=\sum_k\ket{e_{\,k}^{(A)}}\bra{e_{\,k}^{(A)}}$ and $\mathds1_B=\sum_\ell\ket{e_{\,\ell}^{(B)}}\bra{e_{\,\ell}^{(B)}}$, and we define $u_{ik}=\braket{e_{\,i}^{(A)}|U^\dagger|e_{\,k}^{(A)}}$, $v_{\ell j}=\braket{e_{\,j}^{(B)}|V^\dagger|e_{\,\ell}^{(B)}}$. In the new basis, the state (\ref{qi20}) then becomes:

\begin{equation}\label{qi24}
\ket\psi=\sum_{ij}(udv)_{ij}\,\ket{\tilde e_{\,i}^{(A)}}\otimes\ket{\tilde e_{\,j}^{(B)}}\;.
\end{equation}

\noindent
But, for every complex matrix $d$, there always exist unitary transformations $u$ and $v$ such that $udv$ is diagonal which provide the so-called \textit{singular value decomposition} of $d$ with real, non-negative diagonal entries $s_i$ called \textit{singular values} \cite{QI7}. This means that, for any state $\ket\psi$, we can always find local bases $\{\ket{\varphi_{\,i}^{(A)}}\}$ and $\{\ket{\varphi_{\,i}^{(B)}}\}$ in terms of which the decomposition (\ref{qi24}) reduces to

\begin{equation}\label{qi25}
\ket\psi=\sum_i\lambda_i\,\ket{\varphi_{\,i}^{(A)}}\otimes\ket{\varphi_{\,i}^{(B)}}\;,
\end{equation}

\noindent
where $\lambda_i=|s_i|$ are called \textit{Schmidt coefficients}, and the sum is limited by the dimension of the smaller subsystem. Like eigenvalues of a matrix, also singular values are uniquely defined and consequently also the Schmidt coefficients are unique. The number of non-zero Schmidt coefficients $|I|$ is called the \textit{Schmidt rank}. Moreover, since the \textit{Schmidt basis} $\{\ket{\varphi_{\,i}^{(A)}}\otimes\ket{\varphi_{\,i}^{(B)}}\}$ is given by separable states, the Schmidt coefficients encode all information on the entanglement of the state $\ket\psi$. In particular, we see that $\ket\psi$ is separable if and only if the Schmidt rank is one. Indeed, if there is only one non-vanishing Schmidt coefficient, then $\ket\psi$ can be written as a product state, otherwise, when at least two Schimdt coefficients are different from zero (i.e., $|I|>1$), then it is not possible to express $\ket\psi$ in the form (\ref{qi11}) and the state is entangled.\\Let us now focus on the reduced states. For istance, the reduced density matrix of the subsystem $A$ in the Schmidt basis is given by

\begin{equation}\label{qi26}
\begin{split}
\rho_A&=\text{Tr}_B\bigl(\ket\psi\bra\psi\bigr)\\
&=\text{Tr}_B\biggl[\sum_{ij}\lambda_i\lambda_j\bigl(\ket{\varphi_{\,i}^{(A)}}\bra{\varphi_{\,j}^{(A)}}\otimes\ket{\varphi_{\,i}^{(B)}}\bra{\varphi_{\,j}^{(B)}}\bigr)\biggr]\\
&=\sum_{ijk}\lambda_i\lambda_j\bigl(\ket{\varphi_{\,i}^{(A)}}\bra{\varphi_{\,j}^{(A)}}\bigr)\underset{\delta_{ki}}{\underbrace{\braket{\varphi^{(B)}_{\,k}|\varphi^{(B)}_{\,i}}}}\,\underset{\delta_{jk}}{\underbrace{\braket{\varphi^{(B)}_{\,j}|\varphi^{(B)}_{\,k}}}}\\
&=\sum_{i\in|I|}\lambda_i^2\ket{\varphi_{\,i}^{(A)}}\bra{\varphi_{\,i}^{(A)}}\,,
\end{split}
\end{equation}

\noindent
and similarly for $\rho_B$. We then see that the basis vectors of the Schmidt basis are respectively given by the eigenstates of $\rho_A$ and $\rho_B$. Hence, an important property of the reduced density matrices is that they have the same non-vanishing spectrum, which is entirely determined by the Schmidt coefficients. This not only provides a prescription to evaluate the Schmidt coefficients but also a restatement of the above separability criterion for pure states in terms of the degree of mixing of the reduced density matrices. Explicitly:

\begin{equation}\label{qi27}
\begin{matrix}
\;\rho=\ket\psi\bra\psi\;\textbf{separable}&\Leftrightarrow&\rho_{A,B}\;\textbf{pure}&\Leftrightarrow&\text{Tr}\rho_{A,B}^2=1\;\\
& & & &\\
\;\rho=\ket\psi\bra\psi\;\textbf{entangled}&\Leftrightarrow&\rho_{A,B}\;\textbf{mixed}&\Leftrightarrow&\text{Tr}\rho_{A,B}^2<1\;
\end{matrix}
\end{equation}

\noindent
\\Indeed, when $|I|=1$ (i.e., the state is separable), then Eq. (\ref{qi26}) shows that the reduced state is pure while, when $|I|>1$ (i.e., the state is entangled), Eq. (\ref{qi26}) implies that the reduced density matrix is mixed. In particular, a \textit{maximally entangled (pure) state} is a state characterized by maximum Schmidt rank and whose Schmidt coefficients are all equal. In this case we have that the reduced density matrix is maximally mixed.

\subsection{Entanglement Entropy and Mutual Information}

The qualitative distinction between separable and entangled states of the previous sections does not allow to compare the amount of correlations of different states. We therefore need a quantitative description of entanglement. There are different measures of correlation known in the literature and we will now give only the definition of some specific useful quantities and describe their main properties. Given the state of a quantum system in terms of a density matrix $\rho$, the \textit{von Neumann entropy} is defined by

\begin{equation}\label{qi28}
S(\rho)=-\text{Tr}(\rho\log\rho)\;,
\end{equation}

\noindent
where we set $0\cdot\log0\equiv0$. For a statistical mixture

\begin{equation}\label{qi29}
\rho=\sum_ip_i\ket{\psi_i}\bra{\psi_i}\qquad,\qquad\Bigl(p_i\geq0\;,\;\sum_ip_i=1\Bigr)
\end{equation}

\noindent
we have that

\begin{equation}\label{qi30}
S(\rho)=-\sum_ip_i\log p_i\;,
\end{equation}

\noindent
i.e., the von Neumann entropy is equal to the \textit{Shannon information entropy} $H(\{p_i\})$ of the distribution $i\mapsto p_i$ given by the statistical weight by which the pure ensemble described by the state vector $\ket{\psi_i}$ enters the mixture. Then $S(\rho)$ expresses the uncertainty (i.e., the lack of knowledge) about the realization of a particular state $\ket{\psi_i}$ in the mixture.\\The von Neumann entropy satisfies some important properties which we now list without proof:

\begin{enumerate}
\item For any density state $\rho$ we have
\begin{equation}\label{qi31}
S(\rho)\geq0\;,
\end{equation}
where the equality holds if and only if $\rho$ is pure;
\item If $N=dim\mathcal H<\infty$, then we have the following upper bound
\begin{equation}\label{qi32}
S(\rho)\leq\log N\;,
\end{equation}
with equality if and only if $\rho$ is maximally mixed;
\item $S(\rho)$ is invariant under unitary transformations, i.e.
\begin{equation}\label{qi33}
S(U\rho U^\dagger)=S(\rho)\;,
\end{equation}
for any unitary operator $U$ on the Hilbert space;
\item For any set of non-negative numbers $\epsilon_i$ such that $\sum_i\epsilon_i=1$, $S(\rho)$ is a concave functional on the space of density matrices, i.e., it satisfies the following \textit{concavity inequality}
\begin{equation}\label{qi34}
S\Bigl(\sum_i\epsilon_i\rho_i\Bigr)\geq\sum_i\epsilon_iS(\rho_i)\;,
\end{equation}
where the equality holds if and only if all $\rho_i$ with non-zero $\epsilon_i$ are equal to each other. Thus, the entropy of the average over a set of states is at least equal to the average of their individual entropies and is usually larger or, in physical terms, the uncertainty about a mixed state is grater than or equal to the average uncertainty of the states that constitute the mixture;
\item For any bipartite system, the von Neumann entropy obeys the so-called \textit{subadditivity condition}
\begin{equation}\label{qi35}
S(\rho)\leq S(\rho_A)+S(\rho_B)\;,
\end{equation}
and the equality holds if and only if the state of the total system describes an uncorrelated state, i.e., $\rho=\rho_A\otimes\rho_B$. Thus, by tracing over the subsystems we lose information on the correlations between them and, consequently, increase the entropy/uncertainty;
\item For a tripartite system, $S(\rho_{ABC})$ obeys the so-called \textit{strong subadditivity inequality}

\begin{equation}\label{qi36}
S(\rho_{ABC})+S(\rho_B)\leq S(\rho_{AB})+S(\rho_{BC})\;,
\end{equation}

\noindent
which, as will be clear soon, tells us that the subsystems $B$ and $C$ together have more correlations with $A$ then just $B$ by itself does.
\end{enumerate}
Moreover, for a pure state $\rho=\ket\psi\bra\psi$ of a bipartite system, the reduced states $\rho_A$ and $\rho_B$ will each have the same non-negative von Neumann entropies as can be easily derived from the Schmidt decomposition and the isospectral property of $\rho_A$ and $\rho_B$, i.e.

\begin{equation}\label{qi37}
S(\rho_A)=S(\rho_B)=-\sum_{i\in|I|}\lambda_i^2\log\lambda_i^2\geq0\;,
\end{equation}

\noindent
and in particular, according to Eq. (\ref{qi27}), we have

\begin{equation}\label{qi38}
\begin{matrix}
S(\rho_A)=S(\rho_B)=0 & & iff & & \rho\;\textbf{separable}\\
S(\rho_A)=S(\rho_B)>0 & & iff & & \rho\;\textbf{entangled}\\
(S(\rho_A)=S(\rho_B)\;\textbf{max} & & iff & & \rho\;\textbf{max. entangled})
\end{matrix}
\end{equation}

\noindent
The quantity (\ref{qi37}) then measures the degree of entanglement between the two subsystems and is hence called the \textit{entanglement entropy} or simply the \textit{entanglement} between $A$ and $B$ (also denoted by $\mathcal E(A:B)$).\\ \\\textbf{Remark:} Unlike what happens in classical probability theory where the (Shannon) entropy of a system is larger than the entropy of any of its subsystems \cite{QI8}, the von Neumann entropy of a quantum system may be zero while, as a consequence of entanglement, the reduced states both have positive entropy.\\ \\Given two quantum states $\rho$ and $\sigma$, the \textit{relative entropy} $S(\rho\|\sigma)$ is defined by
\begin{equation}\label{qi39}
S(\rho\|\sigma)=\text{Tr}(\rho\log\rho)-\text{Tr}(\rho\log\sigma)\;,
\end{equation}
and satisfies the following properties:

\begin{enumerate}
\item \textit{Klein inequality}
\begin{equation}\label{qi40}
0\leq S(\rho\|\sigma)\leq\infty\;,
\end{equation}

\noindent
with $S(\rho\|\sigma)=0$ iff $\rho=\sigma$ and we set by definition $S(\rho\|\sigma)=\infty$ if the kernel of $\sigma$ has a non-trivial intersection with the image\footnote{The image of a density matrix is defined as the space spanned by the eigenstates with non-zero eigenvalues.} of $\rho$. Thus, the relative entropy is a measure of distinguishability between two quantum states;
\item The relative entropy is invariant under unitary transformations, i.e.:

\begin{equation}\label{qi41}
S(U \rho U^\dagger\| U \sigma U^\dagger)=S(\rho\|\sigma)\;;
\end{equation}

\item For $\rho=\lambda\rho_1+(1-\lambda)\rho_2$ and $\sigma=\lambda\sigma_1+(1-\lambda)\sigma_2$, with $ 0\leq\lambda\leq1$, the relative entropy obeys the inequality
\begin{equation}\label{qi42}
S(\rho\|\sigma)\leq\lambda S(\rho_1\|\sigma_1)+(1-\lambda)S(\rho_2\|\sigma_2)\;,
\end{equation}

\noindent
i.e., it it jointly convex in its arguments;
\item Tracing over a subsystem in both arguments reduces the relative entropy, say:

\begin{equation}\label{qi43}
S(\rho_r\|\sigma_r)\leq S(\rho\|\sigma)\quad,\quad r=A,B\;.
\end{equation}
\end{enumerate}
Finally, the \textit{quantum mutual information} is defined as the relative entropy of the quantum state $\rho$ w.r.t. the corresponding uncorrelated state $\rho_A\otimes\rho_B$, i.e.

\begin{equation}\label{qi44}
I(A:B,\rho)\equiv S(\rho\|\rho_A\otimes\rho_B)=S(\rho_A)+S(\rho_B)-S(\rho)\geq0\;,
\end{equation}

\noindent
\\thus providing a measure of the change of the von Neumann entropy resulting from the tracing over the subsystems, that is, a quantification of the corresponding information loss. In other words, $I(A:B)$ measures the amount of information that $A$ has on $B$ and viceversa (it is symmetric w.r.t. the exchange of the two systems). For a pure state $S(\rho)=0$ and $I(A:B)=2S(\rho_A)=2S(\rho_B)=2\mathcal E(A:B)$. The quantum mutual information physically represents the total amount of correlations (both classical and quantum) in terms of which the strong subadditivity inequality reads as

\begin{equation}\label{qi45}
I(A:B)\leq I(A:BC)\;,
\end{equation}

\noindent
\\thus making explicit the interpretation of $Eq. (\ref{qi36})$ given before. On the other hand, the entanglement entropy defines a measure of purely quantum correlations between the systems $A$ and $B$. Therefore, we can quantify the amount of classical correlations as the difference between the mutual information and entanglement, i.e.

\begin{equation}\label{qi46}
\mathscr C(A:B,\rho)=I(A:B,\rho)-\mathcal E(A:B,\rho)\;,
\end{equation}

\noindent
\\where, just for completeness we mention that for the generic case of a mixed state $\rho$, we define the so-called \textit{entanglement of formation} as the minimal average of the entanglement of the pure states entering the mixture over all possible pure state decompositions of $\rho$ \cite{QI9}

\begin{equation}\label{qi47}
\mathcal E(A:B,\rho)=\min_{\{\ket{\psi_i}\}}\sum_ip_i\mathcal E(A:B,\ket{\psi_i}\bra{\psi_i})\;,
\end{equation}

\noindent
\\and $\mathcal E(A:B,\rho)=S(\rho_A)=S(\rho_B)$ for a pure state.

\section{Examples of Entanglement for Spin Networks}

Following \cite{F1}, in this section we discuss the explicit calculation of the entanglement entropy for some simple example of spin network states. We will start by considering the case of a Wilson line state which allows us to easily visualize what we are doing and familiarize with the calculations. Then, we will apply the same procedure to the dipole graph case.

\subsection{An Introductory Example: Wilson Lines}

The simplest spin network state is provided by the Wilson loop state $\bigl|\gamma,j\bigr>$ whose wave function is given by:

\begin{equation}\label{wl}
\psi_{\gamma, j}[h(A)]=\bigl<h\bigl|\gamma,j\bigr>=\text{Tr}\bigl[D^{(j)}(h_\gamma(A))\bigr]=D^{(j)}_{ab}(h_\gamma(A))\delta^{ab}\;,
\end{equation}
\\where:
\begin{itemize}
\item the graph $\Gamma$ consists of a single curve $\gamma$ such that $\gamma(0)=\gamma(1)$;
\item $j\in\mathbb{N}/2$ is the spin which labels the SU(2) representation associated to the single edge of the Wilson loop graph;
\item there is a single vertex to which we attach the intertwiner $i=\frac{\mathds1_j}{\sqrt{2j+1}}$ with $\mathds1_j$ the identity map in $\mathcal V^{(j)}$.
\end{itemize}

\noindent
\\Here for each node $n$ we are assuming to have normalized intertwiners $i_n$, i.e., $\text{Tr}(i_ni^*_n)=1$. This essentially means that we normalize also the Wigner matrix $D^{(j)}$ by multiplying it by a factor $\sqrt{2j+1}$. Then, similarly, the \textit{Wilson line} state $\bigl|\gamma, j, a, b\bigr>$ is defined by:

\begin{equation}
\psi_{\gamma, j, a, b}[h(A)]=\bigl<h\bigl|\gamma, j, a, b\bigr>=\sqrt{2j+1}\,D^{(j)}_{ab}(h_\gamma(A))
\end{equation}

\noindent
\\in such a way that when we close the two endpoints of the line $\gamma$ to form the vertex of the loop and we contract with the above intertwiner $i$ given by the normalized identity map, we get exactly the expression (\ref{wl}) for the Wilson loop wave function. We can compute explicitly the orthonormality relation for the Wilson line states. Indeed, by using the orthogonality of Wigner matrices

\begin{equation}
\int_{SU(2)}dg\,D^{(j)}_{ab}(g)\overline{D^{(k)}_{cd}(g)}=\frac{1}{(2j+1)}\delta_{jk}\delta_{ac}\delta_{bd}\;,
\end{equation}

\noindent
we have

\begin{equation}\label{ort}
\begin{split}
\bigl<\gamma, j, a, b\bigl|\gamma, j, c, d\bigr>&=(2j+1)\int_{SU(2)}dg\,D^{(j)}_{ab}(g)\overline{D^{(j)}_{cd}(g)}\\
&=(2j+1)\frac{1}{(2j+1)}\delta_{ac}\delta_{bd}\\
&=\delta_{ac}\delta_{bd}
\end{split}
\end{equation}

\noindent
and also for the Wilson loop states:

\begin{equation}
\begin{split}
\bigl<\gamma, j\bigl|\gamma, j\bigr>&=\frac{1}{(2j+1)}\sum_{a,b,c,d=1}^{2j+1}\delta_{ab}\delta_{cd}\bigl<\gamma, j, a, b\bigl|\gamma, j, c, d\bigr>\\
&=\frac{1}{(2j+1)}\sum_{a,c=1}^{2j+1}\bigl<\gamma, j, a, a\bigl|\gamma, j, c, c\bigr>\\
&=\frac{1}{(2j+1)}\sum_{a,c=1}^{2j+1}\delta_{ac}\delta_{ac}\\
&=\frac{1}{(2j+1)}\text{Tr}(\mathds1_j)\\
&=1
\end{split}
\end{equation}

\noindent
In order to compute the entanglement entropy of a Wilson line state we consider the curve $\gamma$ as the composition $\gamma_1\circ\gamma_2$ of two paths $\gamma_1$ and $\gamma_2$ such that $\gamma_1(0)=\gamma(0)$, $\gamma_2(1)=\gamma(1)$ and $\gamma_1(1)=\gamma_2(0)$. Thus we have the following decomposition of the Hilbert space:\\
\begin{equation}
\mathcal{H}_\gamma\subseteq\mathcal{H}_{\gamma_1}\otimes\mathcal{H}_{\gamma_2}\;.
\end{equation}
\\The Wilson line state $\bigl|\gamma, j, a, b\bigr>$ can be therefore written in such a decomposition by inserting the resolution of the identity availble in $\mathcal{H}_{\gamma_1}\otimes\mathcal{H}_{\gamma_2}$, i.e.:

\begin{equation}\label{wlwf}
\begin{split}
\psi_{\gamma_1\circ\gamma_2,j,a,b}[h(A)]&=\bigl<h\bigl|\gamma_1\circ\gamma_2, j, a, b\bigr>=\sqrt{2j+1}D^{(j)}_{ab}(h_{\gamma_1\circ\gamma_2}(A))\\
&=\sqrt{2j+1}\sum_{c=1}^{2j+1}D^{(j)}_{ac}(h_{\gamma_1}(A))D^{(j)}_{cb}(h_{\gamma_2}(A))\\
&=\frac{1}{\sqrt{2j+1}}\sum_{c=1}^{2j+1}\psi_{\gamma_1,j,a,c}[h(A)]\psi_{\gamma_2,j,c,b}[h(A)]\\
&=\frac{1}{\sqrt{2j+1}}\sum_{c=1}^{2j+1}\bigl<h\bigl|\gamma_1,j,a,c\bigr>\bigl<h\bigl|\gamma_2,j,c,b\bigr>
\end{split}
\end{equation}

\noindent
from which it follows that:

\begin{equation}\label{schmidec}
\bigl|\gamma, j, a, b\bigr>=\frac{1}{\sqrt{2j+1}}\sum_{c=1}^{2j+1}\bigl|\gamma_1,j,a,c\bigr>\otimes\bigl|\gamma_2,j,c,b\bigr>\;.
\end{equation}

\noindent
According to the orthogonality of the Wilson line states (\ref{ort}), Eq. (\ref{schmidec}) provides a Schimdt decomposition of the state $\bigl|\gamma, j, a, b\bigr>$. The Schmidt rank is:

\begin{equation}
|I|=(2j+1)\;,
\end{equation}

\noindent
while the Schmidt coefficients are all equal and are given by:

\begin{equation}\label{coeff}
\lambda_i=(2j+1)^{-\frac{1}{2}}\quad\forall i\in|I|\;.
\end{equation}

\begin{figure}[t!]
\centering
\includegraphics[scale=0.40]{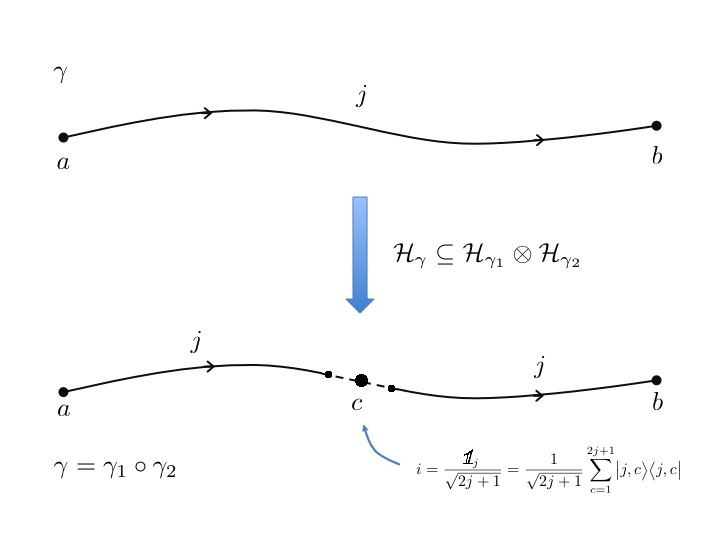}
\caption{\textit{Splitting of a Wilson line into two Wilson lines labelled by the same SU(2) representation $j$ and glued in a bivalent vertex with associated intertwiner $i$.}}
\label{wlglued}
\end{figure}

\noindent
Moreover, by looking at Eq. (\ref{wlwf}), we have the contraction of two Wilson line wave functions with the intertwiner $i$ given by the normalized identity map, i.e.:

\begin{equation}
\psi_{\gamma_1\circ\gamma_2,j,a,b}[h(A)]=(2j+1)D^{(j)}_{ac}(h_{\gamma_1}(A))D^{(j)}_{c'b}(h_{\gamma_2}(A))\frac{\delta^{cc'}}{\sqrt{2j+1}}\;,
\end{equation}

\noindent
where the sum over repeated indices is understood. In other words, what we find is that by gluing two Wilson lines into a bivalent vertex with the prescription showed in Fig. \ref{wlglued} that they are labelled by the same SU(2) representation $j$ and that the intertwiner associated to the new vertex is
\begin{equation}\label{2inter}
i=\frac{\mathds1_j}{\sqrt{2j+1}}=\frac{1}{\sqrt{2j+1}}\sum_{c=1}^{2j+1}\bigl|j, c\bigr>\bigl< j, c\bigr|\;,
\end{equation}
then the resulting Wilson line state is a maximally entangled state of $\bigl|\gamma_1, j, a, c\bigr>$ and $\bigl|\gamma_2, j, c, b\bigr>$. This is coherent with the result discussed in Sec. 1.7, according to which the general group averaging procedure to implement gauge-invariance at the new vertices created by gluing open spin networks amounts to insert bivalent intertwiners like (\ref{2inter}). Thus, there is a close relationship between entanglement and gauge-invariance. It is actually the gauge-invariance requirement to be responsible for the appearence of entanglement in gluing open spin network states. However, it should be stressed that the correspondence ``\textit{gauge invariance = maximizing entanglement}'' holds for basis states. Indeed, one can consider gauge invariant superpositions of spin networks, in particular those corresponding to generic cylindrical functions. In this case, the presence of the modes implies that the states are gauge invariant, but they do not maximize entanglement.\\ \\Finally, having the Schmidt decomposition (\ref{schmidec}), we can now compute the entanglement entropy of the Wilson line state. Indeed, the reduced density matrices will be:
\begin{equation}
\rho_1=\frac{1}{|I|}\sum_c\bigl|\gamma_1, j, a, c\bigr>\bigl<\gamma_1, j, a, c\bigr|
\end{equation}

\begin{equation}
\rho_2=\frac{1}{|I|}\sum_c\bigl|\gamma_2, j, c, b\bigr>\bigl<\gamma_2, j, c, b\bigr|
\end{equation}

\noindent
\\and using the explicit form of the Schmidt coefficients (\ref{coeff}) we get:\\

\begin{equation}
S(\rho_1)=S(\rho_2)=-\sum_{i\in|I|}\lambda_i^2\log{\lambda_i^2}=\log{(2j+1)}\;.
\end{equation}

\noindent
A straightforward iteration of the above procedure allows us to extend the result also to the case in which we consider a region $\Omega$ whose boundary $\partial\Omega$ intersects the curve $\gamma$ at $N$ points. Indeed, we decompose $\gamma$ as $\gamma_1\circ\dots\circ\gamma_N$ with $\gamma_1, \gamma_3,\dots, \gamma_{N-1}\subset\Omega$ and $\gamma_2,\gamma_4, \dots, \gamma_N\subset\overline{\Omega}$, and then repeating the argument which led us to Eq.(\ref{schmidec}) at each intersection point we obtain the following Schmidt decomposition:

\begin{equation}
\begin{split}
\bigl|\gamma, j\bigr>&=\frac{1}{(2j+1)^{\frac{N}{2}}}\sum_{c_1,\dots,c_N=1}^{2j+1}\bigotimes_{i=1}^N\bigl|\gamma_i, j, c_i, c_{i+1}\bigr>\\
&=\frac{1}{(2j+1)^{\frac{N}{2}}}\sum_{c_1,\dots,c_N=1}^{2j+1}\Biggl(\bigotimes_{i=1,3,\dots}^{N-1}\bigl|\gamma_i, j, c_i, c_{i+1}\bigr>\Biggr)\otimes\Biggl(\bigotimes_{i=2,4,\dots}^{N}\bigl|\gamma_i, j, c_i, c_{i+1}\bigr>\Biggr)
\end{split}
\end{equation}

\noindent
where we decompose the Hilbert space as:

\begin{equation}
\mathcal{H}_{\gamma}\subseteq\mathcal{H}_{\gamma\cap\Omega}\otimes\mathcal{H}_{\gamma\cap\overline{\Omega}}
\end{equation}

\noindent
with:

\begin{equation}
\mathcal{H}_{\gamma\cap\Omega}=\mathcal{H}_{\gamma_1}\otimes\mathcal{H}_{\gamma_3}\otimes\dots\otimes\mathcal{H}_{\gamma_{N-1}}\quad,\quad\mathcal{H}_{\gamma\cap\overline{\Omega}}=\mathcal{H}_{\gamma_2}\otimes\mathcal{H}_{\gamma_4}\otimes\dots\otimes\mathcal{H}_{\gamma_{N}}\;.
\end{equation}

\noindent
\\The Schmidt rank and coefficients are now given by:

\begin{equation}
|I|=(2j+1)^N\qquad,\qquad\lambda_i=(2j+1)^{-\frac{N}{2}}
\end{equation}

\noindent
\\and hence the entanglement entropy of the state $\bigl|\gamma, j\bigr>$ is:

\begin{equation}
S(\Omega)=-\sum_{i=1}^{(2j+1)^N}\frac{1}{(2j+1)^N}\log{\frac{1}{(2j+1)^N}}=N\log{(2j+1)}\;.
\end{equation}

\subsection{The Dipole Graph}

\begin{figure}[h!]
\centering
\includegraphics[scale=0.4395]{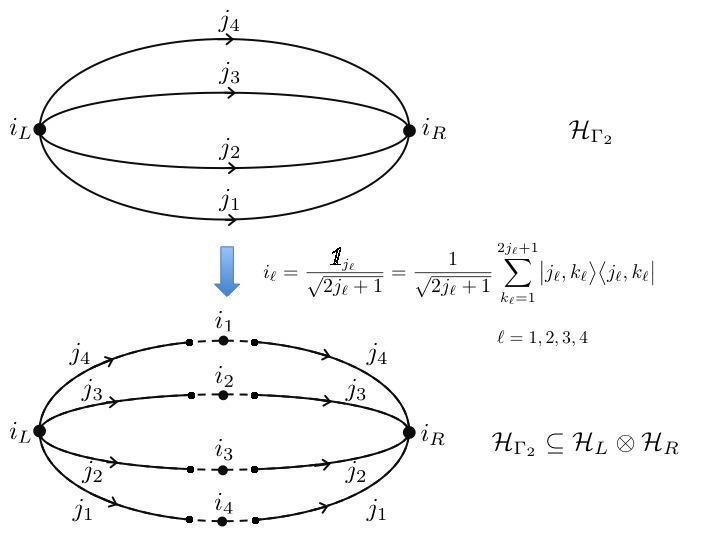}
\caption{\textit{Splitting of a dipole graph into two 4-valent vertices whose links labelled by the same SU(2) representations ($j_1,j_2,j_3,j_4$) are pairwise glued in a bivalent vertex with associated intertwiner $i_\ell$, $\ell=1,2,3,4$.}}
\label{DIPOLE}
\end{figure}

Let us now consider the so-called \textit{dipole graph} $\Gamma_2$ which consists of two 4-valent vertices with four links connecting them. Each link is labelled by a spin $j_{\ell}$ ($\ell=1,2,3,4$) and the intertwiners attached at the two vertices can be written in terms of the $\{3j\}$-symbols as:

\begin{equation}
i_{n}^{m_1m_2m_3m_4}=i^{m_1m_2k}i_k^{m_3m_4}\qquad\qquad(\text{sum over }k)
\end{equation}

\noindent
\\where $n=L,R$ denotes the left and right node in the graph, respectively.\\The gauge invariant wave function is given by:

\begin{equation}
\psi_{\Gamma_2,\vec j,\vec i}[h_1(A),h_2(A),h_3(A),h_4(A)]=\biggl(\prod_{\ell=1}^4\sqrt{2j_\ell+1}\biggr)\bigotimes_{\ell=1}^4D^{(j_\ell)}(h_\ell(A))\cdot\bigotimes_{n=L,R}i_n
\end{equation}

\noindent
\\where we use a shorthand notation $\vec j, \vec i$ respectively for $j_1,j_2,j_3,j_4$ and $i_L,i_R$.\\ \\In order to compute the entanglement entropy we think of $\Gamma_2$ as built up from the gluing of two 4-valent vertex states where all the links are pairwise attached into a bivalent vertex whose intertwiner $i_\ell$ is given by the normalized identity map $\mathds1_{j_\ell}/\sqrt{2j_\ell+1}$ ($\ell=1,2,3,4$) as schematically showed in Fig. \ref{DIPOLE}.

\noindent
\\Similarly to what we have done for the Wilson line state, we now have:

\begin{equation}\label{wfdec}
\begin{split}
\psi_{\Gamma_2,\vec j,\vec i}(h_1,\dots,h_4)&=\biggl(\prod_{\ell=1}^4\sqrt{2j_\ell+1}\biggr)D^{(j_1)}_{m_1n_1}(h_1)\dots D^{(j_4)}_{m_4n_4}(h_4)i_L^{m_1\dots m_4}i_R^{n_1\dots n_4}\\
&=\biggl(\prod_{\ell=1}^4\sqrt{2j_\ell+1}\biggr)D^{(j_1)}_{m_1n_1}(h_1^L\cdot h_1^R)\dots D^{(j_4)}_{m_4n_4}(h_4^L\cdot h_4^R)i_L^{m_1\dots m_4}i_R^{n_1\dots n_4}\\
&=\biggl(\prod_{\ell=1}^4\sqrt{2j_\ell+1}\biggr)\sum_{k_1,k_2k_3,k_4}D^{(j_1)}_{m_1k_1}(h_1^L)D^{(j_1)}_{k_1n_1}(h_1^R)D^{(j_2)}_{m_2k_2}(h_2^L)D^{(j_2)}_{k_2n_2}(h_2^R)\\
&\qquad\qquad D^{(j_3)}_{m_3k_3}(h_3^L)D^{(j_3)}_{k_3n_3}(h_3^R)D^{(j_4)}_{m_4k_4}(h_4^L)D^{(j_4)}_{k_4n_4}(h_4^R)i_L^{m_1m_2m_3m_4}i_R^{n_1n_2n_3n_4}\\
&=\biggl(\prod_{\ell=1}^4\frac{1}{\sqrt{2j_\ell+1}}\biggr)\sum_{k_1,\dots,k_4}\psi^{(L)}_{j_1,\dots,j_4,k_1,\dots,k_4}\bigl(\psi^{(R)}_{j_1,\dots,j_4,k_1,\dots,k_4}\bigr)^*
\end{split}
\end{equation}

\noindent
where $\psi^{(n)}_{\vec j, \vec k}$ with $n=L,R$ are the 4-valent vertex wave functions:

\begin{equation}\label{4valwf}
\psi^{(n)}_{j_1,\dots,j_4,k_1,\dots,k_4}=\sum_{m_1,\dots,m_4}i^{m_1\dots m_4}_n\Bigl(\sqrt{2j_1+1}\,D^{(j_1)}_{m_1k_1}\Bigr)\dots\Bigl(\sqrt{2j_4+1}\,D^{(j_4)}_{m_4k_4}\Bigr)
\end{equation}

\noindent
and in (\ref{wfdec}) we find the complex conjugate because of the orientation of the links coming out from the right vertex.\\ \\Therefore we have the following Schmidt decomposition of the dipole state $\bigl|\Gamma_2, \vec j\bigr>$:

\begin{equation}\label{DP}
\bigl|\Gamma_2, \vec j\bigr>=\biggl(\prod_{\ell=1}^4\frac{1}{\sqrt{2j_\ell+1}}\biggr)\sum_{\vec k}\bigl|\vec j, \vec k, i_L\bigr>\otimes\bigl|\vec j, \vec k, i_R\bigr>^\dagger
\end{equation}

\noindent
\\where again $\vec k$ denotes the 4-tuple $(k_1,k_2,k_3,k_4)$ with $k_\ell=1,\dots,2j_\ell+1\;,\;\forall\;\ell=1,2,3,4$. From Eq. (\ref{DP}) we see that in this case:

\begin{equation}
|I|=\prod_{\ell=1}^4(2j_\ell+1)
\end{equation}

\noindent
\\and

\begin{equation}
\lambda_i=\prod_{\ell=1}^4\frac{1}{\sqrt{2j_\ell+1}}\qquad\forall i\in|I|\;,
\end{equation}

\noindent
\\from which it follows that the entanglement entropy is given by:

\begin{equation}\label{ententrdip}
\begin{split}
S(L)=S(R)&=-\sum_{i\in I}\lambda_i^2\log{\lambda_i^2}\\
&=\log{\biggl(\prod_{\ell=1}^4(2j_\ell+1)\biggr)}\\
&=\sum_{\ell=1}^4\log{(2j_\ell+1)}\;.
\end{split}
\end{equation}

\subsection{Two Tetrahedra Entangled States}

\begin{figure}[h!]
\centering
\includegraphics[scale=0.4]{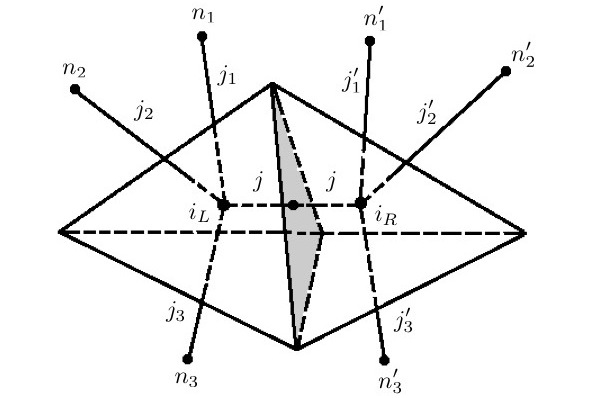}
\caption{\textit{Gluing of two tetrahedra by one of their faces and corresponding dual open spin network given by two 4-valent nodes sharing one link.}}
\label{glutetra}
\end{figure}

The same computation can be repeated for two open 4-valent vertex spin networks and gluing together only some of the links. In this case, the the sum in (\ref{ententrdip}) will be taken only on the spins labelling the glued links. Again, what we find is that entanglement is generated by implementing gauge-invariance at the nodes resulting from the gluing.\\ \\For istance, let us consider explicitly the case of two 4-valent vertices glued by only one link which, as showed in Fig. \ref{glutetra}, corresponds to two tetrahedra sharing one face.

\begin{figure}[h!]
\centering
\includegraphics[scale=0.5]{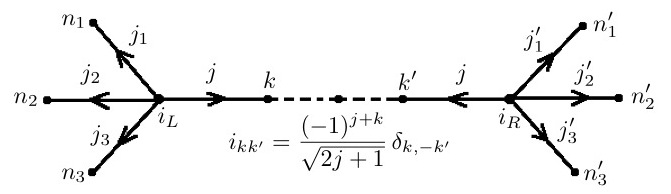}
\caption{\textit{Contraction with a bivalent intertwiner resulting from the gluing of two 4-valent nodes by one of their links.}}
\label{glue2tetr}
\end{figure}

\noindent
In this case the spin network wave function is given by:
\begin{equation}
\begin{split}
\psi(h_1,h_2,h_3,\tilde h, h_1',h_2',h_3')=&\sqrt{2j+1}\biggl(\prod_{\ell=1}^3\sqrt{2j_\ell+1}\sqrt{2j_\ell'+1}\biggr)D^{(j_1)}_{m_1n_1}(h_1)\dots D^{(j_3)}_{m_3n_3}(h_3)\\
&D^{(j)}_{mn}(\tilde h)D^{(j_1')}_{m_1'n_1'}(h_1')\dots D^{(j_3')}_{m_3'n_3'}(h_3')i_L^{m_1m_2m_3m}i_R^{m_1'm_2'm_3'n}\;.
\end{split}
\end{equation}
Therefore, by splitting $\tilde h$ as $h\cdot (h')^{-1}$, we have:
\begin{equation}\label{duetetra}
\begin{split}
\psi=&\sqrt{2j+1}\biggl(\prod_{\ell=1}^3\sqrt{2j_\ell+1}\sqrt{2j_\ell'+1}\biggr)\sum_kD^{(j_1)}_{m_1n_1}(h_1)\dots D^{(j)}_{mk}(h)\\
&\;D^{(j_1')}_{m_1'n_1'}(h_1')\dots D^{(j)}_{kn}((h')^{-1})\,i_L^{m_1m_2m_3m}\,i_R^{m_1'm_2'm_3'n}\;.
\end{split}
\end{equation}
By using now the properties
\begin{equation}
D^{(j)}_{kn}((h')^{-1})=\overline{D^{(j)}_{nk}(h')}=(-1)^{k-n}D^{(j)}_{-n-k}(h')\;,
\end{equation}
\begin{equation}
i_R^{m_1'm_2'm_3'n}=(-1)^{j+n}\;i_R^{m_1'm_2'm_3'-n}\;,
\end{equation}
and renaming the sum over $n$ involved in the contraction with the intertwiner $i_R$, Eq. (\ref{duetetra}) gives
\begin{equation}\label{2tglu}
\psi=\sum_k\frac{(-1)^{j+k}}{\sqrt{2j+1}}\psi_{j_1,j_2,j_3,j,n_1,n_2,n_3,k,i_L}(h_1,h_2,h_3,h)\psi_{j_1',j_2',j_3',j,n_1',n_2',n_3',-k,i_R}(h_1',h_2',h_3',h')\;,
\end{equation}
where the $\psi_{\vec j,\vec n, i_\alpha}$ ($\alpha=L,R$) are the 4-valent vertex wave functions (\ref{4valwf}).\\Thus, we see that the gluing operation amounts to insert a bivalent intertwiner $i_{kk'}=\frac{(-1)^{j+k}}{\sqrt{2j+1}}\delta_{k,-k'}$ which forces the two glued links to have the same spin number $j$ and reverse the direction of one of them (see Fig.\ref{glue2tetr}).

\noindent
Hence, from Eq. (\ref{2tglu}) we have the following Schmidt decomposition for the two-tetrahedra state
\begin{equation}
\ket{\psi}=\sum_k\frac{(-1)^{j+k}}{\sqrt{2j+1}}\ket{j_1,j_2,j_3,j,n_1,n_2,n_3,k,i_L}\otimes\ket{j_1',j_2',j_3',j,n_1',n_2',n_3',-k,i_R}\;,
\end{equation}
whose Schmidt rank and coefficients are given by
\begin{equation}
|I|=2j+1\qquad,\qquad\lambda_k=\frac{(-1)^{j+k}}{\sqrt{2j+1}}\quad(k\in|I|)\;.
\end{equation}
The entanglement entropy is then given by:
\begin{equation}
\begin{split}
S(\rho_L)=S(\rho_R)&=-\sum_{k\in\mathcal I}\lambda_k^2\log{\lambda_k^2}\\
&=\log(2j+1)\;.
\end{split}
\end{equation}
i.e., the only contribution comes from the spin labelling the glued link or, equivalently, the shared face in the dual simplicial picture (the grey surface in Fig. \ref{glutetra}).\\ \\Summarizing, in this chapter we analyse some examples of entanglement on spin network states. The main lesson is that it is actually the gluing of open spin networks to create the entanglement between them. In particular, concerning the basis states, the gauge invariance requirement at the new vertices created by the gluing results into (locally) maximally entangled states. Moreover, in the examples considered here, we find that the entanglement entropy is given by a sum over the intersections of the spin network with the boundary surface dual to the glued open links. Only the spin numbers labelling such links will contribute to the entanglement entropy, just as for the area operator (see Eq. (\ref{qg87})). In other words, the entanglement is a direct consequence of the gauge-invariance at the points on that surface, and in this sense the degrees of freedom responsible for the entanglement entropy are entirely local to the boundary. This argument finds an interesting application to the study of black holes entropy in terms of the entanglement between the region outside the horizon and its complement providing a partition of the horizon surface into elementary cells, each labelled by a spin $j$ \cite{F1,QI10}.

\chapter{\textbf{Geometric Quantum Mechanics and Information Geometry}}

Let us leave for a while the context of Quantum Gravity and introduce now the geometric formulation of Quantum Mechanics in its generality. For the purposes of this thesis we will not deal with the dynamical sector leaving it to future works. We will see how this formulation allows to construct tensorial structures of Riemannian and symplectic geometry on manifolds (and submanifolds) of quantum states and to describe entanglement of composite systems.

\section{Preliminaries}

There are several reasons to motivate the attempt of formulating Quantum Mechanics in geometrical terms as illustrated in \cite{GQM1}. In particular, one of the most appealing reasons is provided by the opportunity of making available ``classical methods'' in a quantum mechanical framework. Moreover, the fact that Classical Mechanics, General Relativity and Gauge Theories has been revealed to have geometric fundamental structures can be regarded as an hint to geometrize also quantum theories in order to better understand the quantum-classical transition or also (and this may be the main ideology underlying the present work) to grasp some tool or intuition to merge Quantum Mechanics and General Relativity into a consistent formulation of a theory of Quantum Gravity.\\ \\But what do we mean by ``\textit{geometric formulation of Quantum Mechanics}'' ?\\ \\According to \cite{GQM2,GQM3,GQM4}, in the modern geometric language, the geometrization program for Quantum Mechanics can be synthesized as the replacement of the usual description of a quantum system in terms of Hilbert spaces with a description in terms of Hilbert manifolds. Such a proposal is very similar to the transition from Special to General Relativity. Indeed, we know that in GR space-time is mathematically described by a Lorentzian manifold and the properties of the (flat) Minkowski space-time of Special Relativity appear locally at the level of the tangent space at each point of the space-time manifold. By doing this, we essentially go from the scalar product $\eta_{\mu\nu}X^\mu X^\nu$ to the Lorentzian metric tensor field $\eta_{\mu\nu}\theta^\mu\otimes\theta^\nu$\footnote{The $\{\theta^\mu\}$ are general 1-forms carrying the information on the non-zero curvature of space-time. In the flat case, they reduce to the $\{dx^\mu\}$.}. Similarly, in the geometrization of QM we go from the scalar product $\bigl<\psi\bigl|\psi\bigr>$ on the Hilbert space to the Hermitian tensor field $\bigl<d\psi\otimes d\psi\bigr>$ on the Hilbert manifold by means of the following steps:

\begin{itemize}
\item[\textbf{1)}] the complex separable Hilbert space $\mathcal{H}$ is replaced by its realification $\mathcal{H}_{\mathbb{R}}:=\mathbb{R}e(\mathcal{H})\oplus\mathbb{I}m(\mathcal{H})$ which is a real differential Hilbert manifold;
\item[\textbf{2)}] the Hermitian inner poduct $\braket{\cdot|\cdot}: \mathcal{H}\times\mathcal{H}\rightarrow\mathbb{C}$ on quantum state vectors is then replaced by an Hermitian tensor on quantum-state-valued sections of the tangent bundle $T\mathcal{H}_\mathbb{R}$ whose real symmetric and imaginary skewsymmetric part define a Riemannian metric tensor and a symplectic structure, respectively. This essentially amounts to identify $\mathcal{H}$ with the tangent space $T_\psi\mathcal{H}$ at each point of the base manifold. 
\end{itemize}

\noindent
It should be stressed that the manifold description in QM comes out naturally already at the standard level of Hilbert spaces. Indeed, as we will discuss in the next sections, the probabilistic interpretation of Quantum Mechanics leads us to identify (pure) states not with vectors in $\mathcal{H}$ but, rather, with rays (i.e., equivalence classes of vectors). The set of rays $\mathcal{R}(\mathcal{H})$ is the complex projective space associated with $\mathcal{H}$ and as such it is a non linear manifold\footnote{Other examples of manifolds of quantum states are provided for istance by coherent states \cite{GQM5} or by the stratified manifold of density states where each stratus contains density states with fixed rank \cite{GQM6}.}.\\In this chapter we first characterize the space of state vectors as a differential manifold and construct the various tensor fields available on it. Then we will focus on the identification of tensor fields through the pull-back to submanifolds of quantum states defined by the orbits of some Lie group acting on the initial Hilbert space. In particular, by considering the case of composite systems, the pull-back of the so-called Fubini-Study metric tensor field from the projective Hilbert space to the orbits of local unitary group (which do not change the entanglement properties of the starting state, i.e., the orbits identify submanifolds of quantum states with the same degree of entanglement) will allow us to give a tensorial characterization of quantum entanglement \cite{GQM7,GQM8}.\\The last part of the chapter is devoted to the connection between Quantum Information and Geometric Quantum Mechanics. The Fubini-Study metric or equivalently its pull-back from the projective Hilbert space to the Hilbert space is actually strictly related to the quantum Fisher information metric \cite{GQM9}. This is quite interesting by itself because it shows how the geometric formulation allows to import the information theoretical setting in the framework of Quantum Mechanics. On the other hand, even more appealing for our purposes, the off-diagonal blocks of the metric tensor can be used to define a measure of entanglement which is naturally connected to a notion of distance between density state operators \cite{GQM10,GQM11}.\\To avoid technicalities and to be sure that all what we are going to do is mathematically well-defined, in what follows we will restrict ourselves to finite-dimensional Hilbert spaces.

\section{Quantum Mechanics on the Space of Rays}

According to Quantum Mechanics, we know that in the standard description of a given quantum system, states are vectors on some complex Hilbert space $\mathcal{H}$ and observables correspond to self-adjoint linear operators on $\mathcal{H}$ \cite{GQM12}. It was in fact the need to take into account the superposition principle of quantum theory to motivate the introduction by Dirac of vector spaces as spaces of states \cite{GQM13}. Moreover, the interpretation of wave functions as probability amplitudes requires the vector space of states to be equipped with an inner product turning it into a Hilbert space. Besides the Hermitian structure $\bigl<\cdot\bigl|\cdot\bigr>$, the complex Hilbert space $\mathcal{H}$ carries also a complex structure which plays a fundamental role in a consistent deduction of the uncertainty relations, another pillar of the quantum world \cite{GQM14}. Usually the presence of a superposition rule and hence a linear structure at the level of the solutions of the dynamical equations requires these equations to be linear.\\However, a simultaneous measurement of a complete set of compatible observables does not define uniquely the state vector $\bigl|\psi\bigr>$ \cite{GQM12}. Indeed the state will be defined up to multiplication by a non-zero complex number and then we are left with an equivalence class of states $\bigl[\bigl|\psi\bigr>\bigr]$ which is usually called a ``ray''. Even if we fix the normalization, an overall phase freedom still remains. In other words, the multiplication by a non-zero complex number $\lambda=|\lambda|e^{i\alpha}$ can be thought of as the composition of a dilation by the modulus $|\lambda|$ followed by a counterclockwise rotation about the origin by an angle $\alpha$. Assuming that $dim_\mathbb{C}\mathcal{H}=n<\infty$, this formally translates into the following double fibration structure:

\begin{equation}\label{gqm1}
\xymatrix{
\mathbb{R}_+\ar[r]&\mathcal{H}_0=\mathcal{H}-\{0\}\ar[d]\\
U(1)\ar[r]&S^{2n-1}\ar[d]\\
&\mathcal{R}(\mathcal{H})\cong\mathbb{C}P^{n-1}
}
\end{equation}

\noindent
where the complex projective space is defined by the following equivalence relation:

\begin{equation}\label{gqm2}
\mathbb{C}P^{n-1}=\biggl{\{}\bigl[\bigl|\psi\bigr>\bigr]\;:\;\bigl|\psi\bigr>\sim\bigl|\psi'\bigr>\Leftrightarrow\bigl|\psi'\bigr>=\lambda\bigl|\psi\bigr>,\;\;\bigl|\psi\bigr>,\bigl|\psi'\bigr>\in\mathcal{H}_0, \lambda\in\mathbb{C}_0\biggr{\}}.
\end{equation}

\noindent
Equivalently, we may argue that the Hilbert space $\mathcal{H}$ acquires the structure of a $\mathbb{C}_0$-bundle with $\mathcal{R}(\mathcal{H})$ as base manifold, i.e.:

\begin{equation}\label{gqm3}
\xymatrix{
\mathbb{C}_0\ar[r]&\mathcal{H}\ar[d]^\pi\\
&\mathcal{R}(\mathcal{H})
}
\end{equation}

\noindent
Let us now try to show briefly that Quantum Mechanics can be reformulated in terms of objects that live in the projective Hilbert space $\mathcal{R}(\mathcal{H})$ (for further details we refer the reader to \cite{GQM15} Ch. 4 and references within). First of all, we recall from Ch.2 that the equivalence class $\bigl[\bigl|\psi\bigr>\bigr]$ can be represented by a rank-one projector also called a ``pure state'':

\begin{equation}\label{gqm4}
\rho_\psi=\frac{\bigl|\psi\bigr>\bigl<\psi\bigr|}{\bigl<\psi\bigl|\psi\bigr>}\,\in\,D^1(\mathcal{H})\;,
\end{equation}

\noindent
which satisfies the following properties:

\begin{equation}\label{gqm5}
\rho_\psi^\dagger=\rho_\psi\qquad,\qquad\text{Tr}\rho_\psi=1\qquad,\qquad\rho_\psi^2=\rho_\psi\;.
\end{equation}

\noindent
Being insensitive to an overall phase, both expectation values of self-adjoint linear operators $A=A^\dagger$ and transition probabilities depend only on the rank-one projectors associated with the states and are respectively given by:

\begin{equation}\label{gqm6}
e_A(\psi)=\bigr<A\bigl>_\psi=\frac{\bigr<\psi\bigl|A\bigl|\psi\bigr>}{\bigl<\psi\bigl|\psi\bigr>}=\text{Tr}(\rho_\psi A)\;,
\end{equation}

\noindent
and

\begin{equation}\label{gqm7}
\frac{\bigl|\bigl<\phi\bigl|\psi\bigr>\bigl|^2}{\bigl<\psi\bigl|\psi\bigr>\bigl<\phi\bigl|\phi\bigr>}=\text{Tr}(\rho_\psi \rho_\phi)
\end{equation}

\noindent
where the trace on the r.h.s. of Eq. (\ref{gqm7}) defines a binary product on pure states.\\As we will see later in this chapter, if $A$ is an observable, i.e., an Hermitian operator defined on the Hilbert space $\mathcal{H}$, we can associate with it a real symmetric function $f_A$ on $\mathcal{H}$ by setting:

\begin{equation}\label{gqm8}
f_A(\psi):=\bigr<\psi\bigl|A\bigl|\psi\bigr>\qquad\forall\psi\in\mathcal{H}\;,
\end{equation}

\noindent
which is related to the expectation value (\ref{gqm6}) by:

\begin{equation}\label{gqm9}
e_A(\psi):=\frac{f_A(\psi)}{\bigr<\psi\bigl|\psi\bigr>}\;\in\mathcal{F}(\mathcal{H}_0)
\end{equation}

\noindent
and coincides with it in the case of normalized states ($\bigl<\psi\bigl|\psi\bigr>=1$). The eigenvectors $\bigl|\tilde{\psi}\bigr>$ of $A$ therefore correspond to the critical points $\tilde\psi$ of the function $e_A(\psi)$, i.e., those points such that $d\,e_A(\psi)=0$, and we find that $e_A(\tilde{\psi})$ is exactly the corresponding eigenvalue. Indeed:

\begin{equation}\label{gqm10}
\bigl|\tilde{\psi}\bigr>\quad:\quad A\bigl|\tilde{\psi}\bigr>=\lambda_A\bigl|\tilde{\psi}\bigr>\quad\Rightarrow\quad e_A(\tilde\psi)=\frac{f_A(\tilde\psi)}{\bigr<\tilde\psi\bigl|\tilde\psi\bigr>}=\frac{\bigr<\tilde\psi\bigl|A\bigl|\tilde\psi\bigr>}{\bigl<\tilde\psi\bigl|\tilde\psi\bigr>}=\lambda_A\;.
\end{equation}

\noindent
According to the spectral decomposition of the operator $A$ and to the identification (\ref{gqm6}), we see that the expectation value $e_A(\psi)$ associates with the observable $A$ a real functional on $\mathcal R(\mathcal H)$ whose critical points and critical values yield the eigenprojectors of $A$ and the corresponding eigenvalues, respectively.\\Despite the fundamental role of the superposition principle for the description of interference phenomena, the identification of states with rays rather than Hilbert space vectors themselves and ultimately with rank-one projectors provides us with the complex projective space (associated with the Hilbert space) as the carrier space of the quantum theory. Such a space is not a linear space anymore but a differential manifold. The non linearity essentially comes with the usual superposition rule of two rank-one projectors which will give, in general, a rank-two operator. Indeed, let $\bigl|\psi_1\bigr>, \bigl|\psi_2\bigr>\in\mathcal{H}$ be two orthonormal states (i.e., $\bigl<\psi_i\bigl|\psi_j\bigr>=\delta_{ij}$ with $i,j=1,2$), the projector $\rho_\psi$ associated with the normalized state vector

\begin{equation}\label{gqm11}
\bigl|\psi\bigr>=c_1\bigl|\psi_1\bigr>+c_2\bigl|\psi_2\bigr>\qquad c_1,c_2\in\mathbb{C}\;,\;|c_1|^2+|c_2|^2=1
\end{equation}

\noindent
is given by

\begin{equation}\label{gqm12}
\rho_\psi=\bigl|\psi\bigr>\bigl<\psi\bigr|=|c_1|^2\rho_1+|c_2|^2\rho_2+(c_1\bar c_2\rho_{12}+\,h.c.)\;,
\end{equation}

\noindent
with

\begin{equation}\label{gqm13}
\rho_i=\bigl|\psi_i\bigr>\bigl<\psi_i\bigr|\;\;(i=1,2)\qquad,\qquad\rho_{12}=\bigl|\psi_1\bigr>\bigl<\psi_2\bigr|\;.
\end{equation}

\noindent
We see from (\ref{gqm12}) that $\rho_\psi$ cannot be expressed directly in terms of the projectors $\rho_1$ and $\rho_2$ associated with the initial states $\bigl|\psi_1\bigr>,\bigl|\psi_2\bigr>$. However, a superposition of rank-one projectors which gives another rank-one projector should be possible in order to maintain the superposition principle as a building block of Quantum Mechanics. This can be achieved in the following way. Considering a third, fiducial vector $\bigl|\psi_0\bigr>$ which is not orthogonal neither to $\bigl|\psi_1\bigr>$ nor to $\bigl|\psi_2\bigr>$ (that is $\text{Tr}(\rho_i\rho_0)\neq0$, $i=1,2$), we can associate normalized vectors $\bigl|\varphi_i\bigr>$ with the projectors $\rho_i$ by setting:

\begin{equation}\label{gqm14}
\bigl|\varphi_i\bigr>=\frac{\rho_i\bigl|\psi_0\bigr>}{\sqrt{\text{Tr}(\rho_i\rho_0)}}\qquad i=1,2\;.
\end{equation}

\noindent
Hence, the projector $\rho$ associated with the linear superposition $\bigl|\varphi\bigr>=c_1\bigl|\varphi_1\bigr>+c_2\bigl|\varphi_2\bigr>$ will be given by:

\begin{equation}\label{gqm15}
\begin{split}
\rho=&\bigl|\varphi\bigr>\bigl<\varphi\bigr|=|c_1|^2\rho_1+|c_2|^2\rho_2+\frac{(c_1\bar c_2\rho_1\rho_0\rho_2+\,h.c.)}{\sqrt{\text{Tr}(\rho_1\rho_0\rho_2\rho_0)}}\\
&=\sum_{i,j=1}^2\frac{c_i\bar c_j\rho_i\rho_0\rho_j}{\sqrt{\text{Tr}(\rho_1\rho_0\rho_2\rho_0)}}\;,
\end{split}
\end{equation}

\noindent
i.e., it is now written entirely in terms of rank-one projectors. Thus, a superposition of rank-one projectors yielding another rank-one projector is possible but it requires the arbitrary choice of the fiducial projector $\rho_0=\bigl|\psi_0\bigr>\bigl<\psi_0\bigr|$.\\To sum up, we argued that the most natural setting for Quantum Mechanics seems to be the projective Hilbert space $\mathcal{R}(\mathcal{H})\cong\mathbb{C}P(\mathcal{H})\cong D^1(\mathcal{H})$ rather than the Hilbert space $\mathcal{H}$ itself. In this spirit, we sketched how the main building blocks of Quantum Mechanics can be rephrased in terms of the elements of the projective Hilbert space. As we will see in the next sections, such a space is not a linear space but it carries a (differential) manifold structure. In this context linear transformations leave the place at the more general notion of diffeomorphisms and the linear structure can be recovered only locally at the level of the tangent space. This implies that all the notions established at the level of the Hilbert space $\mathcal{H}$ which depend on the existing linear structure on $\mathcal H$ should be replaced by tensorial objects which as such maintain their meaning under general transformations and not just linear ones. The tensorial character can be naturally encoded by reformulating Quantum Mechanics with the language of the differential geometry and this will be the topic covered in the rest of the chapter.

\section{From Hilbert Spaces to Hilbert Manifolds: Tensorial Structures on the Space of the State Vectors}

In order to implement the differential-geometric point of view we will consider the (finite-dimensional) Hilbert space of quantum states $\mathcal H$ as a real differential manifold\footnote{Being a real manifold, it carries the usual differential calculus.} and hence the complex structure of the standard Hilbert space will be regarded as an additional structure on that manifold. Specifically, starting with the complex Hilbert space $\mathcal H$, we consider its ``realification'':

\begin{equation}\label{gqm16}
\mathcal{H}_{\mathbb{R}}:=\mathbb{R}e(\mathcal{H})\oplus\mathbb{I}m(\mathcal{H})\;,
\end{equation}

\noindent
such that, if $\bigr\{\bigl|e_k\bigr>\bigl\}_{k=1,\dots,n}$ is a (orthonormal) basis for $\mathcal H$, that is $\mathcal H\cong\mathbb{C}^n$, then $\bigl\{\bigl|e_k\bigr>,i\bigl|e_k\bigr>\bigr\}_{k=1,\dots,n}$ is a basis for $\mathcal{H}_\mathbb{R}\cong\mathbb{R}^{2n}$. Once a basis for $\mathcal{H}$ is chosen, we can introduce complex coordinate functions $\{c^k\}_{k=1,\dots,n}$ by setting:

\begin{equation}\label{gqm17}
\bigl<e_k\bigl|\psi\bigr>=c^k(\psi)\qquad\forall\;\bigr|\psi\bigl>\in\mathcal{H}
\end{equation}

\noindent
which correspond to the real coordinates $\{x^k,y^k\}_{k=1,\dots,n}$

\begin{equation}\label{gqm18}
c^k(\psi)=x^k(\psi)+iy^k(\psi)\qquad\forall\;\bigr|\psi\bigl>\in\mathcal{H}
\end{equation}

\noindent
on $\mathcal H_\mathbb R$, i.e., the corresponding vector in $\mathcal H_\mathbb R$ is represented by the $2n$ real coordinates $(x^1,\dots,x^n,y^1,\dots,y^n)\equiv(x,y)$. By considering the dual basis $\bigr\{\bigl<e_k\bigr|\bigl\}_{k=1,\dots,n}$ we have that $\bar{c}^k(\psi)=:\bigl<\psi\bigl|e_k\bigr>$ ($k=1,\dots,n$) are the complex coordinate functions on the dual space $\mathcal{H}^*$ and by using the inner product we can identify $\mathcal H$ and $\mathcal{H}^*$\footnote{We recall that in finite dimensional cases any vector space is isomorphic (in a basis dependent way) with its dual space.}. By taking now the Hermitian structure

\begin{equation}\label{gqm19}
h(\cdot,\cdot)=\bigr<\cdot\big{|}\cdot\bigl>\,:\,\mathcal H\times\mathcal H\longrightarrow\mathbb C
\end{equation}

\noindent
and separating its real and imaginary parts

\begin{equation}\label{gqm20}
h(\psi,\phi)=\bigr<\psi\bigl|\phi\bigl>=g(\psi,\phi)+i\omega(\psi,\phi)\equiv\mathbb{R}e\,h(\psi,\phi)+i\,\mathbb{I}m\,h(\psi,\phi)\;,
\end{equation}

\noindent
we have that because of the Hermitian structure being a positive-definite, nondegenerate, sesquilinear form (that is linear in the second factor and antilinear in the first one), then $g$ is symmetric, positive and nondegenerate while $\omega$ is antisymmetric and nondegenerate. The explicit expressions in real coordinates on $\mathcal H_\mathbb R\cong\mathbb{R}^{2n}$ are given by:

\begin{equation}\label{gqm21}
\begin{split}
h(\psi,\phi)&=\bigr<\psi\bigl|\phi\bigl>=\sum_{k=1}^n\bigr<\psi\bigl|e_k\bigl>\bigr<e_k\bigl|\phi\bigl>\\
&=\sum_{k=1}^n\bar{c}^k(\psi)c^k(\phi)\\
&=\sum_{k=1}^n(x^k-iy^k)(x'^k+iy'^k)\;,
\end{split}
\end{equation}

\noindent
from which we get:

\begin{equation}\label{gqm22}
g(\psi,\phi)=x\cdot x'+y\cdot y'
\end{equation}

\noindent
and

\begin{equation}\label{gqm23}
\omega(\psi,\phi)=x\cdot y'-y\cdot x'\;.
\end{equation}

\noindent
To turn these entities into tensors we consider $\mathcal H_\mathbb R$ together with its tangent bundle $T\mathcal H_\mathbb R\cong\mathcal H_\mathbb R\times\mathcal H_\mathbb R$. Points in $\mathcal H_\mathbb R$ will be thought of as elements in the first factor while the elements in the second factor will be thought of as tangent vectors at that given point, e.g., the couple $(\phi,\psi)\in\mathcal H_\mathbb R\times\mathcal H_\mathbb R\cong T\mathcal H_\mathbb R$ means that $\bigr|\psi\bigl>$ is a tangent vector at the point $\bigr|\phi\bigl>\in\mathcal H_\mathbb R$, i.e., $\psi\in T_\phi\mathcal H_\mathbb R\cong \mathcal H_\mathbb R$. We can therefore associate with every $\psi\in\mathcal H_\mathbb R$ a vector field

\begin{equation}\label{gqm24}
X_\psi\;:\;\mathcal H_\mathbb R\longrightarrow T\mathcal H_\mathbb R\cong\mathcal H_\mathbb R\times\mathcal H_\mathbb R
\end{equation}

\noindent
by setting

\begin{equation}\label{gqm25}
X_\psi=:(\phi,\psi)\qquad\forall\phi\in\mathcal H_\mathbb R\;.
\end{equation}

\noindent
Eq. (\ref{gqm21}) can be now regarded as the evaluation on tangent vectors at the point $\phi$. Then, $g$ and $\omega$ can be promoted to (0,2)-tensor fields by defining:

\begin{equation}\label{gqm26}
g(\phi)(X_\psi,X_{\psi'})=:g(\psi,\psi')\;,
\end{equation}
\begin{equation}\label{gqm27}
\omega(\phi)(X_\psi,X_{\psi'})=:\omega(\psi,\psi')\;,
\end{equation}

\noindent
which provide us with a Riemannian metric tensor and symplectic structure, respectively. Similarly, by identifying $\mathcal H_\mathbb R$ and $\mathcal H_\mathbb R^*$ by means of the inner product, 1-forms $\alpha: \mathcal H_\mathbb R\longrightarrow T^*\mathcal H_\mathbb R\cong\mathcal H_\mathbb R\times\mathcal H_\mathbb R^*\cong\mathcal H_\mathbb R\times\mathcal H_\mathbb R$ can be thought of as the elements in the second factor. Let now

\begin{equation}\label{gqm28}
\bigl|d\psi\bigr>=:d\bigl|\psi\bigl>=dc^k(\psi)\bigr|e_k\bigl>\qquad(\text{sum over k})
\end{equation}

\noindent
be a vector-state-valued differential form, i.e., a section of the bundle

\begin{equation}\label{gqm29}
\xymatrix{
T^*\mathcal{H}_\mathbb{R}\otimes\mathcal{H}_\mathbb{R}\ar@{<--}[d]^{\ket{d\psi}}\\
\mathcal{H}_\mathbb{R}
}
\end{equation}

\noindent
where we assume that an orthonormal basis $\bigl\{\bigl|e_k\bigl>\bigr\}$ has been selected once and it does not depend on the base point\footnote{To deal with a changing basis (moving frame) we should introduce a connection and the related machinery of covariant differential calculus.}. Then, the Hermitian inner product

\begin{equation}\label{gqm30}
\bigl<\psi\big|\psi\bigr>=\bigl<e_k\big|e_k\bigl>\bar{c}^kc^\ell=\delta_{k\ell}\bar{c}^kc^\ell
\end{equation}

\noindent
\\is promoted to the Hermitian (0,2)-tensor field

\begin{equation}\label{gqm31}
\begin{split}
h&=\bigl<d\psi\otimes d\psi\bigr>=\bigl<e_k\big|e_k\bigl>d\bar{c}^k\otimes dc^\ell=\delta_{k\ell}\,d\bar{c}^k\otimes dc^\ell\\
&=\underbrace{\delta_{k\ell}(dx^k\otimes dx^\ell+dy^k\otimes dy^\ell)}_{\text{euclidean metric } g}+i\,\underbrace{\delta_{k\ell}(dx^k\otimes dy^\ell-dy^k\otimes dx^\ell)}_{\text{symplectic structure } \omega}\;.
\end{split}
\end{equation}

\noindent
Similarly, on $\mathcal H^*$ we have the following Hermitian (2,0)-tensor field:

\begin{equation}\label{gqm32}
\begin{split}
H&=\biggl<\frac{\partial}{\partial\psi}\otimes\frac{\partial}{\partial\psi}\biggr>=\delta_{k\ell}\,\frac{\partial}{\partial\bar{c}_k}\otimes\frac{\partial}{\partial c_\ell}\\
&=\delta_{k\ell}\biggl(\frac{\partial}{\partial x_k}+i\frac{\partial}{\partial y_k}\biggr)\otimes\biggl(\frac{\partial}{\partial x_\ell}-i\frac{\partial}{\partial y_\ell}\biggr)\\
&=\underbrace{\delta_{k\ell}\biggl(\frac{\partial}{\partial x_k}\otimes\frac{\partial}{\partial x_\ell}+\frac{\partial}{\partial y_k}\otimes\frac{\partial}{\partial y_\ell}\biggr)}_{G}+i\underbrace{\delta_{k\ell}\biggl(\frac{\partial}{\partial y_k}\otimes\frac{\partial}{\partial x_\ell}-\frac{\partial}{\partial x_k}\otimes\frac{\partial}{\partial y_\ell}\biggr)}_{\Omega}
\end{split}
\end{equation}

\noindent
where the (2,0) contravariant tensors $G$ and $\Omega$ mapping $T^*\mathcal H_\mathbb R$ to $T\mathcal H_\mathbb R$ are the inverse of the nondegenerate (0,2)-tensors $g$ and $\omega$, respectively. These tensors allow us to define a Jordan bracket $(\cdot,\cdot)$ and a Poisson bracket $\{\cdot,\cdot\}$ on real (smooth) functions on $\mathcal H_\mathbb R$ given by:\\

\begin{equation}\label{gqm33}
(f,f')=:G(df,df')=\frac{\partial f}{\partial x^k}\frac{\partial f'}{\partial x_k}+\frac{\partial f}{\partial y^k}\frac{\partial f'}{\partial y_k}\;,
\end{equation}

\noindent
and

\begin{equation}\label{gqm34}
\{f,f'\}=:\Omega(df,df')=\frac{\partial f}{\partial y^k}\frac{\partial f'}{\partial x_k}-\frac{\partial f}{\partial x^k}\frac{\partial f'}{\partial y_k}\;,
\end{equation}

\noindent
\\for every $f,f'\in\mathcal{F}(\mathcal H_\mathbb R)$. The Riemannian metric tensor and the symplectic 2-form (or equivalently their contravariant counterparts) are related by means of the following (1,1)-tensor field:

\begin{equation}\label{gqm35}
J(\phi)(X_\psi)=:(\phi,J\psi)\;,
\end{equation}

\noindent
where

\begin{equation}\label{gqm36}
J\psi=i\psi\qquad i.e. \qquad J(x,y)=(-y,x)\;.
\end{equation}

\noindent
\\The coordinate expression of such tensor field is:

\begin{equation}\label{gqm37}
J=\delta_{k\ell}\biggl(dx^k\otimes\frac{\partial}{\partial y_\ell}-dy^k\otimes\frac{\partial}{\partial x_\ell}\biggr)\;,
\end{equation}

\noindent
\\and, for every $\phi,\psi$, we find that the following properties hold:

\begin{subequations}
\begin{equation}\label{gqm38a}
J^2=-\mathds 1
\end{equation}
\begin{equation}\label{gqm38b}
g(\phi,\psi)=\omega(J\phi,\psi)
\end{equation}
\begin{equation}\label{gqm38c}
g(J\phi,J\psi)=g(\phi,\psi)\quad,\quad g(J\phi,\psi)+g(\phi,J\psi)=0
\end{equation}
\begin{equation}\label{gqm38d}
\omega(J\phi,J\psi)=\omega(\phi,\psi)\quad,\quad \omega(J\phi,\psi)+\omega(\phi,J\psi)=0
\end{equation}
\end{subequations}

\noindent
\\Therefore, to summarize the whole section, we have replaced the original Hilbert space $\mathcal H$ with the Hilbert manifold $\mathcal H_\mathbb R$, i.e., an even-dimensional real manifold on which we have the following tensor fields:

\begin{equation}\label{gqm39}
\textbf{covariant form}\qquad:\qquad\begin{cases}
g=\delta_{k\ell}(dx^k\otimes dx^\ell+dy^k\otimes dy^\ell)\\
\omega=\delta_{k\ell}(dx^k\otimes dy^\ell-dy^k\otimes dx^\ell)
\end{cases}
\end{equation}

\begin{equation}\label{gqm40}
\textbf{contravariant form}\quad:\quad\begin{cases}
G=\delta_{k\ell}\Bigl(\frac{\partial}{\partial x_k}\otimes\frac{\partial}{\partial x_\ell}+\frac{\partial}{\partial y_k}\otimes\frac{\partial}{\partial y_\ell}\Bigr)\\
\Omega=\delta_{k\ell}\Bigl(\frac{\partial}{\partial y_k}\otimes\frac{\partial}{\partial x_\ell}-\frac{\partial}{\partial x_k}\otimes\frac{\partial}{\partial y_\ell}\Bigr)
\end{cases}
\end{equation}

\noindent
\\together with a (1,1)-tensor field

\begin{equation}\label{gqm41}
J=\delta_{k\ell}\biggl(dx^k\otimes\frac{\partial}{\partial y_\ell}-dy^k\otimes\frac{\partial}{\partial x_\ell}\biggr)
\end{equation}

\noindent
\\which plays the role of a complex structure (Eq.(\ref{gqm38a})). In other words, according to the compatibility conditions (\ref{gqm38b}-\ref{gqm38d}), $\mathcal H_\mathbb R$ becomes in this way a \textit{K\"ahler manifold}.

\section{Tensorial Structures on the Space of Rays}

In the previous sections we said that the probabilistic interpretation of Quantum Mechanics leads us to consider the projective Hilbert space (i.e., the space of rays) $\mathcal{R}(\mathcal H)\cong\mathbb{C}P^{n-1}(\mathcal H)$ as the right stage to describe a quantum system. From the geometrical point of view, the projective Hilbert space carries a natural manifold structure with a well-defined notion of distance between two complex rays which is measured by the so-called \textbf{Fubini-Study metric}\footnote{We recall that if $dim_{\mathbb C}\mathcal H=n$, then the projective Hilbert space $\mathcal R(\mathcal H)$ is homeomorphic to the complex Hilbert space $\mathbb{C}P^{n-1}$ (endowed with a Hermitian form). Such a space consists of equivalence classes of vectors $\mathbf{Z}$ such that $\bf{Z}'\sim\bf{Z}$ iff $\bf{Z}'=\lambda\mathbf{Z}$ with $\lambda\in\mathbb{C}_0$, and it is a K\"ahler manifold equipped with the so-called Fubini-Study metric \cite{GQM16}: $$g_{FS}=\frac{d\bar{\bf{Z}}\otimes_S{d\bf{Z}}}{(\bf{Z}\cdot\bar{\bf{Z}})}-\frac{(d\bf{Z}\cdot\bar{\bf{Z}})\otimes_S(\bf{Z}\cdot d\bar{\bf{Z}})}{(\bf{Z}\cdot\bar{\bf{Z}})^2}$$ together with a symplectic form:$$\omega_{FS}=\frac{d\bar{\bf{Z}}\wedge{d\bf{Z}}}{(\bf{Z}\cdot\bar{\bf{Z}})}-\frac{(d\bf{Z}\cdot\bar{\bf{Z}})\wedge(\bf{Z}\cdot d\bar{\bf{Z}})}{(\bf{Z}\cdot\bar{\bf{Z}})^2}.$$For istance, in the case $n=2$, i.e., $\mathbb{C}P^{1}\cong S^2$, if $Z=x+iy$ is the standard affine coordinate chart on the Riemann sphere and $x=r\cos\theta,y=r\sin\theta$ are polar coordinates on $\mathbb C$, then we have $g_{FS}=\frac{\mathbb{R}e(dZ\otimes d\bar{Z})}{(1+|Z|^2)^2}=\frac{dx^2+dy^2}{(1+r^2)^2}=\frac{1}{4}(d\phi^2+\sin^2\phi d\theta^2)$ which is the round metric on the unit 2-sphere.}. Therefore, the physically relevant distance between two quantum states should be given in terms of the Fubini-Study metric rather than in terms of a Hermitian scalar product. In what follows we will then focus on tensorial structures defined on the projective Hilbert space of complex rays $\mathcal{R}(\mathcal H)$. However, to make formulas explicit and easily computable, it will be convenient to translate the geometric structures on $\mathcal{R}(\mathcal H)$ at the level of the Hilbert space (which we have now turned into a Hilbert manifold). In other words, we are interested in those tensors in $\mathcal H$ defined as the pull-back of tensors defined on $\mathcal{R}(\mathcal H)$.\\Let us start with covariant tensor fields of order zero, i.e., functions. Given a Hermitian operator $A\in\mathfrak{u}^*(\mathcal{H})$ regarded as an element of the dual of the Lie algebra $\mathfrak{u}(\mathcal H)$ of the unitary group $U(\mathcal H)$ (which preserves the triple $(g,\omega,J)$), we recall from Sec. 3.2 that it is possible to associate with $A$ a real symmetric (quadratic) function on $\mathcal H$ given by $f_A(\psi)=\bigl<\psi\big|A\big|\psi\bigr>=\bigl<\psi\big|\psi\bigr>e_A(\psi)$ whose critical points and critical values correspond to eigenvectors and eigenvalues of the operator $A$, respectively. In the previous section we have seen that the punctured Hilbert space $\mathcal H_0=\mathcal H-\{0\}$ is a K\"ahler (and hence symplectic) manifold and thus the map

\begin{equation}\label{gqm42}
\mu\;:\;\mathcal H_0\longrightarrow\mathfrak{u}^*(\mathcal H)\qquad\text{by}\qquad\big|\psi\bigr>\longmapsto\rho_\psi=\frac{\big|\psi\bigr>\bigl<\psi\big|}{\bigl<\psi\big|\psi\bigr>}
\end{equation}

\noindent
\\is a momentum map. By virtue of this map, we notice that $e_A(\psi)$ comes to be the pull-back of a function from $D^1(\mathcal{H})\cong\mathcal R(\mathcal H)$ to $\mathcal H_0$ identified by the relation:

\begin{equation}\label{gqm43}
e_A(\psi)=\rho_\psi(A):=\text{Tr}(\rho_\psi A)\qquad,\qquad\rho_\psi\in D^1(\mathcal H)\;.
\end{equation}

\noindent
\\According to the commutative diagram

\begin{equation}\label{gqm44}
\begin{CD}
\mathcal{H}_0@>\mu>>\mathfrak{u}^*(\mathcal{H})\\
@V\pi VV               @AAi A\\
\mathcal{R}(\mathcal H)@>>\cong>D^1(\mathcal H)
\end{CD}
\end{equation}

\noindent
where $\pi$ denotes the bundle projection in (\ref{gqm3}), the map $\mu$ provides a tool for pulling-back the covariant structures defined on $D^1(\mathcal H)\cong\mathcal{R}(\mathcal H)$ to the Hilbert space $\mathcal H_0=\mathcal{H}-\{0\}$. Explicitly, for a given Hermitian operator $A$, we define the (0,2)-tensor field

\begin{equation}\label{gqm45}
\text{Tr}(A dA\otimes dA)\;,
\end{equation}

\noindent
where the operator-valued differential $dA$ is defined w.r.t. a real parametrization of $\mathfrak{u}^*(\mathcal{H})$, i.e., it is the Hermitian matrix whose elements are the differentials of the elements of (the representation of) $A$. The differential calculus and the covariant structure on the submanifold $\mathcal R(\mathcal H)\cong D^1(\mathcal H)\subset\mathfrak{u}^*(\mathcal H)$ can be inherited from the ambient space $\mathfrak{u}^*(\mathcal H)$. Therefore we get the following (0,2)-tensor field on $D^1(\mathcal H)\cong\mathcal R(\mathcal H)$:

\begin{equation}\label{gqm46}
\text{Tr}(\rho_\psi d\rho_\psi\otimes d\rho_\psi)\;.
\end{equation}

\noindent
By using now the Leibniz rule

\begin{equation}\label{gqm47}
d\Biggl(\frac{\big|\psi\bigr>\bigl<\psi\big|}{\bigl<\psi\big|\psi\bigr>}\Biggr)=\frac{\big|d\psi\bigr>\bigl<\psi\big|+\big|\psi\bigr>\bigl<d\psi\big|}{\bigl<\psi\big|\psi\bigr>}-\frac{d\bigl<\psi\big|\psi\bigr>}{\bigl<\psi\big|\psi\bigr>^2}\;,
\end{equation}

\noindent
together with the fact that

\begin{equation}\label{gqm48}
0=d\bigl<\psi\big|\psi\bigr>\qquad\Longrightarrow\qquad\bigl<d\psi\big|\psi\bigr>=-\bigl<\psi\big|d\psi\bigr>\;,
\end{equation}

\noindent
a straightforward calculation gives us the following $\mu$-induced pull-back tensor field on $\mathcal H_0$:

\begin{equation}\label{gqm49}
\text{Tr}(\rho_\psi d\rho_\psi\otimes d\rho_\psi)=\frac{\big<d\psi\otimes d\psi\big>}{\bigl<\psi\big|\psi\bigr>}-\frac{\big<\psi\big| d\psi\big>}{\bigl<\psi\big|\psi\bigr>}\otimes\frac{\big<d\psi\big| \psi\big>}{\bigl<\psi\big|\psi\bigr>}:=h_{\mathcal H_0}\;.
\end{equation}

\noindent
By introducing complex coordinates $c^k(\psi)=\bigl<e_k\big|\psi\bigr>$, Eq. (\ref{gqm49}) gives the degenerate covariant tensor field:

\begin{equation}\label{gqm50}
h_{\mathcal H_0}=\frac{d\bar{c}^k\otimes dc^k}{\sum_k|c^k|^2}-\frac{c^k d\bar{c}^k\otimes\bar{c}^k dc^k}{\bigl(\sum_k|c^k|^2\bigr)^2}
\end{equation}

\noindent
which is identified as the pull-back of the Fubini-Study tensor field from the space of rays $\mathcal R(\mathcal H)\cong\mathbb{C}P^{n-1}$ to $\mathcal H_0\cong\mathbb{C}_0^n$. The above covariant tensor field decomposes into a real symmetric and an imaginary antysymmetric part:

\begin{equation}\label{gqm51}
h_{\mathcal H_0}=g_{\mathcal H_0}+i\,\omega_{\mathcal H_0}
\end{equation}

\noindent
\\which are the pull-back of a Riemannian and a symplectic structures from the complex projective space to $\mathcal H_0$, respectively. Such a decomposition may either be induced by cartesian coordinates $c^k=x^k+iy^k$ or also by polar coordinates, that is $c^k=p^ke^{iW^k}$. The latter case will be considered later in this chapter when we discuss the connection with the Fisher Information metric.\\Let us stress again that the above covariant structures derived as pull-back tensor fields are degenerate. This implies that they are not invertible and we cannot associate to each of them a corresponding contravariat structure. However, we can define contravariant structures on $\mathcal H_0$ which are projectable on $\mathcal R(\mathcal H)$. Indeed, at the infinitesimal level, the action of the group $\mathbb{C}_0$ is generated by the vector fields:

\begin{equation}\label{gqm52}
\Delta=x^k\frac{\partial}{\partial x^k}+y^k\frac{\partial}{\partial y^k}\qquad(\mathbb{R}_0^+\text{-dilatations})\;,
\end{equation}

\begin{equation}\label{gqm53}
\Gamma=J(\Delta)=y^k\frac{\partial}{\partial x^k}-x^k\frac{\partial}{\partial y^k}\qquad(U(1)\text{-phase transformations})\;.
\end{equation}

\noindent
\\These vector fields commute and they generate an involutive distribution. Referring to (\ref{gqm1}), we can then generate the ray space $\mathcal R(\mathcal H)$ in a Hermitian-structure-independent way by going to the quotient w.r.t. the foliation of $\mathcal H_\mathbb R-\{0\}$ associated with this distribution. Contravariant tensor fields $\mathcal T$ on the Hilbert space will be projectable onto $\mathcal R(\mathcal H)$ if $L_\Delta\mathcal T=0$ and $L_\Gamma\mathcal T=0$, i.e., if they are homogeneous of degree zero and invariant under multiplication of vectors by a phase. The expectation value functions $e_A(\psi)=\text{Tr}(\rho_\psi A)$ hence pass to the quotient. As for the tensors, the complex structure $J$ given in Eq. (\ref{gqm41}) is projectable as it is phase invariant and homogeneous of degree zero. Instead, the tensors $G$ and $\Omega$ in Eq. (\ref{gqm40}) are phase invariant but they are homogeneous of degree -2 and so they will not be projectable. To turn them into projectable objects, we need to modify them by a conformal factor \cite{GQM17}, i.e. :

\begin{equation}\label{gqm54}
\tilde G(\psi)=\bigl<\psi\big|\psi\bigr>G-(\Delta\otimes\Delta+\Gamma\otimes\Gamma)\;,
\end{equation}

\begin{equation}\label{gqm55}
\tilde\Omega(\psi)=\bigl<\psi\big|\psi\bigr>\Omega-(\Delta\otimes\Gamma-\Gamma\otimes\Delta)\;,
\end{equation}

\noindent
\\whose expressions in coordinates are given by:

\begin{equation}\label{gqm56}
\begin{split}
\tilde G(z,\bar{z})=&\Biggl[(\bar{z}^jz_j)\biggl(\frac{\partial}{\partial x^k}\otimes\frac{\partial}{\partial x_k}+\frac{\partial}{\partial y^k}\otimes\frac{\partial}{\partial y_k}\biggr)\Biggr]+\\
&-\Biggl[\biggl(x^k\frac{\partial}{\partial x^k}+y^k\frac{\partial}{\partial y^k}\biggr)\otimes\biggl(x^\ell\frac{\partial}{\partial x^\ell}+y^\ell\frac{\partial}{\partial y^\ell}\biggr)+\\
&+\biggl(y^k\frac{\partial}{\partial x^k}-x^k\frac{\partial}{\partial y^k}\biggr)\otimes\biggl(y^\ell\frac{\partial}{\partial x^\ell}-x^\ell\frac{\partial}{\partial y^\ell}\biggr)\Biggr]\;,
\end{split}
\end{equation}

\begin{equation}\label{gqm57}
\begin{split}
\tilde\Omega(z,\bar{z})=&\Biggl[(\bar{z}^jz_j)\biggl(\frac{\partial}{\partial y^k}\otimes\frac{\partial}{\partial x_k}-\frac{\partial}{\partial x^k}\otimes\frac{\partial}{\partial y_k}\biggr)\Biggl]+\\
&-\Biggl[\biggl(y^k\frac{\partial}{\partial y^k}+x^k\frac{\partial}{\partial x^k}\biggr)\otimes\biggl(y^\ell\frac{\partial}{\partial x^\ell}-x^\ell\frac{\partial}{\partial y^\ell}\biggr)+\\
&-\biggl(y^\ell\frac{\partial}{\partial x^\ell}-x^\ell\frac{\partial}{\partial y^\ell}\biggr)\otimes\biggl(y^k\frac{\partial}{\partial y^k}+x^k\frac{\partial}{\partial x^k}\biggr)\Biggr]\;.
\end{split}
\end{equation}

\noindent
Such tensor fields define respectively a symmetric and a Poisson bracket on pull-back functions from the manifold $\mathcal R(\mathcal H)$. On the other hand, for covariant tensor fields $\tau$ to be the pull-back of tensor fields on $\mathcal R(\mathcal H)$ the conditions $L_\Delta\tau=0, L_\Gamma\tau=0$ together with $i_\Delta\tau=0, i_\Gamma\tau=0$ must hold. This was actually the case for the tensors $g$ and $\omega$ and an explicit computation of the pull-back was provided before through the momentum map $\mu$.

\section{Pull-back Tensor Fields on (Sub)manifolds of Quantum States}

One interesting aspect of the construction discussed before which will turn very useful for the description of entangled systems is the possibility to induce covariant tensorial structures on a given submanifold of quantum states by means of a pull-back procedure. Indeed, for a given embedding of a manifold $\mathcal M$ of quantum states

\begin{equation}\label{gqm58}
i_\mathcal M \;:\;\mathcal M \hookrightarrow\mathcal H
\end{equation}

\noindent
the induced covariant rank-2 Hermitian tensor on $\mathcal M$ is defined via the pull-back $i_\mathcal M^*$ of the Hermitian tensor $\bigl<d\psi\otimes d\psi\bigr>$ from $\mathcal H$ to $\mathcal M$. The real and imaginary parts of this tensor field will define a covariant Riemannian metric tensor $g_\mathcal M$ and a symplectic structure $\omega_\mathcal M$ respectively given by:

\begin{equation}\label{gqm59}
i_\mathcal M^*\bigl(\bigl<d\psi\otimes d\psi\bigr>\bigr)=\underbrace{\mathbb{R}e\Bigl(i_\mathcal M^*\bigl(\bigl<d\psi\otimes d\psi\bigr>\bigr)\Bigr)}_{g_\mathcal M}+i\,\underbrace{\mathbb{I}m\Bigl(i_\mathcal M^*\bigl(\bigl<d\psi\otimes d\psi\bigr>\bigr)\Bigr)}_{\omega_\mathcal M}\;.
\end{equation}

\noindent
In what follows we are interested in the case in which the manifold $\mathcal M$ admits a Lie group structure. When this is the case, we may identify submanifolds of states with the orbits generated by the action of the group on the Hilbert space via a unitary representation. Let us therefore discuss this construction in detail. Let $\mathbb G$ be a Lie group acting on a vector space $V$, which for us will be the Hilbert space $\mathcal H$, and let $\phi$ be its action on $\mathcal H$, i.e., a map:

\begin{equation}\label{gqm60}
\phi\;:\;\mathbb G\times\mathcal H \longrightarrow \mathcal H\;,
\end{equation}

\noindent
\\or equivalently a Lie group homomorphism:

\begin{equation}\label{gqm61}
\phi_g\;:\;\mathbb G\longrightarrow Aut(\mathcal H)\;.
\end{equation}

\noindent
\\By choosing now a normalized fiducial state

\begin{equation}\label{gqm62}
\ket{0}\in\mathcal H\;,
\end{equation}

\noindent
\\then the main idea is to consider the map

\begin{equation}\label{gqm63}
\phi_0\equiv\phi_{\ket{0}}\;:\;\mathbb G\longrightarrow\mathcal H
\end{equation}

\noindent
\\as an embedding of $\mathbb G$ into $\mathcal H$. In this way we can pull-back to $\mathbb G$ the algebra of functions $\phi_0^*(\mathcal F(\mathcal H))\subset\mathcal F(\mathbb G)$ and, according to the relation $d\phi_0^*=\phi_0^*d$ between the exterior differential on the two spaces, we can also pull-back all the algebra of exterior forms. This enables us to construct covariant tensors on $\mathbb G$ out of covariant tensors on $\mathcal H$. Indeed, by considering a unitary representation $U(g)$ ($g\in\mathbb G$) of the Lie group $\mathbb G$, the orbits of the action of $\mathbb G$ on $\mathcal H$ with $\bigr|0\bigl>$ as starting point are identified with the quotient space

\begin{equation}\label{gqm64}
\mathcal O\cong\mathbb G\bigl/\mathbb G_0=\bigl\{\bigl| g\bigr> \bigr\}_{g\in\mathbb G}\Bigl/\sim\qquad\text{with}\qquad \bigl| g\bigr>=U(g)\bigl| 0\bigr>
\end{equation}

\noindent
where $\mathbb G_0$ is the isotropy group of the fiducial state $\bigl| 0\bigr>$. In other words, the homogeneous space $\mathbb G/\mathbb G_0$ admits a smooth embedding via the unitary action of the Lie group $\mathbb G$ as orbit manifold $\mathcal O$ in the Hilbert space. Hence, all the information about the Hermitian tensor on the submanifold of state (orbit) is embodied into the tensor defined on the Lie group as schematically summarized in the following commutative diagram:

\begin{equation}\label{gqm65}
\begin{CD}
\mathbb{G}@>\phi_0>>\;\mathcal H\;\\
@V\pi_0 VV               @AAi_\mathcal O A\\
\mathbb G\bigl/\mathbb G_0@>>\cong>\;\mathcal O\;
\end{CD}
\end{equation}

\noindent
where $\pi_0$ is the canonical projection of $\mathbb G$ onto the quotient $\mathbb G/\mathbb G_0$ and $i_\mathcal O$ is the inclusion map of the orbit $\mathcal O$ on the Hilbert space $\mathcal H$.\\To compute explicitly the pull-back of the Hermitian tensor field $\bigr<d\psi\otimes d\psi\bigl>$ from $\mathcal H$ to $\mathbb G$ we remark that the unitary representation of the Lie group $\mathbb G$ defines a representation $R$ of its Lie algebra $\mathfrak g$ which is defined by means of the tangent map of the representation map \cite{GQM18}. Indeed, by considering the tangent bundle construction, the action $\phi$ of $\mathbb G$ on $\mathcal H$ induces an action $T\phi$ of $T\mathbb G$ on $T\mathcal H\cong\mathcal H\times\mathcal H$

\begin{equation}\label{gqm66}
T\phi\;:\;T\mathbb G\times T\mathcal H \longrightarrow T\mathcal H\;,
\end{equation}

\noindent
and, being $T\mathbb G\cong\mathbb G \times \mathfrak g$, the tangent map $T\phi$ requires the existence of a representation on $\mathcal H$ of the Lie algebra $\mathfrak g$. Therefore, if $X^1\,\dots, X^n$ denote a basis of left-invariant vector fields and $\theta_1,\dots,\theta_n$ the corresponding dual basis of 1-forms, i.e., such that $\theta_k(X^j)=\delta_k^j$, then we have:

\begin{equation}\label{gqm67}
R\;:\; T_e\mathbb G\longrightarrow T_eU(\mathcal H)\qquad;\qquad R([X^j,X^k])=[R(X^j),R(X^k)]\;.
\end{equation}

\noindent
Now the left-invariant vector fields X's are the infinitesimal generators of the right action of $\mathbb G$ on itself and, then, by considering the corresponding one-paramenter subgroup

\begin{equation}\label{gqm68}
U(t)=e^{iR(X^k)t}\;,
\end{equation}

\noindent
\\the Hilbert-space-valued 1-forms $\bigl|dg\bigr>$ are given by:

\begin{equation}\label{gqm69}
\begin{split}
\bigl|dg\bigr>&=d\bigl|g\bigr>=dU(g)\bigl|0\bigr>\\
&=dU(g)\,U^{-1}(g)\bigl|g\bigr>\\
&=iR(X^k)\theta_k\bigl|g\bigr>
\end{split}
\end{equation}

\noindent
where in the last line we used the fact that $dU(g)=iR(X^k)\theta_kU(g)$. More easily, the result (\ref{gqm69}) can be immediately derived recalling that $dU(g)U^{-1}(g)$ is a right-invariant 1-form and as such it can be written as $iR(X^k)\theta_k$. Similarly:

\begin{equation}\label{gqm70}
\bigl<dg\bigr|=\bigl<g\bigr|\bigl(dU(g)\,U^{-1}(g)\bigr)^\dagger=\bigl<g\bigr|\bigl(-iR(X^j)\theta_j\bigr)\;.
\end{equation}

\noindent
Hence, the pull-back of the tensor $\bigr<d\psi\otimes d\psi\bigl>$ to the Lie group $\mathbb G$ is given by:

\begin{equation}\label{gqm71}
\begin{split}
\bigr<dg\otimes dg\bigl>&=\bigl<g\bigr|\bigl(dU(g)\,U^{-1}(g)\bigr)^\dagger\otimes\bigl(dU(g)\,U^{-1}(g)\bigr)\bigr|g\bigl>\\
&=\bigl<g\bigr|R(X^j)R(X^k)\bigr|g\bigl>\theta_j\otimes\theta_k\;.
\end{split}
\end{equation}

\noindent
By using the decomposition

\begin{equation}\label{gqm72}
\begin{split}
\theta_j\otimes\theta_k&=\frac{1}{2}(\theta_j\otimes\theta_k+\theta_k\otimes\theta_j)+\frac{1}{2}(\theta_j\otimes\theta_k-\theta_k\otimes\theta_j)\\
&=\frac{1}{2}\theta_j\underset{S}{\otimes}\theta_k+\frac{1}{2}\theta_j\wedge\theta_k
\end{split}
\end{equation}

\noindent
we can exhibit the real and imaginary part of (\ref{gqm71}) as:

\begin{equation}\label{gqm73}
\begin{split}
\mathbb{R}e\bigl(\bigr<dg\otimes dg\bigl>\bigr)&=\frac{1}{2}\bigl<g\bigr|R(X^j)R(X^k)+R(X^k)R(X^j)\bigr|g\bigl>\theta_j\underset{S}{\otimes}\theta_k\\
&=\frac{1}{2}\bigl<g\bigr|[R(X^j),R(X^k)]_+\bigr|g\bigl>\theta_j\underset{S}{\otimes}\theta_k\;,
\end{split}
\end{equation}

\noindent
and

\begin{equation}\label{gqm74}
\begin{split}
\mathbb{I}m\bigl(\bigr<dg\otimes dg\bigl>\bigr)&=\frac{1}{2}\bigl<g\bigr|R(X^j)R(X^k)-R(X^k)R(X^j)\bigr|g\bigl>\theta_j\wedge\theta_k\\
&=\frac{1}{2}\bigl<g\bigr|[R(X^j),R(X^k)]_-\bigr|g\bigl>\theta_j\wedge\theta_k\;,
\end{split}
\end{equation}

\noindent
where $[\cdot,\cdot]_{\pm}$ denote the anticommutator and the commutator respectively, and for the imaginary part we have used the fact that the commutator of Hermitian operators is antihermitian. The real part (\ref{gqm73}) defines a metric tensor on the manifold of quantum states obtained from $\bigl|0\bigr>$ by acting with the Lie group $\mathbb G$. The skewsymmetric part (\ref{gqm74}) instead defines a closed 2-form on $\mathbb G/\mathbb G_0$ (i.e., a presymplectic structure) which becomes a symplectic tensor when its kernel is trivial (i.e., the closed 2-form is also non-degenerate).\\However, as we have already stressed during the chapter, we are mainly interested in tensorial structures derived as pull-back tensors from the space of rays instead of the Hilbert space. This means that our starting point is not the tensor $\bigl<d\psi\otimes d\psi\bigr>$ anymore but the K\"ahlerian tensor on $\mathcal R(\mathcal H)$

\begin{equation}\label{gqm75}
\frac{\big<d\psi\otimes d\psi\big>}{\bigl<\psi\big|\psi\bigr>}-\frac{\big<\psi\big| d\psi\big>}{\bigl<\psi\big|\psi\bigr>}\otimes\frac{\big<d\psi\big| \psi\big>}{\bigl<\psi\big|\psi\bigr>}\;.
\end{equation}

\noindent
Therefore, the pull-back tensor on $\mathbb G$ is now given by:

\begin{equation}\label{gqm76}
\frac{\big<dg\otimes dg\big>}{\bigl<g\big|g\bigr>}-\frac{\big<g\big| dg\big>}{\bigl<g\big|g\bigr>}\otimes\frac{\big<dg\big|g\big>}{\bigl<g\big|g\bigr>}\;,
\end{equation}

\noindent
and again, by means of the embedding action $U(g)\ket{0}=\ket g$ associated with a given unitary representation $U$ and a fiducial state $\ket{0}$, an analogous computation gives:

\begin{equation}\label{gqm77}
\Biggl(\frac{\bigl<g\big|R(X^j)R(X^k)\big|g\bigr>}{\bigl<g\big|g\bigr>}-\frac{\bigl<g\big|R(X^j)\big|g\bigr>\bigl<g\big|R(X^k)\big|g\bigr>}{\bigl<g\big|g\bigr>^2}\Biggr)\theta_j\otimes\theta_k\;.
\end{equation}

\noindent
Thus we see that the closed 2-form is not modified (except for the normalization factor $\braket{g|g}$), i.e., we have

\begin{equation}\label{gqm78}
\frac{1}{2}\Biggl(\frac{\bigl<g\big|[R(X^j),R(X^k)]_-\big|g\bigr>}{\bigl<g\big|g\bigr>}\Biggr)\theta_j\wedge\theta_k\;,
\end{equation}

\noindent
while the metric tensor is modified by an additional terms, i.e.:

\begin{equation}\label{gqm79}
\Biggl(\frac{1}{2}\;\frac{\bigl<g\big|[R(X^j),R(X^k)]_+\big|g\bigr>}{\bigl<g\big|g\bigr>}-\frac{\bigl<g\big|R(X^j)\big|g\bigr>\bigl<g\big|R(X^k)\big|g\bigr>}{\bigl<g\big|g\bigr>^2}\Biggr)\theta_j\underset{S}{\otimes}\theta_k\;.
\end{equation}

\noindent
More generally, we may use any density state, i.e., a positive normalized functional $\rho\in\mathfrak{u}^*(\mathcal H)$ and replace the tensor (\ref{gqm77}) with the analogous tensor on the group manifold associated with the density state $\rho$. It is important to notice that the construction of the pull-back tensor depends on the choice of the fiducial state and on the choice of the representation of the Lie group and then ultimately on the choice of the action on which depends the way we imbed the group in the carrier space we are interested in. For istance, we may consider the co-adjoint action on the space of states, say

\begin{equation}\label{gqm83}
\rho=\rho(g)=U(g)\rho_0U^\dagger(g)\;,
\end{equation}

\noindent
and we have

\begin{equation}\label{gqm84}
\begin{split}
d\rho&=dU(g)\rho_0U^\dagger(g)+U(g)\rho_0dU^\dagger(g)\\
&=dU(g)\rho_0U^\dagger(g)-U(g)\rho_0U^\dagger(g)dU(g)U^\dagger(g)\\
&=dU(g)U^\dagger(g)U(g)\rho_0U^\dagger(g)-U(g)\rho_0U^\dagger(g)dU(g)U^\dagger(g)\\
&=[dU(g)U^\dagger(g),\rho]_-\\
&=U(g)[dU(g)U^\dagger(g),\rho_0]_-U^\dagger(g)
\end{split}
\end{equation}

\noindent
where in the second step we have used the relation $dU^\dagger(g)=-U^\dagger(g)dU(g)U^\dagger(g)$ that can be easily derived from $0=d(U^\dagger(g)U(g))$ by using the Leibniz rule. Therefore, the pull-back of the (0,2)-tensor Tr$(\rho d\rho\otimes d\rho)$ on the group manifold will be:

\begin{equation}\label{gqm85}
\begin{split}
\text{Tr}(\rho d\rho\otimes d\rho)&=\text{Tr}\Bigl\{U\rho_0U^\dagger\Bigl(U[iR(X^j),\rho_0]_-U^\dagger U[iR(X^k),\rho_0]_-U^\dagger\Bigr)\Bigr\}\theta_j\otimes\theta_k\\
&=-\text{Tr}\Bigl(\rho_0[R(X^j),\rho_0]_-[R(X^k),\rho_0]_-\Bigr)\theta_j\otimes\theta_k\\
&=\Bigl\{\text{Tr}\Bigl(\rho_0^3R(X^j)R(X^k)\Bigr)+\text{Tr}\Bigl(\rho_0^2R(X^k)\rho_0R(X^j)\Bigr)\\
&\qquad-2\text{Tr}\Bigl(\rho_0^2R(X^j)\rho_0R(X^k)\Bigr)\Bigr\}\theta_j\otimes\theta_k\;,
\end{split}
\end{equation}

\noindent
where we have used the cyclic property of the trace in the second line and also to reorganize the terms in the last line after having explicited the product of the commutators\footnote{Let us notice that the structure of the tensor in the second line of Eq. (\ref{gqm85}) shows that this tensor is degenerate along the centralizer of $\rho_0$ and thus it is not degenerate on the homogeneous space $\mathbb G/\mathbb G_{\rho_0}$.}. In the case of a pure state $\rho_0$, i.e., $\rho_0^2=\rho_0$, Eq. (\ref{gqm85}) reduces to the following tensor:

\begin{equation}\label{gqm86}
\begin{split}
\mathcal{K}&=\Bigl(\text{Tr}\bigl[\rho_0R(X^j)R(X^k)\bigr]-\text{Tr}\bigl[\rho_0R(X^j)\rho_0R(X^k)\bigr]\Bigr)\theta_j\otimes\theta_k\\
&=\Bigl(\text{Tr}\bigl[\rho_0R(X^j)R(X^k)\bigr]-\text{Tr}\bigl[\rho_0R(X^j)\bigr]\text{Tr}\bigl[\rho_0R(X^k)\bigr]\Bigr)\theta_j\otimes\theta_k\;,
\end{split}
\end{equation}

\noindent
which corresponds to the following tensor on $\mathcal{H}_0=\mathcal H-\{0\}$

\begin{equation}\label{gqm87}
\Biggl(\frac{\bigr<0\big|R(X^j)R(X^k)\big|0\bigr>}{\bigr<0\big|0\bigr>}-\frac{\bigr<0\big|R(X^j)\big|0\bigr>\bigr<0\big|R(X^k)\big|0\bigr>}{\bigr<0\big|0\bigr>^2}\Biggr)\theta_j\otimes\theta_k
\end{equation}

\noindent
whenever we restrict the density state $\rho_0$ to be the pure state associated with the fiducial state $\bigr|0\bigl>\in\mathcal H_0$.\\ \\\textbf{Remarks:}

\begin{itemize}
\item[i)] Let us stress again that the tensor field $\mathcal K$ in (\ref{gqm86}) is defined on the Lie group $\mathbb G$ by pulling-back from the Hilbert space $\mathcal H_0$ the tensor field which previously has been identified as the pull-back of the Fubini-Study tensor from $\mathcal R(\mathcal H)$ to $\mathcal H_0$. Such a degenerate tensor $\mathcal K$ on $\mathbb G$ however contains the full information of the non-degenerate tensor field on the corresponding co-adjoint orbit $\mathcal O$ which is embedded in the projective Hilbert space $\mathcal R(\mathcal H)$. The embedding of the Lie group and its corresponding orbit is related to the co-adjoint action map on all group elements modulo $U(1)$-representations\\

\begin{equation}\label{gqm88}
\tilde\phi\;:\;\mathbb G/U(1)\longrightarrow\mathcal R(\mathcal H)\quad,\quad g\longmapsto U(g)\rho U^\dagger(g)\quad,\quad \rho\in \mathcal R(\mathcal H)
\end{equation}

\noindent
\\which ammounts to say that the action on the fiducial state $\bigl|0\bigr>$ by a phase multiplication gives rise to degeneracy directions for the Hermitian tensor. The U(1)-degeneracy can be taken into account by extending the diagram (\ref{gqm65}) with an enlarged isotropy group $\mathbb G_0^{U(1)}$ as follows:
\begin{equation}\label{gqm89}
\begin{CD}
\mathbb{G}@>\phi_0>>\;\mathcal H_0\;\\
@V\pi_0 VV               @AAi_\mathcal O A\\
\mathbb G\bigl/\mathbb G_0@>\cong>>\;\mathcal O
\end{CD}\qquad\Longrightarrow\qquad\begin{CD}
\mathbb{G}@>\phi_0>>\;S(\mathcal H)\;\\
@VU(1) VV               @VVU(1) V\\
\mathbb{G}/U(1)@>\tilde\phi_0>>\;\mathcal R(\mathcal H)\;\\
@V\pi_0 VV               @AAi_\mathcal O A\\
\mathbb G\bigl/\mathbb G_0^{U(1)}@>\cong>>\;\mathcal O
\end{CD}
\end{equation}
where $S(\mathcal H)\subset\mathcal H_0$ is the unit sphere of normalized state vectors, i.e., $S(\mathcal H)=\bigl\{\bigl|\psi\bigr>\in\mathcal H_0\;:\;\bigl<\psi\big|\psi\bigr>=1\bigr\}$.

\item[ii)] As we have already said, the construction of the pull-back tensor (\ref{gqm86}) depends on the choice of the fiducial state $\rho_0\in\mathfrak{u}^*(\mathcal H)$ and on the choice of the representation of the Lie group (and hence also on the associated Lie algebra representation). However, it should be kept in mind that the left-invariant 1-forms $\theta_k:\mathbb G\rightarrow T^*\mathbb G\cong\mathbb G\times\mathfrak g^*\cong\mathbb G\times T^*_e\mathbb G$ provide a trivialization of the cotangent bundle $T^*\mathbb G$ and do not depend neither on the fiducial state nor on the representation. Therefore, the dependence of the tensor (\ref{gqm86}) on the choice of the fiducial state is embodied only by its coeffiecients.
\end{itemize}

\noindent
Because of all these pull-backs from a space to another one, the reader may be a little bit confused at this point. It should be therefore useful to close this section by pointing out those results that will turn out to be crucial for the following considerations about entanglement. Essentially, the main lessons of the previous procedure can be summarized in the following points:

\begin{itemize}
\item In the geometrical description of the Hilbert space of a given quantum system the scalar inner product of vectors is described by a Hermitian tensor which is now evaluated on vector fields. The real part of this tensor represents a Riemannian metric tensor while the imaginary part represents a symplectic 2-form;
\item The immersion of submanifolds (of states) in the complex projective space associated with the Hilbert space allows to pull-back tensor fields related to the previous ones, via the immersion map. The main achievement of this construction is to make available methods of usual Riemannian and symplectic geometry on the selected manifold of states.
\end{itemize}

\noindent
We have used Lie groups as a tool to identify submanifolds of quantum states with orbits originated from some fiducial state. The pull-back of the Hermitian tensor from $\mathcal R(\mathcal H)$ to the embedded (co-adjoint) orbit $\mathcal O$ is provided by the tensor

\begin{equation}\label{gqm90}
\mathcal K=\mathcal K_{jk}\theta^j\otimes\theta^k\;,
\end{equation}

\noindent
with coefficients\\

\begin{equation}\label{gqm91}
\boxed{\;\mathcal K_{jk}=\text{Tr}\bigl[\rho_0R(X^j)R(X^k)\bigr]-\text{Tr}\bigl[\rho_0R(X^j)\bigr]\text{Tr}\bigl[\rho_0R(X^k)\bigr]\;}
\end{equation}

\noindent
\\which admit a decomposition into a real (symmetric) and imaginary (skewsymmetric) part given by

\begin{equation}\label{gqm92}
\mathcal K_{jk}=\mathcal K_{(jk)}+i\mathcal K_{[jk]}\;,
\end{equation}

\noindent
\\where

\begin{equation}\label{gqm93}
\boxed{\;\begin{split}
\mathcal K_{(jk)}=\frac{1}{2}\text{Tr}\Bigl(\rho_0\bigl[&R(X^j),R(X^k)\bigr]_+\Bigr)-\text{Tr}\Bigl(\rho_0R(X^j)\Bigr)\text{Tr}\Bigl(\rho_0R(X^k)\Bigr)\\
&K_{[jk]}=\frac{1}{2}\text{Tr}\Bigl(\rho_0\bigl[R(X^j),R(X^k)\bigr]_-\Bigr)
\end{split}\;}
\end{equation}

\noindent
\\In particular, when the fiducial state is a pure state $\rho_0=\frac{\ket{0}\bra{0}}{\braket{0|0}}$, this tensor reduces to (\ref{gqm87}) when it is pulled-back to the Hilbert space $\mathcal H_0$ and coincides with the pull-back of the Fubini-Study tensor (seen from the Hilbert space) on the corresponding orbit embedded in $\mathcal H_0$.\\Let us finally try to summarize all the connections between the various spaces involved in the constructions and diagrams throughout the chapter by collecting them into a unique diagram:

\begin{equation}\label{gqm94}
\xymatrix{
\mathbb{G}\ar[r]^-\phi \ar[d]_=&\mathcal{H}_0\ar[d]_{\pi_{\mathbb{R}_+}}\ar[r]^-\mu\ar@/^2pc/[dd]^\pi&\mathfrak{u}^*(\mathcal H)\\
\mathbb G\ar[r]^-\phi \ar[d]_{\pi_{U(1)}}&S(\mathcal{H})\ar[d]_{\pi_{U(1)}}&\\
\mathbb G/U(1)\ar[r]^-{\tilde\phi} \ar[d]_{\pi_0}&\mathcal R(\mathcal H)\ar[r]^-\cong&D^1(\mathcal H)\ar[uu]_i\\
\mathbb G /\mathbb G_0^{U(1)}\ar[r]^-\cong&\mathcal O\ar[u]_{i_\mathcal O}& &
}
\end{equation}

\noindent
\\from which we shall focus on the Hermitian tensors related by the following pull-backs:

\begin{equation}\label{gqm95}
\xymatrix{
\boxed{h_{FS}\;\text{on}\;\mathbb CP^1}\ar@{-->}[d]_-{\pi^*}& &\\
\boxed{\frac{\big<d\psi\otimes d\psi\big>}{\bigl<\psi\big|\psi\bigr>}-\frac{\big<\psi\big| d\psi\big>}{\bigl<\psi\big|\psi\bigr>}\otimes\frac{\big<d\psi\big| \psi\big>}{\bigl<\psi\big|\psi\bigr>}}\ar@{-->}[rr]^-{\phi^*}& &\boxed{\;\frac{\big<dg\otimes dg\big>}{\bigl<g\big|g\bigr>}-\frac{\big<g\big| dg\big>}{\bigl<g\big|g\bigr>}\otimes\frac{\big<dg\big|g\big>}{\bigl<g\big|g\bigr>}\;}\\
\boxed{\text{Tr}\bigl(\rho_\psi d\rho_\psi\otimes\rho_\psi\bigr)}\ar@{-->}[rd]_-{i_{\mathcal O}^*}\ar@{-->}[u]^-{\mu^*}\ar@{-->}[rr]^-{\tilde{\phi}^*}& &\boxed{\mathcal K=\mathcal K_{jk}\theta^j\otimes\theta^k}\ar@{-->}[u]_-{\pi_{U(1)}^*}\\
&\boxed{\mathcal K_{\mathcal O}}\ar@{-->}[ru]_-{\pi_0^*}&
}
\end{equation}

\section{Tensorial Characterization of Entanglement}

Let us now apply the pull-back procedure discussed before to orbits of quantum states of a bipartite composite system whose Hilbert space is given by:

\begin{equation}\label{gqm96}
\mathcal H=\mathcal H_A\otimes\mathcal H_B\cong\mathbb C^n\otimes\mathbb C^n\;.
\end{equation}

\noindent
In this section we want to discuss the information that can be extracted from pulled-back tensor fields when one considers states of a bipartite system acted upon by local unitary groups (i.e., gauge groups) which do not change the entanglement properties of the initial fiducial state. In other words, in what we are going to do the submanifolds of states are identified with orbits of quantum states with the same content of entanglement. The procedure can be also generalized to multipartite systems but in what follows for the sake of simplicity we will consider only bipartite systems. We will see that the Riemannian and the pre-symplectic pulled-back tensors allow to characterize entanglement without the need of computing the Schmidt decomposition explicitly.\\By considering the product representation, we identify $\mathbb G$ with the subgroup of (unitary) transformations which leave invariant the Schmidt coefficients (i.e., the entanglement) of a state and whose action is realized by the map

\begin{equation}\label{gqm97}
\phi\;:\;\mathbb G\equiv U(n)\times U(n)\longrightarrow Aut(\mathcal H_A\otimes\mathcal H_B)\;,
\end{equation}

\noindent
such that

\begin{equation}\label{gqm98}
\begin{split}
g\equiv(g_A,g_B)\longmapsto U(g)&\equiv U_A(g_A)\otimes U_B(g_B)\\
&=(U_A(g_A)\otimes\mathds 1)\cdot(\mathds 1\otimes U_B(g_B))\;.
\end{split}
\end{equation}

\noindent
The corresponding Lie algebra representation

\begin{equation}\label{gqm99}
\mathfrak g\equiv \mathfrak u(\mathcal H_A)\oplus\mathfrak u(\mathcal H_B)
\end{equation}

\noindent
is provided by means of trace-orthonormal and traceless Hermitian generators $\sigma_a\in T_eU(n)$ with $1\leq a \leq n^2-1$ and $\sigma_0=\mathds 1$ tensored by the identity of a subsystem according to the realization

\begin{equation}\label{gqm100}
R(X_j)=\begin{cases}\sigma_j\otimes\mathds 1&\text{for  }1\leq j \leq n^2\\
\mathds 1\otimes\sigma_{j-n^2}&\text{for  }n^2+1\leq j \leq 2n^2
\end{cases}
\end{equation}

\noindent
of the infinitesimal generators of the one-dimensional subgroup of $U(n)\times U(n)$. Thus, let us consider a fiducial pure state\footnote{In this chapter we will not deal with mixed states. The analysis of the mixed case can be found in \cite{GQM8}.}

\begin{equation}\label{gqm101}
\rho_0\in D^1(\mathcal H_A\otimes\mathcal H_B)\;,
\end{equation}

\noindent
and compute the pull-back of the Hermitian tensor $\text{Tr}(\rho d\rho\otimes d\rho)$ from $\mathcal R(\mathcal H)=\mathcal R(\mathcal H_A\otimes\mathcal H_B)$ to the orbit

\begin{equation}\label{gqm102}
\mathcal O_{\rho_0}:=U(n)\times U(n)\bigl/\mathbb G_{\rho_0}\;,
\end{equation}

\noindent
where $\mathbb G_{\rho_0}$ is the isotropy group of $\rho_0$. The steps of the computation are similar to those that led us to the pulled-back Hermitian tensor field $\mathcal K$ given in Eqs. (\ref{gqm90},\ref{gqm91}). Indeed, by considering the co-adjoint action of the Lie group $\mathbb G$ on the space of states, we have:

\begin{equation}\label{gqm103}
\rho=U(g)\rho_0U^\dagger(g)
\end{equation}

\noindent
\\with $U(g)$ defined in (\ref{gqm98}) and $U_\alpha(g_\alpha)=e^{iR_\alpha(X_j)t_\alpha}$ for each subsystem $\alpha=A,B$. Hence by using the decomposition of the exterior differential operator

\begin{equation}
d=d_A\otimes\mathds1_B+\mathds1_A\otimes d_B
\end{equation}

\noindent
acting on a product representation (\ref{gqm98}), we find

\begin{equation}\label{gqm104}
\begin{split}
dU&=d_AU_A\otimes U_B+U_A\otimes d_BU_B\\
&=iR_A(X_j)\theta_A^jU_A\otimes U_B+U_A\otimes iR_B(X_k)\theta_B^kU_B
\end{split}
\end{equation}

\noindent
and hence

\begin{equation}
\begin{split}
dU\,U^{\dagger}&=d_AU_A\,U_A^{\dagger}\otimes\mathds1_B+\mathds1_A\otimes d_BU_B\,U_B^{\dagger}\\
&=iR_A(X_j)\theta_A^j\otimes\mathds1_B+\mathds1_A\otimes iR_B(X_k)\theta_B^k\;,
\end{split}
\end{equation}

\noindent
where the $\theta_{A,B}$ are a basis of left-invariant 1-forms on the copy of the Lie group representation acting on the subsystem $A,B$ respectively. With the same calculations of Eqs. (\ref{gqm84}-\ref{gqm86}) we finally find that the Hermitian pulled-back tensor on the (co-adjoint) orbit (\ref{gqm102}) is given by:

\begin{equation}\label{gqm105}
\mathcal K=\Bigl(\text{Tr}\bigl(\rho_0R(X_a)R(X_b)\bigr)-\text{Tr}\bigl(\rho_0R(X_a)\bigr)\text{Tr}\bigl(\rho_0R(X_b)\bigr)\Bigr)\theta^a\otimes\theta^b\;,
\end{equation}

\noindent
where we have used a short hand notation with indices $a,b$ such that $1\leq a,b\leq n^2$ without distinguish the range of the indices for the two subsystems as in (\ref{gqm100}). Explicitly, we see that the matrix product between two realizations (\ref{gqm100}) yields

\begin{equation}\label{gqm107}
\begin{split}
R(X_j)R(X_k)&=\left(
\begin{array}{c|c}
(\sigma_a\otimes\mathds 1)\cdot(\sigma_b\otimes\mathds 1) & (\sigma_a\otimes\mathds 1)\cdot(\mathds 1\otimes\sigma_b)\\
\hline
(\mathds 1\otimes\sigma_a)\cdot(\sigma_b\otimes\mathds 1) & (\mathds 1\otimes\sigma_a)\cdot(\mathds 1\otimes\sigma_b)
\end{array}
\right)\\
&=\left(
\begin{array}{c|c}
\sigma_a\sigma_b\otimes\mathds 1 & \sigma_a\otimes\sigma_b\\
\hline
\sigma_b\otimes\sigma_a & \mathds 1\otimes\sigma_a\sigma_b
\end{array}
\right)\;\;,
\end{split}
\end{equation}

\noindent
from which follows that

\begin{equation}\label{gqm108}
[R(X_j),R(X_k)]_{\pm}=\left(
\begin{array}{c|c}
[\sigma_a,\sigma_b]_{\pm}\otimes\mathds 1 & \frac{1}{2}(\sigma_a\otimes\sigma_b\pm\sigma_a\otimes\sigma_b)\\
\hline
\frac{1}{2}(\sigma_a\otimes\sigma_b\pm\sigma_a\otimes\sigma_b) & \mathds 1\otimes[\sigma_a,\sigma_b]_{\pm}
\end{array}
\right)\;.
\end{equation}

\noindent
As in (\ref{gqm92}) and (\ref{gqm93}), the Hermitian tensor field (\ref{gqm105}) can be therefore decomposed into
\begin{equation}\label{gqm109}
\mathcal K_{jk}=\mathcal K_{(jk)}+i\mathcal K_{[jk]}
\end{equation}
where the symmetric Riemannian and the skewsymmetric (pre-)symplectic components are respectively given by the following $2n^2\times 2n^2$ block matrices:

\begin{equation}\label{gqm110}
\mathcal K_{(jk)}=\left(
\begin{array}{c|c}
A&C\\
\hline
C&B
\end{array}
\right)\qquad,\qquad\mathcal K_{[jk]}=\left(
\begin{array}{c|c}
D_A&0\\
\hline
0&D_B
\end{array}
\right)
\end{equation}

\noindent
with the $n^2\times n^2$ blocks given by:

\begin{equation}\label{gqm111}
\begin{cases}
A=\frac{1}{2}\text{Tr}(\rho_0[\sigma_a,\sigma_b]_+\otimes\mathds 1)-\text{Tr}(\rho_0\sigma_a\otimes\mathds1)\text{Tr}(\rho_0\sigma_b\otimes\mathds1)\\
B=\frac{1}{2}\text{Tr}(\rho_0\mathds 1\otimes[\sigma_a,\sigma_b]_+)-\text{Tr}(\rho_0\mathds1\otimes\sigma_a)\text{Tr}(\rho_0\mathds1\otimes\sigma_b)\\
C=\text{Tr}(\rho_0\sigma_a\otimes\sigma_b)-\text{Tr}(\rho_0\sigma_a\otimes\mathds1)\text{Tr}(\rho_0\mathds1\otimes\sigma_b)\\
D_A=\frac{1}{2}\text{Tr}(\rho_0[\sigma_a,\sigma_b]_-\otimes\mathds 1)\\
D_B=\frac{1}{2}\text{Tr}(\rho_0\mathds 1\otimes[\sigma_a,\sigma_b]_-)
\end{cases}
\end{equation}	

\noindent
If now the fiducial state $\rho_0$ is separable, i.e., $\rho_0=\rho_A\otimes\rho_B$, then we find that the off-diagonal blocks $C$ of the metric tensor $\mathcal K_{(jk)}$ vanish. Indeed, when $\rho_0=\rho_A\otimes\rho_B$, the expression (\ref{gqm111}) for C gives:

\begin{equation}\label{gqm112}
\begin{split}
C&=\text{Tr}\bigl((\rho_A\otimes\rho_B)\cdot(\sigma_a\otimes\sigma_b)\bigr)-\text{Tr}\bigl((\rho_A\otimes\rho_B)\cdot(\sigma_a\otimes\mathds1)\bigr)\text{Tr}\bigl((\rho_A\otimes\rho_B)\cdot(\mathds1\otimes\sigma_b)\bigr)\\
&=\text{Tr}(\rho_A\sigma_a\otimes\rho_B\sigma_b)-\text{Tr}(\rho_A\sigma_a\otimes\rho_B)\text{Tr}(\rho_A\otimes\rho_B\sigma_b)\\
&=\text{Tr}_A(\rho_A\sigma_a)\text{Tr}_B(\rho_B\sigma_b)-\text{Tr}_A(\rho_A\sigma_a)\text{Tr}_B(\rho_B)\text{Tr}_A(\rho_A)\text{Tr}_B(\rho_B\sigma_b)\\
&=\text{Tr}_A(\rho_A\sigma_a)\text{Tr}_B(\rho_B\sigma_b)\bigl[1-\text{Tr}_B(\rho_B)\text{Tr}_A(\rho_A)\bigr]\\
&=0
\end{split}
\end{equation}

\noindent
where in the last line we have used the fact that $\text{Tr}_A(\rho_A)=\text{Tr}_B(\rho_B)=1$. On the other hand, if the fiducial state $\rho_0$ is maximally entangled, then we find that the diagonal blocks $D_A,D_B$ of the symplectic tensor $\mathcal K_{[jk]}$ vanish, i.e., the symplectic component of the Hermitian tensor vanishes. Indeed, the blocks $D_{A,B}$ of the skewsymmetric matrix of coefficients $\mathcal K_{[jk]}$ given in (\ref{gqm111}) can be written as:

\begin{equation}\label{gqm113}
D_A=\frac{1}{2}\text{Tr}\Bigl(\text{Tr}_A(\rho_0)[\sigma_a,\sigma_b]_-\Bigr)\;,
\end{equation}

\begin{equation}\label{gqm114}
D_B=\frac{1}{2}\text{Tr}\Bigl(\text{Tr}_B(\rho_0)[\sigma_a,\sigma_b]_-\Bigr)\;.
\end{equation}

\noindent
Moreover, as discussed in Sec. 2.1, a pure state is maximally entangled iff its reduced states $\rho_{A,B}=\text{Tr}_{B,A}(\rho_0)$ are maximally mixed, i.e.:

\begin{equation}\label{gqm115}
\text{Tr}_{A,B}(\rho_0)=\frac{1}{n}\mathds1_{B,A}\;.
\end{equation}

\noindent
We then see immediately that if this is the case, Eqs. (\ref{gqm113}) and (\ref{gqm114}) give 

\begin{equation}\label{gqm116}
D_{A,B}\propto\text{Tr}([\sigma_a,\sigma_b]_-)=0
\end{equation}

\noindent
because the trace of the commutator is zero. This proves that ``$\rho_0\;separable\;\Rightarrow\;C=0$ '', while ``$\rho_0\; maximally\;entangled\;\Rightarrow\;D_A=D_B=0$ ''. Viceversa, by writing $\rho_0\in\mathfrak u^*(\mathbb C^n\otimes\mathbb C^n)$ in its Fano form \cite{Bengts,GQM10}, it is easy to prove the opposite statements ($\Leftarrow$) \cite{GQM3,GQM7,GQM8}. Therefore, these arguments essentially prove the following proposition:\\

\begin{prop}
Let $\mathcal H=\mathcal H_A\otimes\mathcal H_B\cong\mathbb C^n\otimes\mathbb C^n$ be the Hilbert space of a composite bipartite system and let $\mathcal K$ be the pull-back of the Hermitian tensor to an orbit $\mathcal O_{\rho_0}$ of quantum states related, by means of local unitary transformations, to a pure state $\rho_0$. The coefficients of such a tensor are given in (\ref{gqm109}-\ref{gqm111}) and we have that:
$$
\boxed{
\begin{split}
&\quad\rho_0\;\,separable\quad\,\Leftrightarrow\quad C=0\\
&\quad\rho_0\;\,max.\;ent.\quad\,\Leftrightarrow\quad D_A=D_B=0
\end{split}
}
$$
i.e., the pulled-back Hermitian tensor $\mathcal K$ decomposes into a direct sum $\mathcal K_A\oplus\mathcal K_B$ of two Hermitian tensors $\mathcal K_A$, $\mathcal K_B$ associated with the two subsystems when $\rho_0$ is separable, while its symplectic component $K_{[jk]}$ vanishes when $\rho_0$ is maximally entangled.
\end{prop}

\noindent
\\We notice that in general, in contrast to the Riemannian part, the symplectic part splits into two symplectic sctuctures associated with the subsystems $A$ and $B$ independently of the separability of the state $\rho_0$, i.e., it behaves in analogy to classical composite systems. Therefore, the geometrical formalism developed here gives us a quite interesting result according to which the information about the separable or entangled nature of the fiducial state $\rho_0$ is encoded into the different blocks of the pulled-back Hermitian tensor (\ref{gqm105}). In particular, the symplectic part carries information on the separability of the state $\rho_0$ while the Riemannian part carries information on the quantum entanglement, i.e., on the non-local correlations between the two subsystems. Indeed, the vanishing of the symplectic tensor for a maximally entangled state $\rho_0$ corresponds to a vanishing separability while the off-diagonal blocks of the Riemannian tensor are responsible for the entanglement degree of the state $\rho_0$. As we will discuss in Sec. 3.8, these block-off-diagonal matrices allows us to define an associated entanglement monotone which identifies an entanglement measure geometrically interpreted as a distance between entangled and separable states.

\section{Information Theory in Geometric Quantum Mechanics}

In section 3.5 we have seen that, for a given embedding $i_\mathcal M:\mathcal M\hookrightarrow\mathcal H_0$ of a general finite dimensional manifold $\mathcal M$ in $\mathcal H_0=\mathcal H-\{0\}$, we find an induced pull-back of the Fubini-Study metric on $\mathcal M$ in terms of the pull-back of the (degenerate) covariant structure given by (\ref{gqm49}). Let us now discuss an explicit derivation of the pull-back when the Hilbert space under consideration is identified with a space of square integrable functions on some configuration space $Q$. We will find that the Hermitian tensor from the Hilbert space to the submanifold $\mathcal M$ of probability densities over $Q$ identified by means of wave functions contains the (classical) Fisher information metric. However, before enter the details of the pull-back construction which relates the Fubini-Study metric with the Fisher information metric, we will give a very brief introduction of the Fisher metric in the general context of the geometry of statistical models (Information Geometry).\\As well as the geometric formulation makes available ``classical'' differential gemetric methods to approach both the conceptual and mathematical foundations of Quantum Mechanics, the relation of the Fubini-Study metric to the Fisher information could shed light on the mathematical foundantions of Quantum Information Theory \cite{GQM19}.

\subsection{The Geometry of Statistical Models: The Fisher Information Metric}

Information geometry is the study of statistical manifolds and of their invariant properties from a geometrical point of view. The crucial role of differential geometry tools in such a context was first pointed out by Rao, Amari and others (see for istance \cite{GQM20} Sec. 1 for a brief review). According to \cite{GQM21}, a statistical manifold is defined as a space $\mathcal M$ where each point is a hypothesis about some state of affairs. Usually in Statistics, hypothesis means a probability distribution and so a statistical manifold comes to be a space in which at each point $x$ there is an associated probability distribution $p(x)$, i.e., a function $p: \mathcal M\rightarrow \mathbb R$ such that $p(x)\geq0\;,\forall x\in \mathcal M$, and $\int_\mathcal M p(x)dx=1$. More formally, we can give the following:

\begin{defn}{\textbf{(Statistical Manifold)}} Let $(\mathcal M, \Sigma, \mu)$ be a measure on an orientable manifold $\mathcal M$, and let $(\mathcal X, \mathcal F, P)$ be a probability space on $\mathcal X=\mathcal M$, with sigma algebra $\mathcal F=\Sigma$ and probability measure $P=\mu$. By taking the sigma algebra $\Sigma$ fixed, the \textbf{statistical manifold} $S(\mathcal M)$ of $\mathcal M$ is defined as the space of all measures $\mu$ on $\mathcal M$.
\end{defn}

\noindent
Note that $S(\mathcal M)$ is an infinite-dimensional space (usually a Fr\'echet space) whose points are probabilities measures defined on a common probability space. Every statistical manifold comes with a measure of distances and angles which is called the \textbf{Fisher information metric}. We will now sketch the definition and the main aspects of this metric tensor\footnote{For a systematic definition of the Fisher metric together with the other geometrical structures available on a statistical manifold we refer to \cite{GQM20}, Sec. 2 and 5.}. The construction of this metric is based on the following steps:

\begin{itemize}
\item[\textbf{1)}] The starting point is to consider the space $\mathscr{P}(Q)$ of all probability distributions over a space of random variables $Q$ and then to submerge a (smooth) manifold $\mathcal M$ into $\mathscr{P}(Q)$. Hence, we need to give an embedding

\begin{equation}\label{gqm117}
\mathcal M\hookrightarrow\mathscr{P}(Q)\;,
\end{equation}

\noindent
i.e., $\mathcal M$ is a submanifold in the space of probability distributions on some sample space $Q$. Being $\mathscr P(Q)\subset L^1(Q)$, $\mathcal M$ can be also thought of as a subspace of the space of $L^1-$functions on $Q$. Usually $\mathcal M$ is called the \textbf{model statistical manifold} or the \textbf{parameter space} while, according to the definition 4.1, $\mathscr P(Q)$ is called the \textbf{statistical manifold}. The embedding (\ref{gqm117}) is realized by selecting a family of probability distributions $p(x;\xi)$ which depend on the random variables $x\in Q$ and on some specified parameters $\xi$ playing the role of coordinates on $\mathcal M$\footnote{Some example of physical interest is provided by Gaussian probability distributions which have the advantage of being completely determined by means of the first two momenta of the distribution. In this case the parameter space $\mathcal M$ is finite dimensional and we have only two parameters respectively given by the the mean value and the standard deviation (i.e., dim$\mathcal M$=2). Another example is that of the so-called exponential families which can be related to the free energy \cite{GQM22}.}, i.e.:

\begin{equation}\label{gqm118}
\xymatrix{
\mathcal M\equiv S_\xi(Q)=\{p_\xi=p(x;\xi)\in\mathscr P(Q)\;|\;\xi=(\xi^1,\dots,\xi^n)\in\mathbb R^n\}
}
\end{equation}

\item[\textbf{2)}] The \textbf{Fisher metric}, also called the \textbf{information metric}, is a Riemannian metric on the parameter space $\mathcal M$ defined by:

\begin{equation}\label{gqm119}
g(\xi)=g_{jk}(\xi)d\xi^j\otimes d\xi^k\;,
\end{equation}

\noindent
with

\begin{equation}\label{gqm120}
g_{jk}(\xi)=\int_Qdx\;p(x;\xi)\frac{\partial\log{p(x;\xi)}}{\partial \xi^j}\frac{\partial\log{p(x;\xi)}}{\partial \xi^k}
\end{equation}

\noindent
where the integration is carried out using the measure over the whole space $Q$ and we are left only with a dependence on the parmeters $\xi$.
\end{itemize}

\noindent
The Fisher information metric can be regarded as an inner product on the tangent space to the model statistical manifold. Indeed, it is well known that the exponential map provides a map from the tangent space to points in the underlying manifold. Thus, the point $p_\xi\in S_\xi(Q)$ can be thought of as the exponential $p_\xi=e^X$ of a vector $X$ in the tangent space at $T_\xi S_\xi(Q)$ . Conversely, roughly speaking, the logarithm gives a point in the tangent space and we can define an inner product on the tangent space formally written as $\bigl<X_1\big|X_2\bigr>_\xi=g_\xi(X_1,X_2)$, which is equivalent to the definition of the metric given in (\ref{gqm120}).\\Let us now consider a smooth convex function $F(\xi)$ defined on an open set of $\mathbb R^n$ where $\xi$ plays the role of a coordinate system. Its second derivative, i.e., the Hessian matrix

\begin{equation}\label{gqm121}
\frac{\partial^2 F(\xi)}{\partial\xi^j\partial\xi^k}\equiv g_{jk}(\xi)
\end{equation}

\noindent
is a positive definite matrix depending on $\xi=(\xi^1,\dots,\xi^n)$. Thus, if we consider two infinitesimally nearby points $\xi$ and $\xi+d\xi$, we can define the square of their distance by:

\begin{equation}\label{gqm122}
ds^2=\bigl<d\xi\big|d\xi\bigr>=\sum_{j,k}g_{jk}(\xi)d\xi^jd\xi^k=g_{jk}(\xi)d\xi^j\otimes d\xi^k\;.
\end{equation}

\noindent
We notice that this is exactly the second order term of the Taylor expansion of $F(\xi+d\xi)$:

\begin{equation}\label{gqm123}
F(\xi+d\xi)=F(\xi)+\frac{\partial F}{\partial\xi^j}d\xi^j+\frac{1}{2}g_{jk}(\xi)d\xi^jd\xi^k+\dots
\end{equation}

\noindent
Therefore, the idea is that it is possible to generate metrics from the Hessian of convex functions $F(\xi)$, i.e.:

\begin{equation}\label{gqm124}
g=g_{jk}(\xi)d\xi^j\otimes d\xi^k\qquad\text{with}\qquad g_{jk}(\xi)=\frac{\partial^2 F(\xi)}{\partial\xi^j\partial\xi^k}\;.
\end{equation}

\noindent
Note that we have fixed a coordinate system $\xi$ to derive the metric from the convex function $F(\xi)$. But in order to be a proper general geometric structure it should be invariant under coordinate transformations and it is possible to prove this. The strategy is to introduce a metric and an affine connection in a manifold (i.e., a geometric structure) with respect to a specific coordinate system $\xi$ and then to extend it to any coordinate system in an invariant manner (for the details of the construction we refer to \cite{GQM22} and Ch. 2,3,4 of \cite{GQM23}).\\In Statistics what people usually do is to consider convex functions on $\mathcal M\times\mathcal M$. This essentially means to consider the following submersion:

\begin{equation}\label{gqm125}
i_\mathcal M\;:\;\mathcal M\longrightarrow\mathcal M\times\mathcal M\qquad by\qquad m\longmapsto(m,\tilde{m})
\end{equation}

\noindent
and therefore, given a convex function $F$ with a non-degenerate isolated critical point (i.e., a point such that $i_\mathcal M^*(dF)=0)$, we can construct a metric by using the Hessian of $F(m,\tilde m)$:
\begin{equation}\label{gqm126}
\frac{\partial^2F}{\partial\xi^\alpha\partial\xi^\beta}=\frac{\partial^2F}{\partial\tilde{\xi}^\alpha\partial\tilde{\xi}^\beta}=-\frac{\partial^2F}{\partial\xi^\alpha\partial\tilde{\xi}^\beta}
\end{equation}
The point is then to interpret these geometrical objects. This is the reason why people usually uses the relative entropy, which has a well-defined meaning, as a generating function of the metric. For istance, the well-known Von Neumann entropy $S(\rho)=-\text{Tr}\rho\log\rho$ can be regarded as a relative entropy with respect to the uniform distribution, i.e., $S(\rho)=S(\rho||\mathds 1)=-\text{Tr}[\rho(\log\rho-\frac{1}{n}\log\mathds1)]$. Another important example is that of Tsallis q-entropy which in recent times has acquired a key role in the general framework of the theory of divergence functions for which we refer to \cite{GQM23} (Ch. 3,4).

\subsection{Fisher Metric in the Geometrical Formulation of Quantum Mechanics}

In classical probability theory the Fisher information metric can be used to characterize the distance between probability distributions. A generalization of the metric is also available in quantum information theory. On the other side we know that the states of a quantum system are described by state vectors in a Hilbert space $\mathcal H$ (and so by wave functions) or by density matrices. The difference between quantum states corresponds to a distance between the state vectors or the density matrices. To introduce a notion of distance we need to construct a metric on the set of quantum states. Essentially the main point is that we describe probability densities $p(x)$ of random variables with values in $Q$ by means of normalized wave functions $\psi(x)$ defined on $Q$ which are identified with probability amplitudes by setting $p(x)=\bar\psi(x)\psi(x)=|\psi(x)|^2$. Formally, we go from integrable functions to square integrable functions on $Q$. This observation allows to use the metric tensor available on $\mathcal H$ and thereof on the space of pure states $\mathcal R(\mathcal H)\cong D^1(\mathcal H)$ (the Fubini-Study metric) and to pull it back to a submanifold $\mathcal M$ of probability densities on $Q$ as showed in the following scheme:\\

\begin{equation}\label{gqm127}
\xymatrix{
\boxed{\;Parameter\;Space\;\mathcal M\;}\;\ar@{=>}[d]\ar@{^{(}->}[rr]^-{i_\mathcal M}& &\ar@{=>}[d]\boxed{\;(Submanifold\;of)\;\mathcal H\;}\\
\boxed{\;Metric\;h_\mathcal M\;on\;\mathcal M\;}& &\ar@{-->}[ll]^-{i_\mathcal M^*}\boxed{\;Metric\;h_\mathcal H\;on\;\mathcal H\;}
}
\end{equation}

\noindent
\\What we find is that the Hermitian tensor field on $\mathcal R(\mathcal H)$ when pulled-back to $\mathcal M$ gives rise to the Fisher quantum information metric tensor which reduces to the classical one when the states satisfy suitable conditions \cite{GQM9}.\\As discussed in the previous sections, the ray space $\mathcal R(\mathcal H)$ is a K\"ahler manifold equipped with a Hermitian tensor $h_{FS}$ whose real symmetric part gives a metric tensor $g_{FS}$ called the Fubini-Study metric and whose imaginary skewsymmetric part gives a symplectic structure $\omega_{FS}$. The pull-back of this Hermitian tensor to $\mathcal H_0=\mathcal H-\{0\}$ along the projection map\\

\begin{equation}\label{gqm128}
\pi\;:\;\mathcal H_0\longrightarrow\mathcal R(\mathcal H)\qquad\text{by}\qquad\bigl|\psi\bigr>\longmapsto\frac{\bigl|\psi\bigr>\bigl<\psi\bigr|}{\bigl<\psi\big|\psi\bigr>}\;,
\end{equation}

\noindent
\\is given by\\

\begin{equation}\label{gqm129}
h_{\mathcal H_0}=\frac{\big<d\psi\otimes d\psi\big>}{\bigl<\psi\big|\psi\bigr>}-\frac{\big<\psi\bigl|d\psi\bigr>\otimes\big<d\psi\big| \psi\big>}{\bigl<\psi\big|\psi\bigr>^2}
\end{equation}

\noindent
\\where, as explained in Eqs.(\ref{gqm28},\ref{gqm29}), $\bigl|d\psi\bigr>$ is a Hilbert-state-valued 1-form given by $\big| d\psi\big>=dc^k(\psi)\bigl|e_k\bigr>$ when a basis $\bigl\{\bigl|e_k\bigr>\bigr\}$ for $\mathcal H$ has been selected once and for all and it does not depend on the point.\\Let us now realize the Hilbert space as a space of square integrable functions on some configuration space $Q$, say $\mathcal H\cong L^2(Q)$. Therefore, abstract vectors $\bigl|\psi\bigr>$ correspond now to wave functions $\psi(x;\xi)$, where we have explicitly considered the case in which the state of the system may depend on some unknown parameters $\xi=(\xi^1,\dots,\xi^n)$, with a scalar product given by:\\

\begin{equation}\label{gqm130}
\bigl<\psi\big|\phi\bigr>=\int_Q dx\;\bar{\psi}(x;\xi)\phi(x;\xi)\;.
\end{equation}

\noindent
\\By considering the polar representation of the wave function:\\

\begin{equation}\label{gqm131}
\psi(x;\xi)=\sqrt{p(x;\xi)}e^{iW(x;\xi)}\;,
\end{equation}

\noindent
\\we see that the normalization $\bigl<\psi\big|\psi\bigr>=1$ implies that $p(x;\xi)\in L^1(Q)$, i.e., it is a probability density. In other words, the $\xi$'s parametrize a family of wave functions $\psi(x;\xi)\in L^2(Q)$ (and hence of associated probability densities $p(x;\xi)\in L^1(Q)$) which identify a submanifold $\mathcal M$ of $\mathcal H_0$ (or $\mathscr P(Q)$) by means of the following embeddings:

\begin{equation}\label{gqm132}
\xymatrix{
\mathcal M\ar@{^{(}->}[rr]^-{i_\mathcal M}\ar@{_{(}->}[rrdd]_-{i_p} & & \mathcal H_0\cong L^2(Q)\ar@{-->}[dd]^-{(\ref{gqm131})}& &\xi\ar@{|->}[rrr]\ar@{|->}[rrrdd]& & &\psi(x;\xi)\ar@{-->}[dd]^-{(\ref{gqm131})}\\
& & & by & & & &\\
 & &\mathscr P(Q)\subset L^1(Q) & & & & &p(x;\xi)
}
\end{equation}

\noindent
\\and $\xi=(\xi^1,\dots,\xi^n)$ play the role of coordinates on the parameter space $\mathcal M$. Thus, by using the following identities:

\begin{equation}\label{gqm133}
\int_Qdx\;p(x;\xi)=1\quad\Rightarrow\quad0=\int_Qdx\;dp(x;\xi)=\int_Qdx\;p(x;\xi)d\log{p(x;\xi)}
\end{equation}

\begin{equation}\label{gqm134}
\bigl<\psi\big|\psi\bigr>=1\quad\Rightarrow\quad0=d\bigl(\bigl<\psi\big|\psi\bigr>\bigr)\quad\Rightarrow\quad\bigl<d\psi\big|\psi\bigr>=-\bigl<\psi\big|d\psi\bigr>\;,
\end{equation}

\noindent
\\where the differential is taken w.r.t. the parameters $\xi$, then for a given embedding (\ref{gqm132}) we have:

\begin{equation}\label{gqm135}
i_\mathcal M^*\bigl<\psi\big|\psi\bigr>=\int_Qdx\;p(x;\xi)=1\;,
\end{equation}

\begin{equation}\label{gqm136}
i_\mathcal M^*\bigl<\psi\big|d\psi\bigr>=-i_\mathcal M^*\bigl<d\psi\big|\psi\bigr>=\int_Qdx\;p(x;\xi)\;d\log{\psi(x;\xi)}\;,
\end{equation}

\begin{equation}\label{gqm137}
i_\mathcal M^*\bigl<d\psi\otimes d\psi\bigr>=\int_Qdx\;p(x;\xi)\Bigl(d\log{\bar\psi(x;\xi)}\otimes d\log{\psi(x;\xi)}\Bigr)\;,
\end{equation}

\noindent
\\from which, by taking into account that

\begin{equation}\label{gqm138}
\begin{split}
d\Bigl(\log{\psi(x;\xi)}\Bigr)&=d\Bigl[\log{\bigl(\sqrt{p(x;\xi)}e^{iW(x;\xi)}\bigr)}\Bigr]\\
&=\frac{1}{2}d\log{p(x;\xi)}+idW(x;\xi)\;,
\end{split}
\end{equation}

\noindent
\\we find that the pull-back of the tensor field in (\ref{gqm129}) on the submanifold $\mathcal M$ is given by:

\begin{equation}\label{gqm139}
\begin{split}
h_\mathcal M=&i_\mathcal M^* h_{\mathcal H_0}=\frac{1}{4}\int_Qdx\;p(x;\xi)\Bigl(d\log{p(x;\xi)}\Bigr)^{\otimes 2}+\int_Qdx\;p(x;\xi)\Bigl(dW(x;\xi)\Bigr)^{\otimes 2}+\\
&-\Biggl(\int_Qdx\;p(x;\xi)\;dW(x;\xi)\Biggr)^2+i\int_Qdx\;p(x;\xi)\Bigl(dW(x;\xi)\wedge d\log{p(x;\xi)}\Bigr)\;.
\end{split}
\end{equation}

\noindent
From Eq. (\ref{gqm139}) we see that the tensor $h_\mathcal M$ decomposes into a real symmetric and a imaginary skewsymmetric part as:

\begin{equation}\label{gqm140}
h_\mathcal M(\xi)=g_\mathcal M(\xi)+i\omega_\mathcal M(\xi)
\end{equation}

\noindent
with

\begin{equation}\label{gqm141}
g_\mathcal M(\xi):=\frac{1}{4}\mathbb E_p[(d\log{p})^{\otimes2}]+\mathbb E_p[dW^{\otimes2}]-(\mathbb E_p[dW])^2\;,
\end{equation}

\begin{equation}\label{gqm142}
\omega_\mathcal M(\xi):=\mathbb E_p[dW\wedge d\log{p}]\;,
\end{equation}

\noindent
where we have introduced the notation $\mathbb E_p$ for the generalized expectation value integral that for a generic tensor field $\mathcal T(x;\xi)$ (including functions) is given by the average w.r.t. the probability density $p(x;\xi)$, i.e.:

\begin{equation}\label{gqm143}
\mathbb E_p[\mathcal T]:=\int_Qdx\;p(x;\xi)\;\mathcal T(x;\xi)\;,
\end{equation}

\noindent
which traces out the $x-$dependence of the tensor field. We notice that the skewsymmetric part $\omega_\mathcal M$ is related to the geometric phase $W$ and the symmetric part can be further decomposed into:

\begin{equation}\label{gqm144}
g_\mathcal M=\frac{1}{4}\mathscr F+\text{Cov}(dW)\;,
\end{equation}

\noindent
where

\begin{equation}\label{gqm145}
\mathscr F:=\mathbb E_p[(d\log{p})^{\otimes2}]
\end{equation}

\noindent
is the classical Fisher information metric tensor field\footnote{This can be easily checked by evaluating $\mathscr F$ on contravariant vectors $\frac{\partial}{\partial\xi^j},\frac{\partial}{\partial\xi^k}$. Indeed, we get:
$$
\mathscr F_{jk}=\mathscr F\biggl(\frac{\partial}{\partial\xi^j},\frac{\partial}{\partial\xi^k}\biggr)=\mathbb{E}_P\biggl[\frac{\partial\log p}{\partial\xi^j}\frac{\partial\log p}{\partial\xi^k}\biggr]\;,
$$
which is exactly the definition (\ref{gqm120}) of the Fisher metric when we explicit $\mathbb E_p$ as in (\ref{gqm143}).
}, and

\begin{equation}\label{gqm146}
\text{Cov}(dW):=\mathbb E_p[dW^{\otimes2}]-(\mathbb E_p[dW])^2
\end{equation}

\noindent
is the phase-covariance matrix tensor field. In particular, when $dW=0$ the Hermitian tensor field $h_\mathcal M$ on $\mathcal M$ coincides with the Fisher classical information metric up to the $1/4$ factor. In the general case in which $dW\neq0$, the pull-back tensor $h_\mathcal M$ provides us with the quantum version of the Fisher information metric. Indeed, if $\rho_\xi$ denotes a generic density state, the Fisher quantum information metric \cite{GQM25} is defined by:

\begin{equation}\label{gqm147}
\mathscr F_q:=\text{Tr}(\rho_\xi\;d_\ell\rho_\xi\otimes d_\ell\rho_\xi)\;,
\end{equation}

\noindent
where $d_\ell\rho_\xi$ is a Hermitian matrix whose elements are differential 1-forms usually called the symmetric logarithmic differential\footnote{Let us observe that the regularization by means of the symmetric logarithmic derivative is not the only way to define a metric on the space of quantum states. In \cite{fisher1}, for istance, the starting point is to use the relative Tsallis entropy as generating function which amounts to the introduction of q-logarithms. The employement of different regularization procedures yields different results. Indeed, as discussed in \cite{fisher1}, in the case of the Tsallis entropy there is only a symmetric part, while when the quantum Fisher information tensor and the symmetric logarithmic derivative are used, both a symmetric and an antisymmetric contribution appear \cite{fisher2,fisher3}.} implicitly defined by the relation:

\begin{equation}\label{gqm148}
d\rho_\xi=\frac{1}{2}(\rho_\xi d_\ell\rho_\xi+d_\ell\rho_\xi\;\rho_\xi)\;.
\end{equation}

\noindent
For the sake of simplicity let us consider the case of a pure state $\rho_\xi$, i.e., $\rho_\xi^\dagger=\rho_\xi,\text{Tr}\rho_\xi=1$ and $\rho_\xi^2=\rho_\xi$. Then

\begin{equation}\label{149}
\rho_\xi d\rho_\xi+d\rho_\xi\;\rho_\xi=d\rho_\xi^2=d\rho_\xi
\end{equation}

\noindent
from which it follows that

\begin{equation}\label{150}
\textit{Tr}(d\rho_\xi)=0\qquad;\qquad\textit{Tr}(\rho_\xi d\rho_\xi)=0\;,
\end{equation}

\noindent
and comparing with the definition (\ref{gqm148}) we find that

\begin{equation}\label{151}
d_\ell\rho_\xi=2d\rho_\xi\;.
\end{equation}

\noindent
Thus, in the case of a pure state Eq. (\ref{gqm147}) gives:

\begin{equation}\label{gqm152}
\mathscr F_q=4\text{Tr}(\rho_\xi d\rho_\xi\otimes d\rho_\xi)\;.
\end{equation}

\noindent
and, as we know from Sec. 4.4, by pulling-back this tensor field to the Hilbert space we find exactly the expression (\ref{gqm129}) up to the constant factor.\\To conclude, for pure states, the quantum information metric contains both the quantum and the classical version but it collapses to the classical Fisher information metric when $dW=0$, i.e., when the phase is constant. This is coherent with the spirit of \cite{GQM2} where it is stressed the coexistence of both quantum and classical-like structures in every quantum system.

\section{Entanglement Measure and Distance}

In the previous section we have seen that there is a relationship between Von Neumann relative entropy and a notion of distance between quantum states. More precisely, the relative entropy provides a unique entanglement measure for pure states, when applied to the corresponding reduced density states \cite{GQM25}. However, even if people usually refer to relative entropy as a distance, it does not satisfy the usual metric property because it is not symmetric, i.e., $S(\rho || \rho')\neq S(\rho' || \rho)$. On the other hand, the (symmetric) Hessian matrix of the relative entropy gives a completely positive map-monotone metric on the space of quantum states known as the Fisher information metric. This then suggest a relation between entanglement measure and distance on quantum states. In the light of the pull-back relationship between the Fisher information and the Fubini-Study metric, it is therefore natural to ask if it is possible to establish such a connection also in the purely geometric framework for Quantum Mechanics discussed so far. The answer is positive and what we find is that the tensorial characterization of entanglement given in Sec. 3.6 allows us to extract an entanglement measure out of the block-off-diagonal coefficient matrices of the metric tensor (\ref{gqm110}), thus providing us both with a qualitative and a quantitative description of entanglement. As already remarked after proposition 3.1, the off-diagonal blocks $C$ of the pull-back Riemannian tensor field on $U(n)\times U(n)$ are those responsible for the entanglement correlations and it is therefore natural in some sense to search for an entanglement measure associated with them. Indeed, according to \cite{GQM26}, the quantity

\begin{equation}\label{gqm153}
\text{Tr}(C^TC)
\end{equation}

\noindent
is an entanglement monotone. Moreover, this entanglement monotone can be directly related to the measure of entanglement proposed in \cite{GQM10}:

\begin{equation}\label{gqm154}
\text{Tr}(R^\dagger R)
\end{equation} 

\noindent
with

\begin{equation}\label{gqm155}
R:=\rho_0-\rho_0^A\otimes\rho_0^B\;.
\end{equation}

\noindent
To prove this we can use with no loss of generality the Bloch representation for the fiducial $n\times n$ bipartite density state $\rho_0\in\mathfrak{u}^*(\mathbb C^n\otimes\mathbb C^n)$, i.e.:

\begin{equation}\label{gqm156}
\rho_0=\frac{1}{n^2}(\sigma_0\otimes\sigma_0+r_a\sigma_a\otimes\sigma_0+s_b\sigma_0\otimes\sigma_b+t_{ab}\sigma_a\otimes\sigma_b)
\end{equation}

\noindent
with

\begin{equation}\label{gqm157}
r_a:=\text{Tr}(\rho_0^A\sigma_a)=\text{Tr}(\rho_0\sigma_a\otimes\mathds1)\;,
\end{equation}

\begin{equation}\label{gqm158}
s_b:=\text{Tr}(\rho_0^B\sigma_b)=\text{Tr}(\rho_0\mathds1\otimes\sigma_b)\;,
\end{equation}

\begin{equation}\label{gqm159}
t_{ab}:=\text{Tr}(\rho_0\sigma_a\otimes\sigma_b)\;,
\end{equation}

\noindent
and reduced density states

\begin{equation}\label{gqm160}
\rho_0^A=\frac{1}{n}(\sigma_0+r_a\sigma_a)\;,
\end{equation}

\begin{equation}\label{gqm161}
\rho_0^B=\frac{1}{n}(\sigma_0+s_b\sigma_b)\;.
\end{equation}

\noindent
From the expression (\ref{gqm111}) we see that:

\begin{equation}\label{gqm162}
C=t_{ab}-r_as_b\equiv C_{ab}\;,
\end{equation}

\noindent
hence:

\begin{equation}\label{gqm163}
\text{Tr}(C^TC)=\sum_{a,b=1}^nC_{ab}^2=\sum_{a,b=1}^n(t_{ab}^2+r_a^2s_b^2-2t_{ab}r_as_b)\;.
\end{equation}

\noindent
On the other hand, by using the expression (\ref{gqm155}) and the Bloch-representations (\ref{gqm156},\ref{gqm160},\ref{gqm161}), we have:

\begin{equation}\label{gqm164}
\begin{split}
\text{Tr}(R^\dagger R)&=\text{Tr}(\rho_0^2)+\text{Tr}((\rho_0^A)^2\otimes(\rho_0^B)^2)-2\text{Tr}(\rho_0(\rho_0^A\otimes\rho_0^B))\\
&=\frac{1}{n^4}\sum_{a,b=1}^n(n^2+r_a^2n+s_b^2n+t_{ab}^2)+\frac{1}{n^4}\sum_{a,b=1}^n(n^2+r_a^2n+s_b^2n+r_a^2s_b^2)+\\
&\quad\,-\frac{2}{n^4}\sum_{a,b=1}^n(n^2+r_a^2n+s_b^2n+t_{ab}r_as_b)\\
&=\frac{1}{n^4}\sum_{a,b=1}^n(t_{ab}^2+r_a^2s_b^2-2t_{ab}r_as_b)\;.
\end{split}
\end{equation}

\noindent
Comparing Eqs. (\ref{gqm163}) and (\ref{gqm164}), we get:

\begin{equation}\label{gqm165}
\text{Tr}(R^\dagger R)=\frac{1}{n^4}\text{Tr}(C^TC)\;.
\end{equation}

\noindent
Moreover, the basis independent measure of entanglement $\text{Tr}(R^\dagger R)$ can be directly related to a notion of distance between density states, or to be precise to the (Euclidean) distance from the pure state $\rho_0$ to a separable state on the vector space of Hermitian matrices $\mathfrak{u}^*(\mathcal H)$. Indeed, by introducing the so-called \textit{purity parameter} of the state $\rho_0$

\begin{equation}\label{gqm166}
\mu_{AB}:=\textit{Tr}(\rho_0^2)\;,
\end{equation}

\noindent
and similarly for the separable state

\begin{equation}\label{gqm167}
\mu_{sep}:=\textit{Tr}((\rho_0^A)^2\otimes(\rho_0^B)^2)=\textit{Tr}((\rho_0^A)^2)\textit{Tr}((\rho_0^B)^2)=\mu_A\mu_B\;,
\end{equation}

\noindent
then we have

\begin{equation}\label{gqm168}
\begin{split}
\text{Tr}(R^\dagger R)&=\text{Tr}(R^2)=\text{Tr}(\rho_0^2)+\text{Tr}((\rho_0^A)^2\otimes(\rho_0^B)^2)-2\text{Tr}(\rho_0(\rho_0^A\otimes\rho_0^B))\\
&=\mu_{AB}+\mu_A\mu_B-2\sqrt{\mu_{AB}\mu_A\mu_B}\cos\theta\;,
\end{split}
\end{equation}

\noindent
which for a pure state $\rho_0$ (i.e., if $\mu_{AB}=1$ and $\mu_A=\mu_B=\mu$) reduces to

\begin{equation}\label{gqm169}
\text{Tr}(R^\dagger R)=1+\mu^2-2\mu\cos\theta\;,
\end{equation}

\noindent
where we have introduced the angle $\theta$ by setting

\begin{equation}\label{gqm170}
\text{Tr}(\rho_0(\rho_0^A\otimes\rho_0^B))=\sqrt{\mu_{AB}\mu_A\mu_B}\cos\theta\;.
\end{equation}

\noindent
This angle parameter allows to give a geometrical meaning to the entanglement measure under discussion. In fact, if we consider a density matrix as a vector whose components are given by the matrix elements, the purity parameter $\mu$ ($0<\mu\leq1$) can then be thought of as the square of the vector length\footnote{Here we are regarding the trace $\text{Tr}(\rho^\dagger\rho')$ as the scalar product between the two vectors associated with $\rho$ and $\rho'$.}. The entanglement measure $\text{Tr}(R^\dagger R)$ coincides therefore with the square of the length of the vector describing the matrix $R$ which in turn, according to the definition (\ref{gqm155}), is the difference of two other vectors corresponding to the entangled state $\rho_0$ and the separable state $\rho_0^A\otimes\rho_0^B$, respectively. Thus, the angle $\theta$ in (\ref{gqm170}) is the angle between these two vectors and the measure of entanglement (\ref{gqm169}) is actually the Carnot formula for the Euclidean distance between them.\\The connection of the entanglement measure $\text{Tr}(R^\dagger R)$ with $\text{Tr}(C^TC)$ proved in (\ref{gqm165}) finally allows to give the latter a geometric interpretation as a distance between entangled and separable states.

\chapter{\textbf{Quantum Metric and Entanglement on Spin Networks}}

In the previous chapter we have seen that the geometrical formulation of Quantum Mechanics allows to construct tensorial structures on the space of quantum states and how these tensors can be used to characterize entanglement. On the other hand, the notion of entanglement provides a tool to characterize the quantum texture of spacetime in terms of the structure of correlations of spin-network states \cite{F1,F2}. In this sense it is natural to try to describe such entanglement in the geometric language developed so far. To this aim, it is necessary to translate the kinematic Hilbert space structure of LQG into the language of Geometric Quantum Mechanics. In order to create a common dictionary for the various formalisms, as a preliminary step we first try to apply this language to the simple case of a single link (Wilson line), to write down the Fubini-Study metric on the corresponding Hilbert manifold and use it to characterize the entanglement resulting from the gluing of two lines into one.\\Our analysis leads to interpret (the presence of) the link as the result of entanglement and to characterize connectivity (i.e., the existence of the link) by means of the entanglement measure $\text{Tr}(C^TC)$ constructed from the off-diagonal block matrices of the pulled-back metric tensor on orbits of unitarily related quantum states. In particular, and this is our main result, we identify the maximally entangled state with the gauge-invariant Wilson loop state and in this case the associated entanglement measure turns out to be proportional to a power of the area.\\Hence, the implementation of the geometric QM language and the related information theoretical setting can help us to explore the possibility to reconstruct the geometry of quantum spacetime by looking at the correlation structure and entanglement properties of spin-network states giving for istance a proper definition of the concept of a ``\textit{quantum distance measure}''.

\section{The Underlying Idea}

Most of the current background-independent approaches to Quantum Gravity (Loop Quantum Gravity, Spin-Foam models and Group Field Theory) share a microscopic picture of spacetime geometry described by discrete, pre-geometric degrees of freedom of combinatorial and algebraic nature. As discussed in Chapter 1, this picture formally translates in a quantum theory whose kinematic Hilbert space is defined by spin-network states corresponding to a superposition of graphs labelled by group or Lie algebra elements.\\Using the shorthand vectorial notation adopted in the previous chapters, $\bigl|\Gamma, \vec j, \vec i\bigr>$ denotes a spin-network state on a given graph $\Gamma$ which is defined by the half-integer spins $\vec j$ determining the irreducible representations of the Lie group SU(2) on each edge and by the intertwiners (invariant singlet states) $\vec i$ on each node or vertex of the graph. Spin-network states diagonalize geometrical operators such as area and volume. Specifically, spins associated with the links of the graph define quanta of area while intertwiners associated with nodes give the quanta of volume (see Ch. 1). This suggests to interpret spin-networks as the quantum states of (spatial) geometry as it appears clear in the dual picture where each node of a spin network can be naturally interpreted as a polyhedron whose faces correspond to the links of the graph. As discussed in Chapter 1, the discreteness of the spectrum for volume and area operators is a powerful argument in favour of the emergence of discreteness of spacetime in Quantum Gravity. In this thesis, we embrace the idea that Quantum Gravity is not just a theory of quantum General Relativity but rather a theory of the microscopic quantum structure of spacetime based on non-spatiotemporal fundamental building blocks. We are then led to regard spacetime itself as an emergent concept. This essentially means that the continuum spacetime scenario and its geometric structure should be derived from the pre-geometric (non-spatiotemporal) building blocks of the full quantum theory.\\In a framework where quantum aspects become fundamental, entanglement is expected to play a crucial role in the study of the pre-geometric quantum texture of spacetime and in the reconstruction of its geometry. The idea that important notions such as geometricity and topological connectivity of the quantum spacetime might be derived from the interplay between gauge symmetry and entanglement structure of the quantum states has recently gained a growing consensus. In recent times, therefore, much efforts concentrate in the attempt to build an information theoretic framework in which geometric and topological properties of spacetime can emerge from the microsocpic quantum structure of the theory. \cite{F2,F3, STent,F4,F5,F6,F7}\\In this spirit, the information-geometrical approach to Quantum Mechanics outlined in the previous chapter seems to be well-suited to attack the problem. We belive indeed that the possibility to extract geometric structures such as a metric tensor and a symplectic form directly from states and the subsequent characterization of entanglement may provide the formal context to understand the way in which the geometry of space(-time) can be reconstructed, at least at the quantum level and, eventually, recover classical geometric structures in some limit. Once the kinematic pregeometric degrees of freedom of the Hilbert space of the theory has been rephrased in the language of Geometric Quantum Mechanics, the goal is to derive a metric structure directly at the level of the space of spin-network states and use it to give a new interpretation of the geometric quantum observables of the theory in terms of entanglement properties of the quantum states.\\In this sense, one can start from the definition of spin-networks as wave functions which should allow us to find a contact point with concepts and tools available in the literature of the mathematical foundations of Quantum Mechanics and Information Theory, especially their geometric formulations, and ultimately import them into the quantum gravity framework. Indeed, as discussed in Sec. 1.5, quantum states of geometry are wave functions in the Ashtekar-Barbero connection representation or, to be precise, they are defined as cylindrical functionals of that connection (i.e., they depend on it only through the holonomies of the connection along the edges of the given graph). As such, these wave functions realize a finite sampling of the connection thus defining probability distributions on the space of (discrete) connections. More precisely, a normalized cylindrical function $\psi_\Gamma$ based on the graph $\Gamma$ with L links and V vertices defines a probability measure on $SU(2)^L/SU(2)^V$ given by:

\begin{equation}\label{f1}
p\bigl(\{g_\ell, \ell\in\Gamma\}\bigr)\prod_\ell\,dg_\ell\equiv|\psi_\Gamma(g_\ell)|^2\prod_\ell\,dg_\ell\;,
\end{equation}

\noindent
such that:

\begin{equation}\label{f2}
\int_{SU(2)^L}\prod_\ell\,dg_\ell\;|\psi_\Gamma(g_\ell)|^2=1\;.
\end{equation}

\noindent
In this way spin-networks also contain information about the parallel transport on the 3-dimensional hypersurface in which thay are embedded. For istance, in Livine and Terno \cite{F3} the cases of the Wilson loop and of the theta graph are discussed and the maxima of the corresponding probability distributions are interpreted as the most probable parallel transport along the links of the graph.\\According to what we have said in the previous chapter, Eqs. (\ref{f1}) and (\ref{f2}) tell us that in the case of spin-network wave functions the group elements (holonomies along the edges) play the role of random variables and we conclude that now our sample space $Q$ is $SU(2)^L/SU(2)^V$. By considering the probability distributions associated to the cylindrical functions without specifying them, we are essentially dealing with the subset $\mathscr P(Q)=\mathscr P(SU(2)^L/SU(2)^V)$ of the space $L^1(SU(2)^L/SU(2)^V)$. As already stressed, in order to construct the Fisher information metric we have to embed a model statistical manifold (or parameter space) $\mathcal M$ into $\mathscr P(Q)$. Such an embedding is realized by specifying a family of probability distributions on $Q$ which is described by some parameter $\xi$. Once such an identification is made, the construction of Sec. 3.7 can start, at least in principle.\\An alternative way, which circumvents the limitation of having to specify a family of probability distributions, i.e., a parameter space, without any deep motivation neither conceptual nor technical for the time being, consists in working at the level of the abstract Hilbert space and construct the pull-back of the Fubini-Study metric directly on quantum states. The specific realization of these states as wave functions and the possible restriction to particular families of them may however turn useful in the process of interpretation by exploiting the connection between Information Theory and Geometric Quantum Mechanics discussed in the previous chapter.

\section{Outlook of the Procedure}

The formalism developed in the previous chapter provides us in some sense with a general algorithmic procedure to construct tensorial geometric structures on the space of states of a given quantum theory. This has some important advantage both from the technical and computational point of view. Indeed, this geometric language not only makes available tools of differential geometry in the framework of Quantum Mechanics but also provides a way to characterize entanglement in a purely tensorial fashion with no need to explicitly compute the Schmidt coefficients or the entanglement entropy which may enter the discussion only in a second time. The procedure can be synthesized in full generality by means of the following steps:

\begin{itemize}
\item[\textbf{1)}] The space of rays $\mathcal R(\mathcal H)$ identified with the complex projective space $\mathbb CP(\mathcal H)$ $\cong D^1(\mathcal H)$ associated to the Hilbert space $\mathcal H$ of the system is recognized to be the proper setting for Quantum Mechanics. Therefore, the relevant structures are those available on the ray space. In particular, we focus on the pull-back to the Hilbert space of the Hermitian structure naturally available on $D^1(\mathcal H)$ (Fubini-Study) along the momentum map

\begin{equation}\label{f3}
\mu\;:\;\mathcal H_0\ni\ket\psi\longmapsto\rho_\psi=\frac{\ket\psi\bra\psi}{\braket{\psi|\psi}}\in D^1(\mathcal H)\subset\mathfrak{u}^*(\mathcal H)\;,
\end{equation}

\noindent
according to the diagram (\ref{gqm44}). The real and imaginary parts of this tensor provide us with a metric tensor and a symplectic structure, respectively.  

\item[\textbf{2)}] We consider a stratification of the Hilbert manifold by means of the orbits with respect to the action of a Lie group $\mathbb G$ on $\mathcal H$. By choosing a fiducial state $\ket{0}\in\mathcal H_0$ and a representation $U(g)$ of $\mathbb G$, the orbit $\mathcal O$ starting from $\ket 0$ identifies a submanifold of quantum states $\ket g=U(g)\ket 0$ when we consider an embedding map via the action of the Lie group. We can thus restrict ourself to the Hermitian tensor on this submanifold by noticing that it is completely described by the pull-back tensor on the Lie group. 

\item[\textbf{3)}] The application of the pull-back strategy of the point 2) to the case of a composite bipartite system ($\mathcal H=\mathcal H_A\otimes\mathcal H_B$) gives us a Hermitian tensor defined on orbits of unitarily related quantum states. Such a tensor turns out to have the interesting feature that its block coefficient matrices fully encode the information about the entanglement or separability of the fiducial state we start with (see Proposition 3.1). In particular, the off-diagonal blocks C can be used to define an entanglement measure $\text{Tr}(C^TC)$ which is interpreted as a distance from the corresponding separable state.

\end{itemize}

\noindent
Strictly speaking, Quantum Gravity is first of all a quantum theory with its own Hilbert space. Hence, we woukd like to reformulate it into the language of Geometric Quantum Mechanics in order to make all the tools and techniques available to describe the entanglement on spin-network states. Our first aim is therefore to construct tensorial structures on these states.

\section{A Dictionary Correspondence via Wilson Line States}

In order to create a correspondence between the two formalisms and a common dictionary to import the information-geometric machinery in the context of Quantum Gravity, we will consider in this perspective the case of a single link (Wilson line) and we try to apply the various steps of the above procedure in this simple example. 

\subsection{Step 1: Metric Tensor on the Space of States}

Before the computation of the metric tensor can take place, a few observations are in order. Let us recall from Chapter 2 that a generic Wilson line state is given by:

\begin{equation}\label{f4}
\ket{\psi_\gamma^{(j)}}=\sum_{mn}c^{(j)}_{mn}\ket{j,m,n}\;\in\;\mathcal H^{(j)}_\gamma\;,
\end{equation}

\noindent
where we assume for simplicity that $j$ is fixed. As pointed out in \cite{F9} (Sec. III), Eq. (\ref{f4}) is the expansion of the single-link state in the spin basis. Indeed, starting from the classical phase space $T^*SU(2)\cong SU(2)\times\mathfrak{su}^*(2)$, the SU(2)-valued holonomies $h(A)$ along the path $\gamma$ and the $\mathfrak{su}(2)$-valued fluxes of the triad fields through the surface crossed by $\gamma$ play the role of canonically conjugate variables. Passing to the quantum level, we consider the group basis given by a complete set of orthonormal states $\{\ket{h}=\ket{h(A)}\;|\;h\in SU(2)\}$, labelled by group elements, such that:

\begin{equation}\label{f5}
\braket{h'|h}=\delta((h')^{-1}h)\qquad,\qquad\int_{SU(2)}dh\ket{h}\bra{h}=\mathds1
\end{equation}

\noindent
\\where $dh$ denotes the Haar measure on the link. Note that we are considering the group elements themselves as operators instead of some coordinate functions on SU(2). By this we mean that $\ket h$ are the eigenstates of the operator $\hat h$ in the sense that, for any coordinate system on the group, $\ket h$ are the eigenstates of the coordinates as guaranteed by the property $f(\hat h)\ket h\equiv f(h)\ket h$, for any function $f\in\mathcal F(SU(2))$. We therefore define the Hilbert space $\mathcal H_\gamma$ to consist of those states $\ket{\psi_{\gamma}}$ which decompose in the group basis as

\begin{equation}\label{f6}
\ket{\psi_{\gamma}}=\int_{SU(2)}dh\,\psi_{\gamma}(h)\ket{h}\;,
\end{equation}

\noindent
\\where

\begin{equation}\label{f7}
\psi_\gamma(h)\equiv\psi_\gamma[h(A)]=\braket{h(A)|\psi_\gamma}\;\in\;L^2(SU(2), dh)
\end{equation}

\noindent
\\is the (cylindrical) wave function which depends on the holonomy $h(A)$ of the connection along $\gamma$. By using now the Peter-Weyl decomposition of functions on $SU(2)$ in terms of spin representations, we define the spin basis states $\ket{j,m,n}$ in $\mathcal{H}_\gamma^{(j)}\cong L^2(SU(2))$ by

\begin{equation}\label{f8}
\braket{h(A)|j,m,n}:=\sqrt{2j+1}\,D^{(j)}_{mn}(h(A))\;,
\end{equation}

\noindent
\\and the orthogonality relations of the Wigner representation matrices $D^{(j)}_{mn}$ ensure the normalization of the basis states

\begin{equation}\label{f9}
\braket{j',m',n'|j,m,n}=\delta_{jj'}\delta_{mm'}\delta_{nn'}\;,
\end{equation}

\noindent
\\together with the decomposition of the identity\footnote{If we do not assume a fixed $j$, then in (\ref{f10}) there will be also a sum over $j$.}

\begin{equation}\label{f10}
\mathds1=\sum_{mn}\ket{j,m,n}\bra{j,m,n}\;.
\end{equation}

\noindent
\\Any state in $\mathcal H_\gamma$ can be therefore expanded in the spin basis as in (\ref{f4}) with:

\begin{equation}\label{f11}
\begin{split}
c_{mn}^{(j)}&\equiv\braket{j,m,n|\psi_\gamma}=\int_{SU(2)}dh\,\psi_\gamma[h(A)]\braket{j,m,n|h(A)}\\
&=\sqrt{2j+1}\int_{SU(2)}dh\,\psi_\gamma[h(A)]\,\overline{D_{mn}^{(j)}(h(A))}\;.
\end{split}
\end{equation}

\noindent
\\ \textbf{Remark:} Let us stress that the states $\ket{j,m,n}$ are different from the usual spherical harmonics states $\ket{j,m}$, even though they are related by means of the definition of the Wigner matrix elements $D^{(j)}_{mn}(h)\equiv\braket{j,m|D^{(j)}(h)|j,n}$. The states $\ket{j,m}$ are not states on the group but elements of the vector space $\mathcal V^{(j)}$ corresponding to the $(2j+1)$-dimensional representation, whereas $\ket{j,m,n}$ can be thought of as elements of $\mathcal H_\gamma^{(j)}\cong\mathcal V^{(j)}\otimes\mathcal V^{(j)*}$. Indeed, the spherical harmonics are given by \cite{F10}

\begin{equation}\label{f12}
\braket{h|j,m}:=Y^j_m(h)\equiv\sqrt{\frac{2j+1}{4\pi}}\overline{D^{(j)}_{m0}(h)}
\end{equation}

\noindent
and, being invariant under the right multiplication by elements of the group U(1), they are functions on the 2-sphere $SU(2)/U(1)\cong S^2$. Therefore, they do not form a complete basis for $L^2(SU(2))$ and we need to use the states $\ket{j,m,n}$ which instead provide us with a full representation of both right and left multiplication. We will come back later on this point in Sec. 4.2.3.\\ \\Now that all these precisations have been done, let the first step of the procedure begin. We choose to work with the spin basis because in this way our results will depend explicitly on the algebraic data of the spin-network graph ($j, m$ and $n$ in this specific situation). Thanks to the parallelizability of Lie groups (SU(2) in our case), such a basis does not depend on the point. Thus, remembering the notations of the previous chapter, we have the following correspondences:

\begin{equation}\label{f13}
\ket{e_a}\;\quad\qquad\qquad\longleftrightarrow\qquad\qquad\quad\;\ket{j,m,n}\equiv\ket{e^{(j)}_{mn}}
\end{equation}

\begin{equation}\label{f14}
c_a(\psi)\quad\qquad\qquad\longleftrightarrow\qquad\qquad\quad c^{(j)}_{mn}\equiv\braket{j,m,n|\psi_\gamma}
\end{equation}

\begin{equation}\label{f15}
\ket{d\psi}=\sum_adc_a\ket{e_a}\;\qquad\longleftrightarrow\qquad\ket{d\psi_\gamma^{(j)}}=\sum_{mn}dc^{(j)}_{mn}\ket{j,m,n}
\end{equation}

\noindent
Hence the constructions of Sec. 3.3 and 3.4 can be now repeated and we find:

\begin{equation}\label{f16}
\braket{d\psi_\gamma^{(j)}\otimes d\psi_\gamma^{(j)}}=d\overline{c}^{(j)}_{mn}\otimes dc_{mn}^{(j)}\qquad(\text{sum over m,n})\;.
\end{equation}

\noindent
Indeed:

\begin{equation}\label{f17}
\begin{split}
\braket{d\psi_\gamma^{(j)}\otimes d\psi_\gamma^{(j)}}&=\sum_{mn}\braket{d\psi_\gamma^{(j)}|j,m,n}\braket{j,m,n|d\psi_\gamma^{(j)}}\\
&=\sum_{mn}\int_{SU(2)}dh\,\braket{d\psi_\gamma^{(j)}|j,m,n}\braket{j,m,n|h(A)}\braket{h(A)|d\psi_\gamma^{(j)}}\\
&=\sum_{mnm'n'}\int_{SU(2)}dh\,\Bigl(\braket{d\psi_\gamma^{(j)}|j,m,n}\braket{j,m,n|h(A)}\cdot\\
&\qquad\qquad\qquad\qquad\quad\cdot\braket{h(A)|j,m',n'}\braket{j,m',n'|d\psi_\gamma^{(j)}}\Bigr)\\
&=(2j+1)\sum_{mnm'n'}\Biggl(\int_{SU(2)}dh\,\overline{D^{(j)}_{mn}(h(A))}D^{(j)}_{m'n'}(h(A))\Biggr)\;d\overline{c}^{(j)}_{mn}\otimes dc_{mn}^{(j)}\\
&=(2j+1)\sum_{mnm'n'}d\overline{c}^{(j)}_{mn}\otimes dc_{mn}^{(j)}\,\frac{\delta_{mm'}\delta_{nn'}}{2j+1}\\
&=\sum_{mn}d\overline{c}^{(j)}_{mn}\otimes dc_{mn}^{(j)}\;.
\end{split}
\end{equation}

\noindent
Similarly:

\begin{equation}\label{f18}
\braket{\psi^{(j)}_\gamma|d\psi^{(j)}_\gamma}=\overline{c}^{(j)}_{mn}\,dc^{(j)}_{mn}\qquad(\text{sum over m,n})\;.
\end{equation}

\noindent
Finally, the pull-back to the Hilbert space of the Fubini-Study Hermitian tensor is given by:

\begin{equation}\label{f19}
\begin{split}
\mathcal K_{\mathcal H_\gamma}&=\frac{\braket{d\psi_\gamma^{(j)}\otimes d\psi_\gamma^{(j)}}}{\braket{\psi^{(j)}_\gamma|\psi^{(j)}_\gamma}}-\frac{\braket{d\psi^{(j)}_\gamma|\psi^{(j)}_\gamma}\otimes\braket{\psi^{(j)}_\gamma|d\psi^{(j)}_\gamma}}{\braket{\psi^{(j)}_\gamma|\psi^{(j)}_\gamma}^2}\\
&=\frac{d\overline{c}^{(j)}_{mn}\otimes dc_{mn}^{(j)}}{\sum_{mn}|c^{(j)}_{mn}|^2}-\frac{d\overline{c}^{(j)}_{mn}\,c^{(j)}_{mn}\otimes\overline{c}^{(j)}_{mn}\,dc^{(j)}_{mn}}{\Bigl(\sum_{mn}|c^{(j)}_{mn}|^2\Bigr)^2}\;.
\end{split}
\end{equation}

\subsection{Step 2: Pull-back on Orbits of Quantum States}

To pull-back the Hermitian tensor (\ref{f19}) on orbit submanifolds of quantum states according to the construction of Sec. 3.5, we need to understand what are the objects entering the diagram (\ref{gqm89}) in the specific case under examination. Let us therefore choose in $\mathcal H_\gamma^{(j)}$ a fiducial Wilson line state given by:

\begin{equation}\label{f20}
\ket{0}\equiv\ket{\psi_\gamma^{(j)}}=\sum_{mn}c^{(j)}_{mn}\ket{j,m,n}\;.
\end{equation}

\noindent
Since we are considering spin basis states $\ket{j,m,n}$ constructed with the common eigenstates of the operator $J^2$ and one of the $J$'s (say $J_z$), i.e., with a fixed orientation (say the $z$-axis) of the magnetic moments at the endpoints of the link\footnote{We may also consider a more general situation in which we have an additional degree of freedom to take into account a different direction of the magnetic moment. As discussed in \cite{F11}, in this case the basis states are given by $\ket{j,\hat m,\hat n}$, where $\hat m$ simply denotes the new direction ($\sin\theta\cos\varphi, \sin\theta\sin\varphi,\cos\theta$) obtained by rotating the direction $\hat z=(0,0,1)$. This kind of states can be used for istance to account a non-completely precise face matching of polyhedra glued along faces dual to the graph edges which will give some torsion thus providing a generalization of Regge geometries as twisted geometries \cite{F12,F13}.}, the only transformations that we can perform on such states are those generated by the operators $J_1,J_2, J_3$ which have a well-defined action on the basis states. The group $\mathbb G$ acting on $\mathcal H_\gamma$ is thus given by the group $SU(2)$. Therefore, the diagram (\ref{gqm89}) which explain the various level at which the (co-adjoint) orbit $\mathcal O$ is embedded in the projective Hilbert space $\mathcal R(\mathcal H_\gamma)$ is now given by:

\begin{equation}\label{f21}
\xymatrix{
SU(2)\ar[d]_-{U(1)}\ar[r]^-{\phi_0}&S(\mathcal H_\gamma)\ar[d]^-{U(1)}\\
SU(2)\bigl/U(1)\ar[r]^-{\tilde\phi_{0}}\ar[d]_-{\pi_0}&\mathcal R(\mathcal H_\gamma)\\
SU(2)\bigl/\mathbb G_0^{U(1)}\ar[r]^-\cong&\mathcal O\ar[u]_-{i_\mathcal O}
}
\end{equation}

\noindent
where we recall that $\mathbb G_0^{U(1)}$ is the isotropy group of the fiducial states $\rho_0$ (corresponding to $\ket{0}\equiv\ket{\psi_\gamma^{(j)}}$) enlarged to take into account the U(1)-degeneracy. By considering the SU(2) action on the fiducial state $\ket 0$, we realize the embedding of the Lie group into $\mathcal{H}_\gamma-\{0\}$ as

\begin{equation}\label{f22}
\phi_0\;:\; SU(2)\ni h \longmapsto \ket{h}=U^{(j)}(h)\ket{0}\in\mathcal{H}_\gamma-\{0\}\;,
\end{equation}

\noindent
where the spin-j-representation is given by

\begin{equation}\label{f23}
U^{(j)}:SU(2)\longrightarrow Aut(\mathcal H_\gamma)\quad,\quad h\longmapsto U^{(j)}(t)=e^{iR^{(j)}(X^k)t_k}\;.
\end{equation}

\noindent
with $R^{(j)}(X^k)\equiv J_k$ denoting the set of Hermitian operators which represent the $SU(2)$ generators. The corresponding embedding of $\mathbb G\equiv SU(2)$ into the space of states is given by the co-adjoint action map

\begin{equation}\label{f24}
\tilde\phi_0\;:\;h\longmapsto U^{(j)}(h)\rho_0 U^{(j)\dagger}(h)\qquad,\qquad \rho_0\in \mathcal R(\mathcal H_\gamma)\;.
\end{equation}

\noindent
As showed in Eq. (\ref{gqm91}), for a pure fiducial state $\rho_0$, the pull-back of the Hermitian tensor (\ref{f19}) to the co-adjoint orbit starting from it is given by:

\begin{equation}\label{f25}
\mathcal K=\mathcal K_{k\ell}\theta^k\otimes\theta^\ell\;,
\end{equation}

\noindent
with coefficients

\begin{equation}\label{f26}
\mathcal K_{k\ell}=\text{Tr}(\rho_0J_kJ_\ell)-\text{Tr}(\rho_0J_k)\text{Tr}(\rho_0J_\ell)\;.
\end{equation}

\noindent
Moreover, by using the explicit expression for the fiducial state in the case of a pure state, i.e.:

\begin{equation}\label{f27}
\rho_0=\frac{\ket 0\bra0}{\braket{0|0}}=\frac{\ket{\psi_{\gamma}^{(j)}}\bra{\psi_\gamma^{(j)}}}{\braket{\psi_\gamma^{(j)}|\psi_\gamma^{(j)}}}\;,
\end{equation}

\noindent
we find the pulled-back tensor (\ref{gqm87}) on the corresponding orbits in the Hilbert space $\mathcal H_\gamma$:

\begin{equation}\label{f28}
\begin{split}
\mathcal K_{k\ell}&=\frac{\braket{0|J_kJ_\ell|0}}{\braket{0|0}}-\frac{\braket{0|J_k|0}\braket{0|J_\ell|0}}{\braket{0|0}^2}\\
&=\frac{\braket{\psi_\gamma^{(j)}|J_kJ_\ell|\psi_\gamma^{(j)}}}{\braket{\psi_\gamma^{(j)}|\psi_\gamma^{(j)}}}-\frac{\braket{\psi_\gamma^{(j)}|J_k|\psi_\gamma^{(j)}}\braket{\psi_\gamma^{(j)}|J_\ell|\psi_\gamma^{(j)}}}{\braket{\psi_\gamma^{(j)}|\psi_\gamma^{(j)}}^2}\\
&=\braket{J_kJ_\ell}_{\psi_\gamma^{(j)}}-\braket{J_k}_{\psi_\gamma^{(j)}}\braket{J_\ell}_{\psi_\gamma^{(j)}}\;.
\end{split}
\end{equation}

\noindent
We then see that the Hermitian tensor on the orbits embedded in $\mathcal H_\gamma$ coincides with the covariance matrix of the SU(2) generators. Indeed, starting from the definition of the covariance matrix whose entry in the $k$th row and $\ell$th column is

\begin{equation}\label{f29}
\text{Cov}(J)_{k\ell}=\braket{(J_k-\braket{J_k})(J_\ell-\braket{J_\ell})}\;,
\end{equation}

\noindent
we have

\begin{equation}\label{f30}
\begin{split}
\braket{(J_k-\braket{J_k})(J_\ell-\braket{J_\ell})}&=\braket{(J_kJ_\ell-J_k\braket{J_\ell}-\braket{J_k}J_\ell+\braket{J_k}\braket{J_\ell})}\\
&=\braket{J_kJ_\ell}-\braket{J_k}\braket{J_\ell}-\braket{J_k}\braket{J_\ell}+\braket{J_k}\braket{J_\ell}\\
&=\braket{J_kJ_\ell}-\braket{J_k}\braket{J_\ell}\;.
\end{split}
\end{equation}

\noindent
The tensor (\ref{f28}) therefore will measure the correlations in the fluctuations of the $J$ operators. The non-commutativity of such operators implies that the covariance matrix (\ref{f29}) is not symmetric, but if we remember the decomposition (\ref{gqm93}) of the Hermitian tensor in its real symmetric and imaginary skewsymmetric part, we find a metric tensor 

\begin{equation}\label{f31}
\mathcal K_{(k\ell)}=\frac{1}{2}\braket{[J_k,J_\ell]_+}_0-\braket{J_k}_0\braket{J_\ell}_0\equiv\mathbb{R}e\bigl[\braket{(J_k-\braket{J_k})(J_\ell-\braket{J_\ell})}\bigr]\;,
\end{equation}

\noindent
and a symplectic structure 

\begin{equation}\label{f32}
\mathcal K_{[k\ell]}=\mathbb{I}m\biggl(\frac{1}{2}\braket{[J_k,J_\ell]_-}_0\biggr)=\frac{1}{2}\braket{\varepsilon_{k\ell r}J_r}_0\;,
\end{equation}

\noindent
where we have used the commutation relations of the Lie algebra $\mathfrak{su}(2)$, i.e.:

\begin{equation}\label{f33}
[J_k,J_\ell]_-=i\varepsilon_{k\ell r}J_r\;.
\end{equation}

\subsection{Step 3: Link as Entanglement of Semi-links}

We are now ready to apply the third step of the procedure outlined in Sec. 4.2. The simplest bipartite system that we can imagine is the one provided by regarding the single link Hilbert space $\mathcal H_\gamma^{(j)}$ (for fixed $j$) as the tensor product Hilbert space

\begin{equation}\label{f34}
\mathcal H_\gamma^{(j)}\cong\mathcal V^{(j)}\otimes\mathcal V^{(j)*}\;,
\end{equation}

\noindent
where $\mathcal V^{(j)}$ denotes the $(2j+1)$-dimensional linear space carrying the irreducible representation of $SU(2)$ and for any $j\in\frac{\mathbb N}{2}$, the system $\{\ket{j,m}\}_{-j\leq m \leq j}$ is orthonormal, i.e.

\begin{equation}\label{f35}
\mathcal V^{(j)}=span\{\ket{j,m}\}_{-j\leq m \leq j}\;,
\end{equation}

\noindent
while $\mathcal V^{(j)*}$ is its dual vector space. Indeed, as already mentioned (see remark in Sec. 4.3.1), the Wilson line state (\ref{f4}) can be regarded as

\begin{equation}\label{f36}
\begin{split}
\ket{\psi_\gamma^{(j)}}&=\sum_{mn}c^{(j)}_{mn}\,\ket{j,n}\otimes\ket{j,m}^*\\
&=\sum_{mn}c^{(j)}_{mn}\,\ket{j,n}\otimes\bra{j,m}\\
&=\sum_{mn}c^{(j)}_{mn}\,\underset{\ket{j,m,n}}{\underbrace{\ket{j,n}\bra{j,m}}}
\end{split}
\end{equation}

\noindent
in such a way that

\begin{equation}\label{f37}
\begin{split}
\psi_\gamma^{(j)}[A]&\equiv\braket{h(A)|\psi_\gamma^{(j)}}=\sum_{mn}c^{(j)}_{mn}\,\braket{h(A)|j,m,n}\\
&=\sqrt{2j+1}\sum_{mn}c^{(j)}_{mn}\,D^{(j)}_{mn}(h(A))\;,
\end{split}
\end{equation}

\noindent
where $D^{(j)}_{mn}(h(A))=\braket{j,m|D^{(j)}(h(A))|j,n}$ are the Wigner D-matrix elements corresponding to the spin-$j$ irreducible representation labelling the link.\\More precisely, the Wilson line cylindrical basis functions correspond to the Weyl symbols of the holonomy operator $\hat{h}(A)$ with respect to the quantization map $\ket{j,n}\bra{j,m}$ defined by \cite{F14}
\begin{equation}\label{f38}
\mathcal W(\hat{h}(A))\equiv\text{Tr}\bigl(\ket{j,n}\bra{j,m}\hat{h}(A)\bigr)=D^{(j)}_{mn}(h(A))\;,
\end{equation}
where, for any $j\in\frac{\mathbb N}{2}$, the set $\{\hat{v}^j_{nm}\}\equiv\{\ket{j,n}\bra{j,m}\}_{-j\leq n,m \leq j}$ of $(2j+1)\times(2j+1)$ linear maps $\ket{j,n}\bra{j,m}\in\mathcal V^{(j)}\otimes\mathcal V^{(j)*}$ is the canonical basis of the algebra $End(\mathcal V^{(j)})$ of endomorphisms of $\mathcal V^{(j)}$, othonormal with respect to the scalar product $\braket{\hat{v}^j_{nm},\hat{v}^j_{n'm'}}=\text{Tr}\bigl((\hat{v}^j_{nm})^\dagger\hat{v}^j_{n'm'}\bigl)$, which describe the spheres $S_j$.\\Let us now try to characterize the entanglement for the bipartite system (\ref{f34}) importing the tensorial structures of Ch. 3 in the case of our interest where, essentially, we are thinking of the Wilson line state as the composite state of two semilink states (roughly, spherical harmonics $Y^j_m(h)=\braket{h|j,m}$). According to the diagram (\ref{f21}), we select a fiducial pure state
\begin{equation}\label{f39}
\rho_0\in D^1(\mathcal V^{(j)}\otimes\mathcal V^{(j)*})\cong\mathcal R(\mathcal V^{(j)}\otimes\mathcal V^{(j)*})=\mathcal R(\mathcal H_\gamma^{(j)})\;,
\end{equation}
and then we consider the product representation
\begin{equation}\label{f40}
\phi_0\,:\,\mathbb G\equiv SU(2)\times SU(2)\longrightarrow Aut(\mathcal V^{(j)}\otimes\mathcal V^{(j)*})\;,
\end{equation}
providing the following embedding map
\begin{equation}\label{f41}
\mathbb G\ni g \longmapsto \rho_g=U(g)\rho_0 U^\dagger(g)\in\mathcal R(\mathcal V^{(j)}\otimes\mathcal V^{(j)*})
\end{equation}
where $U(g)=e^{iR(X_k)t}$ with the infinitesimal generators $R(X_k)$ realized as the tensor products between the identity of a subsystem and the spin operators $J_k$ representing the $\mathfrak{su}(2)$ algebra in terms of selfadjoint operators on the Hilbert space $\mathcal V^{(j)}$ (cfr. Eq. (\ref{gqm100})). Thus, repeating the computations of Sec. 3.6, we find that the pull-back of the Hermitian tensor $\text{Tr}(\rho d\rho\otimes d\rho)$ from $\mathcal R(\mathcal H_{\gamma}^{(j)})=\mathcal R(\mathcal V^{(j)}\otimes\mathcal V^{(j)*})$ to the co-adjoint orbit
\begin{equation}\label{f42}
\mathcal O_{\rho_0}:=SU(2)\times SU(2)/\mathbb G_{\rho_0}\;,
\end{equation}
where $\mathbb G_{\rho_0}$ is the isotropy group of the fiducial state (\ref{f38})\footnote{The topology of the orbit will thus depend on the isotropy group of the selected fiducial state. We refer to \cite{F15,F16} for a general discussion.}, decomposes into a symmetric Riemannian and a skewsymmetric (pre-)symplectic components (cfr. Eqs. (\ref{gqm110},\ref{gqm111}))

\begin{equation}\label{f43}
\mathcal K_{k\ell}=\mathcal K_{(k\ell)}+i\mathcal K_{[k\ell]}=\left(\begin{array}{c|c}
A&C\\
\hline
C&B
\end{array}\right)+i\left(
\begin{array}{c|c}
D_A&0\\
\hline
0&D_B
\end{array}
\right)\;,
\end{equation}

\noindent
\\with $3\times3$ blocks given by

\begin{equation}\label{f44}
\begin{cases}
A_{ab}=\frac{1}{2}\text{Tr}(\rho_0[J_a,J_b]_+\otimes\mathds 1)-\text{Tr}(\rho_0J_a\otimes\mathds1)\text{Tr}(\rho_0J_b\otimes\mathds1)\\
B_{ab}=\frac{1}{2}\text{Tr}(\rho_0\mathds 1\otimes[J_a,J_b]_+)-\text{Tr}(\rho_0\mathds1\otimes J_a)\text{Tr}(\rho_0\mathds1\otimes J_b)\\
C_{ab}=\text{Tr}(\rho_0J_a\otimes J_b)-\text{Tr}(\rho_0J_a\otimes\mathds1)\text{Tr}(\rho_0\mathds1\otimes J_b)\\
(D_A)_{ab}=\frac{1}{2}\text{Tr}(\rho_0[J_a,J_b]_-\otimes\mathds 1)\\
(D_B)_{ab}=\frac{1}{2}\text{Tr}(\rho_0\mathds 1\otimes[J_a,J_b]_-)
\end{cases}
\end{equation}
or equivalently, for a pure state $\rho_0=\ket0\bra0$ with $\braket{0|0}=1$, the block-coefficient matrices of the pulled-back hermitian tensor field on the corresponding orbit submanifold in the Hilbert space $\mathcal H^{(j)}_\gamma$ is given by
\begin{equation}\label{f45}
\begin{cases}
A_{ab}=\frac{1}{2}\braket{0|[J_a,J_b]_+\otimes\mathds 1|0}-\braket{0|J_a\otimes\mathds1|0}\braket{0|J_b\otimes\mathds1|0}\\
B_{ab}=\frac{1}{2}\braket{0|\mathds 1\otimes[J_a,J_b]_+|0}-\braket{0|\mathds1\otimes J_a|0}\braket{0|\mathds1\otimes J_b|0}\\
C_{ab}=\braket{0|J_a\otimes J_b|0}-\braket{0|J_a\otimes\mathds1|0}\braket{0|\mathds1\otimes J_b|0}\\
(D_A)_{ab}=\frac{1}{2}\braket{0|[J_a,J_b]_-\otimes\mathds 1|0}\\
(D_B)_{ab}=\frac{1}{2}\braket{0|\mathds 1\otimes[J_a,J_b]_-|0}
\end{cases}\;.
\end{equation}	
Therefore, we see that if $\rho_0$ is maximally entangled, that is the reduced states are maximally mixed

\begin{equation}\label{f46}
(\rho_0)_A=(\rho_0)_B=\frac{1}{dim\,\mathcal V^{(j)}}\mathds1_{A,B}=\frac{\mathds1_{j}}{2j+1}\;,
\end{equation}

\noindent
then

\begin{equation}\label{f47}
(D_A)_{ab}=\frac{1}{2}\text{Tr}\bigl((\rho_0)_B[J_a,J_b]_-\bigr)\propto\text{Tr}\bigl([J_a,J_b]_-\bigr)=0\;,
\end{equation}

\noindent
and similarly for $(D_B)_{ab}$. On the other hand, if $\rho_0$ is separable, i.e., $\rho_0=(\rho_0)_A\otimes(\rho_0)_B$, then we have

\begin{equation}\label{f48}
\begin{split}
C_{ab}&=\text{Tr}\bigl((\rho_0)_AJ_a\otimes(\rho_0)_BJ_B\bigr)-\text{Tr}\bigl((\rho_0)_AJ_a\otimes(\rho_0)_B\bigr)\text{Tr}\bigl((\rho_0)_A\otimes(\rho_0)_BJ_B\bigr)\\
&=\text{Tr}\bigl((\rho_0)_AJ_a\bigr)\text{Tr}\bigl((\rho_0)_BJ_B\bigr)-\text{Tr}\bigl((\rho_0)_AJ_a\bigr)\underset{1}{\underbrace{\text{Tr}\bigl((\rho_0)_B\bigr)}}\,\underset{1}{\underbrace{\text{Tr}\bigl((\rho_0)_A\bigr)}}\text{Tr}\bigl((\rho_0)_BJ_B\bigr)\\
&=0\;.
\end{split}
\end{equation}

\noindent
Thus, as expected from the general considerations of proposition 3.1, the information about the separability or entanglement of the fiducial state $\rho_0$ is encoded into the different blocks of the pulled-back Hermitian tensors on the orbit of unitarily related states starting from $\rho_0$. In particular, the symplectic part carries information on the separability of the state $\rho_0$ while the Riemannian part carries information on the quantum entanglement between the two subsystems. Indeed, the vanishing of the symplectic tensor for a maximally entangled state $\rho_0$ corresponds to a vanishing separability while the off-diagonal blocks of the Riemannian tensor are responsible for the entanglement degree of the state $\rho_0$ and allow us to define an associated entanglement monotone $\text{Tr}(C^TC)$ which identifies an entanglement measure geometrically interpreted as a distance between entangled and separable states. As we will discuss later in this chapter, since we are regarding the link as resulting from the entanglement of semilinks, such entanglement monotone gives us a measure of the existence of the link itself and so of the graph connectivity.

\section{Two Explicit Cases: Maximally Entangled and Separable States}

In order to visualize the considerations of the previous section, let us focus on the two extreme cases respectively given by a maximally entangled and a separable Wilson line state, and compute explicitly the pull-back of the Hermitian tensor on a orbit having that state as fiducial state.\\To this aim, we start by considering the following Schmidt decomposition of the normalized state (\ref{f36}):

\begin{equation}\label{f49}
\ket{\psi_\gamma^{(j)}}=\sum_k\lambda_k\ket{j,k}\otimes\bra{j,k}\;.
\end{equation}

\noindent
In the maximally entangled case all Schmidt coefficients are equal and, according to the normalization condition $\braket{\psi_\gamma^{(j)}|\psi_\gamma^{(j)}}=1$, they are given by:

\begin{equation}\label{f50}
\lambda_k=\frac{1}{\sqrt{2j+1}}\qquad\forall\,k\in[-j,+j]\;,
\end{equation}

\noindent
thus yielding a maximally entangled state

\begin{equation}\label{f51}
\ket{\psi_\gamma^{(j)}}=\frac{1}{\sqrt{2j+1}}\sum_k\ket{j,k}\otimes\bra{j,k}\;,
\end{equation}

\noindent
which is nothing but the gauge-invariant Wilson loop state $\ket{\psi_{WL}}$. Indeed, as discussed in Sec. 2.2, such a state corresponds to glue the two endpoints of the link in a bivalent vertex and contract their magnetic moments with an intertwiner provided by the normalized identity in $\mathcal V^{(j)}$, i.e.:

\begin{equation}\label{f52}
\ket{\psi_{WL}}=\sum_{k,k'}\frac{\delta_{k,k'}}{\sqrt{2j+1}}\ket{j,k}\otimes\bra{j,k'}\equiv\sum_{k,k'}i_{k,k'}\ket{j,k}\otimes\bra{j,k'}\;.
\end{equation}

\noindent
Therefore, concerning the open single line state regarded as an entangled state of two semilinks, there is a correspondence between maximal entanglement and gauge-invariance which is actually realized by identifying the maximally entangled state (\ref{f51}) with the closed Wilson loop state, i.e.:

\begin{equation}\label{f53}
\mathcal{H}_{max.\,ent.}\equiv\mathcal{H}_{loop}=\text{Inv}_{SU(2)}\bigl[\mathcal V^{(j)}\otimes\mathcal V^{(j)*}\bigr]\subset\mathcal V^{(j)}\otimes\mathcal V^{(j)*}\cong\mathcal{H}_{link}\;.
\end{equation}

\noindent
Let us then take the maximally entangled Wilson loop state as our fiducial state and compute the corresponding pulled-back Hermitian tensor on the orbit starting from it. For the sake of brevity, in what follows we will give only the results leaving a detailed discussion of the explicit computations in Appendix C. The associated pure density matrix $\rho_0\in D^1(\mathcal V^{(j)}\otimes\mathcal V^{(j)*})$ is given by

\begin{equation}\label{f54}
\rho_0=\ket{\psi_{WL}}\bra{\psi_{WL}}=\frac{1}{2j+1}\sum_{k,k'}\bigl(\ket{j,k}\bra{j,k'}\bigr)\otimes\bigl(\ket{j,k'}\bra{j,k}\bigr)\;,
\end{equation}

\noindent
such that the reduced states are diagonal with eigenvalues exactly given by the square of the Schmidt coefficients, e.g.

\begin{equation}\label{f55}
(\rho_0)_A=\text{Tr}_B(\rho_0)=\frac{1}{2j+1}\sum_k\ket{j,k}\bra{j,k}=\frac{\mathds1_j}{dim\,\mathcal V^{(j)}}\;.
\end{equation}

\noindent
Hence, by using Eqs. (\ref{f43},\ref{f44}), the pull-back of the Hermitian tensor $\mathcal K$ on the orbit $\mathcal O_{\rho_0}$ of Eq. (\ref{f42}) has the following form

\begin{equation}\label{f56}
\begin{pmatrix}
\frac{1}{3}j(j+1) & 0 & 0 & \frac{1}{3}j(j+1) & 0 & 0 \\
0 & \frac{1}{3}j(j+1) & 0 & 0 & \frac{1}{3}j(j+1) & 0 \\
0 & 0 & \frac{1}{3}j(j+1) & 0 & 0 & \frac{1}{3}j(j+1) \\
\frac{1}{3}j(j+1) & 0 & 0 & \frac{1}{3}j(j+1) & 0 & 0 \\
0 & \frac{1}{3}j(j+1) & 0 & 0 & \frac{1}{3}j(j+1) & 0 \\
0 & 0 & \frac{1}{3}j(j+1) & 0 & 0 & \frac{1}{3}j(j+1) \\
\end{pmatrix}
\end{equation}

\noindent
\\from which, using the decomposition

\begin{equation}\label{f57}
\mathcal K_{k\ell}=\mathcal K_{(k\ell)}+i\mathcal K_{[k\ell]}\;,
\end{equation}

\noindent
\\we see that the real symmetric part $\mathcal K_{(k\ell)}$ decomposes in the block-diagonal matrices $A,B$ and the two equal block-off-diagonal matrices $C$, according to

\begin{equation}\label{f58}
\mathcal K_{(k\ell)}=\left(\begin{array}{c|c}
A&C\\
\hline
C&B
\end{array}\right)
\end{equation}

\noindent
with

\begin{equation}\label{f59}
A=B=\begin{pmatrix}
\frac{1}{3}j(j+1) & 0 & 0 \\
0 & \frac{1}{3}j(j+1) & 0 \\
0 & 0 & \frac{1}{3}j(j+1) \\
\end{pmatrix}\;,
\end{equation}

\noindent
\\and

\begin{equation}\label{f60}
C=\begin{pmatrix}
\frac{1}{3}j(j+1) & 0 & 0 \\
0 & \frac{1}{3}j(j+1) & 0 \\
0 & 0 & \frac{1}{3}j(j+1) \\
\end{pmatrix}\;,
\end{equation}

\noindent
\\and, according to Prop. 3.1, the imaginary skewsymmetric part $\mathcal K_{[k\ell]}$\\

\begin{equation}\label{61}
\mathcal K_{[k\ell]}=\left(\begin{array}{c|c}
D_A & 0 \\
\hline
0 & D_B
\end{array}\right)\;,
\end{equation}

\noindent
\\with

\begin{equation}\label{f62}
D_A=D_B=\begin{pmatrix}
0 & 0 & 0 \\
0 & 0 & 0 \\
0 & 0 & 0 \\
\end{pmatrix}\;,
\end{equation}

\noindent
\\gives a vanishing symplectic structure, as expected for the maximally entangled case.\\ \\Moreover, by using the off-diagonal blocks (\ref{f60}) of the Riemannian symmetric part, the associated entanglement monotone (interpreted as a distance with respect to the separable state) is given by:

\begin{equation}\label{f63}
\text{Tr}(C^TC)=\sum_{a,b=1}^3\,C_{ab}^2=\frac{1}{3}[j(j+1)]^2\;.
\end{equation}

\noindent
On the other extreme, if we consider a separable state, the two spin states do not talk with each other and may have in general different spins, i.e.:

\begin{equation}\label{f64}
\ket{0}=\ket{j_1,k_1}\otimes\bra{j_2,k_2}\;,
\end{equation}

\noindent
\\The corresponding pure state density matrix is given by

\begin{equation}\label{f65}
\rho_0=\rho_A\otimes\rho_B=\bigl(\ket{j_1,k_1}\bra{j_1,k_1}\bigr)\otimes\bigl(\ket{j_2,k_2}\bra{j_2,k_2}\bigr)\;.
\end{equation}

\noindent
\\Then, the pull-back of the Hermitian tensor $\mathcal K$ on the orbit $\mathcal O_{\rho_0}$ will take the following form

\begin{equation}\label{f66}
\begin{varpmatrix}[\scriptsize]
\frac{1}{2}[j_1(j_1+1)-k_1^2] & \frac{i}{2}k_1 & 0 & 0 & 0 & 0 \\
-\frac{i}{2}k_1 & \frac{1}{2}[j_1(j_1+1)-k_1^2] & 0 & 0 & 0 & 0 \\
0 & 0 & k_1(k_1-k_2) & 0 & 0 & 0 \\
0 & 0 & 0 & \frac{1}{2}[j_2(j_2+1)-k_2^2] & \frac{i}{2}k_2 & 0 \\
0 & 0 & 0 & -\frac{i}{2}k_2 & \frac{1}{2}[j_2(j_2+1)-k_2^2] & 0 \\
0 & 0 & 0 & 0 & 0 & k_2(k_2-k_1) \\
\end{varpmatrix}
\end{equation}

\noindent
\\from which, according to Prop. 3.1, we see that in the separable case we have vanishing off-diagonal block matrices $C$ and a direct sum

\begin{equation}\label{f67}
\underset{\mathcal K_A}{\underbrace{\begin{varpmatrix}[\scriptsize]
\frac{1}{2}[j_1(j_1+1)-k_1^2] & \frac{i}{2}k_1 & 0 \\
-\frac{i}{2}k_1 & \frac{1}{2}[j_1(j_1+1)-k_1^2] & 0 \\
0 & 0 & k_1(k_1-k_2) \\
\end{varpmatrix}}}\oplus\underset{\mathcal K_B}{\underbrace{\begin{varpmatrix}[\scriptsize]
\frac{1}{2}[j_2(j_2+1)-k_2^2] & \frac{i}{2}k_2 & 0 \\
-\frac{i}{2}k_2 & \frac{1}{2}[j_2(j_2+1)-k_2^2] & 0 \\
0 & 0 & k_2(k_2-k_1) \\
\end{varpmatrix}}}
\end{equation}

\noindent
\\of two decoupled Hermitian tensors $\mathcal K_A$ and $\mathcal K_B$ one for each subsystem. Moreover, a further decomposition of the Hermitian tensor (\ref{f66}) as

\begin{equation}\label{f68}
\mathcal K_{k\ell}=\mathcal K_{(k\ell)}+i\mathcal K_{[k\ell]}\;,
\end{equation}

\noindent
gives a symmetric real part

\begin{equation}\label{f69}
\mathcal K_{(k\ell)}=\left(\begin{array}{c|c}
A&C\\
\hline
C&B
\end{array}\right)
\end{equation}

\noindent
\\with

\begin{equation}\label{f70}
A=\begin{pmatrix}
\frac{1}{2}[j_1(j_1+1)-k_1^2] & 0 & 0 \\
0 & \frac{1}{2}[j_1(j_1+1)-k_1^2] & 0 \\
0 & 0 & k_1(k_1-k_2) \\
\end{pmatrix}\;,
\end{equation}\\

\begin{equation}\label{f70b}
B=\begin{pmatrix}
\frac{1}{2}[j_2(j_2+1)-k_2^2] & 0 & 0 \\
0 & \frac{1}{2}[j_2(j_2+1)-k_2^2] & 0 \\
0 & 0 & k_2(k_2-k_1) \\
\end{pmatrix}\;,
\end{equation}\\

\begin{equation}\label{f71}
C=\begin{pmatrix}
0 & 0 & 0 \\
0 & 0 & 0 \\
0 & 0 & 0 \\
\end{pmatrix}\;,
\end{equation}

\noindent
\\and an imaginary skewsymmetric part

\begin{equation}\label{f72}
\mathcal K_{[k\ell]}=\left(\begin{array}{c|c}
D_A & 0 \\
\hline
0 & D_B
\end{array}\right)
\end{equation}

\noindent
with

\begin{equation}\label{f73}
D_A=\begin{pmatrix}
0 & \frac{1}{2}k_1 & 0 \\
-\frac{1}{2}k_1 & 0 & 0 \\
0 & 0 & 0 \\
\end{pmatrix}\qquad,\qquad D_B=\begin{pmatrix}
0 & \frac{1}{2}k_2 & 0 \\
-\frac{1}{2}k_2 & 0 & 0 \\
0 & 0 & 0 \\
\end{pmatrix}\;.
\end{equation}

\noindent
Finally, we have

\begin{equation}\label{f74}
\text{Tr}(C^TC)=0\;,
\end{equation}

\noindent
i.e., as expected, the entanglement measure associated with the block-off-diagonal matrices $C$ is zero in the unentangled case, coherently with its interpretation as a distance with respect to the separable state.

\section{Entanglement of Glued Links}

Let us now proceed a little step further and consider the description of the entanglement resulting from the gluing of two lines into one. Therefore, the bipartite Hilbert space is now given by two copies of a single link Hilbert space with fixed but different spin labels, i.e.

\begin{equation}\label{f75}
\mathcal H=\mathcal H_{\gamma_1}^{(j_1)}\otimes\mathcal H_{\gamma_2}^{(j_2)}\;,
\end{equation}

\noindent
and the fiducial state is chosen to be a Wilson line state (one link state) coming from the gluing of two other links which, as discussed in Sec. 2.2, admits the following expression (cfr. (\ref{schmidec})):

\begin{equation}\label{f76}
\ket{0}\equiv\ket{\psi_\gamma}=\frac{1}{\sqrt{2j+1}}\sum_{m,n,k,\ell}c_{mn}\ket{j,m,k}\otimes\ket{j,\ell,n}\delta_{k,\ell}\;,
\end{equation}

\noindent
where the local $SU(2)$ gauge-invariance requirement at the gluing point $v\equiv\gamma_1(1)=\gamma_2(0)$, implemented by the bivalent intertwiner $\delta_{k,\ell}/\sqrt{2j+1}$ contracting the magnetic numbers of the glued endpoints, forces the two spins to be equal (i.e., $j_1=j_2=j$). In other words, $\ket{0}$ is a locally $SU(2)$-invariant state in $\mathcal H$, that is

\begin{equation}\label{f77}
\ket{0}\;\in\;\mathcal H_\gamma^{(j)}\subset\mathcal H\qquad,\qquad\gamma=\gamma_1\circ\gamma_2\;.
\end{equation}

\noindent
However, in order to compute an entanglement measure (by using the off-diagonal blocks of the metric tensor) which can be interpreted as the distance of our fiducial state from the separable one, we need to consider the action $\phi$ of a Lie group $\mathbb G$ on $\mathcal H$ and not only on the gauge-reduced level $\tilde\phi:\mathbb G/SU(2)\rightarrow\mathcal H_\gamma^{(j)}$.\\With these precisations, the underlying scheme of the construction of the pulled-back Hermitian tensor on the orbit of states with fixed amount of entanglement will be given by the following diagram

\begin{equation}\label{f78}
\xymatrix{
\mathbb G\ar[dd]_-{SU(2)}\ar[r]^-{\phi_0} & \mathcal H_{\gamma_1}^{(j_1)}\otimes\mathcal H_{\gamma_2}^{(j_2)}\ar[dd]^-{SU(2)}\\
\ar@{}[r]^-{``\text{gluing}"} & \\
\mathbb G\bigl/SU(2)\ar[r]^-{\tilde\phi_{0}}\ar[d]_-{\pi_0} & \mathcal H_{\gamma_1\circ\gamma_2}^{(j)}\\
\mathbb G\bigl/\mathbb G_0\ar[r]^-\cong & \mathcal O\ar[u]_-{i_\mathcal O}
}
\end{equation}

\noindent
\\We recall that the group $\mathbb G$ is a group of local unitary transformations which as such do not modify the degree of entanglement along the orbit starting at the selected fiducial state. In the specific case under consideration, the group $\mathbb G$ is $SU(2)$ and its action on the bipartite Hilbert space (\ref{f75}) is realized through a product representation

\begin{equation}\label{f79}
U(\mathcal H)=U(\mathcal H_{\gamma_1})\otimes U(\mathcal H_{\gamma_2})\;,
\end{equation}

\noindent
whose infinitesimal generators are given by the SU(2)-generators tensored by the identity of one of the subsystems. Indeed, each subsystem Hilbert space reads as

\begin{equation}\label{f80}
\mathcal H_{\gamma_i}^{(j)}\cong\mathcal V^{(j_i)}\otimes\mathcal V^{(j_i)*}\qquad\quad (i=1,2)\;,
\end{equation}

\noindent
and so the bipartite Hilbert space (\ref{f75}) can be regarded as

\begin{equation}\label{f81}
\mathcal H\cong(\mathcal V^{(j_1)}\otimes\mathcal V^{(j_1)*})\otimes(\mathcal V^{(j_2)}\otimes\mathcal V^{(j_2)*})\;.
\end{equation}

\noindent
Now, formally speaking, the gluing operation $\gamma=\gamma_1\circ\gamma_2$ corresponds to select the subspace

\begin{equation}\label{f82}
\mathcal V^{(j_1)}\otimes\text{Inv}_{SU(2)}\bigl[\mathcal V^{(j_1)}\otimes\mathcal V^{(j_2)*}\bigr]\otimes\mathcal V^{(j_2)}\;\subset\;\mathcal H\,,
\end{equation}

\noindent
\\which coincides with the space

\begin{equation}\label{f83}
\mathcal V^{(j)}\otimes\mathcal V^{(j)*}\cong\mathcal H_\gamma^{(j)}\qquad,\qquad j=j_1=j_2
\end{equation}

\noindent
since, according to the \textit{Schur's lemma} \cite{LQG31}, when we have only two spin representations the invariant bivalent intertwining operator $\mathcal V^{(j_1)}\rightarrow\mathcal V^{(j_2)}$ is either proportional to the identity if $j_1=j_2$ or zero if $j_1\neq j_2$, i.e., the invariant subspace is trivial.\\We are thus brought back to the situation of the previous section. The pulled-back Hermitian tensor $\mathcal K$ is again given by (\ref{f45}) with a fiducial state now given by (\ref{f76}) and the spin operators $J$'s act non-trivially only at the free endpoints of the resulting new link. Hence, there is no need to repeat our calculations and we only notice that, coherently with the general considerations of Sec. 3.6, we have:

\begin{itemize}

\item In the skewsymmetric part:

\begin{equation}\label{f84}
\begin{split}
(D_A)_{ab}&=\braket{0|[J_a,J_b]_-\otimes\mathds1|0}\\
&=\frac{1}{2j+1}\sum_{mn\ell m'n'\ell'}\overline{c_{m'n'}}c_{mn}\braket{j,m',\ell'|[J_a,J_b]_-|j,m,\ell}\delta_{\ell',\ell}\delta_{n',n}\\
&=\frac{1}{2j+1}\sum_{mn\ell m'}\overline{c_{m'n}}c_{mn}\braket{j,m',\ell|[J_a,J_b]_-|j,m,\ell}\\
&=\sum_{mnm'}\overline{c_{m'n}}c_{mn}\braket{j,m'|[J_a,J_b]_-|j,m}\;,
\end{split}
\end{equation}

\noindent
where in the last step we have used the normalization $\braket{j,\ell|j,\ell}=1$ and the fact that the sum over $\ell$ gives an overall $2j+1$ factor. We then see that when the fiducial state is maximally entangled, i.e., all the Schmidt coefficients are equal, we end up with the trace of the commutator which is zero.

\item By similar arguments, when $\ket0$ is separable, we see that the block-off-diagonal matrices of the symmetric part vanish:

\begin{equation}\label{f85}
\begin{split}
C_{ab}&=\braket{0|J_a\otimes J_b|0}-\braket{0|J_a\otimes\mathds1|0}\braket{0|\mathds1\otimes J_b|0}\\
&=\braket{J_a}\braket{J_b}-\braket{J_a}\braket{J_b}\\
&=0\;.
\end{split}
\end{equation}

\end{itemize}

\section{Discussion and Interpretation}

As stressed in Chapter 1, spin network states are supposed to be the fundamental degrees of freedom representing the quantum structure of spatial three-dimensional geometry\footnote{This point of view is shared by most of the background-independent QG approaches, not only LQG. From the spin foam point of view \cite{LQG8}, for istance, a spin network can be thought of as a trivial case of a spin foam with no spacetime vertices thus representing a static spacetime set-up.}. According to \cite{F17}, the degrees of freedom of a spin network flow through the network itself which can be then considered as a quantum circuit, i.e., the SU(2)-representation vectors living on the edges of the graph evolve along the edges then meet and intertwine at the nodes, the latter playing the role of quantum gates. This is a quite interesting interpretation from a relational point of view. Indeed, as already remarked, in a diffeomorphism invariant context, only relations between objects have physical meaning and the physical content of the theory should be understood in a purely relation way. A quantum space state essentially defines a set of correlations between the (sub-)regions of the spin network and ultimately between regions of space. In the spirit of Penrose's original idea \cite{LQG58,LQG2}, spin networks then come to be networks of quantum correlations (entanglements) between regions of space. The 3-dimensional quantum space can be therefore identified with the set of processes that can occur on it, processes represented by the various quantum channels forming the network, and its properties (e.g., connectivity and geometry) should be reconstructed from the quantum information encoded in the fundamental degrees of freedom. It is therefore crucial to understand and properly characterize the intrinsic informational content of spin networks.\\In this work we have tried to address this issue using the tensorial structures available in the geometric formulation of Quantum Mechanics and their characterization of entanglement. Specifically, we implemented a three-steps procedure to build up these structures and we have applied it on a single link state which may be regarded as the most simple circuit consisting of two ``one-valent nodes'' (semilink states) and a link connecting them. Indeed, even at the level of the basis states we associate a state $\ket{j,m,n}$ with an irreducible representation of the group $SU(2)$ whose Wigner matrix $D^{(j)}_{mn}$ is actually an amplitude $\braket{j,m|\hat h(A)|j,n}$ from an initial to a final spin state given by the transformed state under the group action. Hence, a Wilson line must be intended as the description of the transformation relating the two (quantum) references represented by the endpoints, i.e., it can be regarded as a process connecting two spin states and generating the minimal element of geometry\footnote{Note that the two states are characterized by the same spin $j$ since the action of $SU(2)$ on such states can modify the magnetic moments but not the spin quatum number.}. Such a relational interpretation is well described in our formalism where, for any fixed spin $j$, the single link Hilbert space is regarded as a bipartite space $\mathcal H^{(j)}\cong\mathcal V^{(j)}\otimes\mathcal V^{(j)*}$ of two spin-$j$ irreducible representation Hilbert spaces (the presence of the dual space is dued to the opposite ingoing orientation at the second endpoint), and the link itself is actually an entangled state of the two semilink spin states.\\Now the point is to understand which informations can be extracted from the tensorial structures contructed on the space of states by means of the geometric formalism developed so far. First of all, let us notice that, being tensors on the state space, they depend on the algebraic and combinatorial data carried by the spin-network thus concerning the pregeometric level of the theory. The latter acquires a geometric character when we define gauge-invariant operators (Dirac observables) acting on the quantum states which are then interpreted as the building blocks of a \textit{quantum geometry} in the dual simplicial picture. In a background-independent framework, such a ``microscopic'' quantum geometry is at the root of space(-time) but how its geometry can be reconstructed in general remains to be undestood. It is actually diffeomorphism invariance to make things trickier. For istance, at a purely combinatorial level, a notion of spatial distance has no meaning. Indeed, in order to say that two points or two regions are close or far in space, we need an ambient manifold and use the geometric structure of such space (i.e., a background metric) to define a notion of distance. In some sense, in order to define a spatial distance, we need to localize points in some space.\\Therefore, concerning the single link case, it seems more natural to look at the notion of connectivity rather than of distance. The metric tensor constructed in this work indeed lives on the space of states not on space(-time). More precisely, it is given by the pull-back of the Fubini-Study metric tensor from the ray space to the Hilbert space. From this point of view, the entanglement monotone $\text{Tr}(C^TC)$ involving the off-diagonal blocks of pulled-back tensor can be interpreted as a measure of the existence of the link itself, i.e., as a measure of connectivity in the sense of graph topology. Indeed, as discussed in Sec. 3.8, in general it can be regarded as a distance with respect to the separable case, i.e.:

\begin{equation}\label{f86}
\text{Tr}(C^TC)\propto\text{Tr}(R^\dagger R)\qquad,\qquad R:=\rho-\rho_A\otimes\rho_B\;.
\end{equation}

\noindent
Thus, it is zero when $\rho$ is separable and it is maximum when the state is maximally entangled. By focusing now on the case of a single link reinterpreted as an entangled state of two semilinks, we may conclude that

$$
\boxed{\begin{split}
&\,\text{Tr}(C^TC)=0\quad\leftrightarrow\quad\text{unentangled semilinks/no link}\quad\leftrightarrow\quad\text{not connected}\,\\
&\,\text{Tr}(C^TC)\neq0\quad\leftrightarrow\quad\;\;\text{link as entangled semilinks}\;\;\;\quad\leftrightarrow\quad\;\;\;\text{connected}\,
\end{split}
}
$$

\noindent
\\This is even more clear in the case of two links where the entanglement is directly associated to their gluing into a single link. Hence, we have a relation between entanglement and graph connectivity as we expect from the relational point of view discussed before where the network graph itself is regarded as describing the correlations between its nodes and hence there will be a link/a process connecting two points only if they are correlated (correlations which translate into entanglement at the quantum level).\\Moreover, in the maximally entangled case, which corresponds to the gauge-invariant Wilson loop state, we find that the entanglement measure (\ref{f63}) depends only on the spin number $j$ labelling the link through the eigenvalue $j(j+1)$ of the area operator.\\Therefore, already at the level of the simplest example of a single link state, the analysis developed so far seems to suggest a connection between entanglement on spin networks states and their geometric properties. For istance, the gluing of two link in a gauge-invariant way resulting into a (locally) maximally entangled state ensures the matching of their dual surfaces into a unique surface dual to the single link resulting from the gluing. Pushing this interpretation further we may argue that the \textit{connectivity} and \textit{geometricity} of the fundamental degrees of freedom might emerge from their entanglement properties. In order to check this statement we need to extend our analysis to more complicated spin networks. Next step may be for istance to consider the gluing of 4 Wilson lines into a 4-valent node. Only in the maximally entangled case in which we associate the resulting node with an invariant tensor (intertwiner) corresponds to a well-defined geometric figure (tetrahedron), i.e.:
$$
\boxed{\begin{matrix}
\text{maximally}\\
\text{entangled}
\end{matrix}
}\longleftrightarrow\boxed{\begin{matrix}
\text{gauge-invariant}\\
\text{gluing}
\end{matrix}
}\longleftrightarrow\boxed{\begin{matrix}
\text{closure}\\
\text{condition}
\end{matrix}
}
$$
\\Moreover, since there are more degrees of freedom involved, the analysis may reveal some new connection with other geometric observables (e.g., volume) which are trivial in the single link case.\\We stress again that however we are still at a pregeometric level. By this we mean that the advocated entanglement-geometry connection concerns abstract non-embedded objects and so the resulting geometric features are not yet those of the physical space geometry. Some ideas concerning this and further possibilities will be outlined in the section of conclusions.

\chapter*{\textbf{Conclusions and Outlook}}
\phantomsection
\addcontentsline{toc}{chapter}{Conclusions and Outlook}
\markboth{Conclusions and Outlook}{}

Motivated by the idea that, in the background independent framework of a quantum theory of gravity, entanglement is expected to play a key role in the reconstruction of spacetime geometry, this dissertation is a preliminary investigation towards the possibility of using the formalism of Geometric Quantum Mechanics (GQM) to give a tensorial characterization of entanglement on spin network states. Our analysis focuses on the simple case of a single link graph state for which we define a dictionary to construct a Riemannian metric tensor and a symplectic structure on the space of states. The manifold of (pure) quantum states is then stratified in terms of orbits of equally entangled states and the block-coefficient matrices of the corresponding pulled-back tensors fully encode the information about separability and entanglement. In particular, the off-diagonal blocks $C$ define an entanglement monotone Tr$(C^TC)$ interpreted as a distance with respect to the separable state.\\ \\The main achievements of our constructions are:

\begin{enumerate}
\item A formalism which fits well to a purely relational interpretation of the link as an elementary process describing the quantum correlations between its endpoints;
\item A quantitative characterization of graph connectivity by means of the entanglement monotone Tr$(C^TC)$ which comes to be a measure of the existence of the process/link;
\item A connection between the GQM formalism and the (simplicial) geometric properties of the quantum states through entanglement. In the maximally entangled case, which for the single link corresponds to a gauge-invariant loop, the entanglement monotone is actually proportional to a power of the corresponding expectation value of the area operator. 
\end{enumerate}

\noindent
These results may be intended as the starting point of a long-term program whose final goal should be to understand in full generality how the tensorial structures defined on the space of spin network states can be used to characterize their geometric features and, in particular, how they can provide new insights to reconstruct the (quantum) geometry of spacetime. We have seen that even at the simplest level of a single link we can already grasp some connections between entanglement and geometry. Obviously, being this case so simple, it does not enable us to explore the full spectrum of possibilities and, as already mentioned, we need to extend our construction to more general cases. Let us then close by sketching some future perspective:

\begin{itemize}
\item A generic region of a spin network can be thought as an intertwiner between the N links punturing the dual surface. From the point of view of the surface, a state of geometry of that region is then described by a superposition of the possible N-valent intertwiners \cite{F17}. It has been proposed in \cite{F3} that a notion of distance between two regions of space should be derived in terms of the entanglement between the two regions A and B of the underlying spin network induced by the rest of the network. After a coarse-graining procedure the influence of the rest of the spin network amounts to a virtual link connecting the two intertwiners. Therefore, we propose to regard the whole system as a bipartite system in which each subsystem is a $n$-level system, where the number of levels is determined by the degeneracy of the new intertwiner space resulting from the coarse-graining and as such it will depend also on the spin labelling the virtual link. Even if not directly leading to a notion of distance, by studying this kind of correlations with our tensors, we expect for istance the entanglement monotone $\text{Tr}(C^TC)$ to be now a good candidate for a measure of spatial connectivity \cite{F18}.
\item  We may focus on coherent states and exploit their interpretation as semiclassical states to study the classical limit of the metric tensor. Let us also notice that in this case we are selecting a particular family of states with their own parameter space. This should enable us to exploit also the connection between the Fubini-Study and the Fisher-Rao metrics and the related tools of Information Geometry. 
\item The analysis of entanglement with classical tensors has been extended also to the case of mixed states \cite{GQM8}. The case of Gibbs states, in which the expectation values of (geometric) observables such as area play the role of the parameters of the exponential family of maximally mixed states, would be interesting to study black holes.
\item A further interesting aspect concerns the very interpretation of these tensors in those cases where the space of states is a tensor product of boundary states spaces of a process. The case of the single link, where the Hermitian tensor can be associated with an amplitude from an initial to a final spin state, may be generalized to a full (spin foam) path integral amplitude, meant as a process generating a region of space-time. In this case, the Fubini-Study metric would provide a metric for the space-time region. This setting has interesting formal analogies with the general boundary formalism \cite{F19,F20}.
\end{itemize}

\part*{Appendices}
\addcontentsline{toc}{chapter}{Appendices}

\appendix

\chapter{\textbf{The Theta Graph Spin Network}}
\markboth{A. The Theta Graph Spin Network}{}

Following \cite{LQG7}, in this appendix we present an explicit construction of the gauge invariant spin network state based on a simple graph.\\

\begin{wrapfigure}{l}{0.49\textwidth}
\includegraphics[width=0.49\textwidth]{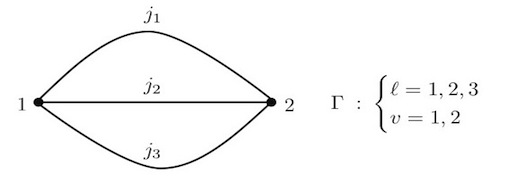} 
\end{wrapfigure}

\noindent
The example considered here is that of the so-called \textit{theta graph} which consists of two 3-valent nodes connected by three links labelled by spin numbers $j_1,j_2,j_3$ as schematically showed in the figure.\\ \\According to the Peter-Weyl decomposition (\ref{qg62}), a generic cylindrical function for this graph can be expressed as:
\begin{equation}
\psi_\Gamma(h_1,h_2,h_3)=\sum_{\vec j, \vec m, \vec n}f_{m_1,m_2,m_3,n_1,n_2,n_3}^{j_1j_2j_3}D^{(j_1)}_{m_1n_1}(h_1)D^{(j_2)}_{m_2n_2}(h_2)D^{(j_3)}_{m_3n_3}(h_3)\;.
\end{equation}
Since a SU(2)-gauge transformation acts only on the group elements, the gauge invariant part of $\psi_\Gamma$ can be determined by considering the gauge-invariant part of the product of Wigner matrices that is implemented via a group averaging, i.e.:

\begin{equation}\label{ga}
\begin{split}
\psi_\Gamma^{(inv)}(h_1,h_2,h_3)&=\sum_{\vec j, \vec m, \vec n}f_{m_1,m_2,m_3,n_1,n_2,n_3}^{j_1j_2j_3}\biggl[D^{(j_1)}_{m_1n_1}(h_1)D^{(j_2)}_{m_2n_2}(h_2)D^{(j_3)}_{m_3n_3}(h_3)\biggr]_{inv}\\
&=\sum_{\vec j, \vec m, \vec n}f_{m_1,m_2,m_3,n_1,n_2,n_3}^{j_1j_2j_3}\int dg_1dg_2\biggl[D^{(j_1)}_{m_1n_1}(g_1h_1g_2^{-1})\\
&\qquad\qquad\qquad D^{(j_2)}_{m_2n_2}(g_1h_2g_2^{-1})D^{(j_3)}_{m_3n_3}(g_1h_3g_2^{-1})\biggr]
\end{split}
\end{equation}

\noindent
By using now the property
\begin{equation}
D^{(j_\ell)}_{m_\ell n_\ell}(g_1h_\ell g_2^{-1})=\sum_{\alpha_\ell,\beta_\ell}D^{(j_\ell)}_{m_\ell\alpha_\ell}(g_1)D^{(j_\ell)}_{\alpha_\ell\beta_\ell}(h_\ell)D^{(j_\ell)}_{\beta_\ell n_\ell}(g_2^{-1})\qquad(\ell=1,2,3)
\end{equation}

\noindent
Eq. (\ref{ga}) can be rewritten as

\begin{equation}
\begin{split}
\psi_\Gamma^{(inv)}(h_1,h_2,h_3)&=\sum_{\vec j, \vec m, \vec n,\vec\alpha,\vec\beta}f_{m_1,m_2,m_3,n_1,n_2,n_3}^{j_1j_2j_3}\mathcal I_{m_1m_2m_3\alpha_1\alpha_2\alpha_3}\\
&\qquad\qquad\mathcal I_{\beta_1\beta_2\beta_3n_1n_2n_3}D^{(j_1)}_{\alpha_1\beta_1}(h_1)D^{(j_2)}_{\alpha_2\beta_2}(h_2)D^{(j_3)}_{\alpha_3\beta_3}(h_3)
\end{split}
\end{equation}

\noindent
where

\begin{equation}
\mathcal I_{m_1m_2m_3\alpha_1\alpha_2\alpha_3}=\int dg_1D^{(j_1)}_{m_1\alpha_1}(h_1)D^{(j_2)}_{m_2\alpha_2}(h_2)D^{(j_3)}_{m_3\alpha_3}(h_3)
\end{equation}

\noindent
is the projector on the gauge invariant space. The integral on the r.h.s. of the expression of the projector can be written in terms of the normalized Clebsch-Gordan coefficients or Wigner $3j$ symbols as \cite{appA}

\begin{equation}
\int dg_1D^{(j_1)}_{m_1\alpha_1}(h_1)D^{(j_2)}_{m_2\alpha_2}(h_2)D^{(j_3)}_{m_3\alpha_3}(h_3)=\begin{pmatrix}
j_1 & j_2 & j_3\\
m_1 & m_2 & m_3
\end{pmatrix}\overline{\begin{pmatrix}
j_1 & j_2 & j_3\\
\alpha_1 & \alpha_2 & \alpha_3
\end{pmatrix}}\;,
\end{equation}

\noindent
and we get

\begin{equation}\label{A7}
\begin{split}
\psi_\Gamma^{(inv)}(h_1,h_2,h_3)&=\sum_{\vec j, \vec m, \vec n,\vec\alpha,\vec\beta}f_{m_1,m_2,m_3,n_1,n_2,n_3}^{j_1j_2j_3}\begin{pmatrix}
j_1 & j_2 & j_3\\
m_1 & m_2 & m_3
\end{pmatrix}\overline{\begin{pmatrix}
j_1 & j_2 & j_3\\
\alpha_1 & \alpha_2 & \alpha_3
\end{pmatrix}}\\
&\quad\begin{pmatrix}
j_1 & j_2 & j_3\\
\beta_1 & \beta_2 & \beta_3
\end{pmatrix}\overline{\begin{pmatrix}
j_1 & j_2 & j_3\\
n_1 & n_2 & n_3
\end{pmatrix}}D^{(j_1)}_{\alpha_1\beta_1}(h_1)D^{(j_2)}_{\alpha_2\beta_2}(h_2)D^{(j_3)}_{\alpha_3\beta_3}(h_3)\\
&=\sum_{\vec j}\prod_\ell D^{(j_\ell)}(h_\ell)\prod_v i_v\sum_{\vec m,\vec n}f_{m_1,m_2,m_3,n_1,n_2,n_3}^{j_1j_2j_3}\begin{pmatrix}
j_1 & j_2 & j_3\\
m_1 & m_2 & m_3
\end{pmatrix}\overline{\begin{pmatrix}
j_1 & j_2 & j_3\\
n_1 & n_2 & n_3
\end{pmatrix}}\\
&=\sum_{\vec j}\tilde f^{j_1j_2j_3}\prod_\ell D^{(j_\ell)}(h_\ell)\prod_v i_v
\end{split}
\end{equation}

\noindent
where the short-hand notation $i_v$ for the $3j$ symbols denotes the invariant tensors in the space of all the spins that enter the node $v$, i.e., $i_v\in\text{Inv}[\bigotimes_{\ell\in v}\mathcal V^{(j_\ell)}]$, and in the last equality we have included the sums over the magnetic numbers $\vec m, \vec n$ into the new coefficients $\tilde f$.\\Thus, in this simple example we explicitly see how the implementation of gauge invariance via group averaging essentially amounts to insert the projector $\mathcal I_v$ at each node $v$ of the graph and ultimately to a gauge-invariant state given by a linear combination of products of the Wigner matrices associated to the holonomies along the links contracted with the intertwiners associated to the nodes.

\chapter{\textbf{Symplectic Reduction and Geometric Quantization}}
\markboth{B. Symplectic Reduction and Geometric Quantization}{}

This appendix is intended as a complement to the discussion about the classical and quantum tetrahedron in 3 dimensions which takes place in Ch. 1. As such, we will review the main aspects of symplectic reduction procedures pointing out their role in the construction of the classical phase space of the tetrahedron. This will be the starting point for the geometric quantization that allows us to construct the Hilbert space of a quantum bivector in 3 dimensions.

\section{Kirillov-Konstant Poisson Structure}

Let us first recall the general definition of symplectic and Poisson reduction \cite{app1}. Let $(\mathbb M, \omega)$ be a symplectic manifold and $\mathbb G$ a Lie group acting on it. If

\begin{equation}\label{B1}
\mathcal J:\mathbb M\longrightarrow\mathfrak g^*
\end{equation}

\noindent
is an $Ad^*$-equivariant momentum map for a canonical $\mathbb G$-action \cite{GQM15} and $0\in\mathfrak g^*$ is a regular value of $\mathcal J$ such that the isotropy group $\mathbb G_0$ of $0$ acts freely and properly on $\mathcal J^{-1}(0)\subset\mathbb M$, then there exists a unique symplectic structure $\omega_0$ on $\mathbb M_0=\mathcal J^{-1}(0)/\mathbb G_0$ whose pullback to $\mathcal J^{-1}(0)$ equals the restriction of $\omega$ to $\mathcal J^{-1}(0)$, i.e.

\begin{equation}\label{B2}
i_0^*\,\omega=\pi^*_0\,\omega_0\;,
\end{equation}

\noindent
where

\begin{equation}\label{B3}
i_0: \mathcal J^{-1}(0)\longrightarrow\mathbb M\qquad,\qquad\pi_0:\mathcal J^{-1}(0)\longrightarrow\mathbb M_0
\end{equation}

\noindent
are respectively the inclusion map and the projection. The quotient $\mathbb M_0=\mathcal J^{-1}(0)/\mathbb G_0$ is called a \textit{symplectic reduction} of $\mathbb M$ by $\mathbb G$. Moreover, if $\mathbb G$ acts freely and properly on $\mathbb M$, then the symplectic leaves of $\mathbb M/\mathbb G$ are the reduced manifolds $\mathcal J^{-1}(\mathcal O)/\mathbb G$, with $\mathcal O$ a coadjoint orbit, according to the following diagram:
\begin{equation}\label{B4}
\xymatrix{
\mathbb{G}\ar[r]^-{Ad^*}\ar[dr]_-\rho&\mathfrak{g}^*&\ar[l]_-{i_\mathcal O}\ar[d]^-{\mathcal{J}^{-1}}\mathcal O \cong\mathfrak g^*/Ad^*(\mathbb G)\\
 & \ar[u]^-{\mathcal{J}}(\mathbb M,\omega)\ar@{-->}[dd]_-{i^*}&\ar[l]_-{i}\mathcal{J}^{-1}(\mathcal O)\subset\mathbb M\ar[d]^-\pi\\
 & & (\mathbb M_\mathcal O=\mathcal{J}^{-1}(\mathcal O)/\mathbb G_\mathcal O, \omega_\mathcal O)\ar@{-->}[d]^-{\pi^*}\\
 & (\mathcal{J}^{-1}(\mathcal O),i^*\omega)\ar@{=}[r]&(\mathcal{J}^{-1}(\mathcal O),\pi^*\omega_\mathcal O)
}
\end{equation}
$\mathbb M_\mathcal O$ inherits a unique Poisson structure provided by that of the reduced symplectic structure $\omega_\mathcal O$. This follows from the fact that $\mathbb G$-invariant functions produce Hamiltonian vector fields $\pi$-related to their reductions. Let us recall very briefly the general construction \cite{app2,app3}. The dual $\mathfrak g^*$ of any Lie algebra $\mathfrak g$ is a vector space with an additional structure that makes it a Poisson manifold, i.e., a manifold with a Poisson bracket on its algebra of functions. Indeed, using the property that $Lin(\mathfrak g^*,\mathbb R)=(\mathfrak g^*)^*\cong\mathfrak g$, we can regard the elements of the Lie algebra $\mathfrak g$ as linear functions on the dual vector space $\mathfrak g^*$ and consequently, $[\ell,m]$ for $\ell,m\in\mathfrak g$ can be thought of as a function on $\mathfrak g^*$. This allows us to define a Poisson bracket (PB) on the linear functions on $\mathfrak g^*$ by considering the bivector field $\Omega$ on $\mathfrak g^*$ such that
\begin{equation}\label{B5}
[\ell,m]=\Omega(d\ell,dm)\;,
\end{equation}
and defining
\begin{equation}\label{B6}
\{f,f'\}=\Omega(df,df')\qquad\forall f,f'\in\mathcal F(\mathfrak g^*)\;.
\end{equation}
Such a PB can be then extended to all smooth functions by means of (\ref{B6}) evaluated on the coordinate functions associated to a basis for the Lie algebra $\mathfrak g$.\\Now, since (\ref{B6}) is linear in $df'$, it can be written as the evaluation of $df'$ on a vector field $X_f$ on $\mathfrak g^*$, i.e.:
\begin{equation}\label{B7}
\{f,f'\}=df'(X_f)=i_{X_f}df'=L_{X_f}f'=X_f(f')\qquad,\qquad X_f\in\mathfrak{X}(\mathfrak g^*)\;.
\end{equation}
In particular, each Lie algebra element $\ell$ determines a vector field $X_\ell$ and the value of $X_\ell m$ at $x\in\mathfrak g^*$ is
\begin{equation}\label{B8}
(X_\ell m)(x)=\braket{x,[\ell,m]}\;,
\end{equation}
where $\braket{\cdot,\cdot}$ denotes the dual pairing. Thus, if $\mathfrak g$ is the Lie algebra of a Lie group $\mathbb G$, the vector fields $X_\ell$ are given by the natural action of $\mathbb G$ on $\mathfrak g^*$, that is the coadjoint action. A 2-form $\omega$ is compatible with the Poisson structure if:
\begin{equation}\label{B9}
\{\ell,m\}=\omega(X_\ell,X_m)\;.
\end{equation}
This however only dermines $\omega$ on the tangent space to the coadjoint orbit (the span of $X_\ell$) and therefore, $\omega$ is defined as a 2-form on each orbit which is called a \textit{symplectic leaf} of the Poisson manifold. The Poisson bivector $\Omega$ is tangent to each leaf and non-degenerate as a bilinear form on the cotangent bundle of each leaf. Its inverse is exactly the symplectic form $\omega$. This shows that the coadjoint orbits of a Lie group are symplectic manifolds on which there exists a non-degenerate closed $\mathbb G$-invariant differential 2-form called the \textit{Kirillov form}.\\In the case $\mathbb G\equiv SO(3)$, the Lie algebra and its dual can be identified with $\mathbb R^3$ with its standard vector cross product, i.e., $\mathfrak{so}(3)\cong\mathfrak{so}(3)^*\cong\mathbb R^3$. Thus, the dual pairing now becomes the Euclidean inner product and we have
\begin{equation}\label{B10}
\{\ell,m\}(x)=\braket{x,[\ell,m]}\;,
\end{equation}
with the r.h.s. now given by the triple scalar product in $\mathbb R^3$. The symplectic 2-form is given by
\begin{equation}\label{B11}
\omega(a,b)=\frac{1}{x^2}\braket{x,[a,b]}\;.
\end{equation}
In this case, the coadjoint action is just the action of rotations on the vector space of angular momenta, and so the symplectic leaves are spheres centered at the origin.

\section{Quantization}

The Hilbert space for each leaf can be constructed by means of geometric quantization, or more precisely, K\"ahler quantization. We will now sketch the procedure without going into the details for which we refer to \cite{LQG45,app4,app5} and references within. Essentially, we first choose a complex structure $J$ on each leaf that preserves the symplectic form $\omega$ thus making the leaf a K\"ahler manifold. Then we choose a holomorphic complex line bundle $\mathbb L$ over the leaf called the \textit{prequantum line bundle}, equipped with a connection whose curvature equals $\omega$, i.e.:
\begin{equation}\label{B12}
\omega(X,Y)s=i\bigl(\nabla_X\nabla_Y-\nabla_Y\nabla_X-\nabla_{[X,Y]}\bigr)s
\end{equation}
for every section $s$ of this bundle. This is possible only if the symplectic leafs satisfy the so-called Bohr-Sommerfeld condition according to which the leafs must be integral, i.e., the closed 2-form $\omega/2\pi$ must define an integral cohomology class \cite{app6}. When this is the case, we define the Hilbert space for the leaf to be the space of square-integrable holomorphic sections of $\mathbb L$ (for non integral leaves the Hilbert space is defined to be zero-dimensional).\\Let us now focus on bivectors. A bivector in $N$ dimensions is an element of $\Lambda^2\mathbb R^N$. A bivector encodes some information on the geometry of a triangle in $\mathbb R^N$. Indeed, as discussed in \cite{LQG5}, a (simple) bivector $E=e_1\wedge e_2=e_2\wedge e_3=e_3\wedge e_1$ given by the wedge product between pairs of edge vectors cycling around an oriented triangle\footnote{The above equalities of wedge products hold because of the condition $e_1+e_2+e_3=0$ which the three edge vectors have to satisfy in order to close a triangle. This means that the bivector $E$ does not depend on the choice of the edges but only on the triangle and its orientation.} determines a 2-dimensional plane in $\mathbb R^N$ with an orientation, and the norm of the bivector in twice the area of the triangle. Furthermore, using the Euclidean metric $\eta$ on $\mathbb R^N$, we may identify bivectors $E=e\wedge f$ with elements of $\mathfrak{so}(N)^*$ by means of the following isomorphism:

\begin{equation}\label{B13}
\beta\;:\;\Lambda^2\mathbb R^N\longrightarrow\mathfrak{so}(N)^*\quad s.t.\quad\beta(e\wedge f)(\ell)=\eta(\ell e, f)
\end{equation}

\noindent
for any $\ell\in\mathfrak{so}(N)$. Therefore, for a tetrahedron in 3-dimensions, a bivector associated to one of its faces can be regarded as elements of $\mathfrak{so}(3)^*$. In the case of $\mathfrak{so}(3)^*$, the integral symplectic leaves are 2-spheres centered at the origin for which the integral of $\omega$ is $2\pi$ times an integer. These are spheres $S_j$ with radii given by non-negative half-integers $j$. Each sphere $S_{j>0}$\footnote{When $j=0$ the sphere reduces to a single point, the prequantum line bundle is trivial, and the corresponding Hilbert space of holomorphic sections is 1-dimensional.} has a complex structure $J$ corresponding to the usual complex structure on the Riemann sphere. Explicitly, using the identification $\mathfrak{so}(3)^*\cong\mathfrak{so}(3)$, we get the following expression for $J$ at a point $x$:

\begin{equation}\label{B14}
J(a)=\frac{1}{|x|}[x,a]\;.
\end{equation}

\noindent
The sphere thus becomes a K\"ahler manifold with Riemannian metric given by:

\begin{equation}
g(a,b)=\omega(a,Jb)=\frac{1}{|x|}\braket{a,b}\;.
\end{equation}

\noindent
We may associate to any Lie algebra element $\ell\in\mathfrak{so}(3)$ a self-adjoint operator $\hat\ell$ on the Hilbert space given by the map

\begin{equation}\label{B15}
\hat\ell\;:\;s\longmapsto-i\hbar\nabla_{X_\ell}s+\ell s\;,
\end{equation}

\noindent
where $\ell$ is regarded as a function on the coadjoint orbit multipling the section $s$ pointwise, and $X_\ell$ denotes the Hamiltonian vector field corresponding to the classical observable $\ell$ (i.e., $i_{X_\ell}\omega=d\ell$). The map (\ref{B15}) gives a representation of $\mathfrak{so}(3)$ which is just the usual spin-$j$ representation $\mathcal V^{(j)}$. Therefore, we can finally obtain the Hilbert space of a quantum bivector in 3-dimensions by taking the sum over all symplectic leaves:

\begin{equation}\label{B16}
\mathcal H=\bigoplus_j\mathcal V^{(j)}\;.
\end{equation}

\noindent
\textbf{Remark:} If $E^k$ denote a basis of $\mathfrak{so}(3)$ satisfying the PB relations $\{E^k,E^r\}=\varepsilon^{kr}_sE^s$ when thought of as coordinate functions on $\mathfrak{so}(3)^*$, then by the above argument we obtain self-adjoint operators $\hat E^k$ on $\mathcal H$ satisfying the usual angular momentum commutation relations $[\hat E^k,\hat E^r]=i\hbar\varepsilon^{kr}_s\hat E^s$. We may interpret these operators as observables measuring the three components of the quantum bivector (as can be seen by geometrically quantizing the coordinate functions on the space of bivectors). Their failure to commute means that the components of a quantum bivector cannot in general be measured simultaneously with complete precision.

\chapter{\textbf{Hermitian Tensor on Orbits of Maximally Entangled and Separable Wilson Line States}}
\markboth{C. Hermitian Tensor on Orbits of Maximally Entangled and Separable Wilson Line States}{}

In this appendix, we will present the explicit calculation of the matrix elements of the pull-back of the Hermitian tensor on the orbit of unitarily related Wilson line states having as fiducial state a maximally entangled and a separable state, respectively.

\section{Case 1: Maximally Entangled State}

As discussed in Sec. 4.4, the maximally entangled Wilson line state corresponds to the Wilson loop state

\begin{equation}\label{C1}
\ket{\psi_{WL}}=\frac{1}{\sqrt{2j+1}}\sum_k\,\ket{j,k}\otimes\bra{j,k}\equiv\ket{0}\;\in\;\mathcal H_\gamma^{(j)}\cong\mathcal V^{(j)}\otimes\mathcal V^{(j)*}\;,
\end{equation}

\noindent
\\with pure state density matrix

\begin{equation}\label{C2}
\rho_0=\ket{0}\bra{0}=\frac{1}{2j+1}\sum_{k,k'}\bigl(\ket{j,k}\bra{j,k'}\bigr)\otimes\bigl(\ket{j,k'}\bra{j,k}\bigr)\;\in\;D^1(\mathcal H_\gamma^{(j)})\;.
\end{equation}

\noindent
Using Eqs. (\ref{f44}), we compute the block-coefficient matrices of the pulled-back Hermitian tensor on the co-adjoint orbit submanifold $\mathcal O_{\rho_0}$ embedded in the ray space $\mathcal R(\mathcal H_\gamma^{(j)})\cong D^1(\mathcal H_\gamma^{(j)})$ as follows.\\ \\Let us first focus on the off-diagonal blocks $C$ of the Riemannian metric tensor whose matrix elements are given by:

\begin{equation}\label{C3}
\begin{split}
C_{ab}&=\text{Tr}(\rho_0J_a\otimes J_b)-\text{Tr}(\rho_0J_a\otimes\mathds1)\text{Tr}(\rho_0\mathds1\otimes J_b)\\
&=\frac{1}{2j+1}\sum_{k,k'}\text{Tr}\Bigl[\bigl(\ket{j,k}\bra{j,k'}J_a\bigr)\otimes\bigl(\ket{j,k'}\bra{j,k}J_b\bigr)\Bigr]+\\
&\quad-\frac{1}{(2j+1)^2}\sum_{k,k'}\text{Tr}\Bigl[\bigl(\ket{j,k}\bra{j,k'}J_a\bigr)\otimes\bigl(\ket{j,k'}\bra{j,k}\bigr)\Bigr]\cdot\\
&\quad\cdot\sum_{k,k'}\text{Tr}\Bigl[\bigl(\ket{j,k}\bra{j,k'}\bigr)\otimes\bigl(\ket{j,k'}\bra{j,k}J_b\bigr)\Bigr]\\
&=\frac{1}{2j+1}\sum_{k,k',k'',k'''}\delta_{k''',k}\braket{j,k'|J_a|j,k'''}\delta_{k'',k'}\braket{j,k|J_b|j,k''}+\\
&\quad-\frac{1}{(2j+1)^2}\sum_{k,k',k'',k'''}\delta_{k''',k}\braket{j,k'|J_a|j,k'''}\delta_{k'',k'}\delta_{k,k''}\cdot\\
&\qquad\cdot\sum_{k,k',k'',k'''}\delta_{k''',k}\delta_{k',k'''}\delta_{k'',k'}\braket{j,k|J_b|j,k''}\\
&=\frac{1}{2j+1}\sum_{k,k'}\braket{j,k'|J_a|j,k}\braket{j,k|J_a|j,k'}+\\
&\quad-\frac{1}{(2j+1)^2}\sum_{k}\braket{j,k|J_a|j,k}\cdot\sum_{k}\braket{j,k|J_b|j,k}\\
&=\frac{1}{2j+1}\sum_{k,k'}\braket{j,k'|J_a|j,k}\braket{j,k|J_b|j,k'}\;,
\end{split}
\end{equation}

\noindent
\\where in the last step we have used the traceless property of the $SU(2)$ generators.\\Now, since the matrix $C$ is symmetric, we need to compute only the diagonal terms and the upper triangle. By using the the following algebraic relations from the theory of angular momentum \cite{appA}

\begin{equation}\label{C4}
\braket{j,k'|J_1|j,k}=\frac{1}{2}\sqrt{j(j+1)-k'k}\,(\delta_{k',k+1}+\delta_{k'+1,k})\;,
\end{equation}

\begin{equation}\label{C5}
\braket{j,k'|J_2|j,k}=\frac{1}{2i}\sqrt{j(j+1)-k'k}\,(\delta_{k',k+1}-\delta_{k'+1,k})\;,
\end{equation}

\begin{equation}\label{C6}
\braket{j,k'|J_3|j,k}=k\,\delta_{k',k}\;,
\end{equation}

\noindent
we then find

\begin{itemize}
\item\begin{equation}\label{C7}
\begin{split}
C_{33}&=\frac{1}{2j+1}\sum_{k,k'}\braket{j,k'|J_3|j,k}\braket{j,k|J_3|j,k'}\\
&=\frac{1}{2j+1}\sum_{k=-j}^{+j}k^2\\
&=\frac{1}{3}j(j+1)\;,
\end{split}
\end{equation}

\noindent
where in the last step we have used the relation

\begin{equation}\label{C8}
\sum_{k=-j}^{+j}k^2=2\sum_{k=1|k=\frac{1}{2}}^{+j}k^2=2\,\frac{j(j+1)(2j+1)}{6}=\frac{1}{3}\,j(j+1)(2j+1)\;,
\end{equation}

\noindent
which holds both for integer and half-integer spins (i.e., $\forall j\in\frac{\mathbb N}{2}$) as explicitly pointed out by the notation $k=1|k=\frac{1}{2}$ in the sum.

\item\begin{equation}\label{C9}
\begin{split}
C_{11}&=\frac{1}{2j+1}\sum_{k,k'}\braket{j,k'|J_1|j,k}\braket{j,k|J_1|j,k'}\\
&=\frac{1}{4(2j+1)}\sum_{k,k'}[j(j+1)-k'k](\delta_{k',k+1}+\delta_{k'+1,k})^2\\
&=\frac{1}{4(2j+1)}\sum_{k}\bigl\{[j(j+1)-k(k+1)]+[j(j+1)-k(k-1)]\bigr\}\\
&=\frac{1}{2(2j+1)}\sum_{k}[j(j+1)-k^2]\\
&=\frac{1}{2(2j+1)}\sum_{k=-j}^{+j}j(j+1)-\frac{1}{(2j+1)}\sum_{k=1}^{+j}k^2\\
&=\frac{1}{2}j(j+1)-\frac{1}{6}j(j+1)\\
&=\frac{1}{3}j(j+1)\;,
\end{split}
\end{equation}

\noindent
where, in the passage from the third-last to the second-last line, we have used the fact that the sum over k in the first term gives a factor (2j+1), and Eq. (\ref{C8}) in the second term.

\item Similarly for $C_{22}$ we have

\begin{equation}\label{C10}
\begin{split}
C_{22}&=\frac{1}{2j+1}\sum_{k,k'}\braket{j,k'|J_2|j,k}\braket{j,k|J_2|j,k'}\\
&=-\frac{1}{4(2j+1)}\sum_{k,k'}[j(j+1)-k'k](\delta_{k',k+1}-\delta_{k'+1,k})(\delta_{k,k'+1}-\delta_{k+1,k'})\\
&=\frac{1}{4(2j+1)}\sum_{k}\bigl\{[j(j+1)-k(k+1)]+[j(j+1)-k(k-1)]\bigr\}\\
&=\frac{1}{3}j(j+1)\;,
\end{split}
\end{equation}

\item\begin{equation}\label{C11}
\begin{split}
C_{31}=C_{13}&=\frac{1}{2j+1}\sum_{k,k'}\braket{j,k'|J_1|j,k}\braket{j,k|J_3|j,k'}\\
&=\frac{1}{2j+1}\sum_{k}k\braket{j,k|J_1|j,k}=0\;,
\end{split}
\end{equation}

\noindent
since $\braket{j,k|J_1|j,k}=0$, as we can see from Eq. (\ref{C4}).

\item\begin{equation}\label{C12}
C_{32}=C_{23}=0\;,
\end{equation}

\noindent
since, similarly to the case of $C_{13}$, we end up with the sum $\sum_k\,k\,\braket{j,k|J_2|j,k}$ which vanishes in virtue of (\ref{C5}).

\item\begin{equation}\label{C13}
\begin{split}
C_{21}=C_{12}&=\frac{1}{2j+1}\sum_{k,k'}\braket{j,k'|J_1|j,k}\braket{j,k|J_2|j,k'}\\
&=\frac{1}{4(2j+1)}\sum_{k,k'}[j(j+1)-k'k](\delta_{k',k+1}+\delta_{k'+1,k})(\delta_{k',k+1}-\delta_{k'+1,k})\\
&=\frac{1}{4(2j+1)}\sum_{k}\bigl\{[j(j+1)-k(k+1)]-[j(j+1)-k(k-1)]\bigr\}\\
&=\frac{i}{2(2j+1)}\sum_{k=-j}^{+j}k\\
&=0\;,
\end{split}
\end{equation}
where in the last step we have used the fact that $\sum_{k=-j}^{+j}k=0$.
\end{itemize}

\noindent
Therefore, collecting the results (\ref{C7}-\ref{C14}) of the above computations, we have that for the maximally entangled fiducial Wilson loop state (\ref{C1}), the off-diagonal blocks of the pulled-back Hermitian tensor are given by:

\begin{equation}\label{C14}
C=\begin{pmatrix}
C_{11} & C_{12} & C_{13} \\
C_{21} & C_{22} & C_{23} \\
C_{31} & C_{32} & C_{33}
\end{pmatrix}=\begin{pmatrix}
\frac{1}{3}j(j+1) & 0 & 0 \\
0 & \frac{1}{3}j(j+1) & 0 \\
0 & 0 & \frac{1}{3}j(j+1)
\end{pmatrix}\;.
\end{equation}

\noindent
\\ \\\textbf{Remark:} The fact that the element $C_{21}\,(=C_{12})$ and $C_{32}\,(=C_{23})$, i.e., those involving the generator $J_2$ only once, are zero should be expected since, according to Eq. (\ref{C5}), the matrix elements of $J_2$ in the sum will give an imaginary number which as suvh it cannot enter the metric tensor, the latter being the real part of the Hermitian tensor.\\ \\Let us compute now the other blocks:

\begin{equation}\label{C15}
\begin{split}
A_{ab}&=\frac{1}{2}\text{Tr}(\rho_0[J_a,J_b]_+\otimes\mathds1)-\text{Tr}(\rho_0J_a\otimes\mathds1)\text{Tr}(\rho_0J_b\otimes\mathds1)\\
&=\frac{1}{2(2j+1)}\sum_{k,k'}\text{Tr}(\ket{j,k}\bra{j,k'}[J_a,J_b]_+)\text{Tr}(\ket{j,k'}\bra{j,k})+\\
&\quad-\frac{1}{(2j+1)^2}\sum_{k,k',k'',k'''}\delta_{k''',k}\braket{j,k'|J_a|j,k'''}\delta_{k'',k'}\delta_{k,k''}\cdot\\
&\qquad\cdot\sum_{k,k',k'',k'''}\delta_{k''',k}\braket{j,k'|J_b|j,k'''}\delta_{k'',k'}\delta_{k,k''}\\
&=\frac{1}{2(2j+1)}\sum_{k,k',k'',k'''}\delta_{k''',k}\braket{j,k'|[J_a,J_b]_+|j,k'''}\delta_{k'',k'}\delta_{k,k''}+\\
&\quad-\frac{1}{(2j+1)^2}\underset{0}{\underbrace{\sum_{k}\braket{j,k|J_a|j,k}}}\cdot\underset{0}{\underbrace{\sum_{k}\braket{j,k|J_b|j,k}}}\\
&=\frac{1}{2j+1}\sum_k\braket{j,k|J_aJ_b|j,k}\\
&=\frac{1}{2j+1}\sum_{k,k'}\braket{j,k|J_a|j,k'}\braket{j,k'|J_b|j,k}\;.
\end{split}
\end{equation} 

\noindent
\\Hence, using the relations (\ref{C4}-\ref{C6}), we find

\begin{itemize}

\item\begin{equation}\label{C16}
A_{33}=\frac{1}{2j+1}\sum_k\braket{j,k|J_3^2|j,k}=\frac{1}{2j+1}\sum_{k=-j}^{+j}k^2=\frac{1}{3}j(j+1)
\end{equation}

\noindent
\\where in the last step we have used Eq. (\ref{C8}).\\

\item\begin{equation}\label{C17}
\begin{split}
A_{11}&=\frac{1}{2j+1}\sum_{k,k'}\braket{j,k|J_1|j,k'}\braket{j,k'|J_1|j,k}\\
&=\frac{1}{4(2j+1)}\sum_{k,k'}[j(j+1)-k'k](\delta_{k,k'+1}+\delta_{k+1,k'})(\delta_{k',k+1}+\delta_{k'+1,k})\\
&=\frac{1}{4(2j+1)}\sum_{k}\{[j(j+1)-k(k+1)]+[j(j+1)-k(k-1)]\}\\
&=\frac{1}{2(2j+1)}\sum_{k=-j}^{j}[j(j+1)-k^2]\\
&=\frac{1}{2}j(j+1)-\frac{1}{2j+1}\sum_{k=1}^{j}k^2\\
&=\frac{1}{2}j(j+1)-\frac{1}{6}j(j+1)\\
&=\frac{1}{3}j(j+1)\;;
\end{split}
\end{equation}

\item\begin{equation}\label{C18}
\begin{split}
A_{22}&=\frac{1}{2j+1}\sum_{k,k'}\braket{j,k|J_2|j,k'}\braket{j,k'|J_2|j,k}\\
&=-\frac{1}{4(2j+1)}\sum_{k,k'}[j(j+1)-k'k](\delta_{k,k'+1}-\delta_{k+1,k'})(\delta_{k',k+1}-\delta_{k'+1,k})\\
&=\frac{1}{4(2j+1)}\sum_{k}\{[j(j+1)-k(k+1)]+[j(j+1)-k(k-1)]\}\\
&=\frac{1}{2(2j+1)}\sum_{k=-j}^{j}[j(j+1)-k^2]\\
&=\frac{1}{3}j(j+1)\;;
\end{split}
\end{equation}

\item Similarly to what we have seen for the block-matrix C, we also have

\begin{equation}\label{C19}
\begin{split}
A_{31}=A_{13}&=\frac{1}{2j+1}\sum_{k,k'}\braket{j,k|J_1|j,k'}\braket{j,k'|J_3|j,k}\\
&=\frac{1}{2j+1}\sum_{k}k\,\underset{0}{\underbrace{\braket{j,k|J_1|j,k}}}\\
&=0\;,
\end{split}
\end{equation}

\begin{equation}\label{C20}
\begin{split}
A_{32}=A_{23}&=\frac{1}{2j+1}\sum_{k,k'}\braket{j,k|J_2|j,k'}\braket{j,k'|J_3|j,k}\\
&=\frac{1}{2j+1}\sum_{k}k\,\underset{0}{\underbrace{\braket{j,k|J_2|j,k}}}\\
&=0\;,
\end{split}
\end{equation}

\begin{equation}\label{C21}
\begin{split}
A_{21}=A_{12}&=\frac{1}{2j+1}\sum_{k,k'}\braket{j,k|J_1|j,k'}\braket{j,k'|J_2|j,k}\\
&=\frac{1}{4i(2j+1)}\sum_{k,k'}[j(j+1)-k'k](\delta_{k,k'+1}+\delta_{k+1,k'})(\delta_{k',k+1}-\delta_{k'+1,k})\\
&=\frac{1}{4i(2j+1)}\sum_{k}\{[j(j+1)-k(k+1)]-[j(j+1)-k(k-1)]\}\\
&=\frac{i}{2(2j+1)}\sum_{k=-j}^jk\\
&=0\;,
\end{split}
\end{equation}
\end{itemize}

\noindent
\\Analogous computations hold for the block $B$. Therefore, collecting the results (\ref{C16}-\ref{C21}), the diagonal-block-matrices of the metric tensor in the maximally entangled case are given by:\\

\begin{equation}\label{C22}
A=B=\begin{pmatrix}
\frac{1}{3}j(j+1) & 0 & 0 \\
0 & \frac{1}{3}j(j+1) & 0 \\
0 & 0 & \frac{1}{3}j(j+1)
\end{pmatrix}\;.
\end{equation}

\noindent
\\Finally, as expected, in the maximally entangled case there is no symplectic structure. Indeed, we have
\begin{equation}\label{C23}
\begin{split}
(D_A)_{ab}&=\frac{1}{2}\text{Tr}\bigl(\rho_0[J_a,J_b]_-\otimes\mathds1\bigr)\\
&=\frac{1}{2(2j+1)}\sum_{k,k'}\text{Tr}\bigl(\ket{j,k}\bra{j,k'}[J_A,J_b]_-\bigr)\text{Tr}\bigl(\ket{j,k'}\bra{j,k}\bigr)\\
&=\frac{1}{2(2j+1)}\sum_{k,k',k'',k'''}\delta_{k''',k}\braket{j,k'|[J_a,J_b]_-|j,k'''}\delta_{k'',k'}\delta_{k,k''}\\
&=\frac{1}{2(2j+1)}\;\underset{\text{Tr}\bigl([J_a,J_b]_-\bigr)=0}{\underbrace{\sum_{k}\braket{j,k|[J_a,J_b]_-|j,k}}}\\
&=0\;,
\end{split}
\end{equation}

\noindent
and similarly for $(D_B)_{ab}$.

\section{Case 2: Separable State}

Let us now consider a separable fiducial state

\begin{equation}\label{C24}
\ket{0}=\ket{j_1,k_1}\otimes\bra{j_2,k_2}\;,
\end{equation}

\noindent
\\whose pure state density matrix is given by

\begin{equation}\label{C25}
\rho_0=\rho_A\otimes\rho_B=\bigl(\ket{j_1,k_1}\bra{j_1,k_1}\bigr)\otimes\bigl(\ket{j_2,k_2}\bra{j_2,k_2}\bigr)\;.
\end{equation}

\noindent  
\\In this case we have that

\begin{equation}\label{C26}
\begin{split}
C_{ab}&=\text{Tr}(\rho_0J_a\otimes J_b)-\text{Tr}(\rho_0J_a\otimes\mathds1)\text{Tr}(\rho_0\mathds1\otimes J_b)\\
&=\text{Tr}_1(\ket{j_1,k_1}\bra{j_1,k_1}J_a)\text{Tr}_2(\ket{j_2,k_2}\bra{j_2,k_2}J_b)\,+\\
&\quad-\text{Tr}_1(\ket{j_1,k_1}\bra{j_1,k_1}J_a)\text{Tr}_2(\ket{j_2,k_2}\bra{j_2,k_2})\text{Tr}_1(\ket{j_1,k_1}\bra{j_1,k_1})\text{Tr}_2(\ket{j_2,k_2}\bra{j_2,k_2}J_b)\\
&=\braket{j_1,k_1|J_a|j_1,k_1}\braket{j_2,k_2|J_b|j_2,k_2}-\braket{j_1,k_1|J_a|j_1,k_1}\braket{j_2,k_2|J_b|j_2,k_2}\\
&=0\;,
\end{split}
\end{equation}

\noindent
\\i.e., as expected, the off-diagonal blocks of the metric tensor vanish in the unentangled case.\\ \\For the diagonal-block matrix A we have

\begin{equation}\label{C27}
\begin{split}
A_{ab}&=\frac{1}{2}\text{Tr}(\rho_0[J_a,J_b]_+\otimes\mathds1)-\text{Tr}(\rho_0J_a\otimes\mathds1)\text{Tr}(\rho_0J_b\otimes\mathds1)\\
&=\frac{1}{2}\braket{j_1,k_1|[J_a,J_b]_+|j_1,k_1}-\braket{j_1,k_1|J_a|j_1,k_1}\braket{j_2,k_2|J_b|j_2,k_2}\;.
\end{split}
\end{equation}

\noindent
\\Thus, if $a=b=3$, Eq. (\ref{C27}) gives

\begin{equation}\label{C28}
\begin{split}
A_{33}&=\braket{j_1,k_1|J_3^2|j_1,k_1}-\braket{j_1,k_1|J_3|j_1,k_1}\braket{j_2,k_2|J_3|j_2,k_2}\\
&=k_1^2-k_1k_2\\
&=k_1(k_1-k_2)\;.
\end{split}
\end{equation}

\noindent
\\Instead, if $a,b=1,2$, then the second term in (\ref{C27}) vanishes since $J_1$ and $J_2$ have zero diagonal elements. Hence, using the relations (\ref{C4}) and $(\ref{C5})$, we find

\begin{equation}\label{C29}
\begin{split}
A_{11}&=\braket{j_1,k_1|J_1^2|j_1,k_1}\\
&=\sum_{k_1'}\braket{j_1,k_1|J_1|j_1,k_1'}\braket{j_1,k_1'|J_1|j_1,k_1}\\
&=\frac{1}{4}\sum_{k_1'}[j_1(j_1+1)-k_1'k_1](\delta_{k_1,k_1'+1}+\delta_{k_1+1,k_1'})(\delta_{k_1',k_1+1}+\delta_{k_1'+1,k_1})\\
&=\frac{1}{4}\bigl\{[j_1(j_1+1)-k_1(k_1+1)]+[j_1(j_1+1)-k_1(k_1-1)]\bigr\}\\
&=\frac{1}{2}[j_1(j_1+1)-k_1^2]\;,
\end{split}
\end{equation}

\noindent
\\and

\begin{equation}\label{C30}
\begin{split}
A_{22}&=\sum_{k_1'}\braket{j_1,k_1|J_2|j_1,k_1'}\braket{j_1,k_1'|J_2|j_1,k_1}\\
&=-\frac{1}{4}\sum_{k_1'}[j_1(j_1+1)-k_1'k_1](\delta_{k_1,k_1'+1}-\delta_{k_1+1,k_1'})(\delta_{k_1',k_1+1}-\delta_{k_1'+1,k_1})\\
&=\frac{1}{4}\bigl\{[j_1(j_1+1)-k_1(k_1+1)]+[j_1(j_1+1)-k_1(k_1-1)]\bigr\}\\
&=\frac{1}{2}[j_1(j_1+1)-k_1^2]\;,
\end{split}
\end{equation}

\noindent
while, using the relation

\begin{equation}\label{C31}
[J_a,J_b]_+=2J_aJ_b-[J_a,J_b]_-=2J_aJ_b-i\varepsilon_{abc}J_c\;,
\end{equation}

\noindent
\\the other elements are given by

\begin{equation}\label{C32}
\begin{split}
A_{12}=A_{21}&=\frac{1}{2}\braket{j_1,k_!|[J_1,J_2]_+|j_1,k_1}\\
&=\braket{j_1,k_1|J_1J_2|j_1,k_1}-\frac{i}{2}\braket{j_1,k_1|J_3|j_1,k_1}\\
&=\sum_{k_1'}\braket{j_1,k_1|J_1|j_1,k_1'}\braket{j_1,k_1'|J_2|j_1,k_1}-\frac{i}{2}\,k_1\\
&=0\;,
\end{split}
\end{equation}

\noindent
\\where we have used the fact that

\begin{equation}\label{C33}
\begin{split}
\sum_{k_1'}\braket{j_1,k_1|J_1|j_1,k_1'}\braket{j_1,k_1'|J_2|j_1,k_1}&=\frac{1}{4i}\sum_{k_1'}[j_1(j_1+1)-k_1'k_1](\delta_{k_1,k_1'+1}+\delta_{k_1+1,k_1'})\cdot\\
&\quad\cdot(\delta_{k_1',k_1+1}-\delta_{k_1'+1,k_1})\\
&=\frac{1}{4i}\bigl\{[j_1(j_1+1)-k_1(k_1+1)]+\\
&\quad-[j_1(j_1+1)-k_1(k_1-1)]\bigr\}\\
&=\frac{i}{2}\,k_1\;.
\end{split}
\end{equation}

\noindent
Similarly

\begin{equation}\label{C34}
\begin{split}
A_{23}=A_{32}&=\frac{1}{2}\braket{j_1,k_1|[J_2,J_3]_+|j_1,k_1}\\
&=\braket{j_1,k_1|J_2J_3|j_1,k_1}-\frac{i}{2}\braket{j_1,k_1|J_1|j_1,k_1}\\
&=k_1\,\underset{0}{\underbrace{\braket{j_1,k_1|J_2|j_1,k_1}}}-\frac{i}{2}\;\underset{0}{\underbrace{\braket{j_1,k_1|J_1|j_1,k_1}}}\\
&=0\;,
\end{split}
\end{equation}

\noindent
and

\begin{equation}\label{C35}
\begin{split}
A_{13}=A_{31}&=\frac{1}{2}\braket{j_1,k_1|[J_1,J_3]_+|j_1,k_1}\\
&=\braket{j_1,k_1|J_1J_3|j_1,k_1}+\frac{i}{2}\braket{j_1,k_1|J_2|j_1,k_1}\\
&=k_1\,\underset{0}{\underbrace{\braket{j_1,k_1|J_1|j_1,k_1}}}+\frac{i}{2}\;\underset{0}{\underbrace{\braket{j_1,k_1|J_2|j_1,k_1}}}\\
&=0\;.
\end{split}
\end{equation}

\noindent
\\Again, we see that the elements involving $J_2$ only once are zero as expected because otherwise, according to Eq. (\ref{C5}), they will take imaginary values which cannot enter the real part of the Hermitian tensor. The calculation of the block $B$ essentially is the same as for $A$ with the replacement of the 1 and 2 subscripts.\\ \\Therefore, collecting the results (\ref{C29}-\ref{C35}), we find that in the case of a separable fiducial state the real symmetric part of the pulled-back Hermitian tensor decomposes in the following way\\

\begin{equation}\label{C36}
\mathcal K_{(k\ell)}=\left(\begin{array}{c|c}
A&0\\
\hline
0&B
\end{array}\right)
\end{equation}

\noindent
\\with\\

\begin{equation}\label{C37}
A=\begin{pmatrix}
\frac{1}{2}[j_1(j_1+1)-k_1^2] & 0 & 0 \\
0 & \frac{1}{2}[j_1(j_1+1)-k_1^2] & 0 \\
0 & 0 & k_1(k_1-k_2)
\end{pmatrix}\;,
\end{equation}

\noindent
\\and\\

\begin{equation}\label{C38}
B=\begin{pmatrix}
\frac{1}{2}[j_2(j_2+1)-k_2^2] & 0 & 0 \\
0 & \frac{1}{2}[j_2(j_2+1)-k_2^2] & 0 \\
0 & 0 & k_2(k_2-k_1)
\end{pmatrix}\;.
\end{equation}

\noindent
\\ \\Finally, let us compute the imaginary symplectic part:

\begin{equation}\label{C39}
\begin{split}
(D_A)_{ab}&=\frac{1}{2}\text{Tr}\bigl(\rho_0[J_a,J_b]_-\otimes\mathds1\bigr)\\
&=\frac{1}{2}\text{Tr}_1\bigl(\rho_A[J_a,J_b]_-)\text{Tr}_2(\rho_B)\\
&=\frac{1}{2}\text{Tr}_1\bigl(\ket{j_1,k_1}\bra{j_1,k_1}[J_a,J_b]_-)\\
&=\frac{1}{2}\sum_{k_1'}\delta_{k_1',k_1}\braket{j_1,k_1|[J_a,J_b]_-|j_1,k_1'}\\
&=\frac{1}{2}\braket{j_1,k_1|[J_a,J_b]_-|j_1,k_1}\\
&=\frac{1}{2}\braket{j_1,k_1|\varepsilon_{abc}J_c|j_1,k_1}\;.
\end{split}
\end{equation}

\noindent
Hence

\begin{align}
&(D_A)_{11}=(D_A)_{22}=(D_A)_{33}=0\;,\\
&(D_A)_{12}=-(D_A)_{21}=\frac{1}{2}\braket{j_1,k_1|J_3|j_1,k_1}=\frac{1}{2}\,k_1\;,\\
&(D_A)_{13}=-(D_A)_{31}=-\frac{1}{2}\braket{j_1,k_1|J_2|j_1,k_1}=0\;,\\
&(D_A)_{23}=-(D_A)_{32}=-\frac{1}{2}\braket{j_1,k_1|J_1|j_1,k_1}=0\;,
\end{align}

\noindent
\\and similarly for $D_B$ replacing $j_1,k_1$ with $j_2,k_2$.\\ \\Therefore, in the separable case, the diagonal-block matrices of the skewsymmetric imaginary part of the pulled-back Hermitian tensor are given by

\begin{equation}\label{C43}
D_A=\begin{pmatrix}
0 & \frac{1}{2}\,k_1 & 0 \\
-\frac{1}{2}\,k_1 & 0 & 0 \\
0 & 0 & 0
\end{pmatrix}\;,
\end{equation}

\noindent
\\and

\begin{equation}\label{C44}
D_B=\begin{pmatrix}
0 & \frac{1}{2}\,k_2 & 0 \\
-\frac{1}{2}\,k_2 & 0 & 0 \\
0 & 0 & 0
\end{pmatrix}\;.
\end{equation}

\newpage

\section*{Acknowledgements}
\thispagestyle{empty}

There are so many people I would like to thank for having made this work possible. I've spent the thesis period at the Max Planck Institute for Gravitational Physics (Albert Einstein Institute) in Golm (Potsdam), where I had the opportunity to meet many people who have helped to make this experience unforgettable.\\First of all, I am deeply grateful to my supervisors, Dr. Goffredo Chirco and Dr. Daniele Oriti, for having accepted me in their group, sharing with me their experience and their knowledge both at the scientific and non-scientific level. A special thank goes to my supervisor in Naples, Prof. Patrizia Vitale, for her kindness, precious advices and support.\\Moreover, I would also like to thank the person I am honored to consider as my mentor, Prof. Giuseppe Marmo, for his willingness and the stimulating discussions during my entire academic career.\\Most of what I've learned comes from discussions with other people. Therefore, I am grateful to all the members of the \textit{Microscopic Quantum Structure and Dynamics of Spacetime} group. In particular, I would like to thank Marco Finocchiaro, Giovanni Tricella, Florian Gerhardt and Alex Kegeles for sharing with me their time and for all the funny moments.\\Now it is time to thank all those who contributed both directly and indirectly to improve myself during the last years, making me a better person. A special thank goes to Marco Laudato. We shared one of the most difficult and exciting moment of our studies and this helped me to know you both as a nice friend and as a great colleague. I am sure that we will continue our collaboration and ``infinite'' discussions in the future...we only need to find a blackboard.\\I cannot avoid to be grateful to Giorgio Nocerino for these years of study and for helping me to understand what a true friend is. Grazie di tutto!\\Being sorry for surely neglecting someone, I would like to thank all the friends and colleagues who have accompanied me in these university years: Daniele, Fo, Luca, Anna, Alessandro e Pierpaolo. Grazie per questi anni indimenticabili!\\Moreover, I would like to thank my parents and all my family for their constant encouragements, support, and for having made me who I am. Vi voglio bene.\\Last but not least, I cannot thank enough my ``second family'': Checco, Domenico, Peppe, Ciro, Davide ed Angelo. You are the most happy and free part of my life and I feel so lucky to have you at my side. DGS per sempre!!!

\begin{flushright}
December 2016,\\
$\mathcal{F}abio$
\end{flushright}

\end{document}